%% file: main.tex
\documentclass[12pt,a4paper,oneside]{book} 
\usepackage[utf8]{inputenc}

\input{setup_stuff.tex}

\begin{document}
% ========================================================
% ========================================================

\pagestyle{empty_page}

% --------------------------------------------------------
% cover page
\input{title_page.tex}
\cleardoublepage 
\setcounter{page}{1}

% --------------------------------------------------------
% table of contents
\tableofcontents
\cleardoublepage
\pagestyle{default_page}

\frontmatter
\input{FRONTMATTER/Introduction.tex}

% --------------------------------------------------------
% input your chapters here
\mainmatter

\input{CHAPTERS/chapter_1.tex}

\cleardoublepage % opens next new chapters in right pages for two-sided printing

\input{CHAPTERS/chapter_2.tex}
\cleardoublepage % opens next new chapters in right pages for two-sided printing

\input{CHAPTERS/chapter_3.tex}

\cleardoublepage % opens next new chapters in right pages for two-sided printing

\input{CHAPTERS/chapter_4.tex}

\cleardoublepage % opens next new chapters in right pages for two-sided printing

\input{CHAPTERS/Conclusions.tex}

\cleardoublepage % opens next new chapters in right pages for two-sided printing

% --------------------------------------------------------

\backmatter

% input final stuff here (appendices, bibliography)
\pagestyle{final_stuff}
\input{CHAPTERS/appendix_A.tex}

\input{CHAPTERS/references.tex}

\cleardoublepage % opens next new chapters in right pages for two-sided printing

\cleardoublepage % opens next new chapters in right pages for two-sided printing

\end{document}

%% file: setup_stuff.tex
\PassOptionsToPackage{hidelinks}{hyperref}
\usepackage[a-1b]{pdfx}   

\usepackage[top=30mm,bottom=35mm,inner=30mm,outer=25mm]{geometry}
\usepackage{setspace}
\usepackage{amsmath}
\usepackage{amsfonts}
\usepackage{amssymb}
\usepackage{amsthm}
\usepackage{chemarr}

\usepackage{pifont}
\usepackage{lettrine}
\usepackage{erewhon}
\usepackage{GoudyIn}
\usepackage{xcolor} 
\definecolor{bostonuniversityred}{rgb}{0.8, 0.0, 0.0}
\definecolor{harvardcrimson}{rgb}{0.79, 0.0, 0.09}
\definecolor{darkmidnightblue}{rgb}{0.0, 0.2, 0.4}

\LettrineTextFont{\itshape}
\usepackage{graphicx}
\usepackage[justification=centering,font={small,singlespacing}]{caption}

\usepackage[Glenn]{fncychap}
\usepackage{caption}
\usepackage{subcaption}
\usepackage{booktabs}
\usepackage{lscape}
\usepackage{float}

\usepackage{natbib}
\newcommand\aap{A\&A}                % Astronomy and Astrophysics
\newcommand\aapr{A\&ARv}             % Astronomy and Astrophysics Review (the)
\newcommand\aj{AJ}                   % Astronomical Journal (the)
\newcommand\ao{Appl. Opt.}           % Applied Optics
\newcommand\apj{ApJ}                 % Astrophysical Journal
\newcommand\apjl{ApJ}                % Astrophysical Journal, Letters
\newcommand\apjs{ApJS}               % Astrophysical Journal, Supplement
\newcommand\apss{Ap\&SS}             % Astrophysics and Space Science
\newcommand\araa{ARA\&A}             % Annual Review of Astronomy and Astrophysics
\newcommand\mnras{MNRAS}             % Monthly Notices of the Royal Astronomical Society
\newcommand\nar{New~Astron.~Rev.}    % New Astronomy Review
\newcommand\nat{Nature}              % Nature
\newcommand\prd{Phys. Rev.~D}        % Physical Review D
\newcommand\prl{Phys. Rev.~Lett.}    % Physical Review Letters
\newcommand\pasp{PASP}               % Publications of the Astronomical Society of the Pacific
\newcommand\pasj{PASJ}               % Publications of the Astronomical Society of Japan
\newcommand\physrep{Phys.~Rep.}      % Physics Reports
\newcommand\ssr{Space Sci. Rev.}     % Space Science Reviews
\usepackage{fancyhdr}
\fancypagestyle{default_page}{
\fancyhead[C]{}
\if@twoside%
    \fancyhead[LE,RO]{\sl CHAPTER \thechapter}
    \fancyhead[RE,LO]{}
\else
	\fancyhead[R]{\ifnum\value{chapter}>0 {\sl CHAPTER \thechapter} \else {\sl INTRODUCTION}\fi}
	\fancyhead[L]{}
\fi

\fancyfoot[L,R]{}
\fancyfoot[C]{\small\thepage}

}

\fancypagestyle{empty_page}{
\fancyhf{}

}

\fancypagestyle{final_stuff}{
\fancyhead[C]{}
\if@twoside%
    \fancyhead[LE,RO]{\sl \leftmark}
    \fancyhead[RE,LO]{}
\else
    \fancyhead[R]{\sl \leftmark}
    \fancyhead[L]{}
\fi
\fancyfoot[L,R]{}
\fancyfoot[C]{{\small\thepage}}

}

\makeatletter
\def\emptypage@emptypage{%
    \thispagestyle{empty}
    \vspace*{\fill}
    \begin{center}
    {\footnotesize This page intentionally left blank.}
    \end{center}
    \vspace{\fill}
    \newpage%    
}%
\def\cleardoublepage{%
        \clearpage%
        \if@twoside%
            \ifodd\c@page%
            \else%
                \emptypage@emptypage%
                
            \fi%
        \fi%
    }%
\makeatother

\usepackage{url}
\usepackage{hyperref}	% Hyperlinks
\hypersetup{colorlinks=true,linkcolor=blue,citecolor=blue,filecolor=blue,urlcolor=blue}
\usepackage{listings}

%% file: title_page.tex
\newgeometry{top=12mm,bottom=15mm,inner=25mm,outer=25mm}
\begin{titlepage}
\begin{spacing}{1.4}
\begin{figure}[ht]
%\centering
\includegraphics[height=0.305\textwidth]{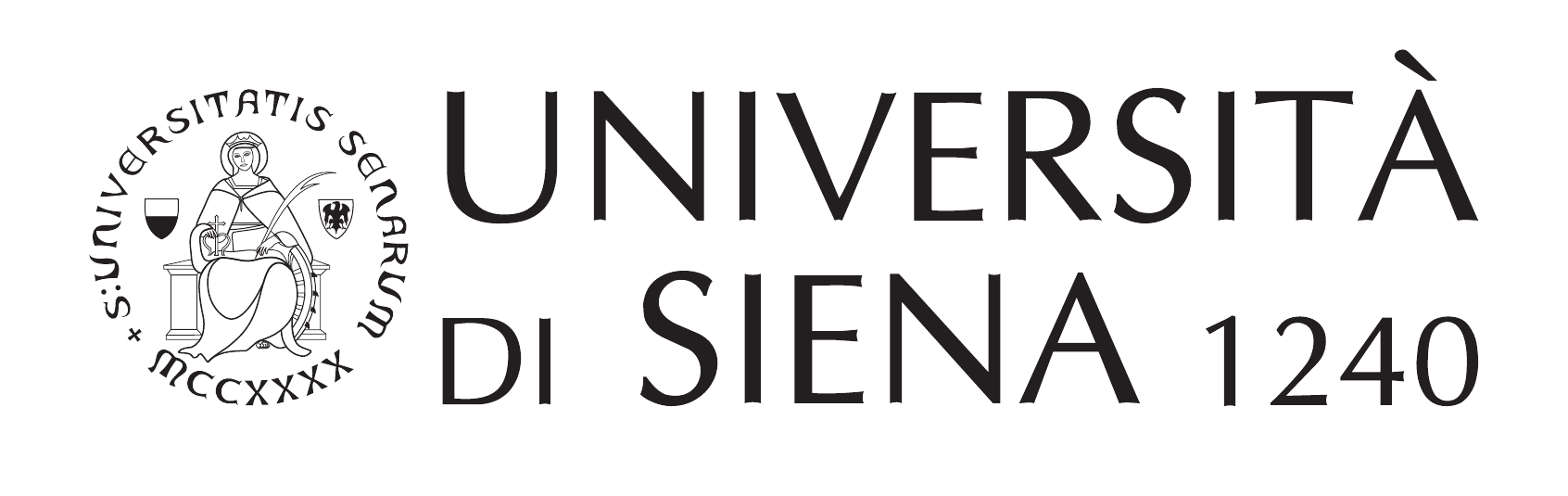}
\end{figure}
\begin{center}
{\scshape \Large
Department of\\Physical Sciences, Earth and Environment
}

\vspace{5mm}

\rule{100mm}{0.1mm}
\rule[5mm]{100mm}{0.4mm}

{\large 
Ph.D Program in\\ 
{ Experimental Physics}
}

\vspace{22mm}

\begin{spacing}{2.0}
{\huge \bf Probing star clusters as cosmic ray factories}
\end{spacing}

\end{center}

\vspace{8mm}
\noindent
\begin{minipage}[t]{0.5\textwidth}
\bf Supervisor:\\
Dr. Elena Amato\\

Co-supervisor:\\
Dr. Giovanni Morlino\\
 
Tutor:\\
Prof. Riccardo Paoletti\\
\end{minipage}
\hfill
\begin{minipage}[t]{0.5\textwidth}\raggedleft
\bf Candidate:\\
Stefano Menchiari\\
Mat. 094422
\end{minipage}

\vfill

\begin{center}
\large Academic Year 2021-2022
\end{center}

\end{spacing}

\end{titlepage}
\restoregeometry

%% file: FRONTMATTER/Introduction.tex
%---------------------------------------------------------
\chapter{Introduction}
%---------------------------------------------------------

\lettrine{S}{tellar} clusters are among the most studied celestial objects in the cosmos. They represent crucial laboratories for understanding a wide range of problems in different subtopics of astrophysics. As they include a vast number of stars with masses spanning a broad interval, star clusters are, for instance, essential in the studies of the stellar Initial Mass Function \citep{Bastian_IMFReview_2010}. In general, all cluster members are born simultaneously and share common primordial properties since they are the offspring of the same progenitor molecular cloud. This makes star clusters an excellent subject for studies of stellar evolution \citep{Kalirai_StellarEvolutionReview_2010}. Some stellar clusters are also gravitationally bound systems, i. e. are held together by the mutual gravitational attraction of their members. From this point of view, they are the perfect target for studies on stellar dynamics \citep{Vesperini_StellarDynamicReview_2010}.

Stellar clusters are fundamental building blocks of galaxies. The analysis of the spatial distribution in the host galaxy of different types of clusters has often been used to determine the galactic structure. One example is the distribution of globular clusters in the Milky Way, which has allowed the estimation of the Milky Way size \citep{Shapley_MilkyWaySize_1918}, the determination of the Galactic Center, and has established the existence of the galactic halo \citep{Bica_GCMWPropertiesReview_2006}. Or similarly, the distribution of young clusters in external galaxies, which is employed to trace star forming regions and the spiral structure in galactic disks \citep{Adamo_YSCTraceSFRReview_2020}.

Stellar clusters are also vital for understanding the star formation mechanism. Indeed, a relevant fraction of stars is born in clustered environments, as demonstrated by \cite{Lada_SCReview_2003}, who showed that the contribution to the local star formation rate from the population of young embedded clusters is similar to that obtained from field stars. This is even more true for massive stars, as almost $\sim$70\% of O-type stars are observed in clusters or associations \citep{Gies_ORunawayStars_1987, Parker_OstarClusters_2007}, and at least $\sim$50\% of the remaining stars are identified as runaways \citep{deWit_OstarsField_2005}. More accurate estimates seem to lower the percentage of massive stars born outside a cluster to $\sim$4\% \citep{deWit_OstarsField_2005}.

Last but not least, stellar clusters play a prominent role in high-energy astrophysics. For a long time, the interest in star clusters by the high-energy astrophysics community was mainly due to the unusual abundance of extreme objects, often referred to as \textit{lusus natur\ae}\footnote{\textit{lusus natur\ae} is Latin for freaks of nature, mutants, or monsters} or stellar exotica. Indeed, globular clusters, because of their old age and relatively homogeneous population, and isolation from their parent galaxies, are known to be rich in unusual extreme objects, such as X-rays binaries \citep{Heinke_XraySourcesInGCs_2010}, pulsars \citep{Ransom_PSRinGC_2008}, black holes \citep{Maccarone_BHinGCs_2007}, etc. However, in recent decades, star clusters, or to be more accurate, young massive star clusters (YMSC), have found themselves at the center of attention not so much as a possible cradle for a future generation of lusis natur\ae, but rather as objects capable of producing and accelerating cosmic rays (CRs). Massive OB-type stars are known to launch powerful winds \citep{Abbott_RadiativelyWinds_1979, Cassinelli_StellarWinds_1979, Kudritzki_WindsHotStars_2000}, and the presence of tens (if not hundreds or, in the most extreme and rare cases, thousands) of massive stars crammed into a small volume can generate favorable conditions for boosting particles to very high-energies. Several CR production mechanisms have been proposed through the years, such as for example, acceleration at the wind termination shock of single massive stars \citep{Casse_CRfromSW_1980, Cesarsky_GammaRaysFromSW_1983} or acceleration by wind-wind interaction \citep{Klepach_WindWindInteractionCR_2000, Reimer_CRfromWWinteraction_2006}. In the case of compact clusters, the winds from individual stars may end up combining, somehow creating a collective cluster wind. In this scenario, particle acceleration may occur at the cluster wind termination shock \citep{Morlino_2021}. As the shock generated by the winds from the stars or by the collective cluster wind interacts with density inhomogeneities of different scales, broad spectra of magnetohydrodynamic fluctuations are generated. In these systems then, second-order Fermi acceleration within the turbulent plasma becomes also a possible efficient mechanism of particle acceleration \citep{Bykov_MSCTurbAcc_2020}. Finally, in the case of YMSCs older than $\sim 5\--10$ Myr, the most massive stars gradually leave the main sequence and rapidly move to their final evolutionary stage. At this point, those stars begin to explode as supernovae, leaving behind supernova remnants (SNRs). In such aged systems, CR acceleration is achieved thanks to the multiple interactions between SNR shocks and stellar winds \citep{Bykov_SuperBubbleSNR_2001}, with a gradual decrease in time of the contribution to the process from the stars' winds \citep{Vieu_MSC+SNR_2022}. The possibility that YMSCs can actually accelerate particles is a fact of extreme relevance in the general panorama of high-energy astrophysics, especially in connection with the problem of the origin of CRs. At present, it is well proven, based both on theoretical arguments \citep{Blandford_SNRsAcceleration_1987, Berezhko_SNRsCRsReview_1988} and empirical observations \citep{Koyama_XraySNR_1995, Reynolds_SNR_2008, Helder_ObservationSNRs_2012}, that SNRs are a class of sources able to produce CRs. Nevertheless, the population of SNRs alone has difficulties accounting for all the observed properties of galactic CRs. More precisely, a single population of SNRs struggles to reproduce two distinct things: the observed CR composition at Earth and the so-called \textit{knee feature} in the CR spectrum. Measurements of CR composition seem to point to an excess in the ratio of some isotopes, such as for example, the $^{22}$Ne to $^{20}$Ne ratio, which is found $5.3 \pm 0.3$ times higher than the solar value \citep{Wiedenbeck_Ne_1981, Binnis_NeInCRs_2008}. YMSCs can easily explain this excess, as these anomalous ratios can result from the winds of Wolf-Rayet and massive stars \citep{Gupta_NeMSC_2020}.

On a very general ground, the observed CR spectrum at the Earth between $\sim 10^{12} \-- 10^{18}$ eV can be adequately described by a broken power law \citep{Workman_PDG_2022}. The change in the slope, referred to, in the literature, as the \textit{knee}, appears to be located at $\sim 3 \times 10^{15}$ eV and has been interpreted as the maximum energy reachable by the population of galactic CR accelerators \citep{Blasi_CRsReview_2013, Amato_GalCROrigin_2014}. Celestial objects able to produce particles up to these energies are commonly called \textit{PeVatrons}. From both the theoretical and observational \citep{Aharonian_MSCs_2019} point of view, SNRs struggle to accelerate CRs up to such energies, unless under specific conditions involving extreme energy releases, at least assuming that the needed magnetic field amplification is well understood \citep{Bell_CRsAccAndEscapeInSNR_2013, Cardillo_SNRIICRspect_2015}. On the contrary, YMSCs seem to be promising PeVatron candidates, as suggested by theoretical considerations \citep{Morlino_2021} and tentative observational hints \citep{Cao_LHAASO12PeVatrons_2021}. The answer to the century-old enigma about the origin of galactic CRs could therefore lie in having multiple populations of galactic accelerators: SNRs could account for the bulk of the observed CRs, while particle acceleration associated with the powerful winds of YMSCs could provide the highest-energy particles. This would in parallel explain the unusual abundances of some elements. But how to confirm or eventually reject this scenario? It is widely known that it is impossible to probe the properties of a given galactic accelerator by observations of CRs at Earth alone. In fact, the propagation of charged particles below a few $10^{18}$ eV in the Galaxy is diffusive because of the scattering with the interstellar magnetic field fluctuations \citep{Strong_CRsDiffusionReview_2007, Amato_CRsDiffusionReview_2018}. The information on the position of the CR sources is totally lost during the propagation process. For this reason, the search for CR sources must rely on somewhat indirect investigation methods based on the interaction between CRs and the interstellar matter (ISM) and radiation field. In practice, one can exploit observations of non-thermal radiation: in this sense $\gamma$-ray emission has a privileged role as it directly traces the presence of high-energy and very-high-energy particles \citep{Tibaldo_GammaTracesCRsReview_2021}. Another promising strategy is to search for indications of a high ionization rate induced by energetic particles in dense clouds close to the accelerator \citep{Gabici_IonRate-GammaCRsTracers_2015}. In summary, in order to assess whether YMSCs are eventually playing a primary role in the origin of galactic CRs, $\gamma$-ray observations and comprehensive studies of the environment close to these objects are of extreme importance.

In this thesis, we investigate the observational properties of YMSCs as very-high-energy sources under the assumption that CR production occurs at the cluster wind termination shock. In particular we will assume CR acceleration and propagation as described by the model of \cite{Morlino_2021} and derive different observational signatures that can help constrain the model and test the efficiency of clusters as CR accelerators. The entire work is divided into three distinct parts.

First, we will try to assess whether and to what extent the employed model of particle acceleration can effectively reproduce the observed $\gamma$-ray emission from a given YMSC. To this purpose, we will focus on the scientific case of Cygnus OB2. This is one of the most iconic cases when discussing YMSCs as CR sources since several experiments have detected diffuse $\gamma$-ray emission in both the high-energy \citep{Ackermann_FermiCocoon_2011} and very-high-energy \citep{Bartoli_ARGOCygnusCocoon_2014, Abeysekara_HAWCCygnusCocoon_2021} bands towards its position. Moreover, the detection of a 1.4 PeV photon from the same region \citep{Cao_LHAASO12PeVatrons_2021} makes it one of the most promising PeVatron candidates in the Galaxy.

In the second part of the thesis, we will rather focus on YMSCs as a population of $\gamma$-ray sources. If YMSCs are indeed particle accelerators, then it is natural to think of them also as $\gamma$-ray emitters. In view of the new and upcoming facilities for $\gamma$-ray observations such as the Cherenkov Telescope Array \citep{CTA_ScienceCTA_2019}, the Astri Mini Array \citep{AstriColl_AstriScience_2022, Vercellone_ASTRIProgram_2022}, and the Southern Wide-field Gamma-ray Observatory \citep{Bakalova_SWGO_2022}, simulating the emission from a synthetic population of galactic YMSCs becomes of fundamental importance. In fact, the comparison between the number of expected versus detected YMSCs can be used to discriminate the capabilities of these sources as particle accelerators and their contribution to the galactic CR sea. Moreover, as we will see, the $\gamma$-ray emission from most of these objects is predicted to be extended and potentially difficult to disentangle from the diffuse. Consequently, YMSCs could also significantly contribute to the galactic diffuse $\gamma$-ray emission, and the estimation of this contribution is of particular importance for all those studies that require a solid modelization of the galactic background emission, such as, for instance, the search for dark matter in the Galactic Center.

In the third and final part of the work, we will focus on the low- energy CRs and their impact on the ISM surrounding the stellar clusters. In addition to highly-energetic particles, YMSCs are expected to produce also a (largely more conspicuous) population of CRs with energies below 1 GeV. Differently from the cluster starlight, these particles are capable of penetrating deep in the core of the dense molecular clouds that are commonly found in the neighborhood of YMSCs, ionizing the cold neutral medium of which the clouds are made. The general aim will then be to evaluate the ionization rate of a molecular cloud located close to a YMSC. This type of investigation is of particular interest for two main reasons. First, the measurement of the ionization rate can be employed as a test to trace the presence of freshly accelerated CRs. This information can be subsequently combined with independent $\gamma$-ray observations to have a self-consistent picture of the CR distribution around the stellar cluster. Secondly, depending on how much the ionization rate is different from the Spitzer value \citep{Spitzer_IonizationMC_1968}, low-energy CRs could be an additional feedback channel for YMSCs to regulate the star formation process in their environment.

The following manuscript is divided into five main chapters. In the first chapter, we give a general review of YMSCs and their main properties, followed by a discussion of their role as particle accelerators. We will review the main acceleration models in stellar clusters and the indirect techniques to observe CRs at the sources. The bulk of the work is then presented in chapters two, three, and four, where we analyze the scientific case of Cygnus OB2, the emission from a synthetic population of YMSCs, and the estimate of the ionization rate in clouds close to a stellar cluster. Finally in chapter five, we summarize and discuss the conclusions of the work.

%% file: CHAPTERS/chapter_1.tex
%---------------------------------------------------------
\chapter{Young Massive Star Clusters as very-high-energy sources}

\lettrine{A}{s} primary players in different fields of astrophysics, from stellar physics to physics and astronomy of star formation, and recently also in the high-energy astrophysics field, the properties of stellar clusters have been widely investigated. In this chapter, we provide an overview of these objects, starting from their classification based on fundamental characteristics, and then briefly reviewing their evolutionary path. Afterward, we describe their capabilities in shaping the ISM around them. We then concentrate on stellar clusters as CR accelerators, giving a comprehensive summary of the most widely considered acceleration mechanism in these systems. 

%---------------------------------------------------------

%---------------------------------------------------------
\section{What is a stellar cluster?}
\label{sec:DefAndTerm}
When it comes to defining a star cluster, several criteria can be given \citep{Krumholz_SCsReview_2019}, and unfortunately, there is no single definition that can be used universally. It follows that the definition of a star cluster has a certain degree of arbitrariness, and it may change from author to author, directly marking their research case. Among the several proposed formulae, one of the most widespread is the one given by \cite{Lada_SCReview_2003}, who define a stellar cluster as a group of stars with a mass density large enough ($\rho_\star \gtrsim 1 M_\odot pc^{-3}$) to withstand tidal disruption in Solar Neighborhood conditions and, in parallel, to have enough members to avoid kinematical evaporation for at least 100 Myr. Another possible criterion is the one proposed by \cite{Zwart_YMSC_2010}, who define a star cluster as a group of stars that are gravitationally bound to one another. According to the virial theorem, a system of stars with density $\rho_\star$ and size $r$ is considered to be gravitationally bound if the velocity dispersion $\sigma$ is such that $\sigma \lesssim r \sqrt{G \rho_\star}$. The prescription suggested by \cite{Lada_SCReview_2003} is somewhat more general as it also includes unbound clusters, and as far as the present work is concerned, it is perhaps the most suitable. In fact, let us now think for a moment in terms of the involved timescales. As we will justify in the next sections, for the aim of this thesis, we are interested in those clusters that are young enough so that the pollution by supernova (SN) explosions is relatively low. This means that we are interested in clusters younger than $\sim$10 Myr, a timescale that is close to the lifetime of a star with mass 20 $M_\odot$. As the number of stars more massive than 20 M$_\odot$ is relatively low\footnote{This statement is somewhat tricky, as the population of stars is potentially correlated with the mass of the stellar cluster, see \S~\ref{subsec:StellarPopInYMSC}} on average, clusters younger than $\sim$10 Myr are expected to have witnessed few SN explosions. The time required for an unbound system to disperse is the crossing time, defined as $t_{cr} = r/\sigma \sim 1/\sqrt{G \rho_\star}$. Considering the threshold in $\rho_\star$ set by \cite{Lada_SCReview_2003}, the crossing time is $t_{cr}\sim 10 Myr$. It follows that unbound clusters must also be considered, as they survive long enough to potentially contribute to CR acceleration. In addition to the threshold in the stellar density, the definition of \cite{Lada_SCReview_2003} includes also a condition on the minimum number of stars in the cluster, due to the requirement of having a kinematic evaporation time less than 100 Myr. For a cluster with $N$ members, the relaxation time is $\tau_{relax} \approx (0.1 N /\ln N) t_{cr}$ \citep{Lada_SCReview_2003}, and the evaporation can be calculated as $\tau_{ev} \approx 100 \tau_{relax}$. Again, even for clusters with a relatively small number of members, the evaporation timescale is well above our threshold of 10 Myr. To sum up, we can safely use \cite{Lada_SCReview_2003} definition of star clusters, with the clarification that, during our work, we will focus specifically on the subcategory of \textit{young clusters} (age less than 10 Myr).

Yet, youth is not the only parameter we require. As the presence of massive stars is a fundamental ingredient for CR acceleration, we also need to consider only those clusters with a significant number of massive stars. Unfortunately, it is not straightforward to formally express this condition. Once the stars initial mass function is known \citep{Salpeter_IMF_1955, Kroupa_IMF_2001}, the number of massive stars clearly depends on the cluster mass. In this way, the condition can be shifted to a lower limit cut on the star cluster mass. Let us now, for the moment, fix this threshold to $10^3$ $M_\odot$, we will then justify a posteriori in \S~\ref{sec:PopYMSCasCR} that this limit is reasonable.

%---------------------------------------------------------

%---------------------------------------------------------
\section{The evolutionary path of a young massive star cluster}
The evolutionary path of a stellar cluster can change drastically from one case to another. Clusters that are not gravitationally bound are expected to disperse in a few crossing times. Even for bound clusters, their long-term fate is not straightforward, as the complex inner stellar dynamics may induce kinematic evaporation on timescales of a few hundred Myr. Despite this, the early evolutionary stages of the various types of star clusters should, on average, be quite similar. On a very general ground, we can divide the initial phases of stellar clusters evolution into two main parts: a first initial stage, where the cluster members are forming as a result of the collapse of dense gas clumps in giant molecular clouds, and a second phase, when the first stars light up, generating stellar feedback on the surrounding gas and stopping the star formation process. In the case of massive star clusters, during the second stage, the presence of a significant number of massive stars with their powerful winds can produce large bubbles filled with hot shocked wind material. In the next two subsections we will describe separately these two evolutionary stages.

\subsection{The birth of a stellar cluster}
\label{subsec:YMSCBirth}
\begin{figure}
     \centering
     \begin{subfigure}{0.3\textwidth}
         \centering
         \includegraphics[width=\textwidth]{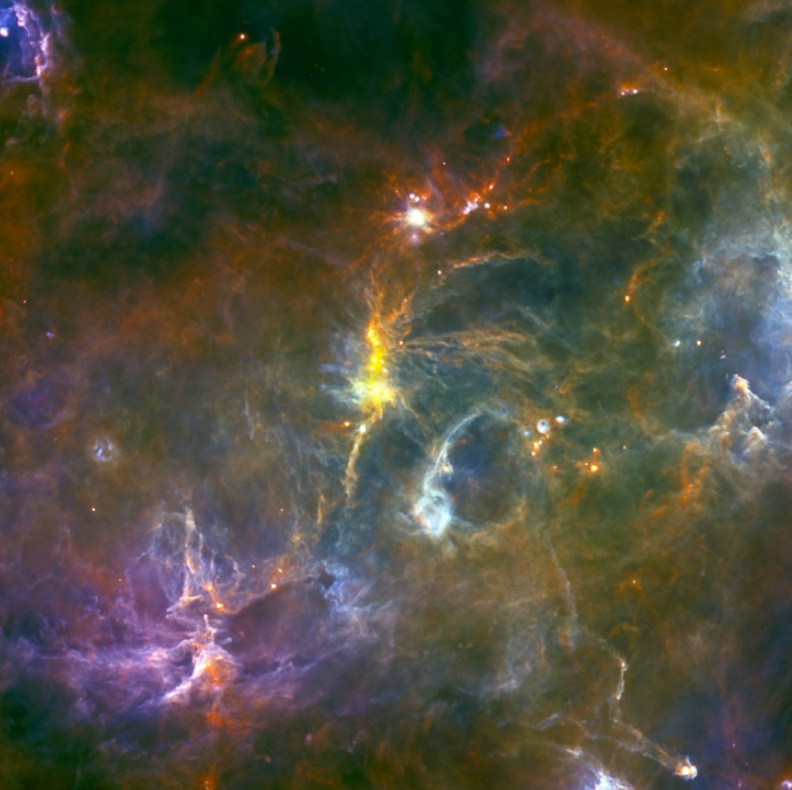}
         \caption{}
         \label{fig:GMC_DR21}
     \end{subfigure}
     \hfill
     \begin{subfigure}{0.3\textwidth}
         \centering
         \includegraphics[width=\textwidth]{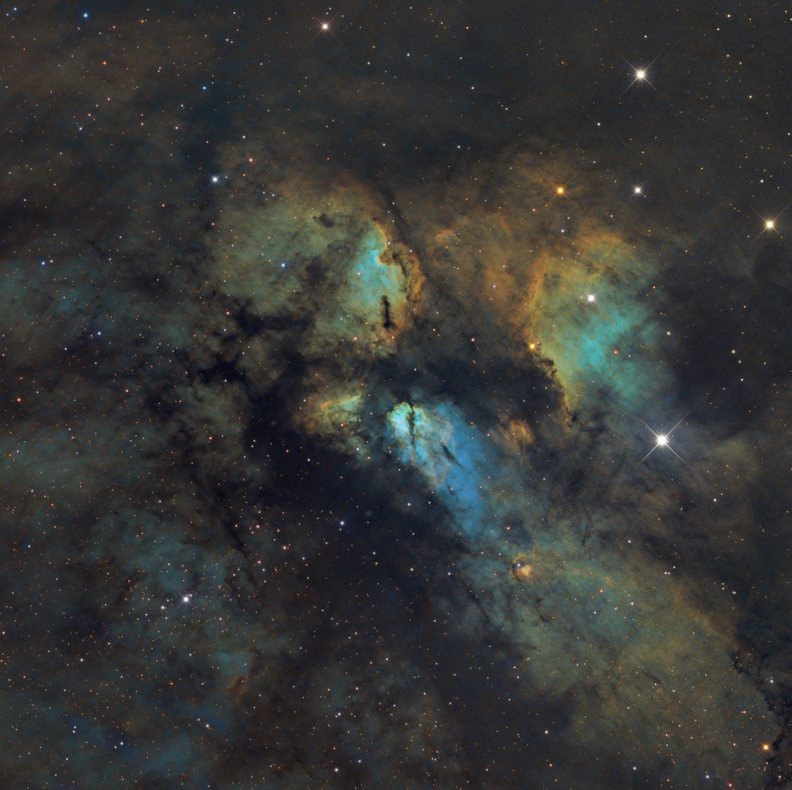}
         \caption{}
         \label{fig:EmbeddedClusterRCW38}
     \end{subfigure}
     \hfill
     \begin{subfigure}{0.3\textwidth}
         \centering
         \includegraphics[width=\textwidth]{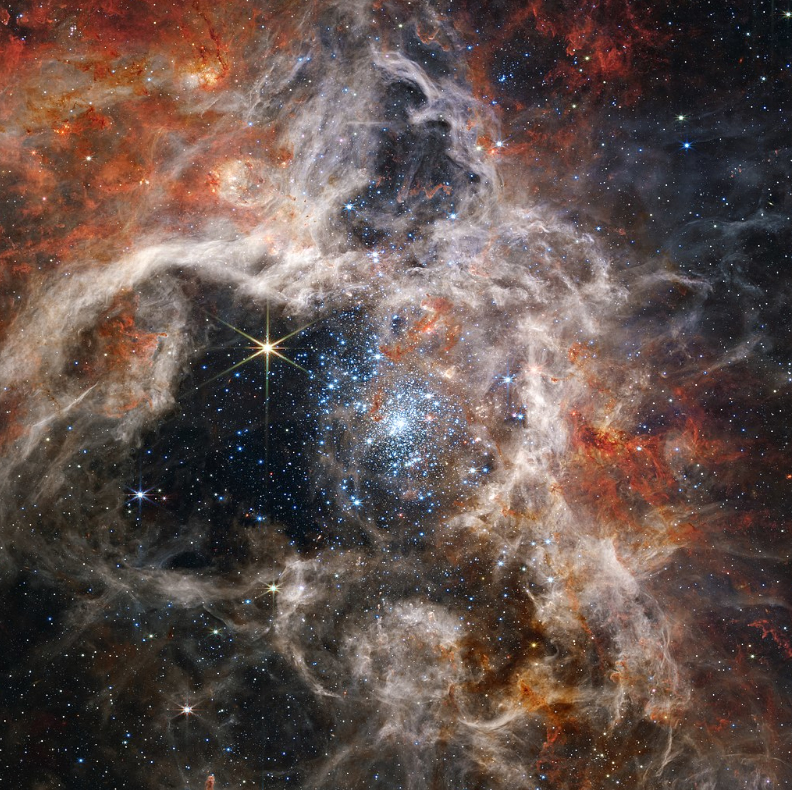}
         \caption{}
         \label{fig:YMSC_30Doradus}
     \end{subfigure}
        \caption{Three different celestial objects representing different evolutionary stages of a massive stellar cluster. (a) Herschel observations of the giant molecular cloud DR21 localized in the Cygnus-X star-forming complex. (b) Hubble observations of the embedded cluster RCW 38. The star cluster itself is invisible at the optical wavelength as it is still deep within its parental cloud. (c) JWST observation of the YMSC 30 Doradus emerging from its GMC as it blows away the surrounding gas.}
        \label{fig:YMSC_EvolutionStages}
\end{figure}

Star clusters are known to form within massive large complexes of cold molecular gas (Fig.~\ref{fig:GMC_DR21}), usually called giant molecular clouds (GMCs). The gas within a GMC is not uniformly distributed. On the contrary, GMCs are characterized by having a self-similar structure over a wide range of scales down to individual protostellar cores, which are aggregated into cluster-forming dense gas clumps \citep{Williams_GMCsStructure_2000, McKee_StarFormTheory_2007} with typical sizes of $0.1 \-- 2$ pc and masses ranging from a few solar masses up to thousands of solar masses \citep{Lada_SCReview_2003}. On average, less than $\sim 10 \%$ of the volume and mass of a GMC end up in the form of dense gas ($>10$ cm$^{-3}$). The clumps where clusters are born are highly localized and occupy a small fraction (a few percent) of the volume of a GMC. The mass distribution of these cluster-forming clumps is well described by a power law with index $\sim -1.7$ \citep{Kramer_ClumpMassDistrib_1998}. Interestingly, with such index, it directly follows that most of the mass of the dense gas component of a GMC is found in its most massive cores. Consequently, as the star formation process is triggered only in dense gas regions, it is not surprising that a significant fraction of stars are born in a clustered environment, coherently with the mass distribution of the GMC clumps. 

As not the whole GMC is in form of dense gas, it follows that not the entire mass of a GMC ($M_{gas}$) ends up into stars ($M_{stars}$). The parameter describing the fraction of gas converted into stars is the star formation efficiency (SFE), defined as $\epsilon_{SFE}=M_{stars}/(M_{gas}+M_{stars})$, which is $\epsilon_{SFE}=1$ if all the gas ends up in the form of stars. For the entire GMC, the  $\epsilon_{SFE}$ is usually of the order of a few percent \citep{Duerr_SFEforGMC_1982}. The value increases up to $\sim 30\%$ when considering single cluster-forming clumps \citep{Lada_SCReview_2003}. The implied low-efficiency at small scales indicates that, at some point, there is something preventing the process of star formation from proceeding \citep{Krumholz_SCsReview_2019}. This is thought to be related mainly to feedback processes associated with the lighting-up of the first stars\footnote{For the sake of completeness, it must be mentioned that a minor contribution to the quenching of the SFE can also be caused by the dynamical properties of the molecular clumps, i.e., if they are not bound \citep{Dobbs_NotBoundMCs_2011} or are highly turbulent \citep{Krumholz_TubulenceRegulatedSF_2005}.}, although some feedback may also arise from the outflows of low-mass protostellar objects \citep{Bally_ProtostellarOutflows_2016}. As the stars enter the main sequence, stellar feedback processes sweep away the gas, causing the cluster to emerge from its cradle and begin the next evolutionary stage.

Before proceeding further, it is interesting to note the following: since the SFE is generally low, a significant fraction of the mass is still in the form of dense gas, which will be swept away when stellar feedback mechanisms kick in. So from the point of view of stellar clusters, the gas expulsion can potentially be a destructive process\footnote{As some sort of cosmic joke, the death of unbound star clusters has been proposed as a mechanism for the birth of the loose aggregations of massive stars known as OB associations. However, this picture has lost relevance over the years due to the growing body of opposing evidence (see \cite{Wright_OBAssociationReview_2022} and references therein for a well made review on the topic)}, as many systems that were once gravitationally bound may find themselves deprived of most of their mass.

Whether a cluster may survive or not from its birth as a bound system can be assessed by comparing its dynamical crossing time with the timescale of gas expulsion ($\tau_{ge}$), which ultimately depends on the dominant feedback process. If $\tau_{ge} \ll t_{cr}$, which seems to be the case for massive star clusters, since OB-type massive stars produce the strongest feedback \citep{Krumholz_SCFFeedback_2014}, the cluster should disperse unless the SFE is not at least as high as $\sim50\%$ \citep{Wilking_SFE50percent_1983}. However, this seems to be in tension with observations.

The formation of clusters and the issue of infant mortality is still not fully understood. The process is still under debate (see \cite{Lada_SCReview_2003} and references therein for a comprehensive review of the problem). Nevertheless, as stated in \S~\ref{sec:DefAndTerm}, assuming that unbound systems can last at least a few crossing times, they live long enough to possibly contribute as CRs accelerators.

\subsection{Emerging from the cradle: creation of a wind bubble}
As soon as stellar feedback kicks in, quickly bringing the star formation process to a halt, the cluster is formally born. At the very beginning, the cluster is still embedded in the dense gas of its parental clump, buried deep within the GMC. Objects in this evolutionary phase are often called embedded clusters (see Fig.~\ref{fig:EmbeddedClusterRCW38}). The embedded phase is usually short-lived as the feedback from the stars rapidly blows away the dense gaseous envelope.

Stellar feedback may come in different flavors. Massive stars are known to produce a significant flux of ionizing photons. The presence of ionizing radiation directly heats the surrounding gas up to $10^4$ K. If not trapped, the heated gas will flow out of the cluster in a breeze called \textit{champagne flow}. This process can effectively remove a certain amount of gas \citep{Williams_OBAssociationInMC_1997}. In parallel, the light from the same zero-age population can also produce significant direct radiation pressure on the surrounding gas. The emitted power (the estimated luminosity to mass ratio is $\sim 1100$ L$_\odot$ M$_\odot^{-1}$) is mostly in the UV band, where the ISM is highly opaque. Depending on the density, this mechanism is one of the most efficient gas removal processes \citep{Fall_StellarFeedbackInMC_2010}. In the case of a very thick envelope, the indirect radiation pressure provided by multiple cycles of absorption and re-emission from the dust may also play an important role \citep{Thompson_IndirectRadPress_2015}.

Last but not least comes the feedback from the powerful winds of massive OB-type stars. Hot stars with surface temperatures higher than a few $10^4$ K are known to blow fast winds with speeds up to several thousand km s$^{-1}$ \citep{Kudritzki_WindsHotStars_2000}. The energy injected through this channel can be considerably high, and slightly less or of the order of supernova explosions \citep{Krause_FeedbackMassiveStarsVishniac_2013}. As the stellar wind impacts the surrounding gas, the wind material gets shocked and can reach temperatures as high as $10^7$ K, entering a regime where radiation cooling becomes practically inefficient. This leads to the formation of hot, expanding bubbles \citep{Castor_InterstellarBubbles_1975, Weaver_1977} that push away all the surrounding cold gas putting formally an end to the embedded phase (see Fig.~\ref{fig:YMSC_30Doradus}). The physics of these bubble-like structures has been comprehensively studied by \cite{Weaver_1977}, who have developed a simple analytical model for their evolution. Although the model by \cite{Weaver_1977} was originally intended for bubbles generated by isolated massive stars, it can be fairly well applied to the case of YMSCs if the cluster is compact enough (we will better state this condition soon) so that the winds from the central stars may combine together to form a collective cluster wind. Understanding these structures is of vital importance for the scope of our work. In the following, we will review the evolution of YMSC bubbles following the work of \cite{Weaver_1977}. 

\subsection{Evolution of the wind bubble}
\label{subsec:WindBubble}
\begin{figure}
\begin{center}
\includegraphics[width=0.7\textwidth]{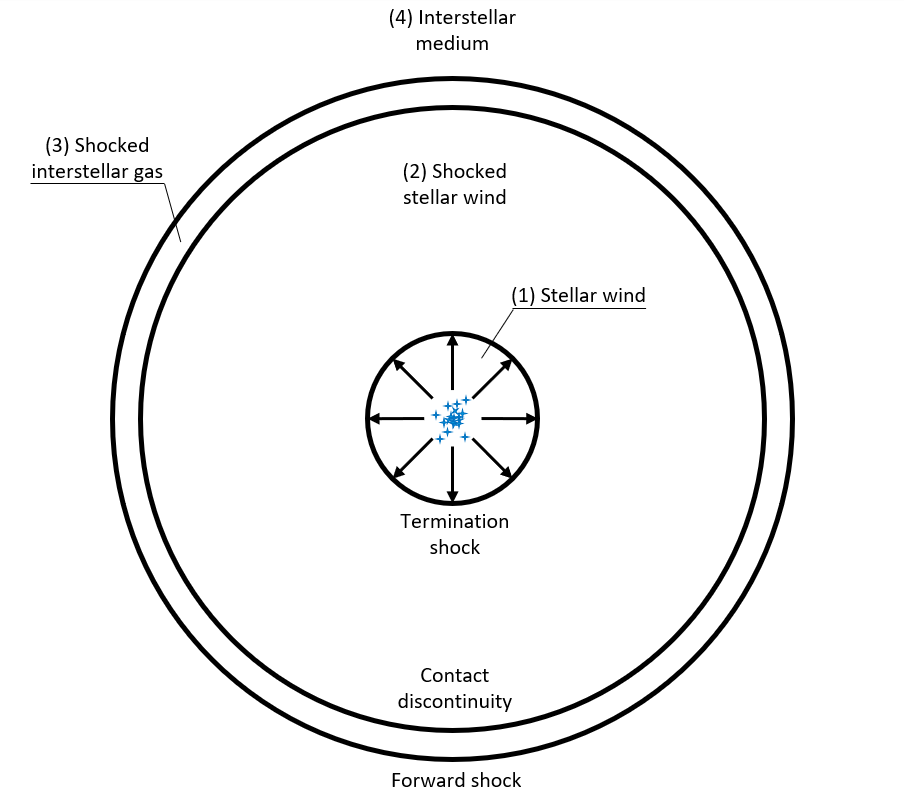}
\caption{Sketch of the bubble structure generated by a YMSC.}
\label{fig:BubbleStructure}
\end{center}
\end{figure}

In a general, ideal, situation, massive stars within the clusters start to blow fast winds as soon as they enter the main sequence. The winds from the individual stars combine to generate a spherically symmetric cluster wind with a speed $v_w$. As a consequence of this wind, the YMSC loses mass at a rate $\dot{M}$. The mechanical wind luminosity $L_w$ is:
\begin{equation}
\label{eq:Lw}
L_w=\frac{1}{2} \dot{M} v_w^2 \ .
\end{equation}
The collision of the supersonic, cold, cluster wind (region 1 in Fig.~\ref{fig:BubbleStructure}) with the surrounding ism generates a shock that propagates in the latter (\textit{forward shock}) and a reverse shock that propagates backward towards the origin of the wind. This reverse shock becomes the wind \textit{termination shock} (TS). The wind material is slowed down and heated up when crossing it, and a bubble of hot gas is formed (region 2 in Fig.~\ref{fig:BubbleStructure}). A contact discontinuity separates the bubble of shocked wind material from the shocked ISM that piles up on top of it (region 3 in Fig.~\ref{fig:BubbleStructure}).
 
 %Fig.~\ref{fig:BubbleStructure} provided a simple sketch of the entire system. 

The dynamical evolution of a wind bubble can be divided into: early, intermediate and late stage, which differ primarily for the importance of losses. The early phase is characterized by a fast adiabatic expansion of the bubble, during which radiative losses do not affect the dynamics of any region. This is not true anymore during the intermediate stage, when radiative losses cause the contraction of the shell of shocked ISM, while the region filled with the hot shocked wind material continues to expand adiabatically. Finally, in the late phase, non adiabatic losses begin to influence the whole hot cavity.

Let us now summarize one by one each evolutionary stage.

\subsubsection{Early stage of a wind bubble}
The early stage of evolution is expected to last a short time (a few 10$^4$ yr), so, in general, it is not particularly relevant in practical terms. Nevertheless, let us now proceed in evaluating the system evolution. The expression of the FS position as a function of time can be easily found through a dimensional analysis, and it is:
\begin{equation}
\label{eq:RfsEarly}
R_{FS}(t)=A L_w^{1/5} \rho_0^{-1/5} t^{3/5}
\end{equation}
where $A$ is a dimensionless constant to be determined. To obtain the evolution of the size of the contact discontinuity ($R_c$), one needs to numerically integrate the equations of continuity of mass and momentum for the warm shocked shell of ISM (region 3) under the assumption of adiabatic expansion. The position of $R_c$ is then set where the shell density profile drops to zero, that is $R_c \simeq 0.86 R_{FS}(t)$. Obtaining the evolution of the cold wind region boundary ($R_{TS}$) is not straightforward, as while the outer shell has a self similar evolution, this is not true for the region filled with hot shocked wind material (region 2). One possible way to obtain an analytical expression for $R_{TS}$ is to consider region 2 as almost isobaric. Under this assumption, which is indeed a good approximation \citep{Weaver_1977}, it is possible to derive from the continuity equation and adiabatic law the profiles of the velocity, density and pressure in region 2. The normalizations of these profiels are obtained using the Rankine-Hugoniot relations for shocks (assuming to known the wind speed). By imposing the regularity conditions at the contact discontinuity, one finds:
\begin{equation}
R_{TS}(t)=0.90 A^{3/2} \dot{M}^{3/10} \rho_0^{-3/10} v_w^{1/10} t^{2/5}
\end{equation}
with $A=0.88$. The early evolutionary stage lasts until the system age becomes comparable to the cooling timescale of the outer shell, at which time the adiabatic approximation cannot be used. The time at which this occurs is \citep{Fall_SWinISM_1975}:
\begin{equation}
\tau_{\rm shell \ collapse}=2.33 \times 10^3 \left( \frac{\dot{M}}{M_\odot yr^{-1}} \right)^{0.32} \left( \frac{v_w^2}{km^2 s^{-2}} \right)^{0.32} \left( \frac{n_0}{cm^{-3}} \right)^{-0.68} yr
\end{equation}  
which for $n_0=1 \ cm^{-3}$ and in the case of an extreme YMSC with $\dot{M}=10^{-4} \ M_\odot yr^{-1}$ and $v_w=3000 \ km s^{-1}$ is $\tau_{\rm shell \ collapse} \simeq 2 \times 10^4$ yr.

\subsubsection{Intermediate stage of a wind bubble}
As the outer shell cools, it contracts to form a thin, cold and dense, isobaric layer. Consequently, in this phase, $R_{FS}  \simeq R_{c}$, and it is then reasonable to assume that the position of the shell marks the total extent of the bubble, that we shall name as $R_b \equiv R_{FS} \simeq R_{c}$. To obtain the evolution of the bubble size (the former forward shock), we can consider the momentum equation for the outer shell:
\begin{equation}
\label{eq:ShellMomentum}
\frac{d}{dt} \left [M_s \frac{dR_{b}}{dt} \right ] = 4 \pi R_{FS}^2 P_2
\end{equation}
where $P_2$ is the pressure in the hot bubble (region 2) and M$_s$ is the mass contained in the swept-up shell, which is $M_s  = (4/3) \pi \rho_0 R_{b}^3$. Eq. (\ref{eq:ShellMomentum}) tells us that the momentum of the outer shell changes due to the work done by the expansion of the hot shocked wind material. This induces a variation of the energy of the hot gas equal to:
\begin{equation}
\label{eq:HotGasEnergyVar}
\frac{dE_2}{dt}=L_w-4 \pi R_{b}^2 P_2 \frac{dR_{b}}{dt}
\end{equation}
with $E_2$ as the energy of the hot gas, defined as:
\begin{equation}
\label{eq:HotGasEnergy}
E_2=\frac{4}{3} \pi R_{b}^3 \frac{P}{\gamma_g-1}=2 \pi R_{b}^3 P
\end{equation} 
where in the last equality we have assumed the case of a monoatomic gas ($\gamma_g=5/3$). Finally, assuming that, similarly to the earlier stage, the forward shock radius can be parametrized as $R_{b}\propto t^\alpha$, combining Eq. (\ref{eq:ShellMomentum}--\ref{eq:HotGasEnergy}) leads to:
\begin{equation}
\label{eq:Rbubble}
R_b (t)=\left ( \frac{250}{308 \pi} \right )^{1/5} L_w^{1/5} \rho_0^{-1/5} t^{3/5} 
\end{equation}
\begin{equation}
\label{eq:PressureInterStage}
P_2(t)= \frac{7}{(3850 \pi)^{2/5}}  L_w^{2/5} \rho_0^{3/5} t^{-4/5}  \ .
\end{equation}
Interestingly, the expression of Eq.~\ref{eq:Rbubble} is equal to that of Eq.~\ref{eq:RfsEarly}, but with $A=0.76$, as a consequence of the contraction of the outer shell. Eq.~\ref{eq:PressureInterStage} can be used to obtain an estimation of the TS position during this intermediate stage, which can be calculated by imposing the balance between the ram pressure of the wind and $P_2$:
\begin{equation}
\label{eq:RamPressEqToBubbPress}
\frac{\dot{M} v_w}{4 \pi R_{TS}^2} = \frac{7}{(3850 \pi)^{2/5}}  L_w^{2/5} \rho_0^{3/5} t^{-4/5}\ ,
\end{equation}
which returns:
\begin{equation}
\label{eq:Rts}
R_{TS}(t) = \sqrt{\frac{(3850 \pi)^{2/5}}{28 \pi}} \dot{M}^{1/2} v_w^{1/2} L_w^{-1/5} \rho_0^{-3/10} t^{2/5}\ .
\end{equation}
Note that in Eq.\ref{eq:HotGasEnergyVar} the contribution of radiation losses is not included. The intermediate stage is considered to last until radiation losses begin to be relevant, and the assumption of adiabatic expansion ceases to be valid. To estimate the cooling timescale and then the expected duration of the intermediate evolutionary phase, one needs to calculate the shocked wind temperature, which is:
\begin{equation}
\label{eq:Tshock}
T_2=\frac{P}{k_B n_2} \simeq 25.5 \times 10^6 \left(\frac{L_w}{10^{37} \rm \ erg \ s^{-1}} \right)^{2/5} \left(\frac{n_0}{10 \rm \ cm^{-3}} \right)^{3/5} \left(\frac{n_2}{10^{-2} \rm \ cm^{-3}} \right)^{-1} \left(\frac{t}{1 \rm \ Myr} \right)^{-4/5}  \rm K
\end{equation}
where $n_2$ is the density in the bubble. The expected temperature is of the order of a few $10^7$ K. However, one needs to account also for the thermal flux due to the heat conductivity between the cold dense shell and the hot shocked plasma, which rapidly cools the bubble temperature to lower values. The heat flux from the hot bubble is: 
\begin{equation}
\label{eq:ThermFlux}
Q = \kappa_S \frac{\partial T}{\partial r}
\end{equation}
where $\kappa_S$ is the thermal conductivity for a fully ionized plasma \citep{Spitzer_IonPlasmaBook_1962}: 
\begin{equation}
\label{eq:HeatFluxSpitzer}
\kappa_S = C T^{5/2} \simeq 1.2 \times 10^{14} \left(\frac{T}{10^8 \rm \ K} \right)^{5/2} \rm erg \ s^{-1} cm^{-1} K^{-1}
\end{equation}
with $C=1.2 \times 10^{-6}$ erg cm$^{-1}$ s$^{-1}$ K$^{-7/2}$. In this way, the conduction timescale ($t_{cond}$) can be estimated as the ratio between the thermal energy and Q:
\begin{equation}
t_{cond}=\frac{1}{2} n_2 k_B \frac{T_2}{Q} R_b \simeq  2 \times 10^{3} \left(\frac{n_2}{10^{-2} \rm \ cm^{-3}} \right) \left(\frac{T}{10^7 \rm \ K} \right)^{-5/2} \left(\frac{R_b}{60 \rm \ pc} \right)^{2} \rm yr
\end{equation}
where we have approximated $\partial T/ \partial r \approx T_2/R_b$. This extremely short timescale indicates that the plasma immediately cools to lower temperatures, transferring heat to the cold, dense shell. Consequently, in a steady-state regime, temperatures reach values such that the heat flow has the same order of magnitude as the mechanical energy flow ($F_m$), which can be written as: 
\begin{equation}
\label{eq:MechFlux}
F_m=\frac{5}{2} P_2 \frac{dR_b}{dt} \ .
\end{equation}
Equating Eq.\ref{eq:ThermFlux} and Eq.\ref{eq:HeatFluxSpitzer} with Eq.\ref{eq:MechFlux} leads to the temperature expression:
\begin{equation}
\label{eq:AvgTbubble}
T_2 \simeq \left(\frac{5 P_2 R_b^2}{2 t C} \right)^{2/7} \simeq 2.3 \times 10^{6} \left(\frac{L_w}{10^{37} \rm \ erg \ s^{-1}} \right)^{8/35} \left(\frac{n_0}{10 \rm \ cm^{-3}} \right)^{2/35}  \left(\frac{t}{1 \rm \ Myr} \right)^{-6/35}  \rm K
\end{equation}
where we used again the approximation $\partial T/ \partial r \approx T/R_b$ and $dR_b/dt \approx R_b/t$, and we also made the dependencies of $P_2$ and $R_b$ explicit using Eq.~\ref{eq:PressureInterStage} and Eq.~\ref{eq:Rbubble} respectively. Due to heat conductivity, the temperature in the bubble is roughly one order of magnitude less than what is expected from a shocked gas, as given by Eq.~\ref{eq:Tshock}. With this temperature in mind, we can now estimate the radiative cooling timescales. A hot plasma with temperatures ranging between $ \sim 10^4 \-- 10^7$ K loses energy as a consequence of collisional excitation of bound electrons. In this regime, the power loss can be approximated as \citep{Draine_ISMPhysics_2011}:
\begin{equation}
\frac{dE}{dV dt} \simeq 1.1 \times 10^{-22} n_2^2 \left(\frac{T}{10^6 \rm \ K} \right)^{-0.7} \rm erg \ cm^{-3} s^{-1}
\end{equation}
which leads to a radiative cooling timescale of 
\begin{equation}
\label{eq:RadCoolingTime}
t_{cool}=\left( \frac{3}{2}n_2 k_B T \right) \left(\frac{dE}{dV dt}\right)^{-1} \simeq 6 \left(\frac{n_2}{0.01 \rm cm^{-3}} \right)^{-1} \left(\frac{T}{10^6 \rm \ K} \right)^{1.7} \rm Myr \ .
\end{equation}
Considering an average temperature of a few $10^6$ K, the radiative cooling should start to be relevant after a few tens of Myrs.

Before proceeding further, it is interesting to point out the following phenomenon. As a direct result of the heat flux from the hot shocked wind region to the swept-up shell, the cold material evaporates into the bubble, carrying a significant amount of mass that increases the gas density. Following the same approach as \cite{Castor_InterstellarBubbles_1975}, a simple way to treat the evaporation process is to assume that the inward gas flux from the shell to the hot cavity is approximated by a stationary plane parallel flow. If so, the mass evaporation rate from the shell ($\dot{M}_{s}$), assuming a constant pressure flow and neglecting radiation losses, is well described by \citep{Zeldovich_GasFlow_1969}:
\begin{equation}
\label{eq:ShellMdotEquation}
\frac{\dot{M}_{s}}{4 \pi R_b^2} \frac{dH}{d\zeta} = \frac{d}{d\zeta} \left( \kappa_S \frac{dT}{d\zeta} \right) 
\end{equation}
where $\zeta=R_b-r$, $H$ is the specific enthalpy $H=5 k_B T / 2 \mu$, with $\mu=0.62 m_p$ as the mean molecular weight, defined as:
\begin{equation}
\mu=\frac{\sum_j n_j A_j}{\sum n_j + n_e}
\end{equation}
where $n_j$ and $A_j$ are the numerical density and the mass number of a given j-th ion respectively, while $n_e$ is the electron density. The mass loss rate can be obtained by integrating Eq.~\ref{eq:ShellMdotEquation} from $\zeta=0$ to $\zeta=R_b$. However, to do so, one needs to know the temperature profile in the bubble. The expression reported in Eq.~\ref{eq:AvgTbubble} can be roughly considered as an average value in the cavity.

The calculation of the temperature profile is a somewhat involved procedure, and has been carried out by \cite{Weaver_1977}. The temperature has a non uniform profile, with an overall decreasing trend towards the cold shell:
\begin{equation}
T(\zeta)=T_2(1-\zeta)^{2/5}
\end{equation}
where $T_2$ is given by Eq.~\ref{eq:AvgTbubble}. Using the latter equation leads to the following expression for the mass evaporation rate:
\begin{equation}
\dot{M}_{s} = \frac{16 \pi \mu}{25 k_B} \kappa_S R_b \ ,
\end{equation}
which can be further expanded by expressing the dependencies of $\kappa_S$ and $R_b$ via Eq.~\ref{eq:HeatFluxSpitzer} and Eq.\ref{eq:Rbubble}:
\begin{equation}
\label{eq:ShellMdot}
\dot{M}_{s} \simeq 2\times 10^{-4} \left(\frac{L_w}{10^{37} \rm \ erg \ s^{-1}} \right)^{27/35} \left(\frac{\rho_0}{10 m_p \rm \ cm^{-3}} \right)^{-2/35} \left(\frac{t}{1 \rm \ Myr} \right)^{6/35} \rm M_\odot yr^{-1}   
\end{equation}
On average, the mass evaporation from the shell is significantly larger than the cluster mass loss rate induced by stellar winds (see \S~\ref{subsec:SpectAnalysis}).

\subsubsection{Late stage of a wind bubble}
As the contribution of radiation losses becomes relevant, the system deviates from the adiabatic expansion solution. To understand the evolution in such a condition, one needs to modify Eqs.~\ref{eq:ShellMomentum},~\ref{eq:HotGasEnergyVar} and~\ref{eq:HotGasEnergy} by including the contribution of radiation losses. In addition, one must take into account the resulting decrease in volume of the hot bubble caused by the expansion of the TS as a result of the lower pressure induced by the decrease in temperature. The new sets of equations that must be solved are then:

\begin{equation}
\label{eq:HotGasEnergyVar_LateStage}
\begin{cases}
\frac{d}{dt} \left [M_s \frac{dR_b}{dt} \right ] = 4 \pi R_b^2 (P-P_3)  \\
\frac{dE}{dt}=L_w-4 \pi R_b^2 P \frac{dR_b}{dt}-L_{rad} \\
E=2 \pi (R_b^3-R_{TS}^3) P
\end{cases}
\end{equation}
where $P_3$ is the pressure of the outer shell, and  
\begin{equation}
L_{rad}= \int_{R_{TS}}^{R_{b}} 4 \pi n_2^2(r)\Lambda(T(r)) r^2 dr \ .
\end{equation}
In principle, the system can be solved numerically by calculating $L_{rad}$ at every instant of time, if the an expression for the temperature is provided \citep{Weaver_1977}. However, such approach is computationally expensive. A cheap alternative to find an approximate description of the evolution is to parametrize the radiation losses as a constant fraction $\xi_{rad}$ of the wind luminosity:
\begin{equation}
L_{rad}= \xi_{rad} L_w \ .
\end{equation}
The new expressions for $R_b$ and $R_{TS}$ are similar to the previous ones but with the wind brightness value rescaled by a factor  $(1-\xi_{rad})$. 

At this point, however, the following should be noted: the wind bubble model developed by \cite{Weaver_1977} was designed for structures generated by single massive stars. In our case, in which we are dealing with YMSCs, the evolution of the bubble during the final stage can be significantly modified by the occurrence of several other physical processes that take place on a shorter time scale. Since the late phase occurs after a few tens of Myr, if not gravitationally bound, the star cluster could have dispersed before entering in the last evolutionary phase, given that the crossing time is of the order of $\sim 10$ Myr. Furthermore, after 10 Myr, a significant number of massive stars should have ended their life cycle, generating a high number of supernova explosions whose feedback can severely alter the dynamic evolution of their surroundings.

\subsubsection{Limitations of the Weaver's wind bubble model}
Despite the simplicity and elegance with which it describes a complex phenomenon, the prescription of \cite{Weaver_1977} still remains extremely simplified model with limitations that we are now going to briefly summarize.

First of all, the model was conceived for wind bubbles around massive stars. Hence, there is the underlying assumption that the central source injecting the wind is pointlike compared to the system overall size. This could not be the case for some YMSCs, which means that the model is still valid if the cluster is compact enough compared to the system size.

On a very general ground, the compactness condition can be translated into the requirement for the existence of a collective wind from the cluster. This can be formalized as $R_{YMSC}\ll R_{TS}$, where $R_{YMSC}$ is the size of the star cluster. After a more critical look, instead of the radius of the cluster core, what really matters is the overall spatial distribution of the most massive stars. This is because the most massive stars are the ones that contribute the most to the wind energy budget, and are hence the main pillars sustaining the TS structure. 

In this regard, it is important to underline the phenomenon of mass segregation in stellar clusters. Several observational pieces of evidence \citep{Lada_EvolutionESCs_1991, Hillenbrand_OrionNebulaCLusterDynamics_1998, Elmegreen_NGC6946_2000, Jiang_M17_2002} have shown that the initial mass function of stars in a cluster is characterized by a spatial dependence described by a power-law with an index that becomes harder towards the center of the cluster. This effect seems to be related to the formation mechanism of stellar clusters \citep{Lada_SCReview_2003, Karam_SCFormation_2022}, and hence, particularly enhanced in young clusters.  As consequence, the most massive stars tend to sprout in a very compact space at the very center of the stellar cluster. So, even if it is not guaranteed that a YMSC can generate a collective wind, the compact spatial distribution of massive stars within a cluster favors this scenario. 

As for other limitations of the model of \cite{Weaver_1977}, it also does not account for other types of feedback, such as for example direct radiation pressure. Indeed it has been estimated that direct radiation pressure feedback can be as efficient as wind feedback (see \cite{Krumholz_SCsReview_2019} and references therein). This fact could affect the dynamics of the shell.

An additional, and potentially significant limitation, is the assumption of evolution in a uniformly distributed medium. While this could be true in the case of isolated massive stars, it is likely not in the case of YMSCs, as they are surrounded by the structure of the parent GMC which is highly non-uniform. The presence of low-density regions generated by the porosity of the GMC structure can induce severe leaks of the hot shocked gas from the bubble. This can dramatically reduce the pressure, thus stopping the expansion. Numerical simulations \citep{Rogers_MSCsWindFeedbackSim_2013} and observations \citep{Lopez_HIIStellarFeedback_2014} seem to point out that, if the gas is not well confined, gas leaks are a common phenomenon.

In parallel to gas leaks, a further process that can seriously affect the evolution of the bubble is the cooling produced by hot-cold gas mixing at the contact discontinuity \citep{Rosen_GoneWindMSCs_2014}. The mixing is induced by both turbulent motions at the interface \citep{Lancaster_StellarWindCooled_2021}, directly generated by the turbulent behavior of the GMC gas, and by the rise of several instability modes at the contact discontinuity, such as thin shell \citep{Vishniac_VishniacInstab_1983} and Rayleight-Taylor-like instabilities \citep{Bucciantini_RTInstab_2004}. As a direct consequence of gas mixing, the density of the bubble increases. Eq.~\ref{eq:RadCoolingTime} shows that the radiative cooling scales with $n_2^{-1}$, so a substantial growth in the density may induce catastrophic radiative cooling. A secondary effect is that the instabilities at the contact discontinuity cause the fragmentation of the shell \citep{Lancaster_StellarWindCooledII_2021}, which leads to a considerable increase of the contact surface between the hot and cold gas. This makes thermal conduction even more efficient and, simultaneously, boosts the mass evaporation rate, causing the density of the bubble to increase further.

Lastly, the solution for the intermediate evolutionary stage given by \cite{Weaver_1977} is not self-consistent. In fact, \cite{Weaver_1977} calculates the evolution of the bubble boundary without accounting for the thermal energy losses due to the heat exchange with the cold shell. To obtain the correct solution, one needs to modify Eq.~\ref{eq:HotGasEnergyVar} in:
\begin{equation}
\frac{dE}{dt}(r)=L_w - 4 \pi R_b^2 \left[ P \frac{dR_b}{dt} + \kappa_S  \frac{\partial T(R_b)}{\partial r} \right]
\end{equation}
As the temperature of the bubble decreases so does the pressure. As a consequence, the size of the TS increases. In parallel, the expansion of the forward shock slows down, and the bubble size becomes smaller. Obtaining a full solution in this situation becomes challenging and non-trivial since now the energy equation has a radial dependence. Nevertheless, similar to what has been done for radiative losses in the late evolutionary stage, the heat conduction can be approximated as a fraction of the overall wind luminosity:
\begin{equation}
\frac{dE}{dt} \simeq L_w - 4 \pi R_b^2 \left( P \frac{dR_b}{dt} + \kappa_S  \frac{T}{R_b} \right) = (1-\xi_{th})L_w -  4 \pi R_b^2 P \frac{dR_b}{dt}
\end{equation}
which means that if $50\%$ of the wind luminosity is lost due to thermal conduction, the forward shock and TS sizes vary by $\sim 15\%$.

%---------------------------------------------------------

\section{Cosmic Rays and the enigma of their origins: Young Massive Star Clusters as Cosmic Ray factories?}
\label{sec:YMSCsCRsAcc}
Discovered more than a century ago by Victor Hess, Cosmic Rays (CRs) are charged particles mainly composed of nuclei (99\%, of which 87\% are protons, and 12\% are alpha particles with a smaller fraction of heavier nuclei) and a minor fraction of electrons and antimatter. CRs are a fundamental component of the Cosmos, affecting the environment of galaxies. Indeed, the energy density associated with CRs in the Milky Way is of the order of $\sim 1$ eV cm$^{-3}$ \citep{Webber_CRsEnergyDensity_1998}, which is similar to that of the interstellar starlight radiation field $\sim 0.5$ eV cm$^{-3}$ \citep{Mathis_StarlightEnDens_1983}, to that of the average ($\sim 3 \ \mu$G) galactic magnetic field \citep{Webber_CRsEnergyDensity_1998}, and to that of the cosmic microwave background $\sim 0.26$ eV cm$^{-3}$ \citep{Workman_PDG_2022}. CRs are often investigated from the high-energy astrophysics prospective, but, as a matter of fact, they are also a vital ingredient for the physics of the interstellar medium and of star formation.

Unquestionably, low-energy CRs are one of the main ingredients regulating star formation in galaxies, providing a negative feedback. One can in this regard, try to evaluate what is the star formation rate in the Milky Way in the assumption of the total absence of regulating mechanisms. To do so, we can approximately estimate the star formation rate as: 
\begin{equation}
\label{eq:SFRestimate}
SFR_{MW} \simeq \frac{M_{MW}}{\tau_{ff}}
\end{equation}
where $M_{MW}\approx 8\times 10^9$ M$_\odot$ is the Milky Way gas mass \citep{Nakanishi_MWmass_2016} and $\tau_{ff}$ is the free fall time describing the timescale for the collapse of a neutral cloud, which can be written as \citep{Spitzer_PhysOfISM_1978}:
\begin{equation}
\tau_{ff}=\sqrt{\frac{3 \pi}{32 G m_p n_{C}}} \ ,
\end{equation}
where $n_C$ is the numerical density of the cloud. If we assume an average density of $n_C = 10$ cm$^{-3}$, the free fall time is $\tau_{ff} \approx 16.3$ Myr and the star formation rate calculated using Eq.\ref{eq:SFRestimate} is $\approx 500$ M$_\odot$ yr$^{-1}$, which is far from the observed value of $SFR_{MW}=2.0 \pm 0.7$ M$_\odot$ yr$^{-1}$ \citep{Elia_MWSFR_2022}.

Accounting for the CR-induced ionization can significantly reduce the expected star formation rate, as partially ionized clouds are more stable against gravitational collapse. This is because during gravitational collapse, the ionized component of a cloud remains frozen to the cloud magnetic field, which opposes resistance to collapse-induced compression. The neutral material is then slowed down thanks to the coupling generated by neutral-ion scattering with the ionized component, which results in a lengthening of the collapse characteristic time scales\footnote{This process is often named ambipolar diffusion \citep{Draine_ISMPhysics_2011}}, making the star formation process less efficient. 

The significance of the CR role is due to their ability to penetrate deeply into the cores of dense molecular clouds, providing ionization where other ionizing sources cannot. This is the case, for example, for X-rays that are usually absorbed in the first layers of a molecular cloud \citep{McKee_XrayAbsorption_1989}. Due to their ability to ionize, as we shall see in \S~\ref{subsec:MCIonization}, low-energy CRs also prove to be crucial in regulating astrochemical processes, indirectly inducing the formation of complex molecules in the interstellar medium \citep{Dalgarno_CRsISMChem_2006}.

\begin{figure}
\begin{center}
\includegraphics[width=0.54\textwidth]{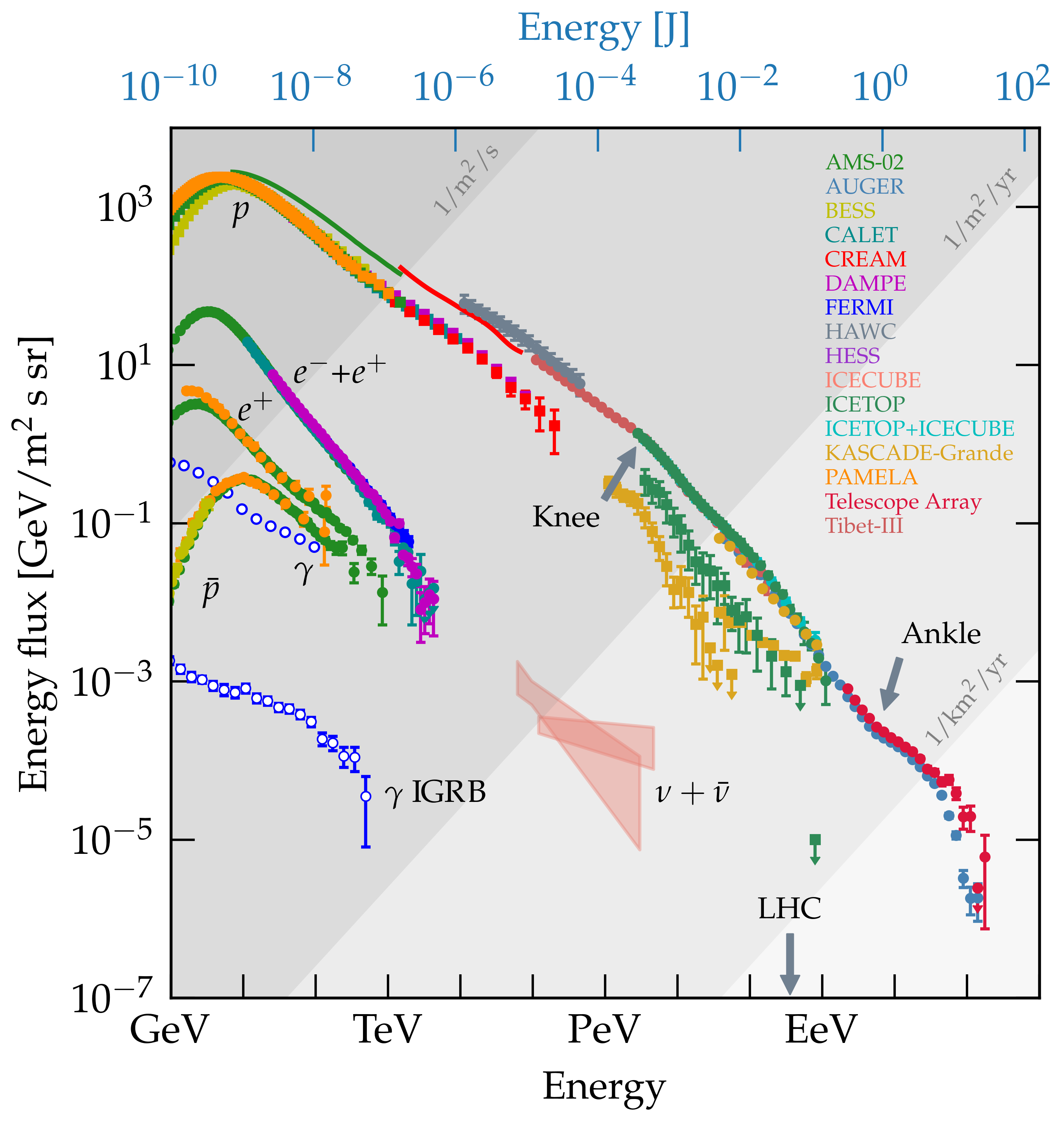}
\caption{CR energy spectrum as observed at the Earth. The plot shows also for comparison the spectra of the galactic diffuse $\gamma$-ray emission, the isotropic diffuse $\gamma$-ray background (IGRB), and the neutrino/anti-neutrino flux. See \cite{Carmelo_CRspectrum_2020} and references therein.}
\label{fig:CRsSpectrmAtEarth}
\end{center}
\end{figure}

From a pure observational point of view, the spectrum of CRs observed at the Earth is well described by a broken power law, starting from $\sim 10$ GeV, with three breaks over almost 12 decades in energy (Fig.~\ref{fig:CRsSpectrmAtEarth}). Below 10 GeV, the CR spectrum is modified by solar modulation, preventing the lowest energy component from penetrating within the Solar System \citep{Gleeson_SolarWindCRModulation_1968, Potgieter_SolarModReview_2013}. Precise measurement of the spectrum at low energy was made possible by the Voyager 1 spacecraft only after exiting the heliosphere \citep{Cummings_Voyager1LECRs_2016}.

At energies above 10 GeV the spectrum is characterized by a spectral index of $\sim -2.7$. The first break appears at $\sim 3\times 10^{15}$ eV, where the spectrum steepens with the spectral index becoming $\sim -3.1$, while the second break occurs instead at $\sim 10^{18.5}$ eV, inducing a hardening of the spectrum and bringing the index back to $\sim-2.7$ \citep{Workman_PDG_2022}. The first and second breaks are respectively known as the \textit{knee} and the \textit{ankle}, given the similarity of the overall spectrum to a human leg.

As far as the ankle is concerned, the general consensus is that this feature is caused by the extragalactic CR population beginning to dominate the Galactic population \citep{Bird_CRsAnkle_1994, Apel_KaskadeAnkle_2013}. Undoubtedly, if the highest energy CRs are protons of extragalactic origin, then, at energies above $\sim 5 \times 10^{19}$ eV, the CR spectrum is expected to be suppressed due to the pion photoproduction mechanism with the cosmic microwave background. This effect is known as the Greisen–Zatsepin–Kuzmin limit (or GZK cutoff) \citep{Greisen_GZK_1966}. The existence of this cut-off has been shown by several ultra-high energy experiments \citep{Abbasi_HiResGZK_2008, Verzi_AugerUHECR_2019}. Although it is still unclear whether we are witnessing the GZK effect or rather observing the maximum energy achievable by cosmic accelerators. The answer to this important question depends on the still uncertain composition of CRs at the highest energies.

Of greater relevance to the topic of this thesis is the origin of the knee. The presence of the knee has been attributed to the maximum energy reachable by galactic sources of CRs. This idea arose from the evidence of a gradual change in the chemical composition of CRs at energies above a few PeV. Above the knee, the fraction of heavy nuclei in CRs appears to increase, and this trend seems to persist up to $10^{17}$ eV \citep{Hoorandel_KneeComposition_2006}. It follows that, in this scenario, the knee could result from the superposition of cutoffs in the spectra of different elements as the large majority of acceleration processes are rigidity\footnote{The rigidity of a particle with charge $Ze$ and momentum $p$ can be defined as $R_g=pc/Ze$} dependent, i.e. proportional to the particle electric charge \citep{Blasi_CRsReview_2013}. This means that if protons are accelerated in the sources to a maximum energy $E_{p, max}\approx 3 \-- 5 \times 10^{15}$ eV, then an iron nucleus will be accelerated to $E_{Fe, max} = 26 E_{p, max} \approx 1 \times 10^{17}$ eV (assuming that during acceleration the iron nuclei are fully ionized, therefore the unscreened charge is $Z = 26$).

Clearly, in order to understand the plausibility of this framework, it is necessary to investigate the properties of the galactic CR sources. Since the first half of the 20th century, supernova remnants (SNR) were suggested as possible candidates as galactic CR accelerators. SNRs are the supersonic ejected material released after the supernova explosion of a massive star\footnote{Collapse supernovae (sometimes also classified as Type II or Type Ib and Ic following the old spectroscopic classification), which occur for stars with masses higher than 8 M$_\odot$ are though to account for 80\% of these explosions. In addition there are Type Ia SNe which arise when an accreting white dwarf exceeds Chandrasekhar limit generating a thermonuclear explosion. See \cite{Vink_SNRsReview_2012, Vink_PhysAndEvolSNR_2020} and references therein for comprehensive reviews on SNRs.}. or of a white dwarf exceeding the Chandrasekhar limit. The suggestion of SNRs as Galactic CR sources was first made by \cite{Baade_CRsSNRsOrigin_1934}, and later motivated by \cite{Syrovatskii_SNR_1963}. The suggestion it is as robust as it is elegant in its simplicity. Whatever the identity of the galactic accelerators, they must be able to sustain the observed CR luminosity, which is: 
\begin{equation}
L_{CR} = \frac{U_{CR}V_{MW}}{\tau_{conf}} \approx 2.5 \times 10^{40} \rm erg \ s^{-1}
\end{equation}
where $U_{CR}\approx 1$ eV cm$^{-3}$ is the galactic CR energy density, $V_{MW}$ is the Milky Way volume\footnote{As a spiral galaxy, the Milky Way volume can be easily approximated as a thick disc with radius $\sim$15 kpc and height $\sim$300 pc.}, and $\tau_{conf} \approx 20$ Myr is the CR confinement timescale in the Galaxy, which can be estimated from measurements of the abundance ratio of unstable isotopes \citep{Connell_CRsConfTime_1998, Hams_MeasureBe10_2004}. In the case of SNRs, the kinetic energy of a supernova explosion is of the order of $E_{SN} \approx 10^{51}$ erg \citep{Carroll_IntroModAstro_1996}. Assuming that the supernova explosion rate in the Milky Way is $\tau_{SN}=1/50$ yr$^{-1}$, then the power injected by supernova explosion is:
\begin{equation}
L_{SN} = \frac{E_{SN}}{\tau_{SN}} \approx 6.3 \times 10^{41} \rm erg \ s^{-1}.
\end{equation}
This implies that if a few percent of the power injected by supernova explosions goes in CR acceleration, then, SNRs can easily account for the observed CR luminosity. But can SNRs actually accelerate CRs to the knee energy?

The supersonic motion of the ejecta produces a collisionless shock wave that propagates in the ISM. Particle acceleration in collisionless shocks is believed to occur through the \textit{first-order Fermi mechanism} (also known as \textit{Diffusive Shock Acceleration}) \citep{Blandford_ParticleAccelerationShocks_1978, Drury_DSA_1983}. We can briefly summarize the process as follows. Let us consider the shock as a discontinuity between two regions with different flow velocities, we define these two velocities in the reference system of the shock as $u_1$ and $u_2<u_1$. We shall refer to the region of unperturbed plasma with speed $u_1$ as \textit{upstream}, and the region of shocked gas with speed $u_2$ as \textit{downstream}. Suppose now that particles, with some velocity $v$ in the rest frame of the shock, enter the upstream, and start to diffuse through scattering with magnetic turbulence. The diffusion causes the particle velocity distribution to rapidly become isotropic in the reference frame of the upstream. After some time, some particles may be able to cross again the shock, entering the downstream. Again, particle will start to diffuse through scattering with magnetic turbulence that is associated with the downstream plasma. On average, particles will see plasma on the other side of the shock moving in their direction with a speed $\sim \lvert u_1 - u_2 \rvert$. Each time the particles cross the shock, they will experience an electric field $\sim \lvert u_1 - u_2 \rvert/c$ times the magnetic field and gain energy. The process is analogous to that of collision with moving wall, and collisions are all head-on, hence inducing a gain in energy. 

It can be shown through that particles will on average increase their energy as $\Delta E \propto \lvert u_1 - u_2 \rvert E/c$. Eventually, particles will return to the upstream side of the shock. The gain occurs at every shock crossing, and several cycles of diffusion back and forth across the shock will lead to a significant increase in particle energy.

To estimate the maximum energy obtained in SNRs, one has to equate the acceleration timescales ($\tau_{acc}$) with the time for which the source is efficiently accelerating particles\footnote{This statement is correct if the energy losses of particles are negligible, which is likely the case for hadrons but typically is not for leptons. In the latter case, the comparison must be made with the Inverse Compton or Synchrotron cooling time scales.}. The acceleration time is clearly related to the timescale with which the upstream-downstream-upstream cycle occurs ($t_{cycle}$), and can be calculated as \citep{Drury_DSA_1983}:
\begin{equation}
\label{eq:t_acc}
\tau_{acc}=\frac{t_{cycle}}{\Delta E/E} = \frac{3}{u_1-u_2} \left( \frac{D_1}{u_1}+\frac{D_2}{u_2}\right) \simeq 8 \frac{D_1}{u_{shock}^2} 
\end{equation}
where $\Delta E/E$ is the average energy gain per cycle, $u_{shock}$ is the shock velocity, while $D_1$ and $D_2$ are the diffusion coefficients in the upstream and downstream respectively. Generally, the diffusion coefficient can be defined as:
\begin{equation}
D=\frac{1}{3} \lambda_{mfp} v_p
\end{equation}
where $\lambda_{mfp}$ is the mean free path and $v_p=\beta_p c$ is the particle speed, parametrized as a fraction $\beta_p$ of the light speed $c$. The diffusion is mediated by the scattering with magnetic field irregularities that have length scales comparable with the particle Larmor radius ($r_L$). In the most extreme case, when the magnetic irregularities $\delta B$ are comparable with the average magnetic field $B$, one has $\lambda_{mfp}=r_L$. This is usually called the \textit{Bohm regime}, and the diffusion coefficient reads:
\begin{equation}
D(E)=\frac{1}{3} r_L(E) \beta_p c = \frac{1}{3} \frac{E}{Z e_0 B} \beta_p c 
\end{equation}
where $E$ and $Ze$ are the particle energy and charge respectively, with $e$ as the electron elementary charge.

SNRs can efficiently accelerate particles during the so-called ejecta-dominated phase (sometimes referred to as the free expansion phase), defined as the evolutionary stage for which most of the explosion kinetic energy is confined in the freely expanding ejecta \citep{Vink_PhysAndEvolSNR_2020}
\begin{equation}
E_{SN}=\frac{1}{2}M_{ej}u_{shock}^2
\end{equation} 
During this phase, the fast shock can easily reach CRs that are diffusing upstream, allowing efficient acceleration. This can be easily seen considering that the shock radius scales as $R_{shock} \propto t$ during the ejecta dominated phase, while the particle diffusion length is proportional to $\propto t^{1/2}$. As the swept-up mass becomes comparable to the ejecta mass ($M_{ej}=4\pi\rho_{ISM} R_{shock}^3/3$), the SNR evolution enters the Sedov-Taylor phase \citep{Vink_PhysAndEvolSNR_2020}. During this stage, the most energetic particles are not caught anymore by the expanding shock ($R_{shock} \propto t^{2/5}$) and are thus free to escape in the upstream region, making the acceleration mechanism no longer efficient. The time at which the Sedov-Taylor phase starts can be estimated as:
\begin{equation}
\tau_{ST}=\frac{R_{shock}}{u_{shock}} \simeq 220 \left( \frac{M_{ej}}{1 \rm \ M_\odot} \right)^{5/6} \left( \frac{E_{SN}}{10^{51} \rm \ erg} \right)^{-1/2} \left( \frac{\rho_{ISM}}{1 m_p \rm \ cm^{-3}} \right)^{-1/3} \rm yr.
\end{equation} 
Finally, equating $\tau_{ST}$ with $\tau_{acc}$ leads to the estimate of the maximum particle energy in SNRs:
\begin{equation}
E_{max, SNRs}=2.5 \times 10^{13} Z \left( \frac{B}{1 \rm \ \mu G} \right) \left( \frac{M_{ej}}{1 \rm \ M_\odot} \right)^{-1/6} \left( \frac{E_{SN}}{10^{51} \rm \ erg} \right)^{1/2} \left( \frac{n_{ISM}}{1 \rm \ cm^{-3}} \right)^{-1/3} \rm eV  
\end{equation} 
In a standard situation, the maximum energy for a proton is expected to be about one order of magnitude below the PeV. This means that, in order to consider SNRs as PeVatrons, a stronger magnetic field is required. 

Several magnetic field amplification processes have been proposed in the literature, such as non-resonant \citep{Bell_MFsAmp2004} and resonant \citep{Skilling_ResStreamInst_1975} streaming instabilities, amplification driven by the dynamics of the shock impinging in density fluctuation of turbulent magnetized plasma \citep{Guo_BAmplTurbMedium_2012}, and acoustic instabilities \citep{Drury_AcusticInstab_2012}. However, even including the most efficient of these amplification mechanisms, the non-resonant streaming instability, a standard SNR can hardly accelerate particles up to PeV energies, unless specific extreme situations with high velocity shocks expanding in dense environments are considered \citep{Cristofari_SNRsPeV_2021}. The difficulty of SNRs to reach PeV energies makes the interpretation of these objects as a single population of Galactic accelerators problematic, and prompts us to consider the possible presence of an additional class of particle accelerators.

The maximum energy of CRs in SNRs is not the only issue related to these sources. Measurements of the CR composition at the Earth have shown an excess in some isotopic ratios, such as the $^{22}$Ne to $^{20}$Ne ratio, which is found $5.3 \pm 0.3$ times higher than the solar value \citep{Wiedenbeck_Ne_1981, Binnis_NeInCRs_2008}. The $^{22}$Ne is copiously produced by Wolf-Rayet (WR) stars \citep{Prantzos_NeWRs_1986, Maeder_IsotopicAnomaliesCRs_1993}, that are the late evolutionary stage of very massive O-type stars ($\gtrsim 25 M_\odot$) \citep{Crowther_WRsReview_2007}. This means that at least some fraction of CR sources must be located in environments close to WR stars, which are very common in YMSCs. The presence of such overabundance then can be considered as a strong clue that points to stellar clusters as possible CR factories. Undoubtedly, if CR acceleration were to occur in these objects, some of the wind material of a WR would end up in the galactic CR population, thus explaining the observed excess. The enigma of the origin of CRs then shifts to the general question of whether and how much YMSCs can actually contribute to the acceleration of CRs in the galaxy. In the past years, different models for particle acceleration in YMSCs have been proposed. We will summarize in the next subsections some of the most popular.

\subsection{Acceleration mechanisms in the core of a YMSCs}
\label{subsec:CoreAcceleration}
In its core, a YMSC may enclose a significant number of massive stars, and each of them can launch a fast wind that collides with the ambient medium and generates a TS. Particle acceleration occurs at the stellar wind TS following the diffusive shock acceleration process. The final distribution of freshly accelerated CRs from the YMSC will then arise as a combination of injected particles from different stellar TSs: such a scenario has been investigated by \cite{Klepach_WindWindInteractionCR_2000}. 

\begin{figure}
\begin{center}
\includegraphics[width=0.7\textwidth]{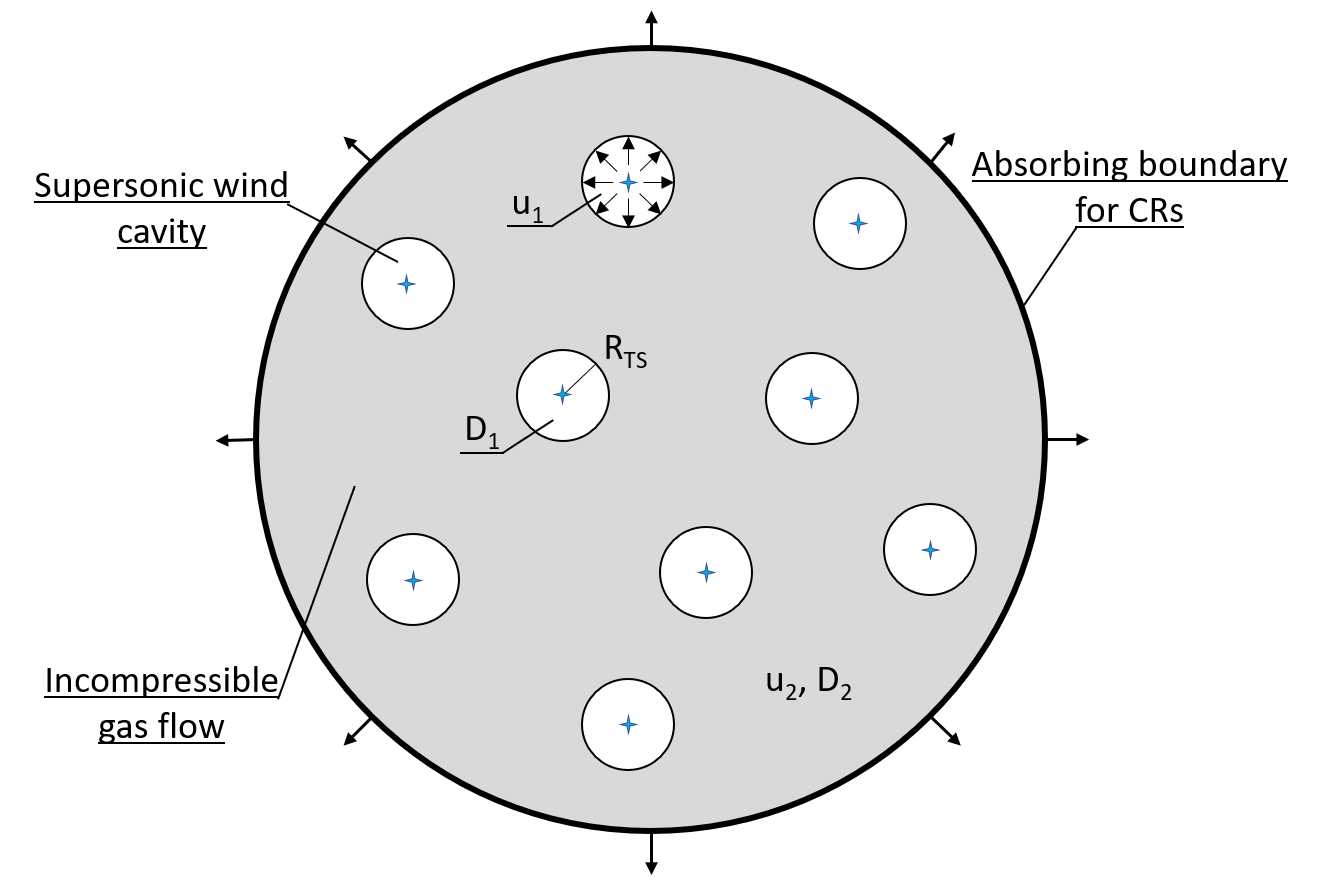}
\caption{System sketch for the multiple TS acceleration mechanism investigated by \cite{Klepach_WindWindInteractionCR_2000}.}
\label{fig:CoreAccSketch}
\end{center}
\end{figure}

Let us consider a system composed of several stellar wind cavities (Fig.~\ref{fig:CoreAccSketch}). Each of the stellar winds is assumed to be highly supersonic, with constant radial velocity $u_1$. The winds impact with the hot shocked material at a distance $R_{TS}$. The time evolution of $R_{TS}$ is somewhat similar to that given by Eq.\ref{eq:Rts}, so it is slowly expanding in time and can be considered as stationary. Following the classical theory of shocks, the gas velocity downstream of the TS is:
\begin{equation}
u_2=\frac{u_1}{\mathcal{R}} \left( \frac{r}{R_{TS}} \right)^{-2} 
\end{equation}
where $\mathcal{R}$ is the TS compression ratio. We will further assume that the overall volume occupied by the wind cavities is much less that the system volume, or equivalently, that $R_{TS}$ is much smaller than the average distance $d$ between the stars. Formally this condition translates into requiring the filling factor $\mathcal{W}=(R_{TS}/d)^3$ to be $\mathcal{W}\ll 1$.

On a very general ground, a CR distribution function $f_{CR}(r,t,p)$ must obey the following transport equation\footnote{This is not the complete transport equation, as some terms are neglected, such as for example, potential energy losses and momentum diffusion, see \S~\ref{subsec:TurbAcc}.}
\begin{equation}
\label{eq:CRsTransportEq}
\frac{\partial f_{CR}}{\partial t} - \vec{\nabla} D \vec{\nabla} f_{CR} + u \vec{\nabla} f_{CR} - \frac{p}{3} \vec{\nabla} u \frac{\partial f_{CR}}{\partial p} = q
\end{equation}
where $p$ is the particle momentum and $q$ is the source term, which describes the injection of freshly accelerated CRs that is assumed to be occurring at the wind TSs. Clearly, the normalization of $f_{CR}$ is determined by the total CR number density $n_{CR}=4\pi \int p^2 f_{CR} dp$.
The CR distribution function ($\langle f_{CR} \rangle$) obtained in such a system will be an average over several $f_{CR}$ calculated from a random distribution of winds. In order to estimate the CR distribution, we need to solve for $\langle f_{CR} \rangle$. The equation for CR injected at a single spherical wind TS, can be written as: 
\begin{equation}
\label{eq:CRsTransportEqSingleWind}
\frac{\partial f_{CR}}{\partial t} - \frac{1}{r^2} \frac{\partial}{\partial r} r^2 D \frac{\partial f_{CR}}{\partial r} + u \frac{\partial f_{CR}}{\partial r} - \frac{r^2}{3} \frac{d (r^2 u)}{dr} p \frac{\partial f_{CR}}{\partial p} = Q \delta (r-R_{TS})
\end{equation}
where $Q(p)$ is the injection term for a single wind TS. The diffusion coefficient is likely to be different in the upstream ($D_1$) and the downstream ($D_2$). \cite{Klepach_WindWindInteractionCR_2000} assume the upstream diffusion coefficient to scale as $D_1 r/R_{TS}$, and that $D_2$ is spatially constant. \cite{Klepach_WindWindInteractionCR_2000} additionally assume that both $D_1$ and $D_2$ are energy independent. Note that since $\mathcal{W}\ll 1$, one will have $f_{CR} \simeq \langle f_{CR} \rangle$ for $r \gg R_{TS}$. To proceed, it is useful to define the following transformations:
\begin{equation}
\Phi(s, r, t)=\int_0^\infty  p^{s-1}f_{CR}(p,r,t) dp
\end{equation}
\begin{equation}
\Psi(s,t)=\int_0^\infty p^{s-1} Q(p, t) dp
\end{equation}
\begin{equation}
X(s,r,t)=\int_0^\infty p^{s-1} q(p, r, t) dp ,
\end{equation}
and to rewrite Eq.~\ref{eq:CRsTransportEq} accordingly:
\begin{equation}
\label{eq:MellinTransfCRsTranspEq}
\frac{\partial \Psi}{\partial t} - \nabla D \nabla \Psi + u \nabla \Psi + s \frac{\nabla u}{3} \Psi = X .
\end{equation}
In principle, the equation for $\langle f_{CR} \rangle$ can be obtained by averaging Eq.~\ref{eq:MellinTransfCRsTranspEq} over the entire system volume and then perform the inverse transformation. Both the averaging and inverse transform are lengthy, nontrivial calculations, and the derivation of $f_{CR}$ may result cumbersome. We here summarize the final outcome under the following assumptions: (1) The gas motion is incompressible downstream of the wind TSs, and in addition, the shocked gas is able to diffuse far away from the TS and to fill the cluster volume such that $\langle \nabla \cdot u \rangle = 3 \mathcal{W} u /  \mathcal{R} R_{TS} $. (2) The solution is calculated assuming a steady state regime ($\partial f_{CR}/ \partial t =0$). (3) Repeated acceleration cycles due to crossing of different TSs may occur. This requires that the particles mean free path must be larger than the TS size: $g_1=R_{TS}u_1/D_1 \ll 1$ and $g_2=R_{TS}u_1/\mathcal{R}D_2 \ll 1$. (4) CRs escape from the cluster boundary with a leakage timescale $T_l$. Under these assumptions, the average CR distribution function is:
\begin{equation}
\langle f_{CR}(p) \rangle = \frac{3 \mathcal{R} Q_0}{u_1(\mathcal{R}-1)[1+(t_{acc}/T_l)(1+G^{-1})^{-1}]} \left(\frac{p}{p_0} \right)^{-S}
\end{equation}
where $Q_0$ is the injection rate at momentum $p_0$, $t_{acc}=\mathcal{R}R_{TS}/3\mathcal{W}u_1$ is the acceleration timescale, and
\begin{equation}
S=\frac{3 \mathcal{R}}{\mathcal{R}-1} \left\{1+ \frac{t_{acc}}{\mathcal{R} T_l \left[1+ G \left(1 +\frac{t_{acc}}{T_l} \right) \right]} \right\}
\end{equation}
\begin{equation}
G=e^{g_2}-1 
\end{equation}
The obtained spectrum is an approximation of the real solution, as in general, the diffusion coefficients depend on energy. Note that for $t_{acc}\ll T_l$, the powerlaw index $S$ becomes $S \simeq 3\mathcal{R}/(\mathcal{R} -1)$ that is the standard result for acceleration at a plane parallel shock \citep{Bell_CRsAccShockFront_1978}. \cite{Klepach_WindWindInteractionCR_2000} also estimate the maximum energy that can be reached with this acceleration mechanism, that is:
\begin{equation}
E_{max} \simeq 2 \times 10^{15} Z \left(\frac{D_2}{D_B} \right)^{-1} \rm \ eV
\end{equation}
where $D_B$ is the diffusion coefficient in the Bohm regime. For standard values of winds from massive stars, the maximum energy is as high as a few $10^{15} eV$ \citep{Klepach_WindWindInteractionCR_2000}, which is encouraging in terms of viewing YMSCs as a possible counterpart for SNRs as PeVatrons.

We may now wonder whether this scenario could actually develop in a YMCs, given that clusters tend to emerge from small clumps with sizes of a few pc and that the entire acceleration process is based on the assumption that $\mathcal{W} \ll 1$. To do so, we can estimate the size of $R_{TS}$ for an isolated star embedded in a hot bubble of shocked wind gas with pressure P given by Eq.\ref{eq:PressureInterStage}. The procedure is identical to that used to obtain the TS radius of the collective wind from a cluster (see Eq:\ref{eq:RamPressEqToBubbPress}), with the difference that now the ram pressure is calculated for the wind of a single star: 
\begin{equation}
\label{eq:RamPressureStars}
\frac{\dot{M_\star} v_{w,\star}}{4 \pi R_{TS}^2} = \frac{7}{(3850 \pi)^{2/5}}  L_w^{2/5} \rho_0^{3/5} t^{-4/5} 
\end{equation}
where $\dot{M_\star}$ and $v_{w,\star}$ are the mass loss rate and the wind velocity for a single star. Note that $L_w$ is the wind luminosity of the entire cluster, as the gas heating is provided by the overall stellar population. The size of the stellar wind TS is then:
\begin{equation}
\begin{split}
\label{eq:RTSStars}
R_{TS} \simeq 1.7 \left(\frac{\dot{M_\star}}{10^{-6} \rm M_\odot yr^{-1}} \right)^{1/2} &
\left(\frac{v_{w,\star}}{2000 \ \rm km \ s^{-1}} \right)^{1/2}  \left(\frac{L_w}{10^{37} \rm erg \ s^{-1}} \right)^{-1/5} \\
&  \left(\frac{\rho_0}{10 m_p \rm cm^{-3}} \right)^{-3/10} \left(\frac{t}{1 \rm Myr} \right)^{2/5} \rm pc .
\end{split}
\end{equation}
For typical values of the parameters $R_{TS} \sim 2$ pc. In general\footnote{From a theoretical pointof view, the cluster mass-size relation is significantly affected by the dominant feedback mechanism operating during the emergence of the YMSC from its dense gas cradle \citep{Krumholz_SCsReview_2019}.}, given the mass of a YMSC ($M_{YMSC}$), the cluster size can be estimated by the following empirical cluster mass-radius relation \citep{Pfalzner_YMSCsMRR_2016}:
\begin{equation}
R_{YMSC} \simeq 1.82 \left(\frac{M_{YMSC}}{10^3 \ \rm M_\odot} \right)^{1/\delta} \ \rm pc
\end{equation} 
whit $\delta=1.71 \pm 0.07$. Note that by cluster size we mean the half-mass radius (the radius within which half of the stellar mass is enclosed). Reasonably, the average stars spacing will be less than, or of the order of, the cluster size. Thus, considering the typical properties of a massive star in terms of wind luminosity, mass loss rate and wind speed (see Eq.~\ref{eq:RTSStars}), $R_{TS}$ has a size similar to the average stellar distance expected in the case of a YMSC. 

It is then clear that this model is likely not a good representation for these objects. Plus, in Eq.\ref{eq:RamPressureStars}, we are not taking into account the effect of pressure reduction due to the cooling induced by the heat flow towards the surrounding ISM, so the $R_{TS}$ calculated in Eq.\ref{eq:RTSStars} is possibly underestimated. Nonetheless, this framework could be valid for older clusters, especially after a few tens of Myr in the case of gravitationally unbound systems. Assuming that the stars velocity dispersion is $\sigma = R_{YMSC}\sqrt{3G M_{YMSC}/4 \pi R_{YMSC}^3}$, after some time $t$ the cluster has expanded by an extent $\Delta R$ of:
\begin{equation}
\Delta R = \sigma t \simeq 7.68 \left(\frac{M_{YMSC}}{10^3 \ \rm M_\odot} \right)^{(\delta-1)/2\delta} \left(\frac{t}{10 \ \rm Myr} \right) \rm pc .
\end{equation}
After a few tens of Myr stars should be sparse enough to validate the mechanism proposed by \cite{Klepach_WindWindInteractionCR_2000}. Eventually, the model could also be well describing the acceleration process in OB associations, which are structures with sizes spanning over a few tens of pc.

\subsection{Acceleration through efficient scattering with turbulence}
\label{subsec:TurbAcc}
The powerful massive star winds, or alternatively, the strong collective cluster wind, interacting with the ISM irregularities generate chaotic magnetohydrodynamic fluctuation with frozen in magnetic field, which end up permeating the entire hot bubble. The scattering with these fluctuations induces both spatial and momentum diffusion if the correlation lengthscales of the turbulent magnetic field are comparable with the particle Larmor radius \citep{Blasi_CRsReview_2013}. The process of momentum diffusion is often called \textit{second-order Fermi acceleration}, as it statistically induces an energy gain \citep{Fermi_Fermi2ndAcc_1949}. However, this mechanism is not particularly efficient as the energy gain scales as $v_a^2/c^2$, where $v_a=B/\sqrt{4 \pi \rho}$ is the plasma Alfvén speed, with $\rho$ the plasma mass density. The momentum diffusion term can be included in the CR transport equation: 
\begin{equation}
\frac{\partial f_{CR}}{\partial t} - \nabla D_{xx} \nabla f_{CR} + u \nabla f_{CR} - \frac{p}{3} \nabla u \frac{\partial f_{CR}}{\partial p} - \frac{1}{p^2} \frac{\partial}{\partial p} \left(p^2 D_{pp} \frac{\partial f_{CR}}{\partial p}\right)= Q
\end{equation} 
where, to avoid confusion, we have renamed the spatial diffusion coefficient $D_{xx}$, while $D_{pp}$ is the momentum diffusion coefficient. Although not generally relevant, second-order Fermi acceleration may play a significant role in the case of strong turbulence, potentially boosting particle energies up to relatively high values \citep{Bykov_MSCTurbAcc_2020}. One can estimate the maximum energy achieved in such a way by comparing the acceleration and the propagation time scales. For second order Fermi acceleration, the characteristic time scale is:
\begin{equation}
\label{eq:tacc2ndFermi}
\tau_{acc}=\frac{p^2}{D_{pp}}\ ,
\end{equation}
where the momentum diffusion coefficient can be written as \citep{Thornbury_CRsReAcc_2017}:
\begin{equation}
D_{pp} \simeq \frac{p^2 v_a^2}{9 D_{xx}} \ .
\end{equation}
In general, the spatial diffusion coefficient is directly linked to the type of plasma turbulence spectrum:
\begin{equation}
\label{eq:SpatialDiffCoeff}
D_{xx} = \frac{1}{3} \frac{\beta c r_L}{\mathcal{F}(1/r_L)} 
\end{equation}
with
\begin{equation}
\label{eq:TurbFk}
\mathcal{F}(k) = \frac{k \mathcal{P}(k)}{B_0^2/8\pi}
\end{equation}
where $k$ is the wave number and $\mathcal{P}(k) \propto k^{-\alpha}$ is the turbulent magnetic field power spectrum. If the power is injected at a characteristic scale $L_{inj}$, then Eq.\ref{eq:TurbFk} can be normalized to the injection scale and rewritten in terms of the magnetic field correlation length scale $l=1/k$:
\begin{equation}
\label{eq:TurbFl}
\mathcal{F}(l) = \mathcal{A} \left(\frac{l}{L_{inj}} \right)^{(\alpha -1)} \ .
\end{equation}
In Eq.\ref{eq:TurbFl} the index $\alpha$ is correlated to the type of plasma turbulence cascade, and reads $\alpha=5/3$ for Kolmogorov-like turbulence \citep{Kolmogorov_K41Turbulence_1941}, $\alpha=3/2$ for a Kraichnan-like cascade \citep{Kraichnan_MHDTurbulence_1965}, and $\alpha=1$ for a flat turbulence spectrum (Bohm diffusion regime). The constant $\mathcal{A}$ is related to the total power in magnetic fluctuation $\delta B_{tot}$, through the condition $\delta B_{tot}^2/B_0^2 = \Lambda_B \mathcal{A}$, where $B_0$ is the unperturbed magnetic field. We consider here the scenario where $\delta B_{tot}=B_0$. The parameter $\Lambda_B$ depends on the turbulence spectral index $\alpha$, and is defined as:
\begin{equation}
\label{eq:fCRNoSea}
\Lambda_B \equiv \int_{k_{inj}}^{k_{max}} \left( \frac{k}{k_{inj}} \right)^{1-\alpha} \frac{dk}{k}=
\begin{cases}
\frac{1}{\alpha-1} \left[ 1 - \left ( \frac{k_{inj}}{k_{max}}\right)^{\alpha-1} \right] & \text{for } \alpha \neq 1  \\
\ln\left(\frac{k_{max}}{k_{inj}} \right) & \text{for } \alpha = 1 
\end{cases}
\end{equation} 
where $k_{inj}=1/L_{inj}$. Generally, the parameter $k_{max}$ corresponds to the inverse of the length scale at which turbulence thermal dissipation occurs. Note that, if this is the case, then one has typically $k_{inj} \ll k_{max}$, and for $\alpha \neq 1$ we have $\Lambda_B\simeq (\alpha-1)^{-1}$. In the case $\alpha=1$, however, it appears unrealistic to extend the assumption of equal turbulent power per decade down to the dissipation scale. We consider this description only appropriate to wave modes within an interval [$k_{inj}$, $k_{max}$] in which it is likely that power injection occurs at all scales. For example, a reasonable guess for such range could be given by the length scale associated with the average distance between the stars, down to the characteristic length scale of wind irregularities.

In \S~\ref{sec:CygOB2fCR} we will show that CR propagation in an expanding hot bubble is in general dominated by advection for particles with energies $\lesssim 1$ TeV, so the propagation time scale can be estimated as:
\begin{equation}
\label{eq:tadv1}
t_{adv} \simeq \frac{R_b}{u_2}\simeq t_{age}
\end{equation}
where $t_{age}$ is the age of the wind-blown bubble (or equivalently the cluster age). Equating Eq.\ref{eq:tacc2ndFermi} with Eq.\ref{eq:tadv1}, leads to 
\begin{equation}
\frac{9D_{xx}}{v_a^2}=  t_{age} \Rightarrow \frac{3 \beta c L_{inj}}{v_a^2} \Lambda_B \left( \frac{r_L}{L_{inj}}\right)^{(2-\alpha)} = t_{age} \  ,
\end{equation}
and the particle maximum energy for the second order Fermi acceleration is readily obtained by considering $r_L=E/eB_0$:
\begin{equation}
E_{max}^{Turb} \simeq \left(\frac{e_0^{2-\alpha}}{12 \pi \rho c} \right)^{1/(2-\alpha)} \Lambda_B^{-1/(2-\alpha)} L_{inj}^{(1-\alpha)/(2-\alpha)}B^{(4-\alpha)/(2-\alpha)} t_{age}^{1/(2-\alpha)}  
\end{equation}
where we have assumed that $\beta\approx 1$. Fixing $\alpha$ to the three previously mentioned values, the maximum energies are:
\begin{equation}
E_{max, K41}^{Turb} \simeq 29 \left(\frac{B_0}{1 \ \rm \mu G} \right)^7
\left(\frac{L_{inj}}{1 \ \rm pc} \right)^{-2}
\left(\frac{t_{age}}{10 \ \rm Myr} \right)^{3}
\left(\frac{\rho_2}{10^{-2}m_p \ \rm cm^{-3}} \right)^{-3} \rm MeV
\end{equation}
\begin{equation}
E_{max, Kra}^{Turb} \simeq 6 \left(\frac{B_0}{1 \ \rm \mu G} \right)^5
\left(\frac{L_{inj}}{1 \ \rm pc} \right)^{-1}
\left(\frac{t_{age}}{10 \ \rm Myr} \right)^{2}
\left(\frac{\rho_2}{10^{-2}m_p \ \rm cm^{-3}} \right)^{2} \rm GeV
\end{equation}
\begin{equation}
E_{max, Bohm}^{Turb} \simeq 5 \frac{1}{\ln (k_{inj}/k_{max})} \left(\frac{B_0}{1 \ \rm \mu G} \right)^3
\left(\frac{t_{age}}{10 \ \rm Myr} \right)
\left(\frac{\rho_2}{10^{-2}m_p \ \rm cm^{-3}} \right)^{-1} \rm TeV .
\end{equation}
The values obtained are far from PeV energies but remain considerably high, especially in the case of Bohm diffusion\footnote{This significantly depends on the value of $k_{max}$, which can potentially be several orders of magnitude above $k_{inj}$.}.

Nevertheless, several caveats may limit the realistic efficiency of such an acceleration process. Acceleration is achieved under the core assumption of particle scattering with resonant fluctuations, and the underlying calculations are performed in the context of quasi-linear theory. However, recent simulations have demonstrated that, by using modern anisotropic magnetohydrodynamic turbulence theories, the wave-particle resonances may be strongly suppressed, severely reducing the efficiency of second order Fermi acceleration (see \cite{Lemoine_MHDTurbAcc_2021} and references therein).

In this section, we have considered the case where acceleration is provided by subsonic turbulence. In older clusters, the combined presence of fast winds and supernova shocks may induce the presence of supersonic turbulence. See the work of \cite{Bykov_MSCTurbAcc_2020} for a comprehensive review of particle acceleration in such a framework.

\subsection{Acceleration at the cluster wind termination shock}
\label{subsec:WindTSAccModel}
YMSCs are probably compact enough to allow single star winds to combine so as to create a collective cluster wind. As we showed in \S~\ref{subsec:WindBubble}, the fast collective cluster wind impacting on the surrounding hot bubble material produces a strong TS at which particle acceleration may occur. This specific case was recently studied by \cite{Morlino_2021}.

Let us now consider a YMSC that has developed a bubble structure and is now in the intermediate evolutionary stage. From the CR point of view, the system can be briefly described as follows: particles are accelerated at the TS via the diffusive shock acceleration mechanism, and subsequently escape from the acceleration site experiencing a combination of advection and diffusion in the hot bubble until they reach the forward shock. From there, CRs are free to leave the system by diffusing in the unperturbed ISM. Given the slow expansion rate of the cavity, the system can be considered as stationary, and if one assumes radial symmetry, and neglects second order Fermi acceleration and particles energy losses, the distribution $f_{CR}(r, E)$ of CRs can be found by solving the following steady-state transport equation\footnote{Note that the system is very similar to that described in \S\ref{subsec:CoreAcceleration} for the acceleration at the wind TS for a single star. The differences are limited to the boundary conditions considered.}
\begin{equation}
\label{eq:TransportEq}
\frac{\partial }{\partial r} \left [ r^2 D(r,p) \frac{\partial f}{\partial r} \right ] - r^2 u(r) \frac{\partial f}{\partial r} + \frac{d [r^2 u(r)]}{dr} \frac{p}{3} \frac{\partial f}{\partial p} +r^2 Q(r,p) = 0
\end{equation}
where $u(r)$ is the plasma speed, and $D(r,p)$ is the spatial diffusion coefficient. The source term $Q(r,p)$ describes the particle injection taking place at the TS:
\begin{equation}
Q(r,p)= \frac{\eta_{inj} n_1 u_1}{4 \pi p^2_{inj}} \delta(p-p_{inj})\delta(r-R_{TS})\ ,
\end{equation}
where $n_1$ is the density immediately upstream of the termination shock, $u_1$ is the speed of the cold wind, and $\eta_{inj}$ is the fraction of particle that are injected in the acceleration process with momentum $p_{inj}$. The global solution of Eq.~\ref{eq:TransportEq} can be found in three main steps: first, the equation must be solved separately in the unperturbed ISM ($r>R_b$) and upstream (in the cold cluster wind, $r<R_{TS}$) and downstream of the TS (in the hot shocked wind bubble, $R_{TS}<r<R_b$). Secondly, the solutions in these three zones are joined together using flux continuity at $r=R_{TS}$ and $r=R_b$. Finally, one needs to specify two boundary conditions. This can be done by assuming no net flux at $r=0$, and requiring that $f_{CR}(r, E)$ at infinity matches the distribution of the galactic CR sea ($f_{gal}$). 

Following the above-mentioned procedure, the CR radial distribution in the three zones is:
\begin{subequations}
\label{eq:CRdistribFunc}
\begin{equation}
\label{eq:FUpstream}
  f_1(r,p)\simeq f_{TS}(p) \cdot exp \left [ -\int_r^{R_{TS}} \frac{u_1}{D_1(r',p)}  dr' \right ]  
\end{equation}
\begin{equation}
\label{eq:FDownstream}
f_2(r,p)= f_{TS}(p) e^{\alpha} \frac{1+\beta(e^{\alpha_B-\alpha}-1)}{1+ \beta(e^{\alpha_B}-1)} + f_{gal}(p) \frac{\beta (e^\alpha-1)}{1+\beta(e^{\alpha_B}-1)} 
\end{equation}
\begin{equation}
f_{ism}(r,p)=f_2(R_b, p)\frac{R_b}{r}+f_{gal}(p) \left ( 1- \frac{R_{TS}}{r} \right )
\end{equation}
\end{subequations}
with
\begin{subequations}
\label{eq:AlphaBetaCRpar}
\begin{equation}
\alpha = \alpha(r, p)=\frac{u_2 R_{TS}}{D_2(p)} \left ( 1 -  \frac{R_{TS}}{r} \right )
\end{equation}
\begin{equation}
\alpha_B=\alpha(r=R_b,p)
\end{equation}
\begin{equation}
\beta = \beta(p) = \frac{D_{ism}(p) R_b}{u_2 R_{TS}^2}
\end{equation}
\end{subequations}
where the subscripts $1$, $2$ and $ism$ refers orderly to values assumed by the variables in the upstream, downstream, and interstellar medium regions. Notice that Eq.~\ref{eq:FUpstream} is a first-order approximation of the full solution presented by \cite{Morlino_2021}, which in principle should be formally obtained by iteratively solving Eq.~\ref{eq:TransportEq} in the upstream. $f_{gal}$ is the average spectrum of the Galactic CR sea, e.g. as inferred from AMS-02 data \citep{Aguilar_AMS02fsea_2015}. Finally, $f_{TS}$ is the distribution of injected particles at the TS.

The formal solution for $f_{TS}$ can be written in the following form 
\begin{equation}
\label{eq:fTSformal}
f_{TS}(p)=s \frac{\eta_{inj}n_1}{4 \pi p_{inj}^3} \left ( \frac{p}{p_{inj}} \right )^{-s} e^{-\Gamma_1(p)}e^{-\Gamma_2(p)}.
\end{equation}
The function is composed of three terms: the first one is the standard power-law spectrum  resulting from particle acceleration in plane shocks. The second term contains the function $\Gamma_1(p)$, which depends itself on $f_{TS}$, implying a non-linear nature of the solution. From a physical point of view, the suppression term $e^{-\Gamma_1(p)}$ can be seen as a modification, induced by the spherical geometry of the system, to the usual energy gain obtained in parallel shocks. The last term finally, $e^{-\Gamma_2(p)}$, describes the cut-off caused by the escape of particles at the bubble boundary. 

The final form of $f_{TS}$ is non-analytical. Nevertheless, it can be approximated with good accuracy using a modified power-law with an exponential cut-off, whose expression slightly changes depending on the model for particle diffusion around the TS:
\begin{equation}
\label{eq:fTSnorm}
f_{TS}(p) \simeq s \frac{\eta_{inj} n_1}{4 \pi p_{inj}^3} \left ( \frac{p}{p_{inj}} \right )^{-s} \left [ 1 + a_1 \left (\frac{p}{p_{max}} \right )^{a_2} \right ] e^{- a_3 (p/p_{max})^{a_4}},
\end{equation}
where $p_{max}$ is related to the maximum achievable momentum in the system, and the parameters $a_i$ depend on the type of magnetohydrodynamic turbulence in the plasma (see Tab.~\ref{tab:fTSparamenters}). The normalization of $f_{TS}$ is determined by $\eta_{inj}$ and $p_{inj}$, which can be usefully expressed in terms of $\epsilon_{CR}$, the fraction of cluster wind luminosity $L_w$ converted into accelerated particles:
\begin{equation}
\label{eq:LCRDef}
\epsilon_{CR} L_w = L_{CR} = 4\pi R_{TS}^2 u_2 \int f_{TS}(p) E_k(p) d^3p ,
\end{equation}
where $L_{CR}$ is the CR luminosity and $E_k=E(p)-m_pc^2$ is the particles kinetic energy. Knowing that $L_w=4 \pi m_p n_1 R^2_{TS} u_1^3$, from Eq.\ref{eq:LCRDef} one can obtain an expression for $\eta_{inj}$, which can be used in Eq.~\ref{eq:fTSnorm} to obtain:
\begin{equation}
\label{eq:fTS}
f_{TS}(p) \simeq \frac{3 n_1 u_1^2 \epsilon_{CR}}{4 \pi \Lambda_p (m_pc)^3 c^2 } \left ( \frac{p}{m_p c} \right )^{-s} \left [ 1 + a_1 \left (\frac{p}{p_{max}} \right )^{a_2} \right ] e^{- a_3 (p/p_{max})^{a_4}}
\end{equation}
where:
\begin{equation}
\Lambda_p =\int_{x_{inj}}^{\infty} x^2 f_{TS}(x) \left( \sqrt{1+x^2}-1 \right ) dx ,
\end{equation}
with $x=p/m_p c$.

\begin{table}
\begin{center}
\begin{tabular}[c]{l c c c c}  
\toprule \toprule
Models  & a$_1$ & a$_2$ & a$_3$ & a$_4$ \\
\midrule
Kolmogorov & 10 & 0.308653 & 22.0241 & 0.43112\\
Kraichnan & 5 & 0.448549 & 12.52 & 0.642666\\
Bohm & 8.94 & 1.29597 & 5.31019 & 1.13245\\
\bottomrule %\bottomrule
\end{tabular}
\caption{Parameter values used to calculate the distribution of injected particles.}
\label{tab:fTSparamenters}
\end{center}
\end{table}  

Interestingly, once the parameters of the star cluster are fixed, $f_{TS}$ is fully described by only two parameters, namely the efficiency of CR production ($\epsilon_{CR}$) and the spectral index of injected particles ($s$). Indeed, it is possible to fix $p_{max}$, or equivalently, the maximum energy of accelerated particles ($E_{max}$), to the intrinsic properties of the YMSC. The maximum momentum of particles can be calculated by equating the particle diffusion length to the size of the TS\footnote{From a formal point of view, this approach to estimating $E_{max}$ is somewhat approximate. In fact, one would have to consider the confinement of the particle in the downstream, since if the particle escapes the bubble it will immediately diffuse into the ISM, effectively terminating the acceleration process. This mechanism is formally included in the term $e^{-\Gamma(p)_2}$ in Eq.~\ref{eq:fTSformal}.}:
\begin{equation}
\label{eq:DiffLRTS}
\frac{D_1(E_{max})}{u_1}=R_{TS} \ .
\end{equation}
This is because the probability of crossing the TS and being further accelerated decreases significantly when the particles have diffusion lengths greater than or comparable to $R_{TS}$. If we consider again the three turbulent cascade models mentioned above, given by the Kolmogorov, Kraichnan and flat (Bohm-like) spectrum, the corresponding diffusion coefficients upstream are easily obtained from Eq.~\ref{eq:SpatialDiffCoeff}, and are respectively:
\begin{equation}
\label{eq:DK41}
D_{K41}(E)=\frac{1}{3}\Lambda_B \beta c r_L^{1/3} L_{inj}^{2/3}
\end{equation}
\begin{equation}
\label{eq:DKra}
D_{Kra}(E)=\frac{1}{3} \Lambda_B \beta c r_L^{1/2} L_{inj}^{1/2}
\end{equation}
\begin{equation}
\label{eq:DBho}
D_{Bohm}(E)=\frac{1}{3} \Lambda_B \beta c r_L.
\end{equation}
where $r_L=E/e \delta B_{tot}$. 

The diffusion coefficients are directly linked to the intensity of the magnetic field fluctuations upstream of the TS. We assume that the total power in magnetic fluctuation $\delta B_{tot}/8\pi = \int \mathcal{P}(k) dk$ is a fraction $\eta_B$ of the wind luminosity, such that:
\begin{equation}
\label{eq:BfieldUpTS}
\frac{\delta B_{tot}^2}{4 \pi}= \eta_B \frac{L_w}{4 \pi u_1 R_{TS}^2}
\end{equation}
we furthermore consider the scenario of strong turbulence, so that $\delta B_{tot}/B_1=1$, with $B_1$ the magnetic field upstream of the TS, and we shall rename for simplicity $\delta B_{tot}=\delta B_1$.
%\begin{equation}
%\label{eq:DeltaB1}
%\delta B_1 (r) \approx \sqrt{\frac{\eta_b \dot{M} u_1}{2 r^2}},
%\end{equation}
%so that at the TS position $\delta B_1 \approx \sqrt{\eta_b \dot{M} u_1 / 2 R_{TS}^2}$.
Solving Eq.~\ref{eq:DiffLRTS} for $E_{max}$ and rewriting the expression for $R_{TS}$ using Eq.~\ref{eq:Rts} and Eq.~\ref{eq:Lw} leads to the following maximum energies:
\begin{equation}
\begin{split}
\label{eq:EmaxK41}
E_{max}^{K41} \simeq 10^{14} \Lambda_B^{-3} \eta_{B}^{1/2} \left( \frac{\dot{M}}{10^{-4} \rm M_\odot yr^{-1}}\right )^{11/10} \left( \frac{u_1}{10^3 \rm \ km s^{-1}} \right )^{37/10} & \left( \frac{\rho_0}{m_p \rm \ cm^{-3}} \right )^{-3/5} \\
\left(\frac{t_{age}}{10 \rm \ Myr}\right )^{4/5} \left( \frac{L_{inj}}{2 \rm \ pc} \right )^{-2} \rm eV
\end{split}
\end{equation}
\begin{equation}
\begin{split}
\label{eq:EmaxKra}
E_{max}^{Kra} \simeq 4 \times 10^{14} \Lambda_B^{-2} \eta_{B}^{1/2} \left( \frac{\dot{M}}{10^{-4} \rm M_\odot yr^{-1}}\right )^{4/5} \left( \frac{u_1}{10^3 \rm \ km s^{-1}} \right )^{13/5} &  \left( \frac{\rho_0}{m_p \rm \ cm^{-3}} \right )^{-3/10} \\
\left(\frac{t_{age}}{10 \rm \ Myr}\right )^{2/5} \left( \frac{L_{inj}}{2 \rm \ pc} \right )^{-1} \rm eV
\end{split}
\end{equation}
\begin{equation}
\begin{split}
\label{eq:EmaxB}
E_{max}^{Bohm} \simeq 7.53 \times 10^{15} \Lambda_B^{-1} \eta_{B}^{1/2} \left( \frac{\dot{M}}{10^{-4} \rm M_\odot yr^{-1}}\right )^{1/2} \left( \frac{u_1}{10^3 \rm \ km s^{-1}} \right )^{3/2} \rm eV
\end{split}
\end{equation}
As we will show in \S~\ref{subsec:StellarPhys}, $u_1$ is expected to be of the order of the speed of the winds from the most massive stars, which ranges between 2000$\--$3000 km s$^{-1}$. This implies that energies of a few PeV are easily reached in the case of protons. For the sake of completeness, it must be noted that $E_{max}$ depends on two parameters, namely $\eta_B$ and $L_{inj}$ that are observationally hard to estimate and are currently largely unknown. For example, $L_{inj}$ can vary by one order of magnitude, depending on whether the turbulence is injected at a characteristic scale of the average distance between stars (a few pc) or at the typical length scale of the TS size ($\sim 10 \-- 20$ pc), and this has a significant impact in the cases of Kraichnan and Kolmogorov-like cascades. For the Bohm case, the parameter $L_{inj}$ is instead substituted by $\Lambda_B$, which accounts for the interval in length scales for which the power spectrum is flat. 

Finally, knowing the relation between $u_1$ and $L_w$ given by Eq.~\ref{eq:Lw}, one can rewrite the equations for the maximum energy in a more handy form, which directly depends on $L_w$. Assuming that $\eta_B \ll 1$, Eqs.~\ref{eq:EmaxK41} --~\ref{eq:EmaxKra} --~\ref{eq:EmaxB} becomes:

\begin{equation}
\begin{split}
\label{eq:EmaxK41Lw}
E_{max}^{K41} \simeq 1.2 \Lambda_B^{-3} \left(\frac{\eta_B}{0.1}\right)^{1/2} \left( \frac{\dot{M}}{10^{-4} \rm M_\odot yr^{-1}}\right )^{-3/4} \left( \frac{L_w}{10^{39} \rm \ erg s^{-1}} \right )^{37/20} & \left( \frac{\rho_0}{20 m_p \rm \ cm^{-3}} \right )^{-3/5} \\
\left(\frac{t_{age}}{3 \rm \ Myr}\right )^{4/5} \left( \frac{L_{inj}}{2 \rm \ pc} \right )^{-2} \rm PeV
\end{split}
\end{equation}
\begin{equation}
\begin{split}
\label{eq:EmaxKraLw}
E_{max}^{Kra} \simeq 2.8 \Lambda_B^{-2} \left(\frac{\eta_B}{0.1}\right)^{1/2} \left( \frac{\dot{M}}{10^{-4} \rm M_\odot yr^{-1}}\right )^{-5/10} \left( \frac{L_w}{10^{39} \rm \ erg s^{-1}} \right )^{13/10} &  \left( \frac{\rho_0}{ 20 m_p \rm \ cm^{-3}} \right )^{-3/10} \\
\left(\frac{t_{age}}{3 \rm \ Myr}\right )^{2/5} \left( \frac{L_{inj}}{2 \rm \ pc} \right )^{-1} \rm PeV
\end{split}
\end{equation}
\begin{equation}
\begin{split}
\label{eq:EmaxBLw}
E_{max}^{Bohm} \simeq 10.07  \Lambda_B^{-1} \left(\frac{\eta_B}{0.1}\right)^{1/2} \left( \frac{\dot{M}}{10^{-4} \rm M_\odot yr^{-1}}\right )^{-1/4} \left( \frac{L_w}{10^{39} \rm \ erg s^{-1}} \right )^{3/4} \rm PeV
\end{split}
\end{equation}
%=========================================================================================

\section{Probing YMSCs as CRs accelerators}
In \S~\ref{sec:YMSCsCRsAcc} we acknowledged that, from the the theory point of view, YMSCs are able to accelerate CRs in different ways. Moreover, the maximum energies achieved in these systems can be as high as a few PeV, making YMSCs conceivable as galactic PeVatron. The general question that one may ask is how to empirically confirm or reject the possibility that of YMSCs actually produce a sizable amount of CRs and accelerate particle up to $10^{15}$ eV.

Unfortunately, it is impossible to probe CR sources directly from the reconstruction of the trajectories of particles arriving at the Earth, since CR propagation in the Galaxy is fully diffusive at 1 PeV. The validity of this statement is readily proven by computing the Larmor radius of a proton in the average Galactic magnetic field ($\sim 3$ $\mu$G):
\begin{equation}
r_L \simeq 0.36 \left(\frac{E}{1 \ \rm PeV} \right) \left(\frac{B}{3 \ \rm \mu G} \right)^{-1} \rm \ pc,
\end{equation}
which means that after a few pc from the source, a CR has deviated significantly from its original escape trajectory.

The study of CR accelerators must then rely on observational techniques that can probe the presence of accelerated particles by means of their interaction with the environment close to the CR source. One possibility is to consider the radiation emitted by CRs. High-energy photons within the $\gamma$-ray range can serve as a direct indicator of the presence of CRs. This is because only energetic particles, via non-thermal processes, can emit radiation with energies higher than a few tens MeV. 

Indeed, during the last decades, several YMSCs have been observed in coincidence with large diffuse $\gamma$-ray emission, both in the high-energy ($\gtrsim 1$ GeV) and very high-energy ($\gtrsim 1$ TeV) bands. Examples are Westerlund 1 \citep{Abramowski_Wd1VHE_2012, Aharonian_HESSWesterlund1_2022}, Westerlund 2 \citep{Yang_Wd2Fermi_2018}, Cygnus OB2\citep{Ackermann_FermiCygnusCocoon_2011, Bartoli_CygOB2Argo_2014, Abeysekara_CygOB2HAWC_2021}, and NGC 3603 \citep{Saha_NGC3603_2020}. The presence of $\gamma$-ray emission has largely strengthened the hypothesis of YMSC as CRs factories. This is even more true considering that in all the detected YMSCs, the observed $\gamma$-ray luminosity, if interpreted as hadronic, is easily explained assuming that a fraction of a few percent of the cluster wind power ends up in accelerated particles.

Let us consider, for instance, the specific case of the YMSC Cygnus OB2. \cite{Aharonian_MSCs_2019} measured a total $\gamma$-ray luminosity towards Cygnus OB2 of $L_\gamma\approx 3.02 \times 10^{34}$ erg s$^{-1}$ for photon energies above 10 GeV. The total energy $W_p$ in terms of hadronic particles to explain such emission can be estimated as \citep{Aharonian_MSCs_2019}:
\begin{equation}
W_p \simeq 2.55 \times 10^{50} \left(\frac{L_\gamma}{10^{34}\ \rm erg s^{-1}} \right) \left(\frac{n_2}{0.1 \ \rm cm^{-3}} \right)^{-1} \ \rm erg ,
\end{equation}
which is $\approx 0.7 \times 10^{51}$ erg for Cygnus OB2. Given the wind power and the age of Cygnus OB2 of $L_w \approx 2 \times 10^{38}$ erg s$^{-1}$ and $t_{age} \approx 3$ Myr respectively (see \S~\ref{sec:CygOB2cluster}), the total injected energy by the stellar cluster is $E_{cluster} = L_w t_{age} \approx 2 \times 10^{52}$ erg, that implies a reasonable efficiency of CR production of the order of $\sim 10 \%$ to account for the observed emission. For the sake of completeness, it must be underlined that this estimation has been carried out assuming that the observed $\gamma$-ray emission was of hadronic nature. In principle, also leptons may generate high-energy radiation, and, often, discriminating the nature of the emission is a challenging task. Nevertheless, a comprehensive study of the morphology and spectrum of the $\gamma$-ray emission from a YMSC can provide essential insight into the properties of the freshly accelerated CRs, revealing the main characteristics of the ongoing particle acceleration processes in such systems. In \S~\ref{subsec:RadMechanisms} a general overview of the principal leptonic and hadronic $\gamma$-ray emission processes is reported. 

Another way to probe the presence of energetic particles is to observe the enhanced ionization degree induced by the low-energy tail of the CR population in dense regions of the ISM close to a CR source. At the beginning of \S~\ref{sec:YMSCsCRsAcc}, we emphasized the importance of low-energy CRs as regulators of the molecular cloud dynamics and in general of the star formation process.

A diffuse HI cloud which is embedded in the Galactic CR sea is foreseen to have a ionization rate of $\zeta_{HI}\gtrsim 6.8 \times 10^{-18}$ s$^{-1}$ \citep{Spitzer_IonizationMC_1968}, while for dense molecular cloud the ionization rate is instead expected to be $\zeta_{H2}\gtrsim 1 \times 10^{-17}$ s$^{-1}$ \citep{Glassgold_IonRateMC_1974}. Close to a CR source, the flux of ionizing CRs is likely to be higher, and one could consider searching for an enhanced ionization rate in cloud close to the acceleration site to confirm the presence of freshly accelerated particles.

This technique has been successfully used in the case of SNRs. A well known case is that of W28, where the presence of an increased ionization rate in nearby molecular clouds was found, in agreement with the idea of SNRs as CR accelerators \citep{Vaupre_W28IonRate_2014}. Interestingly, the information on the ionization rate can be further combined with $\gamma$-ray observations, to eventually help to discriminate the hadronic nature of the emission \citep{Gabici_IonRate-GammaCRsTracers_2015}.

So far, no attempt to extend this approach to YMSC has been made, in spite of the fact that the environment is foreseen to be particularly promising given the large amount of clumped molecular gas that is expected close to a YMSC as a result of the fragmentation of the swept-up shell. See \S~\ref{subsec:MCIonization}, for a review on how to calculate the ionization rate from a population of CRs and how to assess its value from observations based on the detection of specific molecular lines.

%% file: CHAPTERS/chapter_2.tex
%---------------------------------------------------------
\chapter{The scientific case of Cygnus OB2}
\label{ch:CygOB2}
%---------------------------------------------------------
\lettrine{T}{he} understanding of YMSCs as particle accelerators must necessarily pass, to a large extent through $\gamma$-ray observations, as their investigation through direct detection of CRs is severely limited by the diffusive behavior of the accelerated particles, as is the case for all galactic sources. Currently, a handful of YMSCs have been found in coincidence with extended $\gamma$-ray emission, including Cygnus OB2.

Cygnus OB2 represents, perhaps, one of the most intriguing cases for 3 main reasons. First, the detected diffuse $\gamma$-ray emission has been widely investigated from both the spectral and morphological point of view in both the high-energy ($1 < E_\gamma < 100$ GeV) and very-high-energy bands ($0.1 < E_\gamma < 100$ TeV), a fact that allows a robust modelization of the underlying CR distribution, potentially leading to an exhaustive knowledge of the acceleration properties. Secondly, the stellar population of Cygnus OB2 has been extensively studied over the decades, paving the ground for a realistic estimate of fundamental stellar cluster parameters such as the total wind luminosity and the mass loss rate. Last but not least, the recent detection by the LHAASO experiment \citep{Cao_LHAASO12PeVatrons_2021} of a 1.4 PeV $\gamma$-ray in coincidence with the star cluster could indicate the presence of CRs with energies of at least 10 PeV, marking Cygnus OB2 an excellent PeVatron candidate.  

In this chapter, we aim to interpret the observed $\gamma$-ray emission assuming an underlying distribution of CRs described by the model of particle acceleration at the cluster wind TS developed by \cite{Morlino_2021} summarized in \S~\ref{subsec:WindTSAccModel}. The information given by the combination of the morphological and spectral shapes of the extended $\gamma$ emission can be used to constrain, at some level, the propagation mechanism of CRs. This is particularly critical as CR propagation close to the acceleration site is directly related to the type of plasma turbulence in the system, which in turn affects the maximum particle energy achievable by the accelerator.

The chapter is structured as follows: in the first part, we will review the main properties of the YMSC Cygnus OB2, and we will estimate the mass loss rate and wind luminosity considering the population of stars in the cluster. In the second part, knowing the properties of the star cluster, we comment on the morphological properties of the freshly accelerated particle distribution. We furthermore describe how to evaluate the $\gamma$-ray flux under the assumption of pure hadronic emission. In the last part, we compare the spectro-morphological properties of the expected $\gamma$-ray emission with available observations from different experiments. Finally, we comment on the obtained results, discussing on the model validity and its limitations.

%---------------------------------------------------------
\section{The young massive star cluster Cygnus OB2}
\label{sec:CygOB2cluster}

Cygnus OB2 (Cyg OB2) is one of the most massive and compact OB associations in the Milky Way, located towards the center of the Cygnus-X star-forming complex (l$\approx$80.22$^\circ$, b$\approx$0.79$^\circ$), an extended ($\sim$10$^\circ$) radio structure hosting numerous molecular clouds \citep{Schneider_CygnusXSFR_2006}, HII regions \citep{Dickel_CygusHII_1969} and several other OB associations \citep{Uyaniker_CygSuperBubble_2001}. Cyg OB2 harbor hundreds, possibly thousands of massive stars. The first study of its population has been carried out by \cite{Reddish_CygOB2_1966}, who inferred with large uncertainties a total of 400--3000 OB stars, based on star counts on the Palomar Sky Survey plates. Similarly, \cite{Knodlseder_CygOB2_2000} found a compatible result using star counts in the near-infrared, estimating a total population of 2600$\pm$400 OB stars, with 120$\pm$20 being O-type stars. However, because of the problematic background subtraction and highly patchy extinction pattern towards the association, the amount of stars in Cyg OB2 is possibly lower, as noted by \cite{Wright_CygOB2SFH_2010}, who estimated a total star content of $\sim$1200 OB stars, with $\sim$75 O-type stars. The radial stellar distribution from observations seems to follow a compact and peaked profile, with a high stellar density in the core of the association, similar to the YMSCs observed in the Large Magellanic Cloud \citep{Knodlseder_CygOB2_2000}. Due to the peaked morphology of Cyg OB2, a large fraction of the stars is enclosed in the central part of the association, in a region with a radius of $\sim$14 pc. A recent census of this central core has revealed the presence of 169 OB stars, of which 52 are O-type, and 3 are Wolf-Rayet stars \citep{Wright_MassiveStarPopCygOB2_2015}. 

Several estimations of the age of Cyg OB2 have been made through the years. The presence of O-type dwarf stars and high-luminosity blue supergiants in the sample of 85 OB stars selected by \cite{Hanson_CygOB2SuperStarCluster_2003} suggests that Cyg OB2 should not be older than a few Myr, with a likely value of 2$\pm1$ Myr. An investigation of the population of A-type stars in Cyg OB2 indicated the presence of a group of 5--7 Myr old stars, located mainly in the southern part of the association \citep{Drew_CygOB2AStars_2008}. In parallel, X-ray analysis of low-mass stars seems to point to an age of 3--5 Myr \citep{Wright_CygOB2SFH_2010}. \cite{Wright_MassiveStarPopCygOB2_2015} found a typical age of 2--3 Myr and 4--5 Myr by applying respectively non-rotating and rotating evolutionary stellar models to the selected sample of 169 OB stars. The latter results seem to agree with a scenario describing an overall continuous star formation activity, starting $\sim$7 Myr ago and going on until $\sim$1 Myr ago, with a possible peak of star formation around 4--5 Myr. This is compatible with the results found by \cite{Comeron_NewCygOB2Members_2012}, that were also suggestive of a continuous star-forming activity in the region for the last 10 Myr. 

The distance of Cyg OB2 is a subject still under debate in the community. Right after the discovery of the association, \cite{Johnson_ObscuredOBAssociation_1954} measured the distance of Cyg OB2 using spectroscopic observations of 11 stars, finding a value of $\sim$1500 pc. In the first comprehensive investigation of the Cyg OB2 population, \cite{Reddish_CygOB2_1966} estimated a distance of 2100 pc. Independent studies in the early 90s based on the method of spectroscopic parallax resulted in a distance of $\sim$1700 pc \citep{TorresDodgen_PhotometryCygOB2_1991, Massey_MassiveStarsCYGOB2_1991}. Perhaps the most commonly adopted value, at present, is the one measured by \cite{Hanson_CygOB2SuperStarCluster_2003}, who inferred a distance of 1400$\pm$80 pc after analyzing the absolute magnitude and extinction of 14 OB stars. This measure is reasonably compatible with the position of some molecular clouds in the Cygnus-X region, whose distance has been calculated using maser parallaxes \citep{Rygl_MaserParallCygX_2012}. Moreover, this value is also in agreement with the results of a recent work exploiting parallax based distances calculated using eclipsing binaries, from which a distance of 1330 $\pm$ 60 pc was evaluated \citep{Kiminki_GAIAParCygOB2_2015}. Finally, a comprehensive study using parallaxes from the second data release by Gaia seems to point out that the association is actually composed of two main subgroups, the first located at a distance of $\sim$1350 pc and the second at $\sim$1755 pc \citep{Berlanas_DisentanglingSubstructCygOB2_2019}.

\subsection{Cygnus OB2 wind luminosity and mass loss rate}
\label{subsec:CygOB2LwMdot}
As we will detail later in this chapter, two fundamental parameters regulating multiple aspects of the CR distribution properties are the cluster mass loss rate and the cluster wind luminosity. In order to calculate these parameters for Cyg OB2, we need to compute the mass loss rate $\dot{M}_i$ for every i-th member of Cyg OB2. We then consider the stars at the core of the association belonging to the sample studied by \cite{Wright_MassiveStarPopCygOB2_2015}. We use two different recipes to calculate $\dot{M}_i$. The first one is a theoretical formula given by \cite{Yungelson_EvolutionFateMassiveStars_2008}
\begin{equation}
    \dot{M}_i=\frac{L_{i}}{v_{\infty, i} c} \frac{1}{(1-\Gamma)^{(\alpha-0.5)}}
\end{equation}
where $L_i$ is the bolometric stellar luminosity of the i-th star of the sample, $c$ is the speed of light, $\Gamma=L_i/L_{\rm Edd}$ with $L_{\rm Edd}$ the Eddington luminosity, $\alpha$=0.25 \citep{Yungelson_EvolutionFateMassiveStars_2008}, and $v_{\infty,i}$ is the wind terminal velocity of the i-th star. The latter is defined as \citep{Kudritzki_WindsHotStars_2000}:
\begin{equation}
\label{eq:Vinfty}
v_{\infty,i}=C(T_{\rm eff})\left [\frac{2 G M_i (1-\Gamma)}{R_{\star, i}} \right ]^{1/2}
\end{equation}
with $R_{\star, i}=0.85 (M_i/M_\odot)^{0.67} R_\odot$ \citep{Demircan_StarsMRR_1991} the stellar radius, G the gravitational constant, $M_i$ the stellar mass, and $C(T_{\rm eff})$ a parameter that depends on the star temperature $T_{\rm eff}$, which is $C(T_{\rm eff})=2.65$ for $T_{eff}>21000$ K \citep{Kudritzki_WindsHotStars_2000}.

The second equation we use is an empirical relation valid for stars with temperature\footnote{Note that this condition is respected for all stars in the sample of \cite{Wright_MassiveStarPopCygOB2_2015}, whose masses are greater than 20 $M_\odot$} 27500 K$<T^i_{\rm eff}<50000$ K given by \cite{Vink_OBwind_2000}:
\begin{equation}
\label{eq:MdotRenzo}
    \begin{split}
        \textrm{log}_{10} \left (\frac{\dot{M}_i}{M_\odot yr^{-1}} \right)=-6.668(80)+2.210(31)  \textrm{log}_{10} \left (\frac{L_i}{10^5 \ L_\odot} \right ) \\
        -1.339(68) \textrm{log}_{10} \left (\frac{M_i}{30 \ M_\odot} \right )-1.601(55) \textrm{log}_{10} \left (\frac{v_{\infty, i}}{2 v_{ esc}} \right ) \\
        +1.07(10)  \textrm{log}_{10} \left (\frac{T_{\rm eff, i}}{40000  \ \textrm{K}} \right )+0.85(10) \textrm{log}_{10} \left (\frac{Z_i}{Z_{\odot}} \right )
    \end{split}
\end{equation}
where $Z_i$ is the stellar metallicity and $v_{\rm esc}=\sqrt{2GM_i (1-\Gamma)/R_{\star,i}}$ is the escape velocity. When using equation~\ref{eq:MdotRenzo} we will assume solar metallicity. For every star in their sample, \cite{Wright_MassiveStarPopCygOB2_2015} provide an estimation of the stellar parameters $\tilde{L}_i$, $\tilde{T}_{\rm eff, i}$, and $\tilde{M}_i$ (we will refer to the quantities estimated by \cite{Wright_MassiveStarPopCygOB2_2015} using the $\sim$ diacritic symbol) together with the associated parameter uncertainty ($\delta \tilde{L}_i$, $\delta \tilde{T}_{\rm eff, i}$, and $\delta \tilde{M}_i$).

\begin{figure}
     \centering
     \begin{subfigure}{0.7\textwidth}
         \centering
         \includegraphics[width=\textwidth]{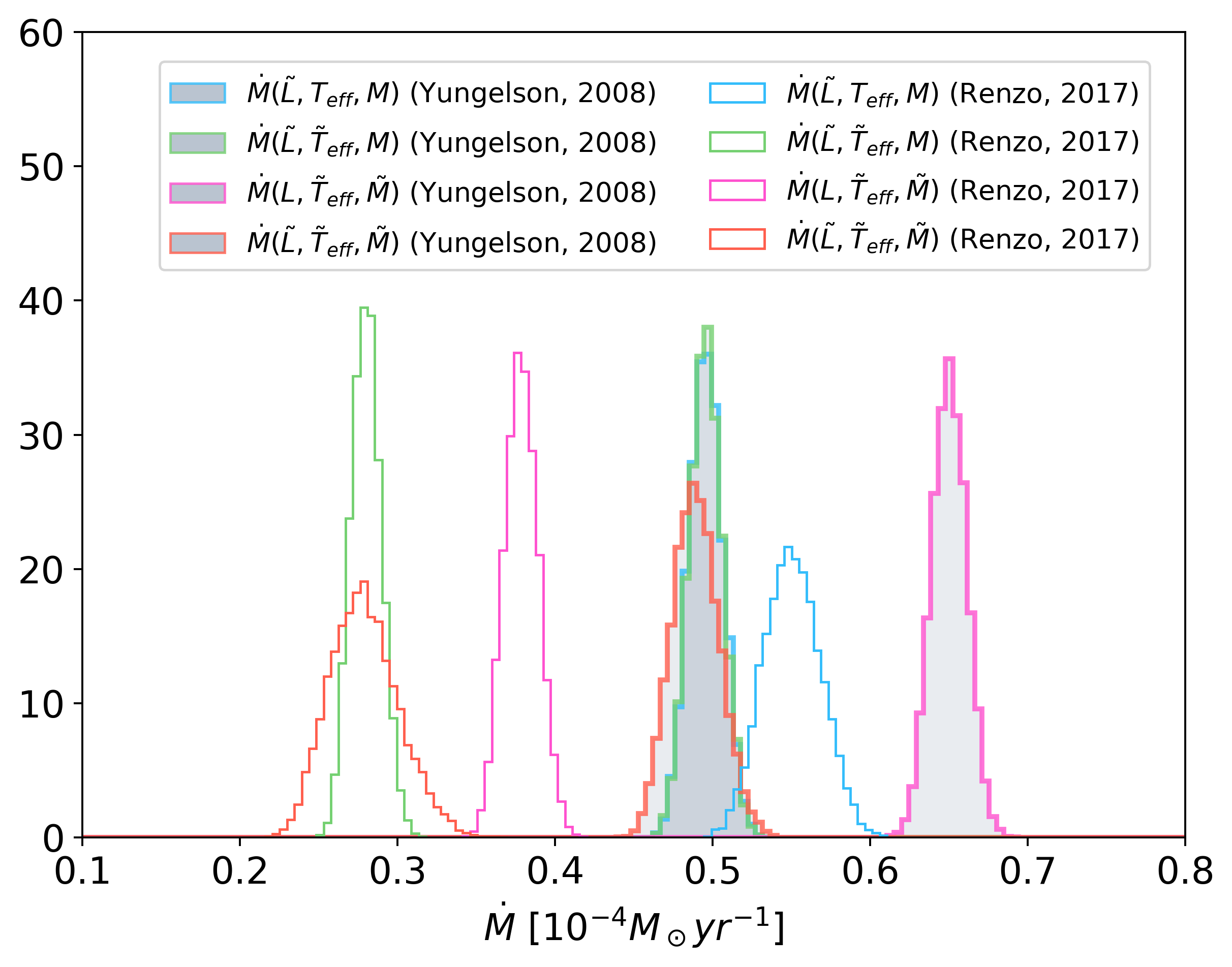}
         \caption{}
         \label{fig:MdotCygOB2}
     \end{subfigure}
     %\hfill \\
     \begin{subfigure}{0.7\textwidth}
         \centering
         \includegraphics[width=\textwidth]{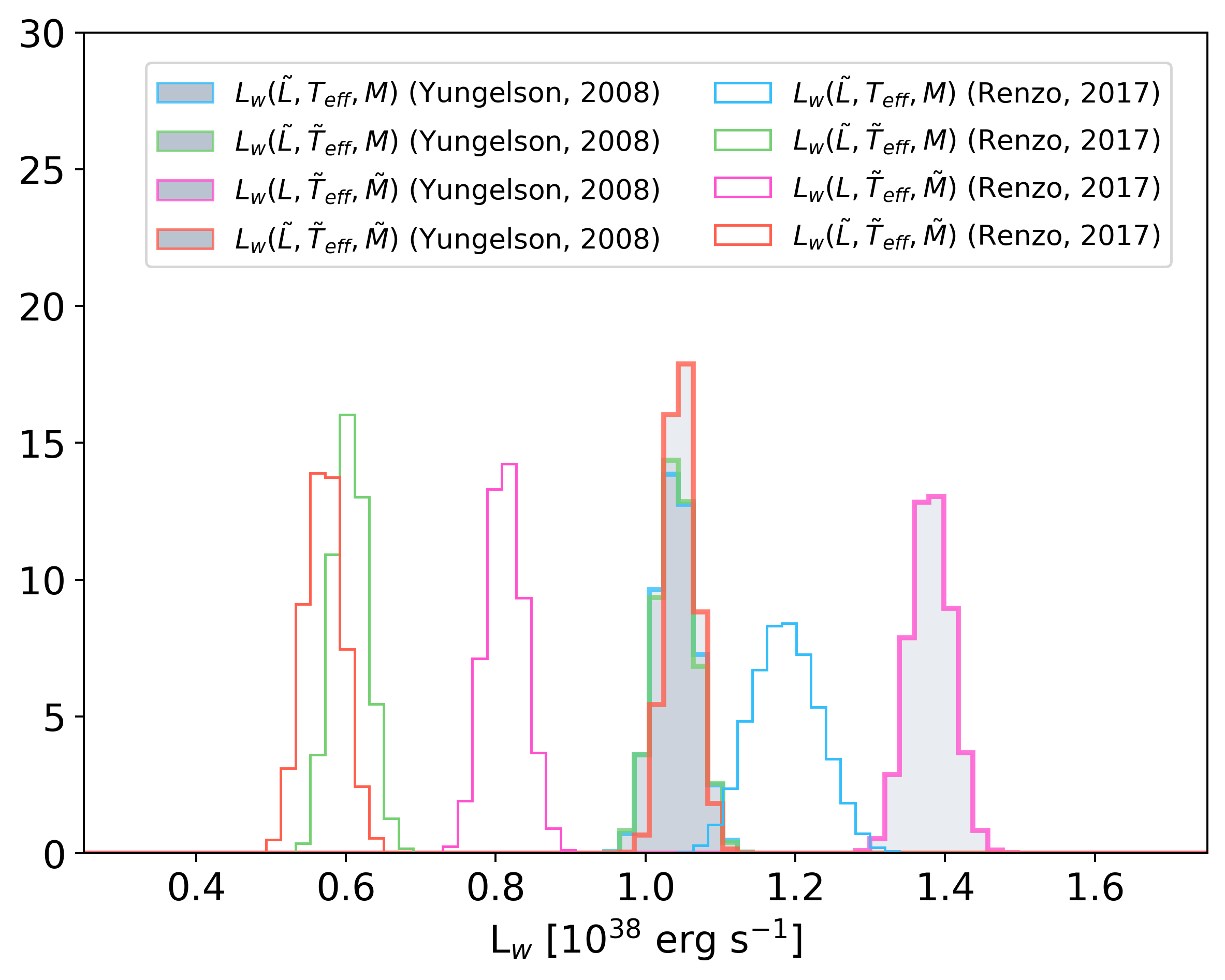}
         \caption{}
         \label{fig:LwCygOB2}
     \end{subfigure}
        \caption{(a) Mass loss rate distribution for Cygnus OB2 considering the sample of stars investigated by \cite{Wright_MassiveStarPopCygOB2_2015}. The mass loss rate is estimated using two different recipes, a theoretical one \citep{Yungelson_EvolutionFateMassiveStars_2008} and an empirical one \citep{Vink_OBwind_2000}, using diverse stars measured quantities (see text). The plot only shows the contribution from single star systems and does not include WR stars. (b) Wind luminosity of Cygnus OB2 inferred using the estimated mass loss rates.}
        \label{fig:YMSC_EvolutionStages}
\end{figure}

We compute four diverse estimation of $\dot{M}$, using different combinations of the measured parameters $\tilde{L}_i$, $\tilde{T}_{\rm eff, i}$, and $\tilde{M}_i$. More precisely,
\begin{itemize}
    \item as a first trial, we calculate the total mass loss rate using all the measured parameters.
    \item as a second trial, we use only $\tilde{L}_i$, while $T_{\rm eff, i}$ and $M_i$ are calculated using respectively the Stefan-Boltzmann law and the inverted Luminosity-Mass scaling relation given by \cite{Yungelson_EvolutionFateMassiveStars_2008}:
        \begin{equation}
            T_{\rm eff, i}=\left (\frac{\tilde{L}_i}{4 \pi R_{\star,i}^2 \sigma_b} \right )^{1/4}
        \end{equation}
        \begin{equation}
        \label{eq:StellarMFromL}
            \frac{M_i}{M_\odot}=\left [ 10^{-3.48} \left( \frac{\tilde{L}_i}{L_\odot} \right ) \right ]^{0.75} \ 
        \end{equation}
        where $\sigma_b$ is the Stefan-Boltzmann constant.
    \item as third trial, we use  $\tilde{L}_i$ and $\tilde{T}_{\rm eff, i}$, while $M_i$ is obtained using equation~\ref{eq:StellarMFromL}.
    \item lastly, we use $\tilde{T}_{\rm eff, i}$ and $\tilde{M}_{i}$ while $L_i$ is calculated inverting equation~\ref{eq:StellarMFromL} \citep{Yungelson_EvolutionFateMassiveStars_2008}.
\end{itemize}
%Finally, for every trial, we also take into account the uncertainties in the estimation of  $\tilde{L}_i$, $\tilde{T}_{\rm eff, i}$, and $\tilde{M}_i$. We do so by using a Montecarlo approach, generating $10^4$ different star samples for which the used measured parameters ($\tilde{L}_i$, $\tilde{T}_{\rm eff, i}$, and $\tilde{M}_i$) are gaussian-fluctuated by a quantity corresponding to the associated error ($\delta \tilde{L}_i$, $\delta \tilde{T}_{\rm eff, i}$, and $\delta \tilde{M}_i$).
In each trial, we account for the uncertainties in the measuraments of $\tilde{L}_i$, $\tilde{T}_{\rm eff, i}$, and $\tilde{M}_i$ by utilizing a Montecarlo method. This involves creating $10^4$ different samples of stars where the measured parameters ($\tilde{L}_i$, $\tilde{T}{\rm eff, i}$, and $\tilde{M}_i$) are randomly fluctuated following a Gaussian distribution with a width equal to the associated error ($\delta \tilde{L}_i$, $\delta \tilde{T}_{\rm eff, i}$, and $\delta \tilde{M}_i$). Fig.~\ref{fig:MdotCygOB2} shows the result for the $\dot{M}$ calculation after considering only single stars (no binary systems) with $M_i>20 M_\odot$ and without the contribution of Wolf-Rayet stars. The value of $\dot{M}$ lies between $\sim 0.2\--0.7 \times 10^{-4}$ M$_\odot$ yr$^{-1}$. To account for the contribution of the three Wolf-Rayet (WR) stars in the sample, we consider an ad-hoc empirical relation reported by \cite{Renzo_SystemSurveyMdot_2017}: 
\begin{equation}
%\label{eq:MdotRenzo}
    \begin{split}
        \textrm{log}_{10} \left (\frac{\dot{M}_{\textrm{WR},\ i}}{M_\odot yr^{-1}} \right)=-11.0+1.29(14)  \textrm{log}_{10} \left (\frac{L_{\textrm{WR}, \ i}}{10^5 \ L_\odot} \right ) \\
        +1.73(42)  \textrm{log}_{10} \left (\frac{Y_{\textrm{WR}, \ i}}{Y_\odot} \right )+0.47(09) \textrm{log}_{10} \left (\frac{Z_{\textrm{WR}, \ i}}{Z_{\odot}} \right ) \ .
    \end{split}
\end{equation}
Adopting for $Y_{\textrm{WR},i}$ (helium fraction) and $Z_{\textrm{WR},i}$ (metallicity) the solar values, we find a contribution to the mass loss rate from WR stars of $\dot{M}_\textrm{WR}(\tilde{L}_{\textrm{WR}, \ i})\sim 0.3 \times 10^{-4}$ M$_\odot$yr$^{-1}$. 

The star sample of \cite{Wright_MassiveStarPopCygOB2_2015} includes also 9 known binary systems, whose contribution to $\dot{M}$ is not straightforward to quantify. However, we can make a rough estimation assuming that the observed luminosity $\tilde{L}_i$ of the system is equally distributed between the two companions. By doing so, and by following the same approach as for single star systems for the calculation of $\dot{M}_i$ using only $\tilde{L}_i$, we find $\dot{M}_{\textrm{Binary}}\sim 0.22 \-- 0.3 \times 10^{-4}$ M$_\odot$yr$^{-1}$. By summing up all the contributions, Cyg OB2 mass loss rate should lie in a conservative range of $\dot{M}\simeq 0.7 \-- 1.3 \times 10^{-4}$ M$_\odot$yr$^{-1}$.
Once $\dot{M}_i$ and $v_{\infty,i}$ are known, it is then possible then to estimate the wind luminosity ($L_i^w$) for each star as:
\begin{equation}
\label{eq:LwindStar}
L_i^w = \frac{1}{2} \dot{M_i} v_{\infty, i}^2 .
\end{equation}
Consequently, the cluster wind luminosity is easily found as:
\begin{equation}
\label{eq:LwindCluster}
L_w=\sum_i L_i^w 
\end{equation}
The contribution of single, non Wolf-Rayet stars, accounting for both the values of $\dot{M}_i$ inferred using the theoretical and empirical recipes of \cite{Yungelson_EvolutionFateMassiveStars_2008} and \cite{Renzo_SystemSurveyMdot_2017}, ranges in the range $L_w^{NBS} \simeq 0.4 \-- 1.6 \times 10^{38}$ erg s$^{-1}$ (see Fig.~\ref{fig:LwCygOB2}). In addition to this, we also estimate the wind luminosity contribution from the three Wolf-Rayet stars included in the sample of \cite{Wright_MassiveStarPopCygOB2_2015} assuming an average wind speed of 2500 km s$^{-1}$: this turns out to be $L_w^{\rm WR} \simeq 0.6 \times 10^{38}$ erg s$^{-1}$. Finally, the contribution of the 9 binary systems using the previously computed mass loss rate, is $L_w^{\rm BS} \simeq 0.55 \-- 0.7  \times 10^{38}$ erg s$^{-1}$. Accounting for all contributions, Cyg OB2 wind luminosity should lie between $L_w \simeq 1.55 \-- 2.9 \times 10^{38}$ erg s$^{-1}$. It is worth noticing that the obtained value of $L_w$ is compatible with other estimations done by different authors. For example, \cite{Ackermann_FermiCygnusCocoon_2011} calculate a wind luminosity of $L_w \simeq 2 \-- 3 \times 10^{38}$ erg/s considering a different sample of stars that include the presence of 17 stars with $M_\star > 35 M_\odot$ and 5 Wolf-Rayet stars.

One last point worth mentioning is that the values obtained for the theoretical and empirical prescription for $\dot{M}$ are approximately in good agreement within a factor of a few, with the trial based on the measured bolometric luminosity becoming the one with the better agreement. There are several possible recipes given in the literature \citep{Renzo_SystemSurveyMdot_2017}, and a discrepancy of a factor 2--3 is reasonably expected.

\subsection{The wind blown bubble of Cygnus OB2}
As a young (<7 Myr) and massive stellar cluster, it is reasonable to expect Cyg OB2 to be surrounded by a wind-blown bubble. Over the years, several attempts were made to find this structure: one of the first suggestions was to identify the Cyg OB2 wind bubble with the extended X-ray source known as Cygnus Superbubble \citep{Cash_CygSupBubbXray_1980, Uyaniker_CygSuperBubble_2001}. However, this hypothesis encountered several problems, such as the fact that Cyg OB2 is significantly offset from the geometrical center of the superbubble. Moreover, various portions of the superbubble itself appeared to be uncorrelated, prompting the idea that the Cygnus Superbubble is actually a combination of several structures along the line of sight \citep{Uyaniker_CygSuperBubble_2001}. Another attempt to search for the wind bubble was performed through direct search of the cold shell of dense swept-up material surrounding the expanding hot shocked gas \citep{Lozinskaya_CygOB2Shell_2002}. Unfortunately, the unlucky position of Cyg OB2 at Galactic longitudes close to 90$^\circ$ prevents a robust determination of the gas location by using kinematic distances. The detection of the Cyg OB2 bubble still remains nowadays a matter of debate.

\begin{landscape}
\begin{figure}
\begin{center}
\includegraphics[width=1.5\textwidth]{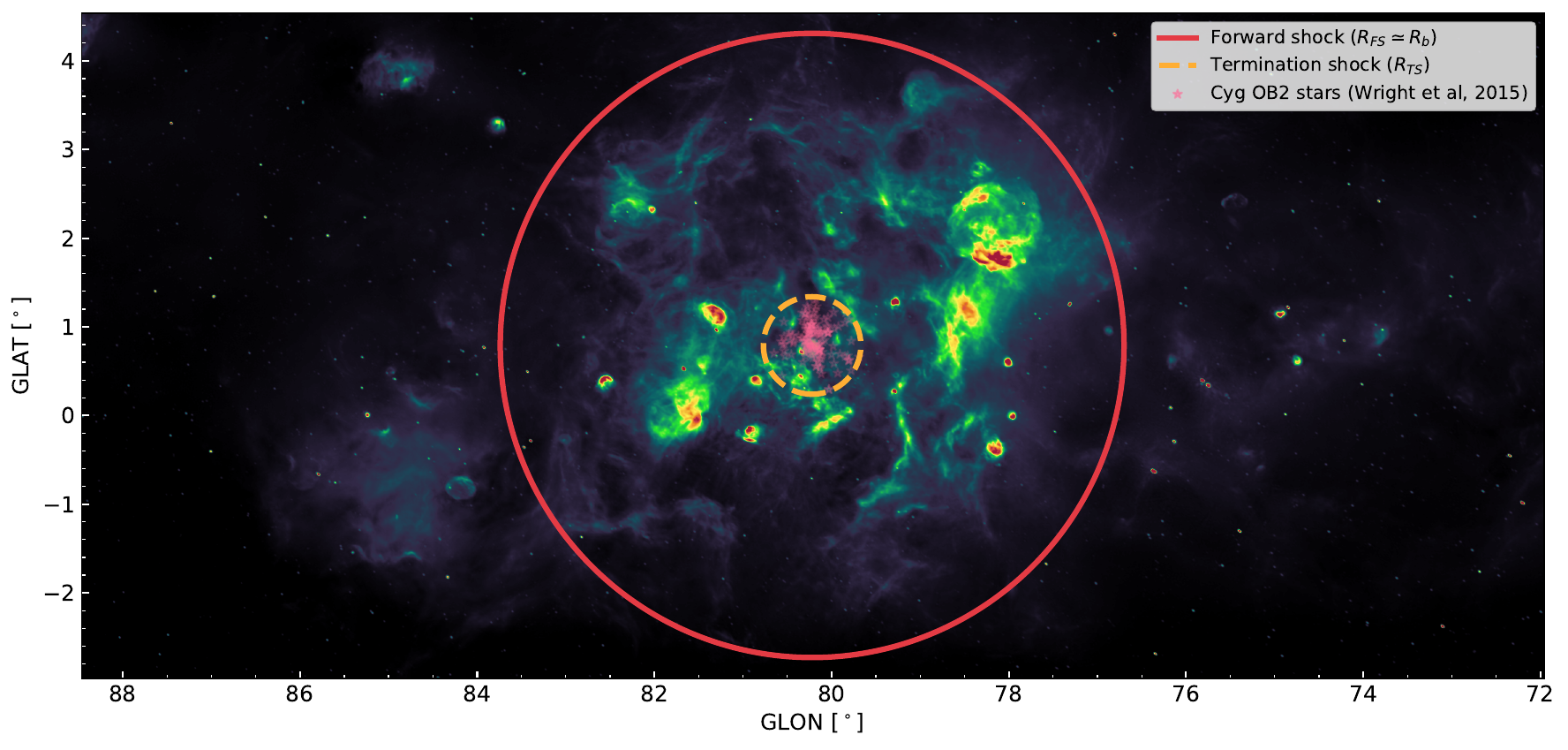}
\caption{Continuum observations at 1.42 GHz from the Canadian Galactic Plane Survey \citep{Taylor_CGPS_2003}. A large fraction of this emission is generated by thermal free-free radiation of material possibly ionized by Cyg OB2. Data points indicates the position of Cyg OB2 stars taken from the sample of \cite{Wright_MassiveStarPopCygOB2_2015}. Continuous and dashed circles represent the position of the forward and TS respectively. Color scales are given in arbitrary units.}
\label{fig:CygX_21cmContinuum}
\end{center}
\end{figure}
\end{landscape}

For the sake of curiosity, we can try to estimate the expected dimension of the wind bubble considering the theory of \cite{Weaver_1977}, introduced in \S~\ref{subsec:WindBubble}. If we assume the following reasonable values for the parameters of Cyg OB2: L$_w=2 \times 10^{38}$ erg s$^{-1}$ (compatible to what has been calculated in \S~\ref{subsec:CygOB2LwMdot}), cluster age of 3 Myr, and $\rho_0=20 m_p$ cm$^{-3}$, the size of the forward shock position tracing the location of the swept-up shell of dense material is (Eq.~\ref{eq:Rbubble}) $R_b\approx 86$ pc. Considering a distance of 1.4 kpc, the projected size of the forward shock is of the order of a few degrees in radius. Interestingly, this is slightly larger, but in good agreement with the dimension of the diffuse, filamentary, 21 cm continuum radio emission of the Cygnus-X star-forming region (Fig.~\ref{fig:CygX_21cmContinuum}). 

Knowing that a significant fraction of the continuum emission is of thermal nature \citep{Xu_ThermalEmissionCygX_2013}, the overall filamentary structure observed at 21 cm could be tracing what is left of the former (now fragmented) shell of dense material. Thermal emission could then be produced in this scenario by the ionized HII region trapped within the fragmented dense shell.

\begin{figure}
\begin{center}
\includegraphics[width=0.7\textwidth]{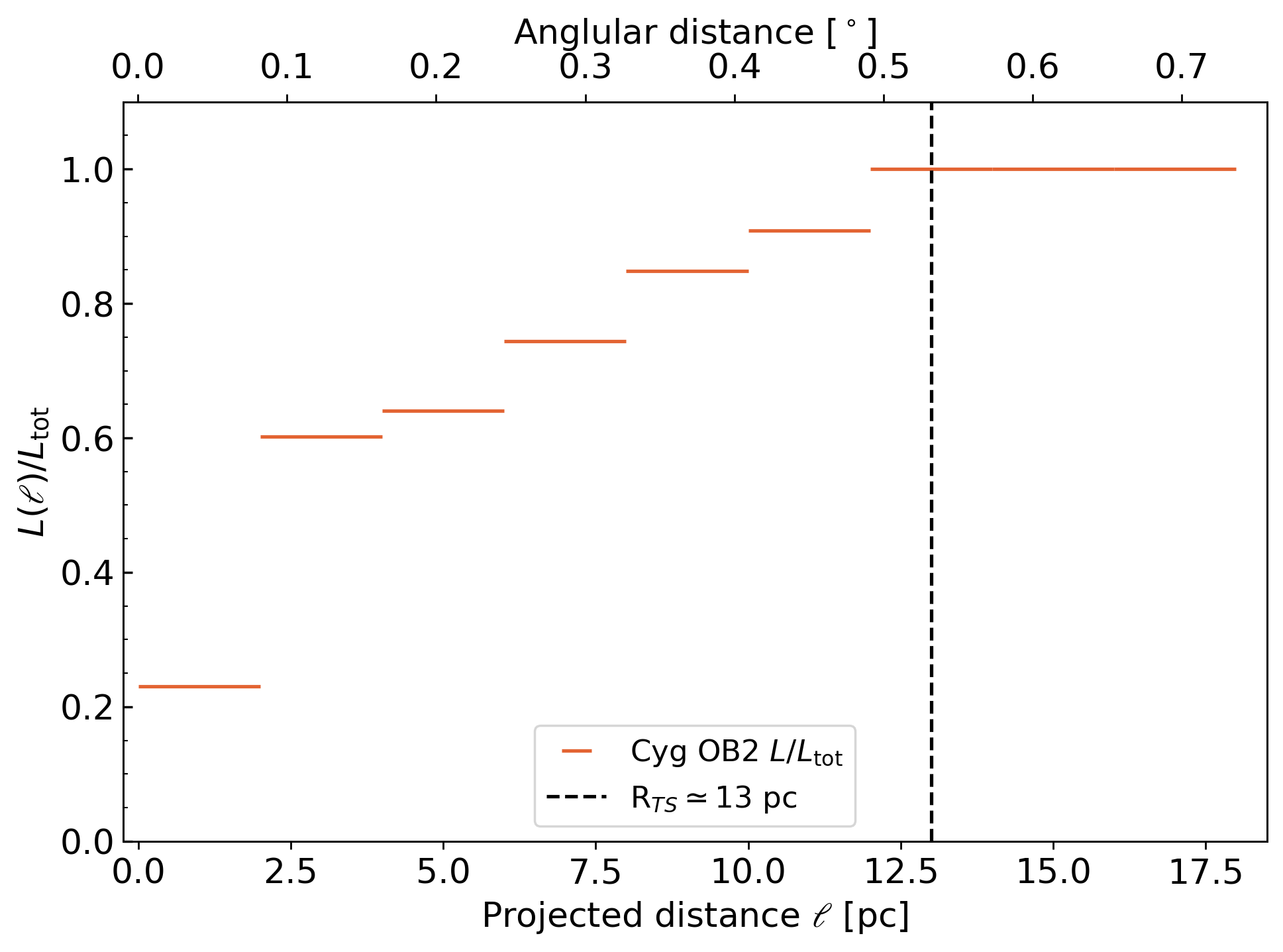}
\caption{Cumulative radial distribution of Cyg OB2 stellar luminosity calculated using the star sample of \cite{Wright_MassiveStarPopCygOB2_2015}. }
\label{fig:CygOB2LumCumDistrib}
\end{center}
\end{figure}

Currently, no evidence for a collective wind TS has ever been searched for. It is therefore fair to ask whether Cyg OB2 fulfills the conditions to develop a collective wind. Using the same parameters for the estimation of R$_b$, and considering a star cluster mass loss rate $\dot{M}=10^{-4}$ M$_\odot$ yr$^{-1}$, the TS shock radius can be calculated using Eq.~\ref{eq:Rts}, from which we obtain R$_{TS} \approx$13 pc. This value is similar to the overall size of the Cyg OB2 association core studied by \cite{Wright_MassiveStarPopCygOB2_2015}. The half-mass radius is, however, smaller by a factor $\sim 0.4$ (R$_{hm}\approx 5.2$ pc) \citep{Pfalzner_UniYMSCSeq_2009}, which is compatible with the half-luminosity radius R$_{hl}\approx 2.5$ pc obtained from the sample of \cite{Wright_MassiveStarPopCygOB2_2015} (see Fig.~\ref{fig:CygOB2LumCumDistrib}). Since the most massive stars are also the brightest, R$_{hl}$ should roughly delimit the region containing the most massive stars, i.e., those that contribute most to the creation of the collective cluster wind. Reasonably assuming that the average distance between the stars is a fraction of R$_{hl}$, the hypothesis that a collective wind exists becomes rather concrete (see \S~\ref{subsec:WindBubble}).
%---------------------------------------------------------

%---------------------------------------------------------
\section{The distribution of CRs in Cygnus OB2}
\label{sec:CygOB2fCR}
Given the compactness of Cyg OB2 and the possibility of developing a wind TS, it is likely that the acceleration mechanism in this system, and the resulting distribution of CRs ($f_{CR}$), is the one described in \S~\ref{subsec:WindTSAccModel}, appropriate for particles acceleration at the wind TS. In such a scenario, Eq.~\ref{eq:CRdistribFunc} describes the distribution of CRs. Note that $f_{CR}$ is composed of a contribution of two populations, the freshly accelerated particles escaping from the acceleration site and the 
population of Galactic CRs that may end up penetrating in the system. Let us now rewrite Eq.~\ref{eq:CRdistribFunc} accounting only for the contribution of freshly accelerated CRs, i. e., considering $f_{gal}=0$:
\begin{equation}
\label{eq:fCRNoSea}
f_{CR}(r, p) = 
\begin{cases}
f_{TS}(p) \cdot exp \left [ -\int_r^{R_{TS}} \frac{u_1}{D_1(r',p)}  dr' \right ] & \text{for } r \leq R_{TS}  \\
f_{TS}(p) e^{\alpha} \frac{1+\beta(e^{\alpha_B-\alpha}-1)}{1+ \beta(e^{\alpha_B}-1)} & \text{for } R_{TS} \leq r \leq R_{b} \\
f_{TS}(p) \frac{e^{\alpha_B}}{1+ \beta(e^{\alpha_B}-1)} \frac{R_b}{r} & \text{for } r \geq R_b
\end{cases}
\end{equation}
where $\alpha$, $\alpha_B$ and $\beta$ are defined in Eq.~\ref{eq:AlphaBetaCRpar}, and $f_{TS}$ is given by Eq.~\ref{eq:fTS}. 

\begin{figure}
\begin{center}
\includegraphics[width=0.6\textwidth]{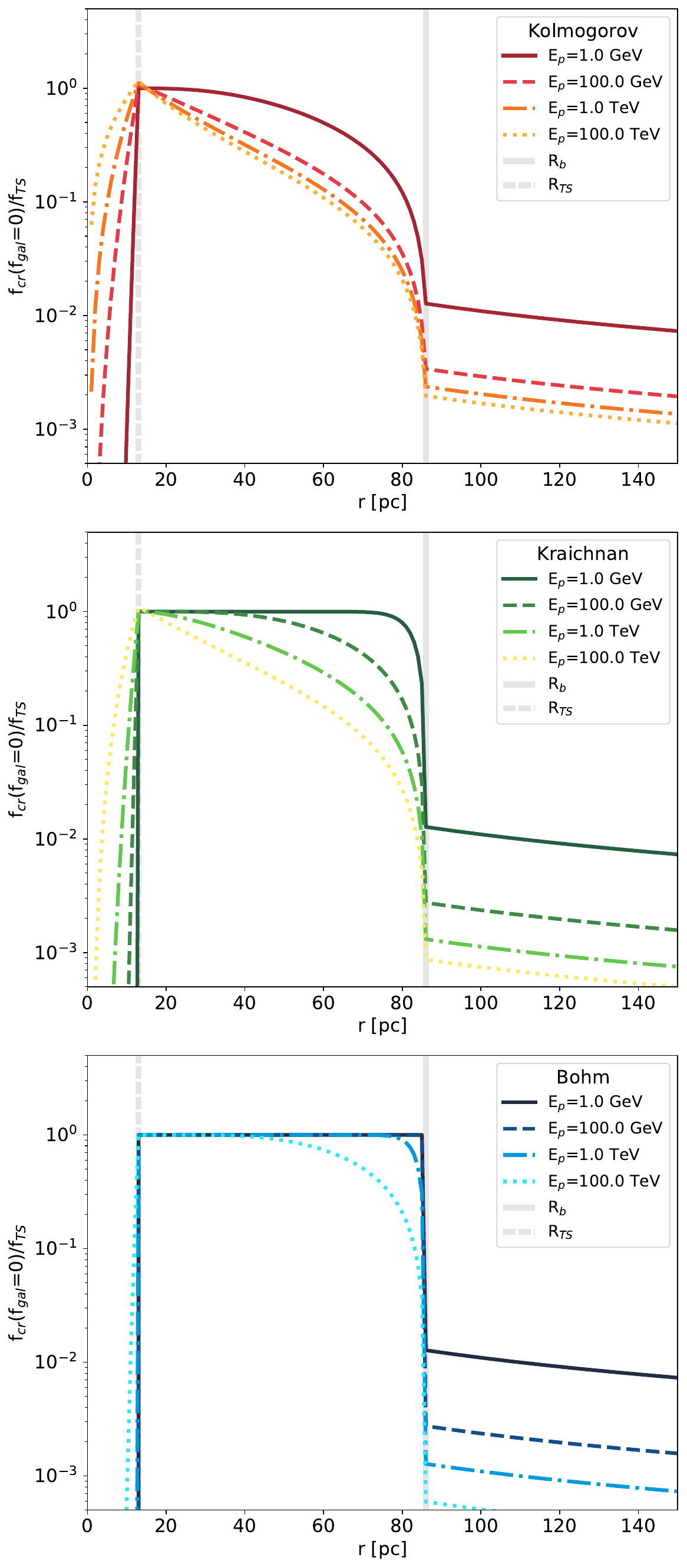}
\caption{Radial CR distribution normalized to its value at the termination shock in the case of Kolmogorov (top panel), Kraichnan (central panel), and Bohm (bottom panel) turbulence. In these plots, the contribution of the galactic CR sea to $f_{CR}$ has not been considered.}
\label{fig:CygOB2fCR}
\end{center}
\end{figure}

\begin{figure}
\begin{center}
\includegraphics[width=0.7\textwidth]{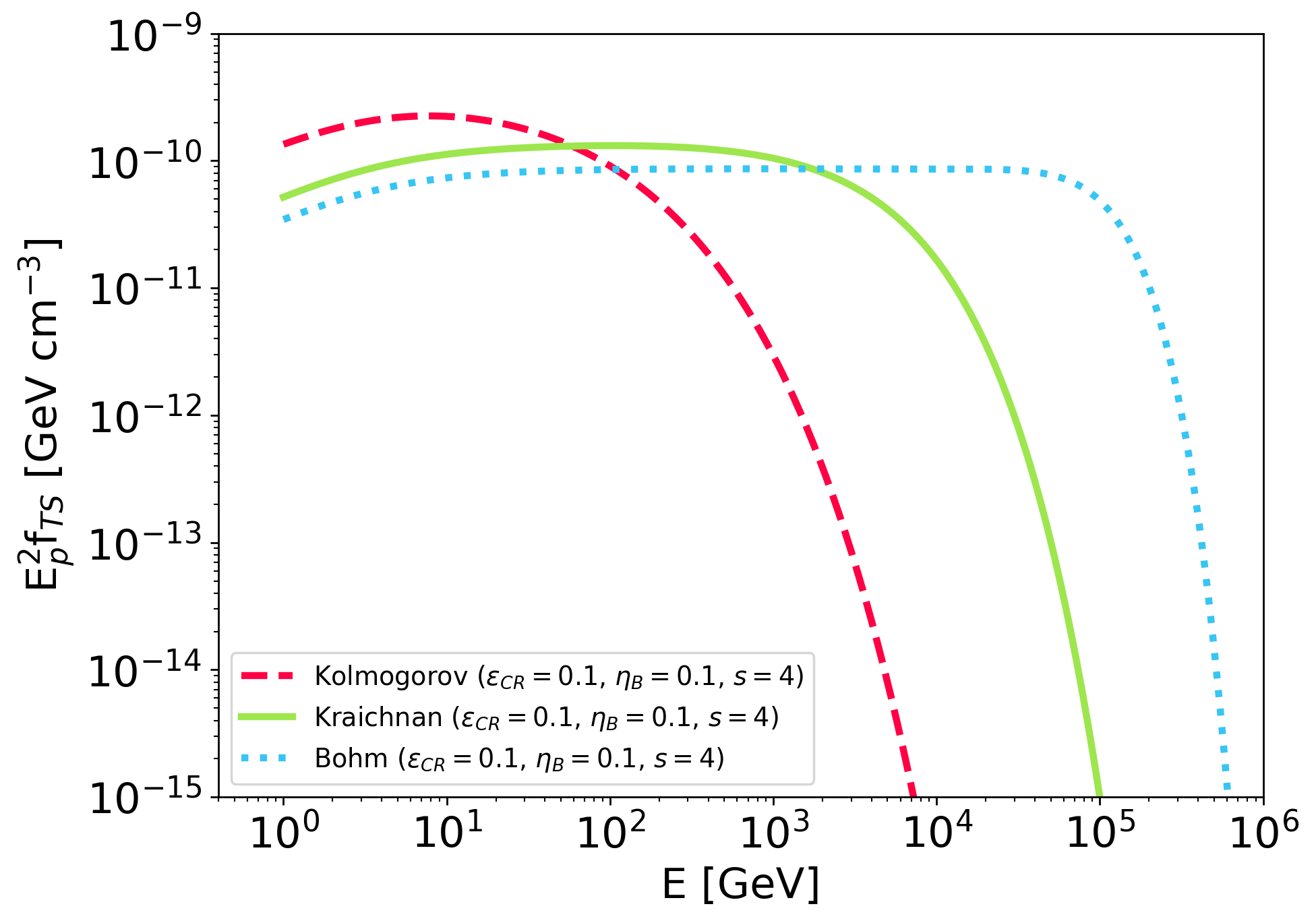}
\caption{Injected particles spectrum at the TS with same cluster and turbulence injection properties but with different diffusion coefficient in the system.}
\label{fig:CygOB2fTS}
\end{center}
\end{figure}

Clearly, both the spectral and morphological characteristics of $f_{CR}$ depend more or less directly on the properties of Cyg OB2, but also on the type of diffusion in the system, which is related to the (unknown) turbulence spectrum in the bubble. We therefore proceed to evaluate Eq.~\ref{eq:fCRNoSea} in the specific case of Cyg OB2, considering three different cases of turbulence spectrum: Kolmogorov-like, Kraichnan-like and flat spectrum. The following values are used for Cyg OB2: $L_w=2 \times 10^{38}$ erg s$^{-1}$, $\dot{M}=10^{-4}$ M$_\odot$ yr$^{-1}$, cluster age of 3 Myr, and $\rho_0=20 m_p$ cm$^{-3}$ (related to the mean density of the GMC from which Cyg OB2 was formed, unknown in fact and totally hypothetical). We model the plasma turbulence in the three cases considering that the total power in magnetic turbulence is a fraction $\eta_B=0.1$ of $L_w$, and that the power injection scale of the turbulence is $L_{inj} = 2$ pc, as it is reasonable to assume it to be of the order of the cluster size $R_{hl}$. For the Bohm case, we also assume that the turbulence is injected down to scales of $k_{min}=10^{-5}$ pc$^{-1}$. Finally, we additionally assume that a fraction $\epsilon_{CR}=0.1$ of $L_w$ goes into acceleration of CRs, and that the slope of the injected particle spectrum at the TS is $s=4$. The results obtained evaluating Eq.~\ref{eq:fCRNoSea} under these assumptions are shown in Fig.~\ref{fig:CygOB2fCR} and Fig.~\ref{fig:CygOB2fTS}, which show respectively the radial shape of $f_{CR}$ at different energies and the spectra of injected particles at the TS for the three cases under analysis. 

By looking at the radial shape of $f_{CR}$, one can readily see how the three cases correspond to different spatial profiles: the Kolmogorov turbulence produces a more peaked shape, while a Bohm diffusion induces a flat profile. The Kraichnan case instead, generates a profile that is intermediate between Kolmogorov and Bohm, with a flat distribution at low energies and a peaked profile at energies higher than $\sim$100 TeV. Actually, the morphology is energy-dependent in all the tree models, with an increasingly peaked profile at higher energies. This change in shape is caused by the onset of dominance of diffusive propagation over advection at high energies. 

\begin{figure}
\begin{center}
\includegraphics[width=0.7\textwidth]{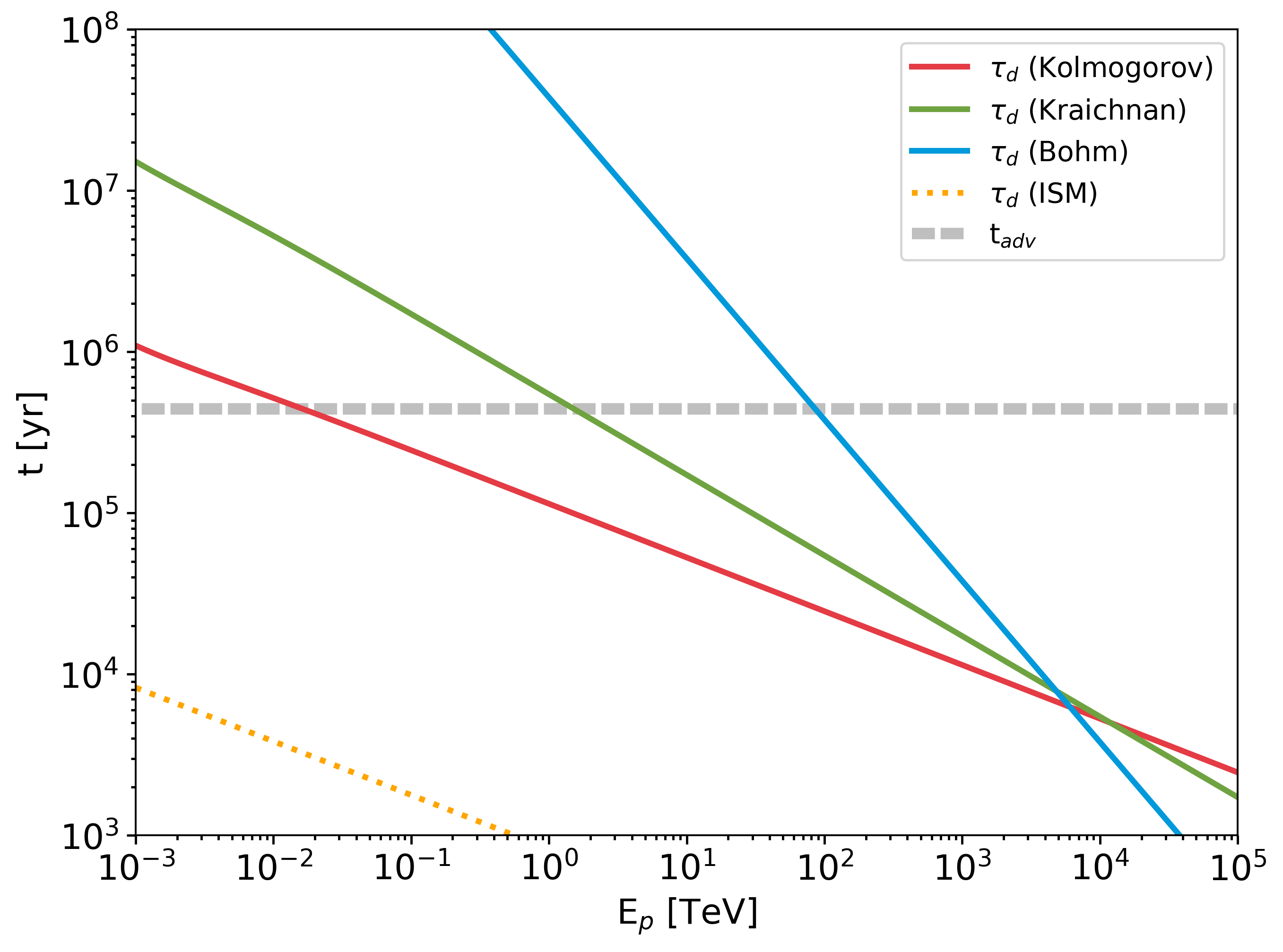}
\caption{Comparison between diffusion and advection time scales as a function of the particle energy. For showing purpose, we also included the diffusion time scale obtained considering the standard diffusion coefficient in the ISM (dotted line) \citep{Strong_CRsDiffusionReview_2007}.}
\label{fig:DiffVsAdvTimeBubble}
\end{center}
\end{figure}

It is possible to find at what energy the two transport processes are equally important by equating the advection time scale $t_{adv}$, with the diffusion time $\tau_{d}$, defined as:
\begin{equation}
t_{adv}=\int_{R_{TS}}^{R_b} \frac{dr}{u_2(r)}
\end{equation}
\begin{equation}
\label{eq:TauDiffBubble}
\tau_{d}(E)=\frac{(R_b-R_{TS})^2}{2 D_2(E)},
\end{equation}
where the advection timescale is obtained under the assumption of a strong shock considering a velocity in the downstream of $u_2 (r) = (u_1/4) (r/R_{TS})^{-2}$. Using the standard parameters for Cygnus OB2, we find that $t_{adv}\approx 1.8$ Myr. In Eq.~\ref{eq:TauDiffBubble}, the diffusion coefficients for the three cases under analysis ($D_2$) are given by Eqs.~\ref{eq:DK41} --~\ref{eq:DKra} --~\ref{eq:DBho}. We calculate them considering the turbulent magnetic field in the downstream, that is $\delta B_2 = \sqrt{11} \delta B_1(R_{TS})$, with $\delta B_1$ given by Eq.~\ref{eq:BfieldUpTS}. Fig.~\ref{fig:DiffVsAdvTimeBubble} shows the energy at which $t_{adv}=\tau_{d}(E)$. This energy is $\sim$20 GeV, $\sim$2 TeV, and $\sim$100 TeV for Kolmogorov, Kraichnan, and Bohm respectively. It is crucial to underline that these numbers are average values. If one takes into account the velocity profile, the effect of advection should be stronger (i.e., shifting to higher energies the former values) at distances closer to the TS. This can be clearly seen in Fig.~\ref{fig:CygOB2fCR}, where the profile becomes flatter close to the TS.

In addition to the propagation properties, the diffusion coefficient also regulates the confinement of particles around the acceleration site; hence, it directly affects the maximum energies that CRs can reach, and the spectral shape of injected particles. Both effects are noticeable in Fig.~\ref{fig:CygOB2fTS}: in spite of begin calculated with the same parameters for Cyg OB2, the particle spectra at the TS present very different shapes of the cut-off, and the position itself of the cut-off is different for the three considered diffusion regimes. Kolmogorov turbulence, for example, is not very efficient in confining particles close to the wind TS. As a consequence, the maximum energy will be less if compared to a harder turbulent cascade spectrum, as for example the flat spectrum behind Bohm-like diffusion. 

%---------------------------------------------------------

%---------------------------------------------------------
\section{Modeling $\gamma$-ray emission from Cygnus OB2}
\label{sec:CygGammaEmission}
During the escape from the acceleration site, the hadronic component of CRs is expected to interact with the surrounding medium causing the creation of neutral pions. As described in \S~\ref{subsec:RadMechanisms}, the $\pi^0$s subsequently decay with the emission of $\gamma$-rays. The observed spectrum and morphology will strongly depend on both the distribution of CRs and the target medium. The $\gamma$-ray flux from $\pi^0$ production is described by Eq.~\ref{eq:GammaFluxGeneral}. 

In our specific case, $f_{CR}(E_p,r)$ is the radial distribution of CRs given by Eq.~\ref{eq:fCRNoSea}, while $n(r)$ represents the number density distribution of the target medium in the vicinity of Cyg OB2, which is largely unknown and must be assumed a priori (see \S~\ref{subsec:CygOB2ISM}). Having in mind the spherical geometry of the system, Eq.\ref{eq:GammaFluxGeneral} can be simplified by considering the volume integral in terms of cylindrical coordinates. Knowing that $r=\sqrt{\ell'^2+z^2}$, where $z$ is the direction along the line of sight and $\ell'$ the projected distance on the sky,  Eq.~\ref{eq:GammaFluxGeneral} can be rewritten as
\begin{equation}
\label{eq:GammaFluxRadial}
\phi_\gamma(\ell, \Delta \ell, E_\gamma)= \int_\ell^{\ell+\Delta \ell} \overline{n}(\ell') \xi(\ell', E_\gamma)  \ell' d\ell' , 
\end{equation}
where $\overline{n}(\ell')$ is the average ISM density profile that can be inferred from observations (see \S~\ref{subsec:CygOB2ISM}) and $\xi(\ell', E_\gamma)$ defined as
\begin{equation}
\xi(\ell', E_\gamma) =\frac{1}{d^2} \iint c f_{CR}(\ell',z, E_p) \frac{d \sigma (E_p, E_\gamma) }{dE_p} dE_p dz.
\end{equation}
By varying the limits of integration, Eq.~\ref{eq:GammaFluxRadial} can be used to estimate the total $\gamma$-ray flux from a particular area of the sky (for spectral analysis) and also to obtain the $\gamma$-ray radial profile (for morphological analysis).

\subsection{The interstellar medium close to Cygnus OB2}
\label{subsec:CygOB2ISM}
The distribution of the ISM is a crucial parameter that directly affects the morphology of the observed $\gamma$-ray emission. On a very general ground, the ISM can be divided into three main gas phases: ionized, neutral atomic gas, and molecular gas. The latter two phases are typically the densest and the most massive, and hence are the ones that contribute the most to the production of $\gamma$-ray emission. The neutral atomic gas is usually observed using the emission of the 21 cm line of HI, while molecular H$_2$ must be traced indirectly using $^{12}$CO, since direct observation of H$_2$ is not possible due to the symmetry of the molecular structure which forbids dipole emission. 

\begin{figure}
\begin{center}
\includegraphics[width=0.5\textwidth]{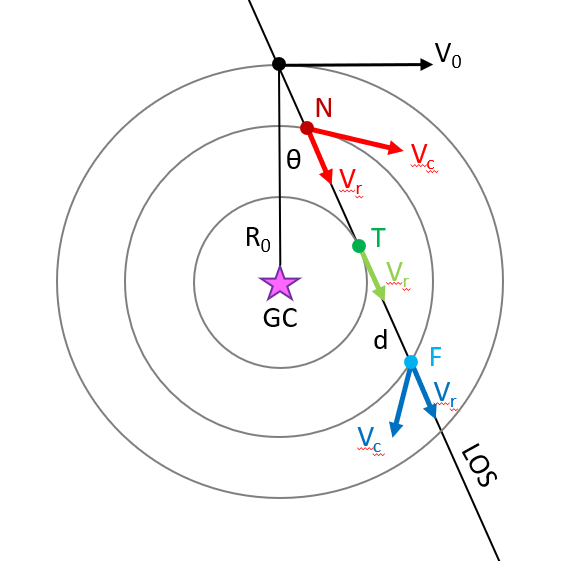}
\caption{Scheme for calculating the kinematic gas distance. The Doppler shift of an atomic or molecular transition depends on the component of the gas velocity due to the galactic revolution ($v_c$) projected along the line of sight ($v_r$). This produces an ambiguity in position (near and far position, N and F points respectively). The case where the radial velocity of the gas is totally aligned with the direction of the line of sight defines the tangent point (T point).}
\label{fig:KinDist}
\end{center}
\end{figure}

Clearly, due to the projection effect, not all the molecular or atomic gas observed in the vicinity of a given source is actually in its vicinity. The location of the gas along the line of sight can be estimated using the method of kinematic distances \citep{Roman-Duval_GasKinDist_2009}. Kinematic distances are based on the Doppler shift of a certain transition (atomic or molecular) induced by Galactic differential rotation. Knowing the Galactic rotation curve, a specific Doppler shift value identifies three points along a given line of sight with Galactic longitude $\theta$. These three points correspond to three different positions returning the same projected radial velocity $V_r$ (see Fig.~\ref{fig:KinDist}), and are known as Near distance, Tangent points, and Far distance. The distance ($d$) of these point can be calculated as:
\begin{equation}
\label{eq:KinDist}
d=R_0 \cos(\theta) \pm \sqrt{r_{gal}^2 - R_0^2 \sin(\theta)}
\end{equation}
where $R_0$ is the Galactocentric distance of the Sun, $V_0$ is the Sun orbital velocity, and $r_{gal}$ is the Galactocentric distance of the gas:
\begin{equation}
r_{gal}=R_0 \sin(\theta) \frac{V(r_{gal})}{V_r+V_0 \sin(\theta)}
\end{equation}
with $V(r_{gal})$ as the Galactic rotation curve. In Eq.~\ref{eq:KinDist}, the tangent point is obtained for $r_{gal}=R_0 \sin(\theta)$, and so $d=R_0 \cos(\theta)$.
It is impossible to discern between near, far and tangent points based on kinematics alone. This issue is generally known as \textit{kinematic distance ambiguity}. In general, additional pieces of information are required to remove the ambiguity, as for example, dust absorption or parallax measures. Note that this problem exists only for $r_{gal}< R_0$, as in the outer Galaxy the radial velocity along the line of sight decreases monotonically.

In principle, to model the gas distribution near Cyg OB2, one could use specific kinematic cut on the gas velocity. Unfortunately, the Cygnus-X region is located at Galactic longitudes where the differential galactic rotation up to $\sim$4 kpc results in low radial velocities, with values compatible with the typical gas motion dispersion, thus preventing a robust 3D modelization of the ISM profile along the line of sight. Even if the kinematic ambiguity prevents a small-scale spatial modelization of the gas, we consider anyway a velocity cut between -20 km s$^{-1}$ and 20 km s$^{-1}$, selecting in such a way the gas associated with the Cygnus-X star forming complex \citep{Schneider_CygnusXSFR_2006} while removing the gas contribution of both the Perseus and Outer arms. Once the gas is kinematically selected, we assume the most straightforward case where all the observed gas is uniformly distributed along the line of sight in a range $\Delta z=\pm400$ pc around Cyg OB2 position. The choice of $\Delta z$ stems from the fact that the total extent of 800 pc is inferred from the distribution of dust towards the Cygnus-X star forming complex \citep{Green_DustMap3D_2019}. Note that this specific gas model is perfectly consistent with the expression for the $\gamma$-ray emission in Eq.\ref{eq:GammaFluxRadial}, which was derived by implicitly assuming a constant gas profile along the line of sight.

\begin{figure}
\centering
\includegraphics[width=0.7\columnwidth]{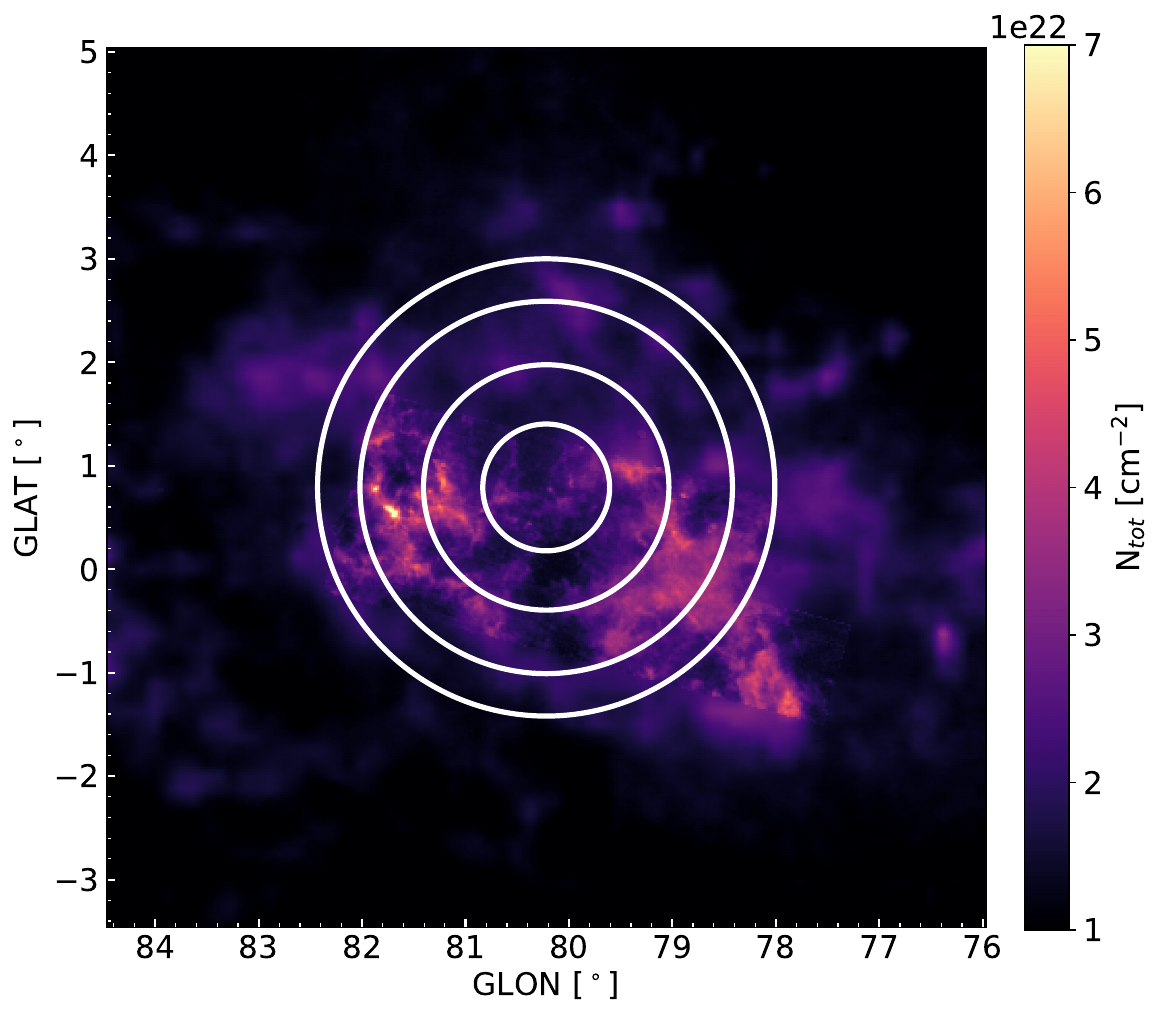}
\caption{Total target column density considering the interstellar medium in the neutral and molecular phases. The column density of HI has been evaluated using the 21 cm line observation from the CGPS, while for the molecular gas, we used a combination of $^{12}$CO observation from the Nobeyama radio telescope and the data from the $^{12}$CO galactic plane survey by \cite{Dame_12COSurvey_2001}. All the observations are kinematically cut selecting only the gas between -20 and 20 km/s. White rings represent the regions used by \cite{Aharonian_MSCs_2019} and \cite{Abeysekara_CygOB2HAWC_2021} to build the $\gamma$-ray radial profile. The circle radii correspond to angular sizes of 0.61$^\circ$, 1.19$^\circ$, 1.8$^\circ$ and 2.21$^\circ$.}
\label{fig:ISM_Map}
\end{figure}

To quantify the amount of neutral hydrogen, we use 21 cm line data from the Canadian Galactic Plane Survey (CGPS) \citep{Taylor_CGPS_2003}. For the molecular component, we use high-resolution observations of $^{12}$CO J(1--0) spectral line from the Nobeyama radio telescope \citep{Takekoshi_NobeyamaCygC18O_2019} in combination with the data from the composite galactic survey of \cite{Dame_12COSurvey_2001}. The neutral hydrogen column density is estimated using the approach described by \cite{Wilson_ToolsRadAstr_2009}:
\begin{equation}
\left [ \frac{N_{\rm HI}}{\rm cm^{-2}} \right ] = - 1.8224 \times 10^{18} \left[ \frac{T_s}{\rm K} \right] \int_{-20}^{20} \log \left( 1 - \frac{T^{\rm HI}_B (v)}{T_s-T_{\rm BG}} \right) \left[ \frac{dv}{\rm km / s} \right],
\end{equation}
where $T^{\rm HI}_B$ is the observed line brightness temperature, $T_s$ is the spin temperature, assumed to be 150 K, and $T_{BG}=2.66$ K is the brightness temperature of the cosmic microwave background at 21 cm. For the molecular hydrogen, we calculate the column density using the standard $X_{\rm CO}$ conversion factor:
\begin{equation}
    N_{\rm H_2} = X_{\rm CO} \int_{-20}^{20} T_B^{\rm CO}(v) \left[ \frac{dv}{\rm km / s} \right], 
\end{equation}
with $X_{\rm CO}=1.68 \times 10^{20}$ mol. cm$^{-2}$ km$^{-1}$ s  K$^{-1}$ as found by \cite{Ackermann_FermiCygnusCocoon_2011}. Finally, we can write the total target column density as $n(r)=(N_{\rm HI}+2N_{\rm H_2})/\Delta z$ (see fig.~\ref{fig:ISM_Map}).

%---------------------------------------------------------

%---------------------------------------------------------
\section{Comparison between expected and observed $\gamma$-ray emission}
To test the validity of the model described in \S~\ref{subsec:WindTSAccModel}, we compare the spectral and morphological properties of the expected $\gamma$-ray emission obtained in \S~\ref{sec:CygGammaEmission} with currently available observations by several experiments. The general idea is to find the best values of the parameters in terms of $L_w$, $s$, and $\epsilon_{CR}$ that can adequately describe the observed $\gamma$-ray spectrum. Then, a posteriori, we check if the best values found are reasonable, by comparing for example the obtained wind luminosity with the estimates in \S~\ref{subsec:CygOB2LwMdot}. For the sake of simplicity, when not specified otherwise, we keep all other parameters describing Cyg OB2 (i.e. age, $\dot{M}$, distance of Cyg OB2, $\rho_0$, $\eta_B$, $L_{inj}$, $k_{min}$) fixed to the values used in \S~\ref{sec:CygOB2fCR}. 

%In the search for the best parameters, we use the following approach. We start by considering a wide range of possible wind luminosities, in the range $10^{37} \-- 5\times 10^{39}$ erg/s, spanning one order of magnitude around the estimate in \S~\ref{subsec:CygOB2LwMdot}. Then, for every value of $L_w$, we find the best value of $\epsilon_{CR}$ and $s$ by fitting, through a $\chi^2$ minimization approach, the $\gamma$-ray spectrum extracted from a region of 2.2$^\circ$ centered on the stellar cluster (corresponding to a projected radius of $\sim 54$ pc).
In the search for the best parameters, we use the following approach. We fit through $\chi^2$ minimization the observed spectral energy distribution extracted from a region of 2.2$^\circ$ centered on the stellar cluster (corresponding to a projected radius of $\sim 54$ pc). We define the $\chi^2$ as:
\begin{equation}
\chi^2 = \sum_k \frac{[(E_k^2 \phi_{\gamma, k})_{data} - (E_k^2 \phi_\gamma(E_k))_{model}]^2}{\sigma_k^2}
\end{equation}
where $k$ is the index associated to the k-th spectral point, and $\sigma_k$ is the error of the k-th spectral energy distribution point. During the fit procedure, the parameters are left free to vary in a range of $10^{37} \-- 5\times 10^{39}$ erg s$^{-1}$ for $L_w$, $1.8 \-- 2.6$ for $s$ and $10^{-3} \-- 10^{-1}$ for $\epsilon_{CR}$. In terms of datasets, we consider the Cygnus Cocoon flux points measured by Fermi-LAT in the 4FGL (4FGL J2028.6+4110e) \citep{Abdollahi_Fermi4FGL_2020}, and the very-high-energy observations carried out by ARGO (ARGO J2031+4157) \citep{Bartoli_CygOB2Argo_2014} and HAWC (HAWC J2030+409) \citep{Abeysekara_CygOB2HAWC_2021}.

All the employed spectral points are rescaled in order to account only for the observed flux coming from a region of 2.2$^\circ$. This is done by considering that in all the cases the emission towards Cyg OB2 is modeled using a 2D symmetric Gaussian profile with different sizes: 2.0$^\circ$ for 4FGL J2028.6+4110e, 1.8$^\circ$ for ARGO J2031+4157 and 2.13$^\circ$ for HAWC J2030+409, which leads to rescaling factors of 0.45, 0.53 and 0.41 respectively. In addition, for the calculation of the $\chi^2$, we do not account for the highest energy datapoint by HAWC, as it differs, by almost one order of magnitude from the flux measured by LHAASO (LHAASO J2032+4102) at 100 TeV \citep{Cao_LHAASO12PeVatrons_2021}. The flux measured by LHAASO J2032+4102 is also not considered as the size of the source is not provided, thus the flux cannot be rescaled accordingly to the analyzed sky region size. Finally, note that the expected $\gamma$-ray emission is calculated using Eq.~\ref{eq:fCRNoSea}, hence, without accounting for the contribution of $f_{gal}$. This is because in principle, all the employed flux points should be background subtracted. In general, the background emission by Galactic CRs is non-negligible in the Fermi-LAT band. However, the Fermi-LAT data does not include this contribution. This is because, in the analysis procedure, a template model that follows the gas distribution is fitted along with the source model to account for the background emission.

Before proceeding further, a note of caution is mandatory. The implemented analysis approach is relatively naive, so the final best fit value should be handled with care. This is even more true for the associated confidence intervals. Clearly, to obtain a precise estimation of the parameters, a robust multi-instrument joint analysis accounting for the systematics between different experiments is required. However, even with this simplified approach, we can still obtain a rough estimation of our parameters of interest. We do not however return confidence intervals for the parameters, as they could be significantly affected by the systematics between the experiments.

Once we obtain the CR distribution that best describes the observed spectrum, we investigate the corresponding expected $\gamma$-ray radial profile to understand if one specific type of propagation model among the ones implemented can best reproduce the observed morphology. For this purpose, we calculate the total $\gamma$-ray luminosity in four different rings centered on Cyg OB2, with projected sizes of 0$\--$15 pc, 15$\--$29 pc, 29$\--$44 pc, and 44$\--$54 pc (see Fig.~\ref{fig:ISM_Map}). The $\gamma$-ray luminosity is defined as:
\begin{equation}
\label{eq:GammaLum}
L_\gamma=4 \pi d_{OB2}^2 \int_{E_{-}}^{E_{+}} E_\gamma \phi_\gamma(E_\gamma) dE_\gamma    
\end{equation}
where $d_{OB2}$ is the distance of Cyg OB2 and $\phi_\gamma (E_\gamma)$ is the spatial integrated flux from eq.~\ref{eq:GammaFluxRadial}. we then divide the obtained value of the luminosity by the rings surfaces, and compare the result with the value estimated for the same sky regions by \cite{Aharonian_MSCs_2019} using Fermi-LAT data and by HAWC \citep{Abeysekara_CygOB2HAWC_2021}. Consequently, we set the limit of integration in Eq.~\ref{eq:GammaLum} to account only for the luminosity in the energy ranges for which the measurements refer, that are respectively $E_-=10$ and $E_+=300$ GeV for the Fermi-LAT band and $E_-=1$ and $E_+=250$ TeV for the HAWC observations. 

In the following subsections, we will separately discuss the analysis outcomes for the three different transport models implemented, then, we will proceed to discuss the implications of the obtained results in the next section.

\subsection{Kolmogorov case}
\label{subsec:K41Case}
\begin{figure}
\centering
\includegraphics[width=0.7\columnwidth]{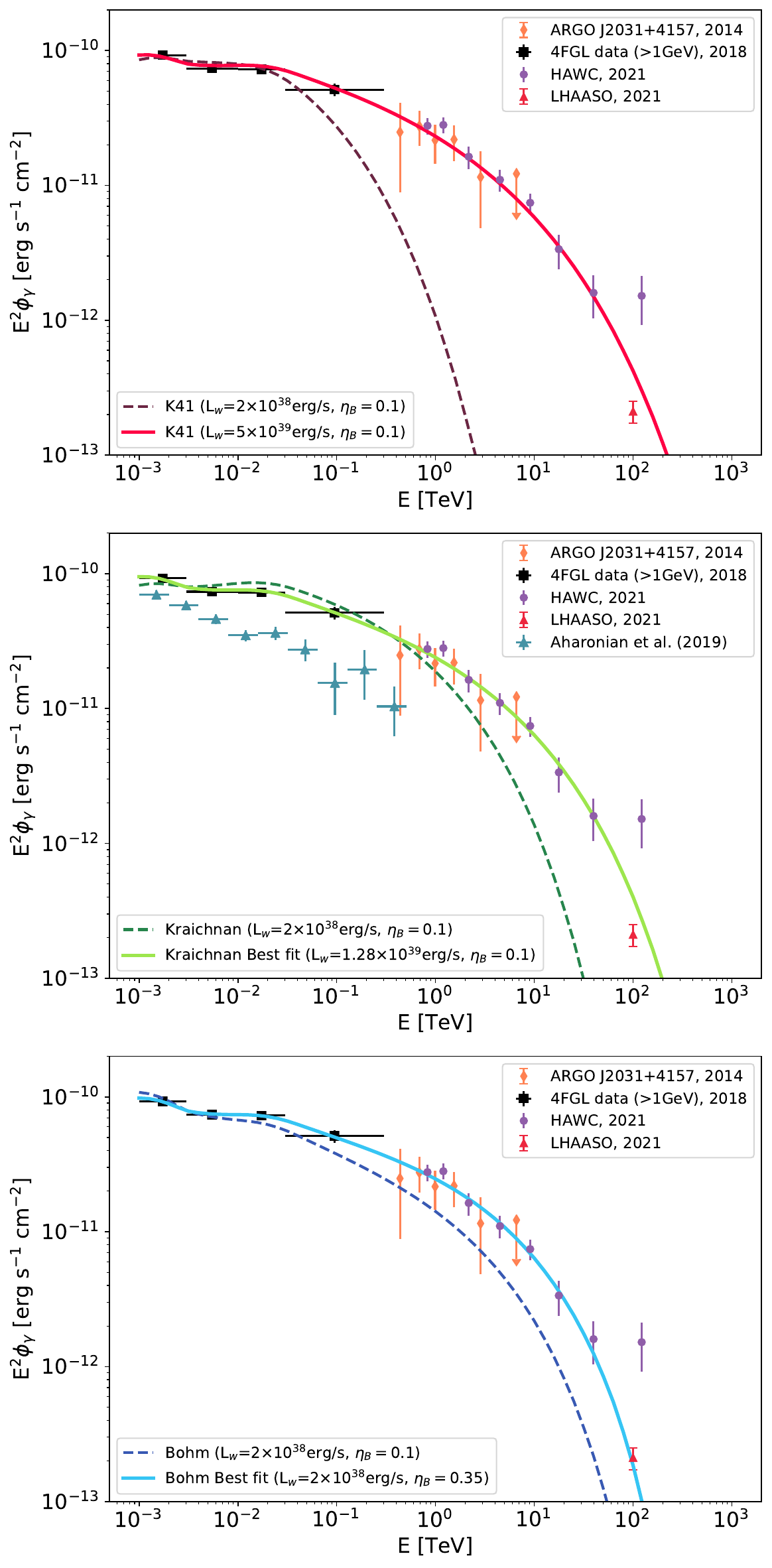}
\caption{Spectral energy distribution of the expected hadronic $\gamma$-ray emission extracted from a region of 2.2$^\circ$ centered on Cyg OB2. Results are shown for the Kolmogorov (top panel), Kraichnan (middle panel), and Bohm diffusion (bottom panel). See text for more details. }
\label{fig:E2GammaFlux}
\end{figure}

We start by studying the case in which the CR distribution is calculated assuming a Kolmogorov like diffusion. We found that $\chi^2$ as a function of $L_w$ has an overall decreasing trend with increasing $L_w$ in the considered wind luminosity interval, with no clear sign of a minimum. This trivially translates into a global $\chi^2$ minimum for $L_w=5 \times 10^{39}$ erg s$^{-1}$. With this value of $L_w$, the best fit values for the spectral index of injected particles and the efficiency of CR production are $s \simeq 4.17$ and $\epsilon_{CR} \simeq 4 \times 10^{-3}$. Although the position of the global $\chi^2$ minimum does not fall in the considered luminosity interval, the value of $L_w=5 \times 10^{39}$ erg s$^{-1}$ should not be far from the true minimum position, which is probably a factor of a few higher. This can be seen in the top panel of Fig.~\ref{fig:E2GammaFlux}, where the spectrum corresponding to $L_w=5 \times 10^{39}$ erg s$^{-1}$ describes fairly well the observations. It follows that to adequately reproduce the observed spectrum, a wind luminosity equal or higher than $5 \times 10^{39}$ erg s$^{-1}$ is needed, which is more than one order of magnitude above the range of values of $L_w$ estimated in \S~\ref{subsec:CygOB2LwMdot}. The reason for such a high luminosity is to be found in a combination of two factors: first, the low efficiency of Kolmogorov turbulence in confining particles at the acceleration site, causes low maximum energies even for high wind luminosities; second, the large diffusion coefficient produced by Kolmogorov turbulence in the bubble is inefficient in trapping particles at high energy, hence suppressing the emission at very-high energy. One can notice these two effects by considering the $\gamma$-ray spectrum obtained by fixing the wind luminosity to the value expected from the star population of Cyg OB2, $L_w=2 \times 10^{38}$ erg s$^{-1}$, and by fitting $s$ and $\epsilon_{CR}$. In this case, the maximum energy of particles is too low, and the expected spectrum fails to reproduce the observed flux at very high-energies.

Even taking in to account the large uncertainties on the estimated wind luminosities in \S~\ref{subsec:CygOB2LwMdot}, such a high $L_w$ is largely in disagreement with our expectations, a fact that strongly disfavors the model with diffusion resulting from Kolmogorov turbulence. %, consequently, we do not investigate further the properties of the $\gamma$-ray emission.\\ 

As a final consideration, it is worth noticing that the maximum reachable energy for $L_w=5 \times 10^{39}$ erg/s is $E_{max}\simeq 23$ PeV. Since the problem is, in part, linked to the value of $E_{max}$, one could consider decreasing the required $L_w$ to obtain such value by considering a higher value for $\eta_B$, as $E_{max}$ is directly proportional to the square root of $\eta_B$ (see Eq.~\ref{eq:EmaxK41Lw}). However, even by increasing $\eta_B$ to an unrealistic value of 50\%, the required luminosity only decreases by $\sim$40\%, which is still in strong disagreement with the expected values. Given the difficulties of this model at reproducing the observed spectrum, we do not investigate further the properties of the $\gamma$-ray emission for the Kolmogorov case. %The thin lines in top panel of  Fig.~\ref{fig:E2GammaFlux} shows the expected flux for the two extreme luminosity values obtained in section~\ref{sec:CygOB2} but with  $\eta_B$ set to 90\%. 

\subsection{Kraichnan case}
\label{subsec:KraichnanCase}
\begin{figure}
\centering
\includegraphics[width=0.7\columnwidth]{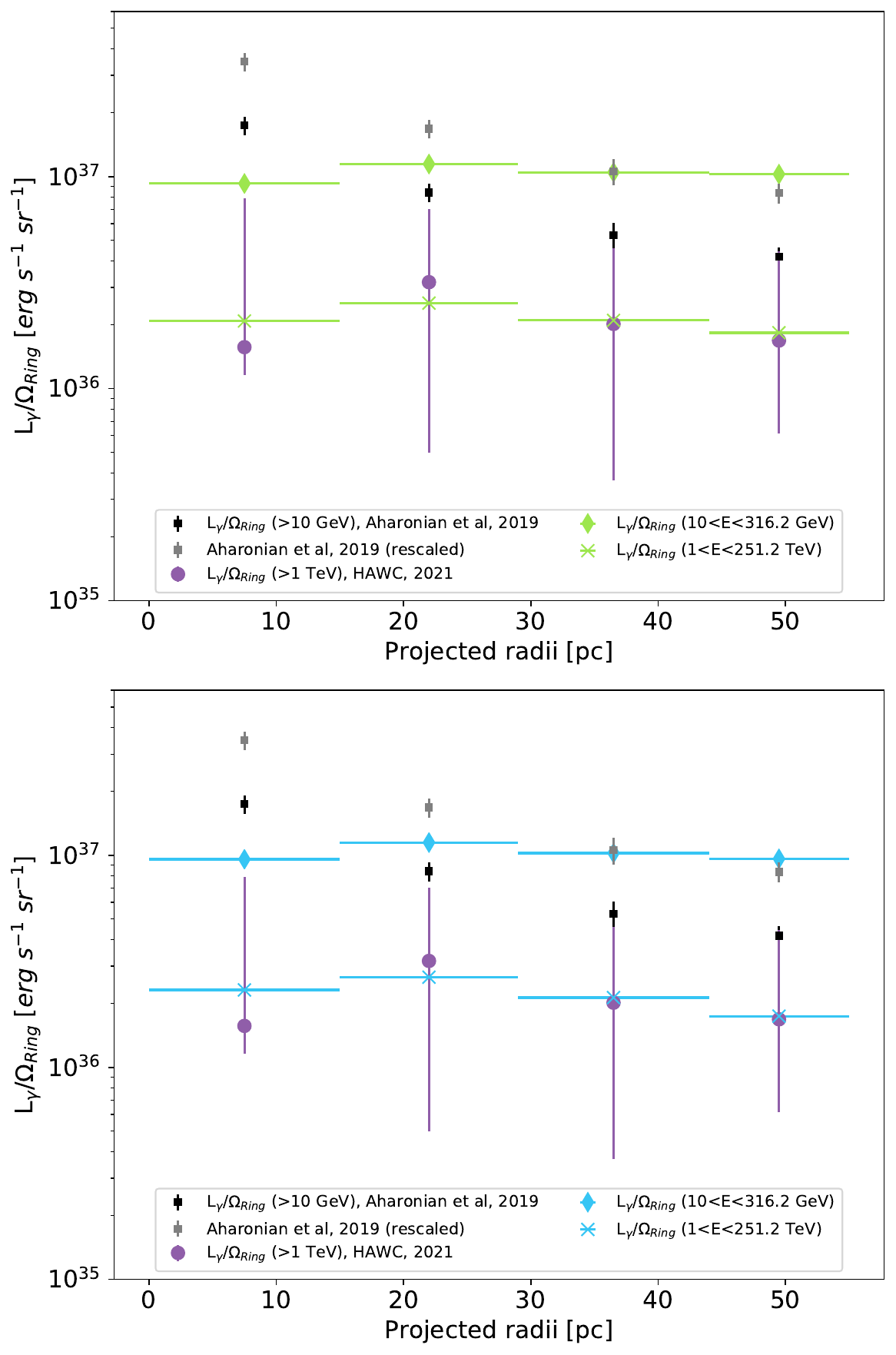}
\caption{Comparison between the expected $\gamma$-ray surface brightness radial profile assuming Kraichnan (upper panel) or Bohm (bottom panel) diffusion. Black squares show the measurements by Fermi-LAT \citep{Aharonian_MSCs_2019}, while purple circles indicate the surface brightness observed by HAWC \citep{Abeysekara_CygOB2HAWC_2021}. Grey squares represent instead Fermi-LAT measurements rescaled by a factor of 2 (see text). Diamonds and asterisks indicate the expected surface brightness values from the models at energies of 10 GeV$<E_\gamma<316$ GeV and 1 TeV$<E_\gamma<251$ TeV respectively.}
\label{fig:GammaRadProfile}
\end{figure}

Let us now consider the results for the case of Kraichnan turbulence. After running the analysis procedure, we found the global minimum for the $\chi^2$ for a wind luminosity of $L_w=1.28 \times 10^{39}$ erg s$^{-1}$. The corresponding best fit values for the spectral index and the CR efficiency are $s \simeq 4.23$ and $\epsilon_{CR} \simeq 7 \times 10^{-3}$. The required $L_w$ is lower compared to the Kolmogorov case, but still a factor of $\sim$4 higher than the maximum estimated luminosity of $2.9 \times 10^{38}$ erg s$^{-1}$. %However, given the large uncertainties in the calculation of $L_w$, a discrepancy of a factor of 2 does not represent a piece of strong evidence to a priori exclude the model.\\ %Interestingly, with this choice for $L_w$, the maximum energy reachable by CR in the system is $\sim$3.97 PeV. \\
Although $L_w$ is at odds with the estimates made for the wind luminosity, let us anyway proceed with our analysis without discarding the Kraichnan case, and compare the predicted $\gamma$-ray radial profile with current observations. The top panel of Fig.~\ref{fig:GammaRadProfile} shows this comparison for both the high-energy and very high-energy bands. Starting from the high-energy band, one can distinctly see that, excluding the inner ring, the normalization of the expected $\gamma$-ray luminosity per ring is, on average, a factor of $\sim$2 higher than the one estimated by \cite{Aharonian_MSCs_2019}. This discrepancy is caused by the different flux normalization between the 4FGL datapoints (used in our spectral analysis) and the spectrum measured by \cite{Aharonian_MSCs_2019}, as shown in the middle panel of Fig.~\ref{fig:E2GammaFlux}. Moreover, even if we account for this difference in normalization by rescaling the observed luminosity by a factor of 2, we can see that the predicted profile shape is still not in agreement with the observations. In fact, we expect a flat profile, contrary to observations that seem to show a peaked morphology. The situation is different at very high energy, where the morphology observed by HAWC is in good agreement with our model, which still predicts a flat profile.

\subsection{Bohm case}
\label{subsec:BohmCase}
Bohm-like turbulence is highly efficient in confining particles around the acceleration site, producing higher maximum energies with small values of $L_w$. If we, in fact, fix the wind luminosity value to the usual $L_w=2 \times 10^{38}$ erg s$^{-1}$, and we perform the fit varying $s$ and $\epsilon_{CR}$, the resulting spectrum is close to the observed one, with the exception of the very high energy part (see bottom panel of Fig.~\ref{fig:E2GammaFlux}). Since this wind luminosity provides a reasonable fit to the spectrum, we decided to use a different approach, consisting in fixing the wind luminosity to $L_w=2 \times 10^{38}$ erg s$^{-1}$ and performing the fit in $s$, $\epsilon_{CR}$ and $\eta_B$. Following this method, we found $\eta_B=0.35$, with an associated spectral index and CR efficiency of $s \simeq 4.27$ and $\epsilon_{CR} \simeq 0.022$. The corresponding maximum energy for particle is $E_{max} \simeq 466$ TeV.

The bottom panel of Fig.~\ref{fig:GammaRadProfile} shows the corresponding predicted $\gamma$-ray radial profile. The result is analogous to that obtained for the Krainchnan case, with an overall flat trend in both energy bands.

\section{Discussion}
Table~\ref{tab:BestFit} summarizes the results obtained in the analysis for all the considered cases. We now proceed to a deeper analysis of the results related to the Kraichnan and Bohm cases, while the Kolmogorov model was ruled out because of the high wind luminosity needed to reproduce the $\gamma$-ray spectrum.

Note that in principle, the Kraichnan case, similarly to the Kolmogorov model, also requires a wind luminosity value that is inconsistent with the estimates calculated in \S~\ref{subsec:CygOB2LwMdot}. However we keep the Kraichnan case for the purpose of discussion. In fact, the value of $L_w$ required to reproduce the spectrum decreases as we consider turbulence models with gradually harder spectra, at the cost of increasing by a factor of a few the fraction of power in magnetic turbulence. We may then conclude that the true turbulence spectral index lies between the cases of Bohm and Kraichnan-like cascades. In this scenario, we can keep the two models as limiting cases, bearing in mind that the most realistic scenario, assuming that CR acceleration occurs at the wind TS, will have intermediate properties between the two diffusion models.

The discussion of the results is organized as follows: first, we will comment on the results obtained from the spectral analysis, and later concentrate on the morphological analysis. Finally, we will discuss which, among the considered cases, is the one that can best reproduce the observed emission properties.

\begin{table*}
\begin{center}
\begin{tabular}[c]{l c c c c c c c c}  
\toprule \toprule
Models  &  $L_w$ & $s$ & $\epsilon_{CR}$ & $\eta_B$ & $E_{\rm max}$ & $R_{TS}$ & $R_b$ & $\bar{\chi}^2_{\min}$ \\
  & [erg s$^{-1}$] &  & [\%] &  & [PeV] & [pc] & [pc] & \\

\midrule

Kolmogorov & $5 \cdot 10^{39}$ & 4.17 & $0.4 $ & 0.1 (fixed) & 23 & 16 & 163 & 0.66\\
Kraichnan & $1.28 \cdot 10^{39}$ & 4.23 & $0.7 $ & 0.1 (fixed) & 3.97 & 14 & 124 & 0.39\\
Bohm & $2 \cdot 10^{38}$(fixed) & 4.27 & $2.2$ & 0.35 & 0.47 & 13 & 86 & 0.25\\

\bottomrule %\bottomrule
\end{tabular}
\caption{Best fit parameters values and main system properties for three different models. For the Kolmogorov and Kraichnan cases, the parameters varied during the fit are $L_w$, $s$, and $\epsilon_{CR}$. For the Bohm case we fix the wind luminosity and vary $s$, $\epsilon_{CR}$ and $\eta_B$.}
\label{tab:BestFit}
\end{center}
\end{table*}

\subsection{Discussion on the spectral analysis results}
\label{subsec:SpectAnalysis}
Starting from the spectral analysis results, one can notice how both Kraichnan and Bohm cases require a particle injection slope at the termination shock $s>4$. This is softer than the standard spectral index predicted from diffusive shock acceleration at strong shocks, as one would expect the TS of a compact cluster to be. A possible reason for this softening could be related to the plasma cooling in the bubble caused by the heat transmission to the cold shell of swept material (see \S~\ref{subsec:WindBubble}). The temperature drop in the downstream can potentially affect the TS, resulting in an effective weakening of the shock itself and causing a softening in the spectra of accelerated particles. 

A significant difference between the two turbulence models is the maximum energy reached by the particles. For the Kraichnan case $E^{Kra}_{max}\simeq 4$ PeV, which is almost an order of magnitude higher than in the Bohm case, $E^{Bohm}_{max} \simeq 0.5$ PeV. Interestingly, despite the different maximum energies, the $\gamma$-ray spectra of the two cases are remarkably similar, with a cutoff feature that in both cases is located in the range 10--100 TeV. As the maximum energy in the Kraichnan case is higher, one could naively expect the cutoff in the $\gamma$-ray spectrum to be positioned at energies slightly higher than 100 TeV. This does not happen because of two distinct factors. The first one is related to the different spectral cutoff shapes of the emitting particle spectrum for the two considered cases. As explained in the paper by \cite{Morlino_2021}, the maximum energy calculated in \S~\ref{subsec:WindTSAccModel} does not represent the exact location of the cutoff of the injected particles spectrum. This can be readily seen by computing $E_{max}$ for the standard parameters of Cyg OB2 given in \S~\ref{sec:CygOB2fCR} and by comparing the results with the cutoff positions shown in Fig.~\ref{fig:CygOB2fTS}. The deviation from a power law and the beginning of the cutoff region usually starts at energies lower than $E_{max}$. On a general ground, the energy shift is related to the spherical symmetry of the TS, which affects primarily the most energetic particles, whose diffusion length is of the order of the TS radius. The magnitude of the shift directly depends on the diffusion coefficient in the upstream region: the hardest (the softest) the momentum dependence of the diffusion coefficient, the smaller (the larger) the energy shift. The second factor is related to the different propagation properties in the downstream region. Particles propagating in Kraichnan turbulence are not as effectively trapped inside the bubble as they are in Bohm-like turbulence, and as a result, they enter the diffusion regime at lower energies. This leads to a reduction in the number of particles at high energies, causing the $\gamma$-ray cutoff to shift towards lower energies.

Noticeably, both the Kraichnan and Bohm cases require a low fraction, of order a few percent, of the wind luminosity to be converted into CR production. One might then think that the process is inefficient in accelerating particles. This is, however, not true: the best fit value found for $\epsilon_{CR}$ should be considered as a sort of lower limit, with true values potentially higher by a factor of $\sim$ 10. The reason for this lies in the method implemented in \S~\ref{subsec:CygOB2ISM} to model the distribution of the ISM around Cyg OB2, which consists of a uniform density profile along the line of sight. Under the assumption of a constant distribution, the average numerical particle density in the region is $\sim 8$ cm$^{-3}$. However, the density in the downstream, where most of the $\gamma$-ray emission is produced, is expected to be lower.

An estimate of the density can be obtained assuming that the entire bubble is only filled with the material provided by the cluster wind, that is $n_2 \simeq \dot{M}t/(4/3 \pi R_b^3)\approx 0.005$ cm$^{-3}$. This value is extremely low, and in fact, not realistic. As explained in \S~\ref{subsec:WindBubble}, due to the heating of the cold swept-up shell, part of the shell material evaporates into the downstream region, causing an increment of the density.

Considering the parameters of Cyg OB2, using Eq.~\ref{eq:ShellMdot}, we can estimate the density in the bubble as: $n_2 \simeq \dot{M}_s t/(4/3 \pi R_b^3)\approx 0.1$ cm$^{-3}$. The latter value could be even higher by a factor of a few, if one accounts for the shell fragmentation, that potentially boosts the evaporation rate \citep{Lancaster_StellarWindCooledII_2021}. In the end, we can reasonably expect $n_2$ to lie between 1--0.1 cm$^{-3}$. If these are the density values in the downstream, the observed spectrum would require $\epsilon_{CR}$ a factor of $\sim$10 larger.

\subsection{Discussion on the morphological analysis results}
\label{subsec:MorphAnalysis}
\begin{figure}
\begin{center}
\includegraphics[width=0.7\textwidth]{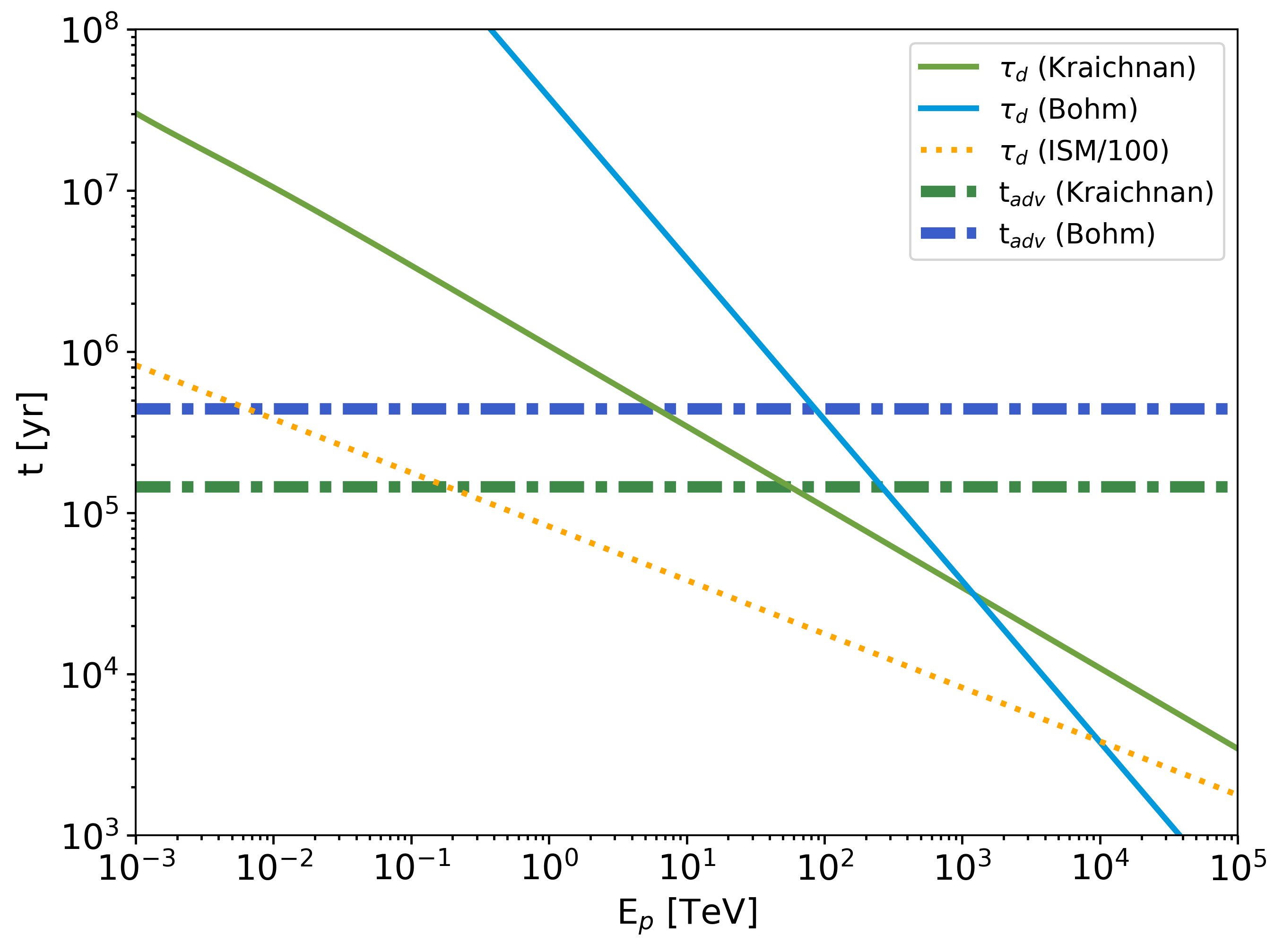}
\caption{Same as Fig.~\ref{fig:DiffVsAdvTimeBubble} but now both diffusion and advection timescales are calculated only considering a region of 54 pc, corresponding to the projected size of 2.2$^\circ$ that is the extent of the outermost ring used for the morphological analysis. Note that Kraichnan and Bohm advection times (dot dashed lines) are different as the best fit L$_w$ are not equal. The dotted orange line represents the diffusion timescale considering the standard interstellar medium diffusion coefficient suppressed by two orders of magnitude.}
\label{fig:DiffVsAdvTime54pc}
\end{center}
\end{figure}

As briefly mentioned in the previous section, both cases considered show a flat morphology of the $\gamma$-ray emission profile, regardless of the energy band. The uniformity of the profile is a result of the advection dominated transport of CRs, making their distribution uniform in space. While such result is not unexpected in the case of Bohm-like turbulence, in the Kraichnan case, one could expect that the diffusion would dominate in the HAWC energy band. This does not happen because we are probing the morphology of the $\gamma$-ray emission in a projected sky area of about 54 pc. This region is very small compared to the size of the forward shock, which for our best-fit, in the Kraichnan case, is $R_b \simeq 126$ pc. As described in \S~\ref{sec:CygOB2fCR}, since the advection velocity scales as $u_2 \propto r^{-2}$, the vicinity of the TS will be characterised by high advection velocities, making it the main propagation mechanism. This can be easily understood by looking at Fig.~\ref{fig:DiffVsAdvTime54pc}, where the timescales of advection and diffusion are compared taking into account the size of the region under analysis. For the Kraichnan case, in a region of 54 pc, advection dominates over diffusion up to energies of almost a hundred TeV. The same is also true for Bohm-like turbulence. 

Actually, to be accurate, one must also consider the effect of projection along the line of sight of the $\gamma$-ray emission, so that the observed emission does not come only from a spherical volume of radius 54 pc. However, during the calculation of the radial profile shown in Fig.~\ref{fig:GammaRadProfile}, this effect is already taken into account. Nevertheless, a flat trend is consistent with the HAWC observation. Yet this is not the case for the radial morphology observed by Fermi-LAT, which shows a peaked profile in the $\gamma$-ray luminosity. This is clearly in tension with the prediction of our model, where at lower energy, the effect of advection is expected to become even stronger.

According to \cite{Aharonian_MSCs_2019}, the observed $\gamma$-ray morphology is consistent with the expected emission generated by a pure diffusive distribution of CRs continuously injected by Cyg OB2. However, this type of profile cannot be achieved in the picture where YMSCs are surrounded by an expanding bubble of hot gas. In order to reproduce the size of the observed emission, \cite{Aharonian_MSCs_2019} require a diffusion coefficient at 10 TeV which is $\sim 5 \times 10^{25}$ cm$^2$ s$^{-1}$. With this normalization of the diffusion coefficient, if the energy dependence is $\propto E^{\delta}$, with $\delta>0$, the diffusion time scale will always higher than the advection time at energies below 10 TeV. \cite{Aharonian_MSCs_2019} also provide an upper limit to the diffusion coefficient based on the efficiency of particle acceleration at 10 TeV. The upper limit is found to be two orders of magnitude below the Galactic diffusion coefficient. Assuming that the diffusion coefficient scales at low energy as $D \propto E^{1/3}$ (hence following a Kolmogorov-like diffusion), even in this case the diffusion time is always larger or comparable with the advection timescale (see Fig.~\ref{fig:DiffVsAdvTime54pc}). Thus, the CR distribution should be flatter than a pure diffusive 1/r profile.

%One possible solutions for this problem consists trying to explain the increasing $\gamma$-ray luminosity towards the cluster through the leptonic inverse Compton emission (IC) from an electron population located in a thin shell around the TS. We will discuss the validity of this scenario in the following paragraph. 

\section{Cygnus OB2 as a cosmic ray accelerator?}
Throughout this chapter, we tried to interpret the extended $\gamma$-ray radiation detected in coincidence with Cyg OB2 in terms of hadronic emission from a population of freshly accelerated CR under the assumption of particle acceleration at the wind TS. We tested three different CR distributions, considering different models of particle transport based on three specific diffusion coefficients calculated in quasi linear theory for the following plasma turbulence spectra: Kolmogorov-like, Kraichnan-like, and flat spectrum. Before drawing conclusions and arguing about the capability of Cyg OB2 to accelerate particles, let us first summarize the main results obtained for the three different models:
\begin{itemize}
 \item We have rejected the hypothesis that the acceleration and propagation of CRs are governed by Kolmogorov-like turbulence, since the resulting $\gamma$-ray emission would disagree with observations unless invoking unreasonable values of the cluster wind luminosity compared to the one estimated from the stellar population (see \S~\ref{subsec:CygOB2LwMdot}).
 \item The Kraichnan model is disfavored as it also requires wind luminosities at least a factor of a few higher than the estimates based on the cluster star population.
 \item In contrast to the two previous cases, if one considers Bohm-like diffusion, the resulting CR distribution is able to account for the $\gamma$-ray spectrum with a reasonable $L_w$ and an efficiency of the ordrer of $\sim30$\% of turbulent magnetic field production.
\end{itemize}
Taking into account the results obtained for Kraichnan and Bohm models, it is plausible to consider the existence of a turbulence spectrum midway between the two, which is capable of reproducing the $\gamma$-ray spectrum with reasonable values of both $L_w$ and $\eta_B$. In this scenario, the analyzed Kraichnan and Bohm models can be treated as limit cases, with the actual solution leading to spectral and morphological properties intermediate between the two. However, since both models considered are able to reproduce the spectrum equally well (as testified by the almost equivalent $\chi^2$ value), and both return a predominantly flat morphology, it is natural to expect that also the true solution will adequately reproduce the spectrum as well, and will be likely characterized by a flat radial profile of $\gamma$-ray emission.

Having this in mind, we expect that the maximum energy of accelerated particles will be around 1 PeV. A model characterized by a flat morphology, both in high-energy and very high energy $\gamma$-ray, is at odds with the peaked profile observed in the Fermi-LAT energy band by \cite{Aharonian_MSCs_2019}. One possible solution for this problem consists in trying to explain the increasing $\gamma$-ray luminosity towards the cluster center through the leptonic inverse Compton emission from an electron population located in a thin shell around the TS. This scenario will be explored in a future work.
%Recently, \cite{Astiasarain_CygCocoon_2023} performed a new analysis of the Fermi-LAT data of the Cygnus Cocoon, modeling the observed emission as a combination of two components, namely FCES G78.74+1.56 and FCES G80.00+0.50. The first source is well described by a 2D Gaussian and has an extent of $\sim 4.4^\circ$, while the second component is modeled using a template with an overall size of $\sim 2^\circ$ derived from the ionized gas emission as traced by Plank 408 MHz observations. The latter source is the one accounting for the $\gamma$-ray emission in our region of interest of $2.2^\circ$. Although \cite{Astiasarain_CygCocoon_2023} do not provide the radial profile for the observed emission below $\sim 2^\circ$, we can reasonably presume (since the template is well fitting the data) that the radial observed morphology follows the radial profile of the Plank 408 MHz observations. This emission is actually tracing photodissociation regions, which in turn are similarly bright in the IR. The radial profile of the ionized gas template used by \cite{Astiasarain_CygCocoon_2023} should not be far from the radial morphology of the IR radiation field, showed in Fig.~\ref{fig:ISM_Profiles}. If this is the case, then also the results of \cite{Astiasarain_CygCocoon_2023} are likely not consistent with the peaked $\gamma$-ray morphology found by \cite{Aharonian_MSCs_2019}.

\begin{figure}
\centering
\includegraphics[width=0.7\columnwidth]{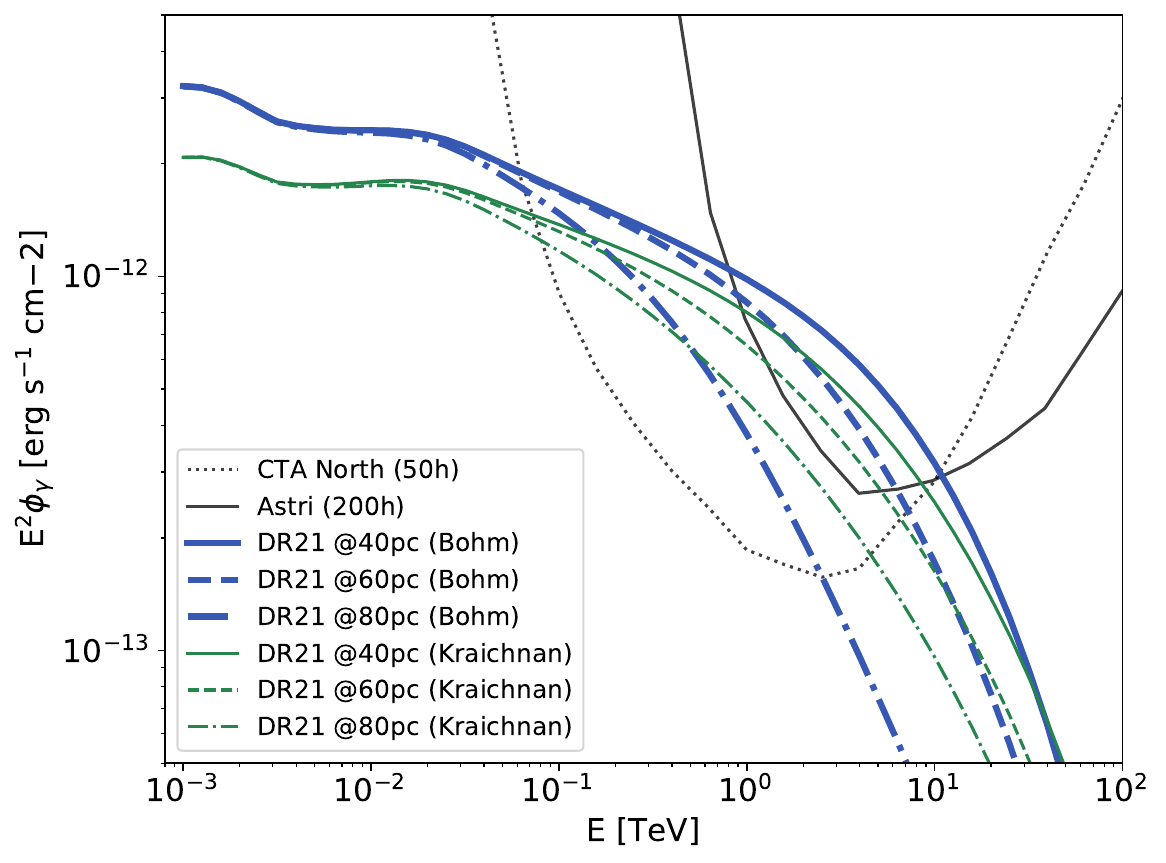}
\caption{Expected hadronic $\gamma$-ray spectra from the molecular cloud DR21 considering the two possible CR distributions given by the Kraichnan (thin lines) and Bohm (thick lines) cases. The spectra are calculated considering different relative distances from Cyg OB2: 40 pc (continuous line), 60 pc (dashed line), and 80 pc (dot-dashed line). The dotted line shows the CTA North point source sensitivity for 50h of exposure considering the Alpha Configuration \citep{CTA_Performances}, while black solid line is the Astri sensitivity for a point-like source for 200h of exposure \citep{Lombardi_AstriPerformance_2022}.}
\label{fig:DR21Emission}
\end{figure}

To understand what the actual turbulence spectrum in the system is, a more detailed study would have to be carried out considering a multi-instrument analysis of the $\gamma$-ray emission. In a not so far future, this might be possible with the use of new-generation $\gamma$-ray telescopes, such as the Cherenkov Telescope Array and the ASTRI Mini Array. In this regard, additional information can be used during the analysis to better constrain the diffusion mechanism in the bubble. First, one can consider the $\gamma$-ray spectrum originating from molecular clouds that are possibly found inside the forward shock. Thanks to precise maser parallaxes measurements, we know the existence of a few massive molecular clouds (i.e. DR21, W75N, and DR20) in the Cygnus-X region, located in close proximity to Cyg OB2. Unfortunately, the exact distance between the clouds and the star cluster is not easy to estimate, as the position of Cyg OB2 is still not well constrained. Fig.~\ref{fig:DR21Emission} shows the expected $\gamma$-ray spectrum from the molecular cloud DR21, considering different possible distances from Cyg OB2. It is interesting to note that, while below $\sim$100 GeV the spectra obtained from the Kraichnan and Bohm models are similar, except for a normalization factor, in the very-high-energy range, the spectral shape in the two cases can be significantly different. The emission at very-high-energy can be hence used to constrain the diffusion coefficient.

\begin{figure}
\centering
\includegraphics[width=0.7\columnwidth]{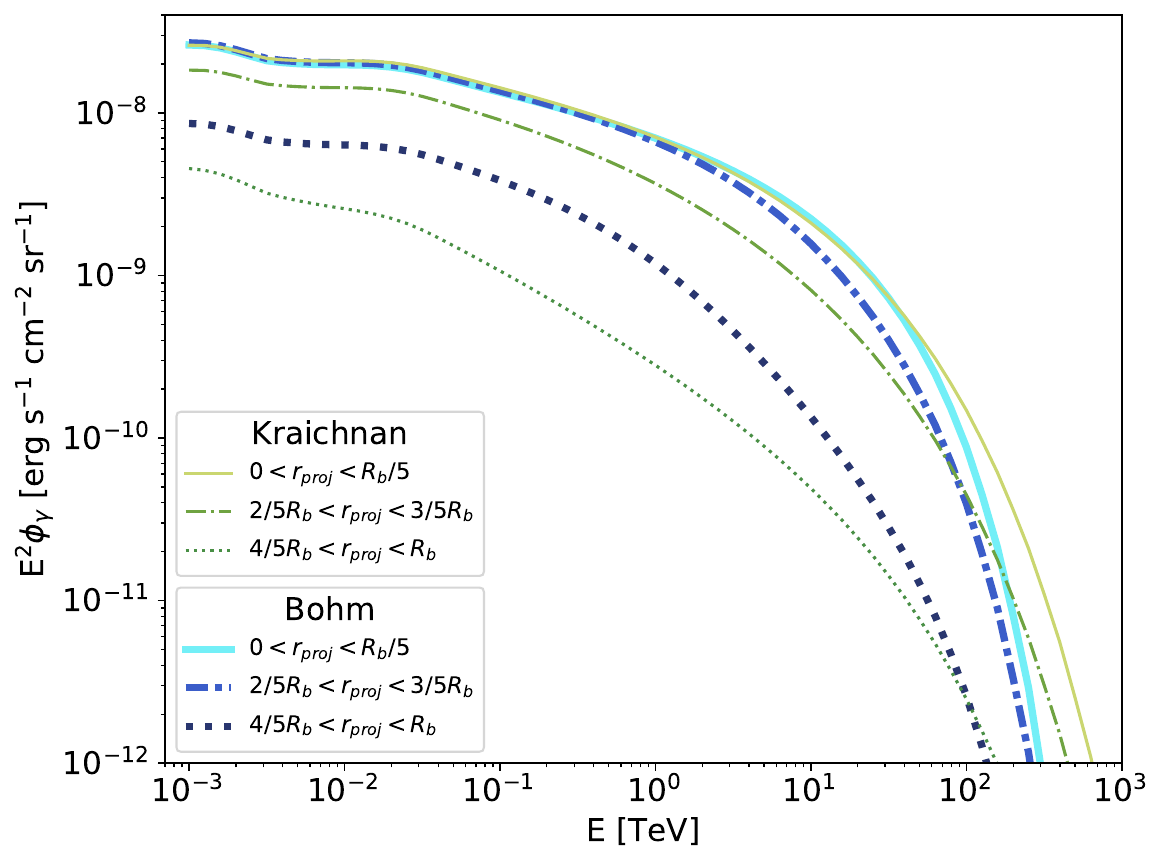}
\caption{Expected hadronic $\gamma$-ray spectra extracted from rings at projected distances of 0<r<$\frac{1}{5}R_b$ (continuos lines), $\frac{2}{5}R_b$<r<$\frac{3}{5}R_b$ (dash-dotted lines), and $\frac{4}{5}R_b$<r<$R_b$ (dotted lines), for the Kraichnan (thin lines) and Bohm (thick lines) cases.}
\label{fig:SpectraPerRegion}
\end{figure}

A second possible way to constrain the diffusion coefficient is to analyze the $\gamma$-ray spectrum at different projected distances from Cyg OB2. Fig.~\ref{fig:SpectraPerRegion} shows the spectra extracted from different regions of the system. The main difference between the two limit cases is observed in the cutoff part of the spectrum. Similarly to the Bohm case, a hard turbulence spectrum will produce sharper cutoff shapes. In the vicinity of the forward shock, the discrepancy between the two limit cases starts to become appreciable even at lower energies (>10 GeV), with the Kraichnan case producing softer emission. Indeed, modeling of spatially resolved $\gamma$-ray spectra appears as a promising way to constrain particle transport and contribute to unveil the mechanism of particle acceleration in YMSCs. 

%% file: CHAPTERS/chapter_3.tex
%---------------------------------------------------------
\chapter{The contribution of YMSCs to the diffuse $\gamma$-ray emission}
\label{Ch:chap3}
%---------------------------------------------------------
\lettrine{G}{iven} the ability of YMSCs to produce CRs, it is natural to expect them to also be $\gamma$-ray emitters. In addition to the case of Cygnus OB2, a dozen of other clusters have been associated with diffuse $\gamma$-ray emission detected by different telescopes (see Tab.~\ref{tab:ClusterList}), namely: Westerlund 1 \citep{Abramowski_Wd1VHE_2012, Aharonian_HESSWesterlund1_2022} (observed by HESS and Fermi-LAT), Westerlund 2 \citep{Yang_Wd2Fermi_2018} (observed with HESS and Fermi-LAT), NGC 3603 \citep{Saha_NGC3603_2020} and NGC 6618 \citep{Liu_NGC6618_2022} (observed with Fermi-LAT).

At first glance, one might think that the number of cases is rather low compared to the expected population of galactic YMSCs. The reason for such a low number of observed YMSCs is probably due to a combination of two factors. First, most YMSCs are less massive than the objects listed above, which should be considered somewhat extreme cases. Therefore, a large fraction of Galactic YMSCs is expected to have lower $L_w$. This has a direct impact on the luminosity in terms of CRs, which affects the $\gamma$-ray emission. Secondly, as we saw in Ch.~\ref{ch:CygOB2}, most of the $\gamma$-ray emission comes from particles propagating within the wind blown bubble. Consequently, the projected size of the emission region can be considerably large, especially if compared to the resolution of current $\gamma$-ray telescopes. As we will show in \S~\ref{sec:PopYMSCasCR}, the typical size of a bubble is of the order of tens of parsecs (10--100 pc), which corresponds to an angular size of:
\begin{equation}
\theta \simeq 0.57 \left(\frac{R_b}{10 \rm \ pc} \right) \left(\frac{d}{1 \rm \ kpc} \right)^{-1} \ \rm deg
\end{equation}
with $d$ the distance of the cluster from the Sun. The fact that they are potentially extended sources makes detection even more difficult, especially given the problems associated with studying extended sources in the $\gamma$-ray band. 

\begin{table}[]
\begin{tabular}{cccccc}
\toprule \toprule
Name                                                                      & $\log M/$M$_\odot$                                                   & \begin{tabular}[c]{@{}c@{}}$r_c$\\ {[}pc{]}\end{tabular}      & \begin{tabular}[c]{@{}c@{}}$D$\\ {[}kpc{]}\end{tabular} & \begin{tabular}[c]{@{}c@{}}Age\\ {[}Myr{]}\end{tabular} & \begin{tabular}[c]{@{}c@{}}$L_w$\\ {[}erg s$^{-1}${]}\end{tabular} \\ \hline
Westerlund 1                                                              & $4.6 \pm 0.045$                                                      & $1.5$                                                         & $4$                                                     & $4\--6$                                                 & $10$                                                               \\
Westerlund 2                                                              & $4.56 \pm 0.035$                                                     & $1.1$                                                         & $2.8 \pm 0.4$                                           & $1.5 \-- 2.5$                                           & $2$                                                                \\
Cygnus OB2                                                                & $4.7 \pm 0.3$                                                        & $5.2$                                                         & $1.4$                                                   & $2\--7$                                                 & $2$                                                                \\
NGC 3603                                                                  & $4.1 \pm 0.1$                                                        & $1.1$                                                         & $6.9$                                                   & $2\--3$                                                 & -                                                                  \\
BDS 2003                                                                  & $4.39$                                                               & $0.2$                                                         & $4$                                                     & $1$                                                     & -                                                                  \\
W40                                                                       & $2.5$                                                                & $0.44$                                                        & $0.44$                                                  & $1.5$                                                   & -                                                                  \\
RSGC 1                                                                    & $4.48$                                                               & $1.5$                                                         & $6.6$                                                   & $10 \-- 14$                                             & -                                                                  \\
MC 20                                                                     & $\sim 3$                                                             & $1.3$                                                         & $3.8 \-- 5.1$                                           & $3\-- 8$                                                & $\sim 4$                                                           \\
NGC 6618                                                                  & -                                                                    & $3.3$                                                         & $\sim 2$                                                & $<3$                                                    & -                                                                  \\
\begin{tabular}[c]{@{}c@{}}30 Dor (LMC)\\ NGC 2070 / RCM 136\end{tabular} & \begin{tabular}[c]{@{}c@{}}$4.8 \-- 5.7$\\ $4.34 \-- 5$\end{tabular} & \begin{tabular}[c]{@{}c@{}}multiple\\ subcluster\end{tabular} & $50$                                                    & \begin{tabular}[c]{@{}c@{}}$1$\\ $5$\end{tabular}       & -                                                                  \\ \bottomrule
\end{tabular}
\caption{List of YMSCs for which a diffuse $\gamma$-ray emission has been detected in their coincidence. References of each cluster: Westerlund 1 \citep{Abramowski_Wd1VHE_2012, Aharonian_HESSWesterlund1_2022}, Westerlund 2 \citep{Yang_Wd2Fermi_2018}, Cygnus OB2 \citep{Bartoli_CygOB2Argo_2014, Abeysekara_CygOB2HAWC_2021, Astiasarain_CygCocoon_2023}, NGC 3603 \citep{Saha_NGC3603_2020}, BDS 2003 \citep{Albert_BDS2003_2021}, W40 \citep{Sun_W40_2020}, RSGC 1 \citep{Sun_RSGC1_2020}, MC 20 \citep{Sun_MC20_2022}, NGC 6618 \citep{Liu_NGC6618_2022}, and the Large Magellanic Clusters \citep{HESS_30Doradus_2015}. Values marked by a '-' are not provided in the literature.}
\label{tab:ClusterList}
\end{table}

Given the difficulty of studying single sources, we may study the contribution of multiple overlapping $\gamma$-ray halos arising from a population of YMSCs, to estimate whether this may result in a non-negligible contribution to the large-scale diffuse emission along the Galactic plane. The estimation of this emission and subsequent comparison with the data is a key element that can be used to constrain YMSCs as CR factories. In addition, calculating the emission from a population of YMSCs is also useful in view of the next generation of gamma telescopes, since it can be used to estimate the number of potentially observable YMSCs. 

In this chapter, we will calculate the diffuse $\gamma$-ray emission arising from a synthetic population of Galactic YMSCs and compare it with available observations from Fermi-LAT. The structure of the chapter traces the workflow implemented to reach our objectives. First, we describe how to generate for each cluster a stellar population from known initial mass functions and how to calculate the fundamental properties of stars such as luminosity, radius, and temperature. We also describe the recipes used to model stellar winds. Next, we discuss how to simulate a population of Galactic YMSCs, illustrating our choices for the distribution in mass, galactic position, and age. 

Once we have all the pieces of the puzzle, we proceed to investigate the properties of the resulting Galactic YMSC population in terms of particle acceleration. To this goal, we estimate the expected $\gamma$-ray emission for each cluster, and we compute the diffuse contribution from the entire population. Finally, we compare the results with existing observations and discuss our findings. We anticipate that our results will provide a lower limit to the possible contribution to the $\gamma$-ray sky in that we are neglecting the role of SN exploding inside stellar clusters.

%---------------------------------------------------------
\section{Modeling the star population inside YMSCs}
\label{sec:ModStarPop}
For any given YMSC, characterized by a specific mass and age, we need to build a stellar population that is consistent with its properties. The general approach is the following: knowing the mass and age of the cluster, we build a population of stars given the initial stellar mass function (IMF). Then, based on the cluster age, we remove all the stars that are expected to have exploded as supernovae. Afterward, we compute the intrinsic characteristics of all the stars that are left, such as luminosity, radius, and temperature. To this purpose, we do not use any stellar model, but rather the observed mass-luminosity, mass-radius, and mass-temperature relations. The usage of empirical relations is preferred to the usage of more robust stellar models as computational time is significantly decreased. Finally, for every star, we calculate the main parameters of the stellar wind, such as the wind luminosity, the wind speed, and the mass loss rate.

In the next subsection we describe in details all the ingredients implemented in the afore mentioned procedure.

\subsection{Mass distribution of stars inside clusters}
\label{subsec:StellarPopInYMSC}
\begin{figure}[ht]
\begin{center}
\includegraphics[width=0.7\textwidth]{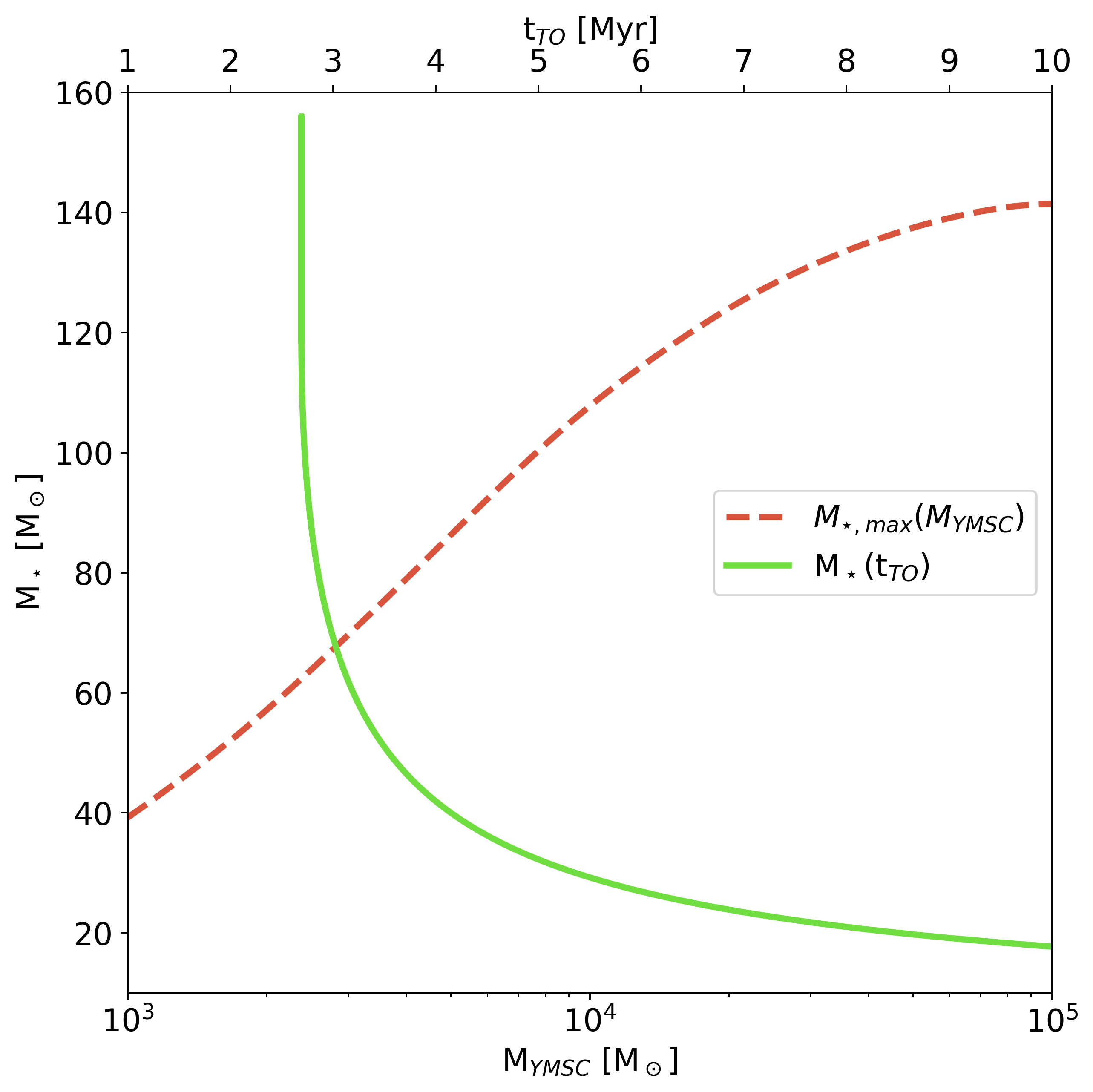}
\caption{Relationship between the maximum stellar mass that can be generated in a cluster and the mass of the host cluster (dashed line, lower x-axis). The graph also reports the mass of stars leaving the main sequence at a given time t$_{TO}$ (solid line, upper x-axis), given by inverting Eq.~\ref{eq:tTO}.}
\label{fig:MStarmaxVsMymscVsTOtime}
\end{center}
\end{figure}

The number of stars formed as a function of their mass ($M_\star$) is generally referred to as the \textit{stellar initial mass function} ($f_\star(M_\star)$). Generally such a function is parameterized as a set of broken power laws. Above $\sim$10 M$_\odot$ there is a general consensus that the IMF follows the Salpeter law $\propto M^{-2.3}$. At lower mass the IMF is flatter, but its exact shape is still debated. Here we have decided to use the expression given by \cite{Kroupa_IMF_2001}:
\begin{equation}
\label{eq:StellarIMF}
f_\star(M_\star) \propto \frac{dN_\star}{dM_\star} = 
\begin{cases}
 M_\star^{-0.3} & \text{for } M_\star < 0.08 \text{ M}_\odot  \\
 0.08 M_\star^{-1.3} & \text{for } 0.08 \text{ M}_\odot \leq M_\star \leq 0.5 \text{ M}_\odot \\
 0.04 M_\star^{-2.3} & \text{for } M_\star > 0.5 \text{ M}_\odot
\end{cases}
\end{equation}
Each cluster with mass $M$ will initially include a total number of stars equal to:
\begin{equation}
N_\star(M)=M \frac{\int_{M_{\star,min}}^{M_{\star,max}} f_\star(M_\star)dM_\star}{\int_{M_{\star,min}}^{M_{\star,max}} M_\star f_\star (M_\star) dM_\star}
\end{equation}
where $M_{\star,min}$ and $M_{\star,max}$ are respectively the minimum and maximum stellar masses that can be generated in a cluster. We fix the value of $M_{\star,min}$ to 0.08 M$_\odot$, which is the minimum theoretical mass to support significant nuclear burning \citep{Carroll_IntroModAstro_1996}. The choice of maximum mass, on the other hand, turns out to be an extremely delicate problem\footnote{See \cite{Bastian_UniversalIMF_2010} for a comprehensive description of the topic.}. As a matter of fact, $M_{\star,max}$ represents a crucial parameter for the purposes of this work, as massive stars are the ones that contribute the most to the YMSC wind luminosity. Clearly, this value cannot be arbitrarily high, and a first limit to $M_{\star,max}$ is given by $M_{\star,max}=150$ M$_\odot$. This value seems to be widely recognized as a fundamental mass upper limit for stars with zero metallicity that form in clusters \citep{Weidner_MaxStellarMassInYMSC_2004, Figer_SGR180620_2005, Oey_StarMassUL_2005, Koen_StarULinLMC_2006}, and it is also the maximum value of stellar mass observed in our Galaxy. In general, it is reasonable to consider the existence of a relation between $M_{\star,max}$ and $M$. Indeed, this must be true for low mass stellar clusters since, for example, a 100 M$_\odot$ cluster cannot include stars with masses equal or greater than 100 M$_\odot$. However, for massive stellar clusters, the matter is presently still under debate. \cite{Weidner_StarMaxMvsClusterM_2010} seem to prove the existence of this relationship through a comprehensive study of the literature, showing that it is extremely unlikely to reproduce the observed stellar populations with a random sampling of the initial mass function that does not account for the clusters masses. On the other side, a parallel analysis of published data performed by \cite{Maschberger_MaxStarMassVsClusterM_2008} using sophisticated unbiased selection criteria for stellar clusters shows that even low mass clusters do possess populations of massive stars. Given the importance of the parameter $M_{\star,max}$ and its current uncertainties, we decide to run our analysis considering two different scenarios. In the first we use a constant maximum mass for all YMSCs fixed to the 150 $M_\odot$ limit. In the other case, we consider a maximum mass that depends on the cluster mass, following the relation provided by \cite{Weidner_MaxStellarMassInYMSC_2004} and reported in Fig.~\ref{fig:MStarmaxVsMymscVsTOtime}.

After generating the stellar population, depending on the YMSC age, we remove all those stars that exploded as supernova. We do so by considering that a star with a given mass $M_\star$ will leave the main sequence (and soon after explode as a supernova) at a turn-off time ($t_{TO}$) approximately given by the following relation \citep{Buzzoni_TOtime_2002}:
\begin{equation}
\label{eq:tTO}
\log \left(\frac{t_{TO}}{1 \rm \ yr} \right) = 0.825 \log^2 \left(\frac{M_{\star}}{120 \rm \ M_\odot} \right) + 6.43
\end{equation}
which is obtained fitting several of sets of theoretical models. Fig.~\ref{fig:MStarmaxVsMymscVsTOtime} shows the resulting inverse relation. Note that no stars explode before $\sim 2.8$ Myr, and that, for the maximum cluster age we are interested in (10 Myr), only stars with less than 20 M$_\odot$ can survive.

\subsection{Modeling stellar parameters}
\label{subsec:StellarPhys}
For each generated star, we estimate stellar luminosity and radius using empirical relations. The choice of this method, expected to provide estimates that are less robust than those based on stellar models, comes for two needs: first, considering a larger mass interval than that usually covered by stellar models. Second, to minimize computation time. The latter turns out to be an extremely relevant aspect since, eventually, we intend to calculate multiple synthetic stellar populations to obtain an estimate of the diffuse $\gamma$-ray emission together with an appropriate guess of its statistical fluctuations. Note that, from now on, we will consider in our analysis only stars with masses larger than 2.75 M$_\odot$, roughly corresponding to the lower mass limit for a B-type star. We do so as the contribution to the cluster wind power of stars with masses below $\sim 3$ M$_\odot$ is believed to be negligible.

\subsubsection{Mass-Luminosity relation}
Various mass-luminosity relationships (MLRs) have been proposed over the years. In general, several authors provide MLRs that are valid only in specific mass ranges. \cite{Eker_MLR_2018}, for example, model the MLR as a series of power laws valid from 0.179--31 M$_\odot$. For the range of masses in which we are interested, the expression is:
\begin{equation}
\label{eq:Eker18MLR}
\log \left(\frac{L_\star}{\rm L_\odot} \right) =  
\begin{cases}
3.967 \log \left(\frac{M_\star}{\rm M_\odot} \right) + 0.093 & \text{for } 2.4 < \frac{M_\star}{\rm M_\odot} < 7  \\
2.865 \log \left(\frac{M_\star}{\rm M_\odot} \right) + 1.105 & \text{for } 7  \leq \frac{M_\star}{\rm M_\odot} \leq 31 
\end{cases}
\end{equation}
In the case of very massive stars (M$_\star$>100 M$_\odot$), the MLR is provided by \cite{Yungelson_EvolutionFateMassiveStars_2008}, whose expression has been used in \S~\ref{subsec:CygOB2LwMdot} for the calculation of Cygnus OB2 parameters. The formal equation is obtained inverting Eq.~\ref{eq:StellarMFromL}:
\begin{equation}
L_\star=10^{3.48} \left(\frac{M_\star}{\rm M_\odot} \right)^{1.34} \ \rm L_\odot
\end{equation} 
No MLR is defined over a broad enough range of masses for our purposes, as we are interested in a wide interval ranging from 2.75 to 150 M$_\odot$. To overcome this problem, we decided to merge the two recipes given by \cite{Eker_MLR_2018} and \cite{Yungelson_EvolutionFateMassiveStars_2008}. We do so by implementing a set of smoothed broken power-laws, defined as:
\begin{equation}
\label{eq:MergedMLR}
L_\star =  
\begin{cases}
L_{b1} \left(\frac{M_\star}{M_{b1}} \right)^{\alpha_1} \left[ \frac{1}{2}+ \frac{1}{2} \left( \frac{M_\star}{M_{b1}} \right )^{1/\Delta_1}\right ]^{(-\alpha_1+\alpha_2)\Delta_1} & \text{for } 2.4 \leq \frac{M_\star}{\rm M_\odot} < 12  \\
\mathcal{K} L_{b2} \left(\frac{M_\star}{M_{b2}} \right)^{\alpha_2} \left[ \frac{1}{2}+ \frac{1}{2} \left( \frac{M_\star}{M_{b2}} \right )^{1/\Delta_2}\right ]^{(-\alpha_2+\alpha_3)\Delta_2} & \text{for } M_\star \geq 12 \text{ M}_\odot 
\end{cases}
\end{equation}
where $L_{b1}=3191$ L$_{\odot}$ and $L_{b2}=368874$ L$_{\odot}$ are the luminosity values calculated using Eq.\ref{eq:Eker18MLR} at the mass break points $M_{b1}=7$ M$_\odot$ and $M_{b2}=36.089$ M$_\odot$ respectively. The value of $M_{b2}$ is the intersection point between the two MLRs of \cite{Yungelson_EvolutionFateMassiveStars_2008} and \cite{Eker_MLR_2018}. The power law indexes are $\alpha_1=3.97$, $\alpha_2=2.86$, and $\alpha_3=1.34$ respectively. The parameters $\Delta_1$ and $\Delta_2$ are used to smooth the junction between the power law components. We fix the two parameters to 0.01 and 0.15 respectively. Finally, $\mathcal{K}=0.817$ is a normalization constant providing continuity at 12 M$_\odot$
Note that in the interval 2.4--12 M$_\odot$ the MLR is an adaption of Eq.\ref{eq:Eker18MLR} in the form of a smoothed power law. We decided to use this form rather than Eq.\ref{eq:Eker18MLR} as we found that Eq.\ref{eq:Eker18MLR} is not continuous at $M_{b1}=7$ M$_\odot$. To check whether the extension of the MLR given by \cite{Yungelson_EvolutionFateMassiveStars_2008} to masses less than 100 M$_\odot$ is valid, we compare Eq.~\ref{eq:MergedMLR} with archival data of massive stars where both bolometric luminosities and masses are given. For this purpose, we consider massive stars observed in different YMSCs, such as Cygus OB2 \citep{Wright_MassiveStarPopCygOB2_2015}, R136 \citep{Brands_R136_2022}\footnote{For every star \cite{Brands_R136_2022} provide different estimated properties after fitting stellar models with a different number of free parameters. The choice of the number of parameters ultimately depends on the available spectroscopy for each star (UV and optical or only optical).}, and 30 Doradus \citep{Schneider_30Dor_2018}. We additionally include eclipsing binary stars from the DebCAT catalog \citep{Southworth_DebCAT_2015} and the sample of stars used by \cite{Eker_MLR_2015} in their first work for the estimation of the MLR. Top panel of Fig.~\ref{fig:MLRTR} shows the result of this comparison: the MLR given by Eq.~\ref{eq:MergedMLR} is in fair agreement with the observations, with the underlying implication that the relationship of \cite{Yungelson_EvolutionFateMassiveStars_2008} is valid even at smaller masses, down to $M_{b2}$ (eventually the relation can be considered still valid down to about $\sim$20 M$_\odot$).
%\begin{figure}[ht]
%\begin{center}
%\includegraphics[width=0.8\textwidth]{FIGURES/Chapter_3/MLR.png}
%\caption{Mass-luminosity relationship for stars above 2 M$_\odot$. Thick dashed line shows the relation used in this work (Eq.~\ref{eq:MergedMLR}). Thick continuous and dotted lines represent the relations provided by \cite{Eker_MLR_2018} and \cite{Yungelson_EvolutionFateMassiveStars_2008} respectively. For comparison purposes, we also plot an additional MLR given by \cite{Salaris_StarPopEvol_2005}. The different data points report observed stellar luminosities and computed masses for several stars in various clusters, such as Cygus OB2 \citep{Wright_MassiveStarPopCygOB2_2015}, R136 \citep{Brands_R136_2022}, and 30 Doradus \citep{Schneider_30Dor_2018}. We also include stars from the DebCAT catalog \citep{Southworth_DebCAT_2015} and the previous work on MLRs of \cite{Eker_MLR_2015}.}
%\label{fig:MLR}
%\end{center}
%\end{figure}

\subsubsection{Mass-Radius relation}
Mass-radius relations (MRRs) are in general less constrained and much broader than MLRs. Similarly to what we have done in \S~\ref{subsec:CygOB2LwMdot}, we employ the relation provided by \cite{Demircan_StarsMRR_1991}: 
\begin{equation}
\label{eq:MRR}
R_{\star}=0.85 \left(\frac{M_\star}{M_\odot} \right)^{0.67} R_\odot \ .
\end{equation} 
Eq.~\ref{eq:MRR} is plotted in the middle panel of Fig.~\ref{fig:MLRTR}, where a comparison with archival data is made showing that the MRR is able to adequately reproduce the observations. To this purpose, we used the same dataset employed for the MLR, with the exception of the Cygnus OB2 stellar cluster where the information on stellar radii is not provided. For the sake of completeness, we also report in Fig.~\ref{fig:MLRTR} an additional MRR obtained from the work of \cite{Yungelson_EvolutionFateMassiveStars_2008}. We found the latter relation in strong disagreement with both Eq.~\ref{eq:MRR} and data from the literature, although the authors claim that the relation should be valid for the mass range 25--115 M$_\odot$. All stellar radii employed for the comparison are obtained using stellar models calibrated to other observable stellar parameters (see references for details). The only exception is that of the DebCAT data, which are based on direct measurements of the stellar radii using detached eclipsing binaries.

%\begin{figure}[ht]
%\begin{center}
%\includegraphics[width=0.8\textwidth]{FIGURES/Chapter_3/MRR.png}
%\caption{Mass-radius relationship for stars above 2 M$_\odot$. Continuous line shows the relation used in this work (Eq.~\ref{eq:MRR}) and obtained from the work of \cite{Demircan_StarsMRR_1991}. Dotted line represent instead the MRR provided by \cite{Yungelson_EvolutionFateMassiveStars_2008}. The references for the several data points are reported in the caption of Fig.\ref{fig:MLR}.}
%\label{fig:MRR}
%\end{center}
%\end{figure}

\subsubsection{Mass-Temperature relation}
\begin{figure}[H]
\begin{center}
\includegraphics[width=0.6\textwidth]{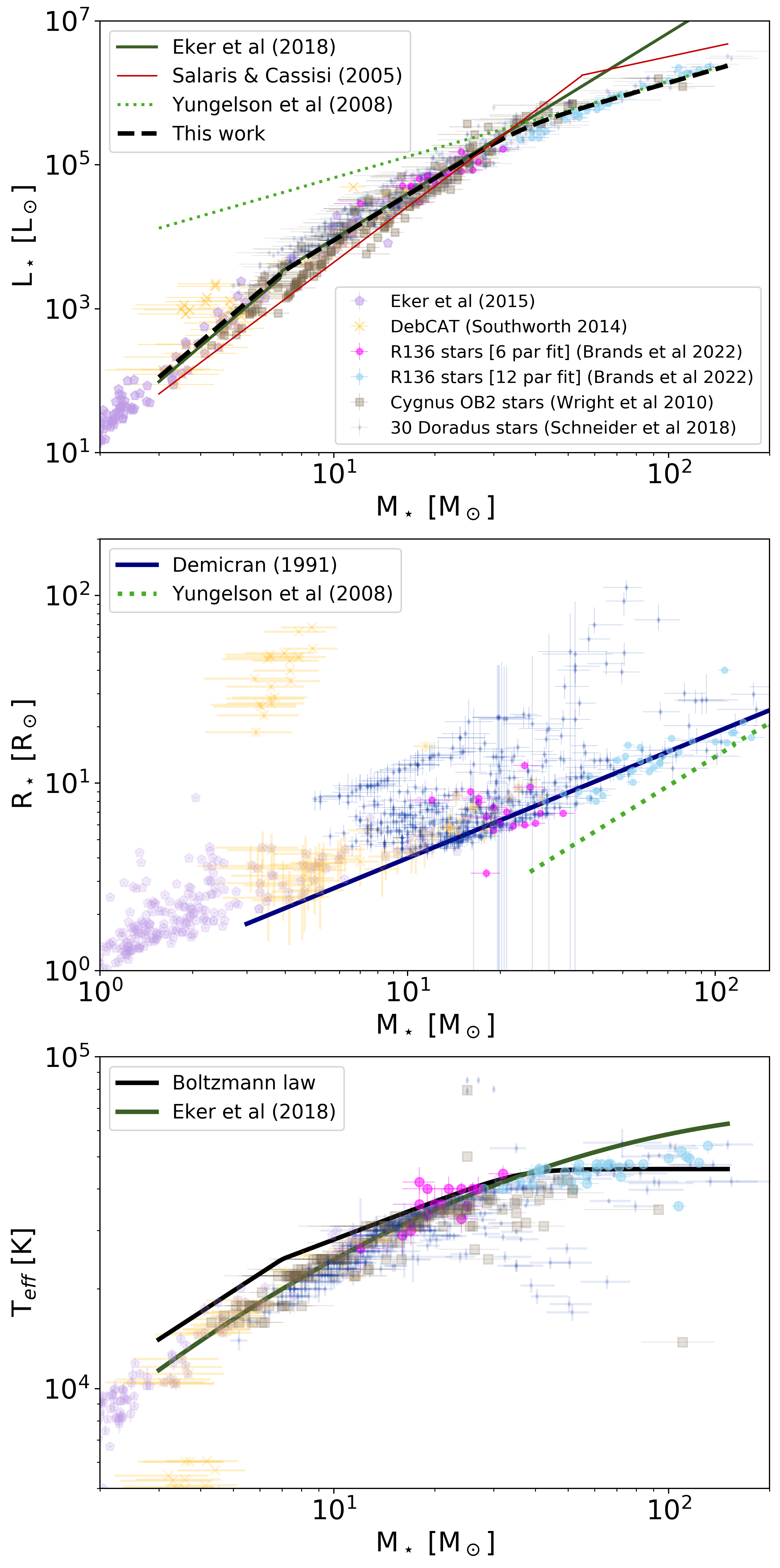}
\caption{\textit{Top panel}: Mass-luminosity relationship used in this work (Eq.~\ref{eq:MergedMLR}) (dashed black line) together with the relations provided by \cite{Eker_MLR_2018} and \cite{Yungelson_EvolutionFateMassiveStars_2008}. For comparison, we also plot the well known MLR given by \cite{Salaris_StarPopEvol_2005}. \textit{Central panel}: Mass-radius relationship used in this work \citep{Demircan_StarsMRR_1991} compared to the MRR provided by \cite{Yungelson_EvolutionFateMassiveStars_2008}. \textit{Bottom panel}: Comparison of empirical mass-temperature relation given by \cite{Eker_MLR_2018} with the relation obtained via Boltzmann's law (Eq.~\ref{eq:BoltzmannLaw}). In all panels the various data points report observed stellar properties and computed masses for several stars in different clusters, such as Cyg OB2 \citep{Wright_MassiveStarPopCygOB2_2015}, R136 \citep{Brands_R136_2022}, and 30 Doradus \citep{Schneider_30Dor_2018}. We also include stars from the DebCAT catalog \citep{Southworth_DebCAT_2015} and the previous work on MLRs of \cite{Eker_MLR_2015}.}
\label{fig:MLRTR}
\end{center}
\end{figure}

Even though empirical mass-temperature relations (MTRs) exist in the literature, once the MLR and MRR are given, we can readily calculate stars effective temperatures using Boltzmann law:
\begin{equation}
\label{eq:BoltzmannLaw}
T_{\rm eff}=\left [\frac{L_\star(M_\star)}{4 \pi R_{\star}(M_\star)^2 \sigma_b} \right ]^{1/4} \ .
\end{equation}
The bottom panel of Fig.~\ref{fig:MLRTR} shows the difference between the empirical relation given by \cite{Eker_MLR_2018}, and the resulting MTR obtained from Eq.~\ref{eq:BoltzmannLaw}. For comparison, we also display the mass and temperature of massive stars. Once again, we consider the same set of stars extracted from catalogs that are used for the study of the MLR. With our choice of MLR and MRR, Eq.~\ref{eq:BoltzmannLaw} is consistent with observed data, although, at masses below 10 M$_\odot$, the temperature is slightly overestimated (of the order of few tens percent). Note that for $M_\star \gtrapprox M_{b2}$, the effective temperature becomes independent from the mass. This is a direct consequence of the specific combination of the power index of our chosen MLR and MRR. From Eq.~\ref{eq:MergedMLR} we have that $L_\star (M_\star \gg M_{b2}) \propto M_\star^{1.34}$, while from Eq.~\ref{eq:MRR}, $R_\star \propto M_\star^{0.67}$. As from Eq.~\ref{eq:BoltzmannLaw} one has that $T_{\rm eff} \propto L_\star^{1/4} R_\star^{-1/2}\propto M_\star^0$, for $M_\star \gg M_{b2}$, the resulting temperature is independent of the stellar mass. 

%\begin{figure}[ht]
%\begin{center}
%\includegraphics[width=0.8\textwidth]{FIGURES/Chapter_3/MTR.png}
%\caption{Comparison of empirical mass-temperature relation given by \cite{Eker_MLR_2018} with the relation obtained via Boltzmann's law (Eq.~\ref{eq:BoltzmannLaw}). Temperatures and masses of different stars obtained from various catalogs (see caption of Fig.~\ref{fig:MLR} for references) are also plot for comparison.}
%\label{fig:MTR}
%\end{center}
%\end{figure}

\subsection{Modeling stellar winds}
Once the stellar parameters are known, we can readily estimate the wind speed, mass loss rate, and wind power for each star. From these we can afterward evaluate the cluster wind properties. We here use for the wind speed ($v_{\star, \infty}$) and wind luminosity ($L_{\star, w}$) the same expressions implemented in \S~\ref{subsec:CygOB2LwMdot}, that are Eq.~\ref{eq:Vinfty} and Eq.~\ref{eq:LwindStar} respectively, which we again report below for convenience:
\[
v_{\star, \infty}=C(T_{\rm eff})\left [\frac{2 G M_\star (1-\Gamma)}{R_{\star}} \right ]^{1/2}
\]
\[
L_{\star, w} = \frac{1}{2} \dot{M_\star} v_{\star, \infty}^2 .
\]
For what concerns the mass loss rate ($\dot{M}_{\star}$), we use a different recipe from that of \cite{Vink_OBwind_2000} and \cite{Yungelson_EvolutionFateMassiveStars_2008}, employed for the case of Cygnus OB2. This is because both expression are expected to be accurate only for very massive star, while here we need a prescription valid for a broader range of masses. The empirical formula provided by \cite{Vink_OBwind_2000} works only for stars with $27500$ K$<T_{\rm eff}<50000$ K, which, considering Eq.~\ref{eq:BoltzmannLaw}, implies stars with masses larger than $\sim 20$ M$_\odot$. Similarly, the expression provided by \cite{Yungelson_EvolutionFateMassiveStars_2008} is in principle valid for $M_\star > 60$ M$_\odot$, although the result obtained including also stars of mass about to 20 M$_\odot$ is approximately correct within a factor of a few, as demonstrated in \S~\ref{subsec:CygOB2LwMdot}. 

Given our plan to consider also less massive stars, we use the expression provided by \cite{Nieuwenhuijzen_Mdot_1990}, which reads:
\begin{equation}
\label{eq:MdotNieu}
\log \left( \frac{\dot{M_\star}}{\rm M_\odot yr^{-1}} \right) = -14.02+ 1.24 \log \left( \frac{L_\star}{\rm L_\odot} \right) + 0.16 \log \left( \frac{M_\star}{\rm M_\odot} \right) + 0.81 \left( \frac{R_\star}{\rm R_\odot} \right)
\end{equation}
Eq.~\ref{eq:MdotNieu} is valid for stars with $T_{\rm eff}>5000$ K, hence it is adequate for all the considered mass range. 

\begin{figure}[H]
\begin{center}
\includegraphics[width=0.8\textwidth]{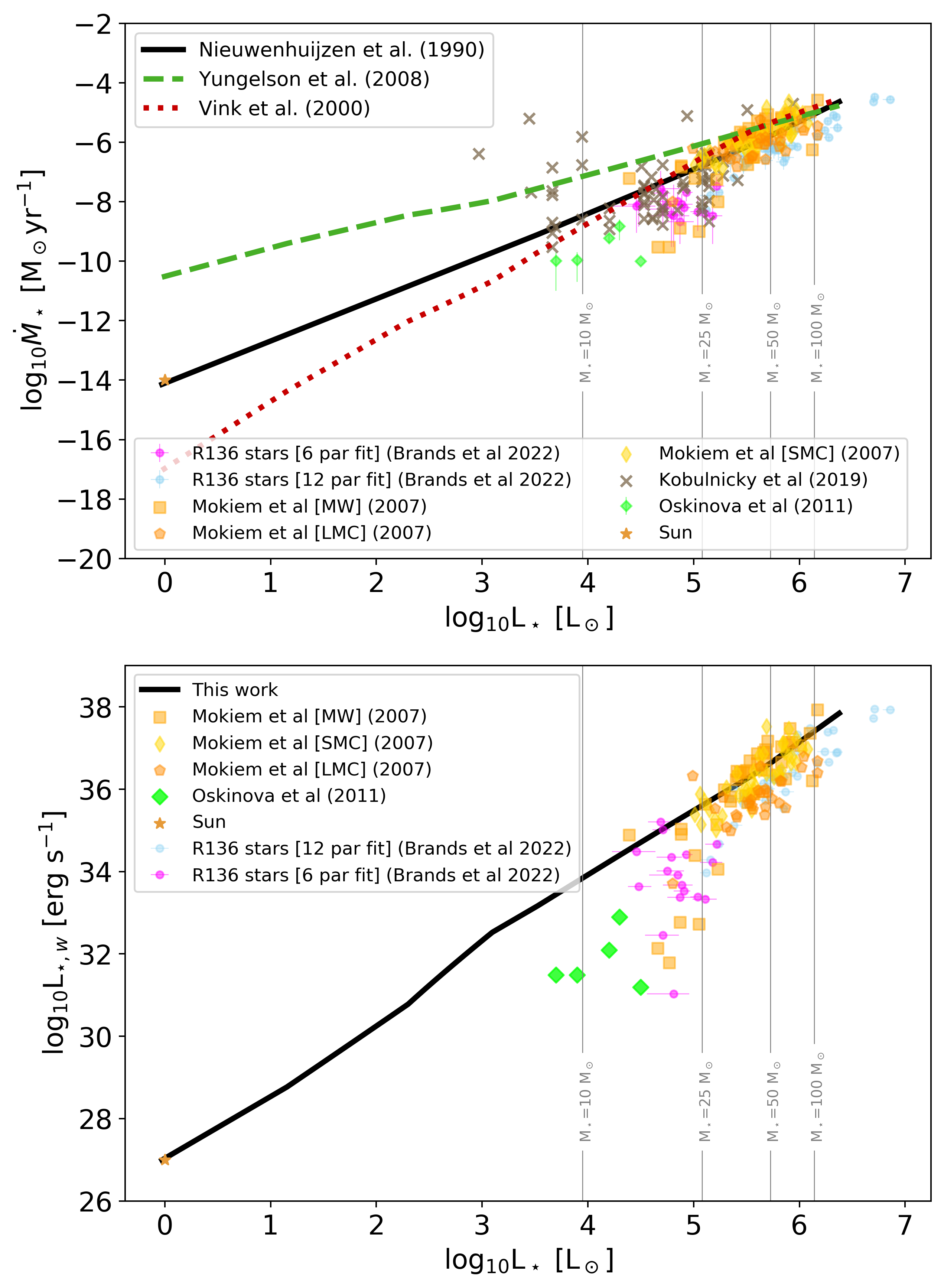}
\caption{\textit{Top panel}: Expected stellar mass loss rates as a function of $L_\star$ compared with observations \citep{Mokiem_MdotMetallicity_2007, Brands_R136_2022}. The Sun mass loss rate is reported by \cite{Carroll_IntroModAstro_1996}. The continuous line shows the relation used in this work, while dashed and dotted lines are the mass loss rates provided by \cite{Yungelson_EvolutionFateMassiveStars_2008} and \cite{Vink_OBwind_2000} respectively, which have been used in \S~\ref{subsec:CygOB2LwMdot}. \textit{Bottom panel}: Wind power as a function of stellar luminosity computed using the mass loss rate given by \cite{Nieuwenhuijzen_Mdot_1990}. The solar wind luminosity is calculated considering a wind speed of 400 km s$^{-1}$ \citep{Carroll_IntroModAstro_1996}.}
\label{fig:StarsLwMdot}
\end{center}
\end{figure}

As a final consistency check, we compare stellar wind luminosity and mass-loss rates with data available in the literature. \cite{Brands_R136_2022} provide these data for a subsample of stars in the YMSC R136. Similarly, \cite{Mokiem_MdotMetallicity_2007} provide them for a set of stars located both in the Milky Way and in the Small and Large Magellanic Clouds. Fig.~\ref{fig:StarsLwMdot} shows the result of this comparison. In this regard, it is interesting to note the following: first, the expression (Eq.~\ref{eq:MdotNieu}) used for $\dot{M}_\star$ is in good agreement with the data. Second, for masses below $\sim 25$ M$_\odot$ the stars wind luminosity appears to be highly overestimated, by almost 1\--2 orders of magnitude on average. However, we expect that this error will not significantly affect the final result. In fact, even if heavily overestimated, the contribution to the cluster wind luminosity of tens of stars under 25 M$_\odot$ is negligible compared to the wind power of a single star of mass greater than $\sim 30$ M$_\odot$.

%---------------------------------------------------------

%---------------------------------------------------------
\section{Generating a synthetic population of YMSCs}
In order to simulate a population of stellar clusters, the core ingredient from which one must start is the \textit{cluster distribution function}:
\begin{equation}
\xi_{SC}(M, t, r, \theta)=\frac{dN}{dM dt dr d\theta}
\end{equation}
defined such that the total number of clusters in the Milky Way with masses ranging in a given interval [$M_{min}$, $M_{max}$], and age [$t_{min}$, $t_{max}$] is:
\begin{equation}
\label{eq:NYMSC}
N_{SC}=\int_{M_{min}}^{M_{max}} \int_{t_{min}}^{t_{max}} \int_0^{R_{MW}} r \xi_{SC}(M, t, r)dM dt dr d\theta
\end{equation}
where $R_{MW}$ is the Milky Way radius. The true form of $\xi_{SC}$ is not known, however, assuming that the cluster distribution is factorized in mass, time, and space, $\xi_{SC}$ can be written as:
\begin{equation}
\xi_{SC}(M, t, r)= f(M)\psi(t) \rho(r, \theta)
\end{equation}
where $f(M)$, $\psi(t)$, and $\rho(r, \theta)$ are the cluster initial mass function, the cluster formation rate and the cluster spatial distribution respectively. In the next subsections, we discuss each single term separately. 

\subsection{Cluster age and mass distribution}
It is possible to infer both the cluster formation rate and cluster mass function from observations. One of the seminal works for the study of these two functions is the analysis done by \cite{Piskunov_GalSCGlobSurvIV_2018} on the Milky Way Star Cluster Survey (MWSCS). \cite{Piskunov_GalSCGlobSurvIV_2018} used 2242 stellar clusters from this survey for their analysis, all within 2.5 kpc from the Sun. Actually, the sample can be considered complete only up to 1.8 kpc. Hence the result from \cite{Piskunov_GalSCGlobSurvIV_2018} should be considered as local. 

In their paper, \cite{Piskunov_GalSCGlobSurvIV_2018} model the cluster initial mass function as a broken power law:
\begin{equation}
\label{eq:ClusterIMF}
f(M) = \frac{dN}{dM} = 
\begin{cases}
 k_1 M^{-(x_1+1)} & \text{for } M_{min} \leq M \leq M_b \\
k_2 M^{-(x_2+1)} & \text{for }  M_b \leq M \leq M_{max}
\end{cases}
\end{equation}
where $M_b=100$ M$_\odot$ is the mass at which the break occurs, and $x_1$ and $x_2$ are parameters used to fit the observed cluster mass distribution. Note that Eq.~\ref{eq:ClusterIMF} coincides with the observed cluster mass distribution only in the case of young clusters. As explained in \S~\ref{subsec:YMSCBirth}, some of the clusters may not survive the initial stage of gas expulsion triggered by the appearance of stellar feedback mechanisms. This eventually ends up affecting the mass distribution of currently observed clusters, hence, in order to obtain the initial cluster mass function, one has to account for this effect. Numerical N-body simulations show that the cluster survival depends on the percentage of gas filling the cluster Roche lobe \citep{Ernst_RocheLobeSC_2015}. When it comes to inferring $f(M)$, \cite{Piskunov_GalSCGlobSurvIV_2018} consider three different scenarios associated with underfilled, filled, and overfilled Roche lobes. Ultimately, this produces differs values for the parameters $x_1$ and $x_2$. The scenario that best agrees with the observations is the underfilled case, for which $x_1=0.39$ and $x_2=0.54$. Finally, $k_1$ and $k_2$ are two constants obtained by requiring both continuity at $M_b$ and normalization of the distribution $ \int_{M_{min}}^{M_{max}} f(M) dM = 1$, which are calculated as:
\begin{subequations}
\label{eq:k1k2}
\begin{equation}
k_2 = \left [ M_b^{(x_1-x_2)} \int_{M_{min}}^{M_b}M^{-(x_1+1)}dM - \int_{M_b}^{M_{max}}M^{-(x_2+1)}dM \right ]^{-1}
\end{equation}
\begin{equation}
k_1 = k_2 M_b^{(x_1-x_2)} \ .
\end{equation}
\end{subequations}
In the work carried out by \cite{Piskunov_GalSCGlobSurvIV_2018}, the minimum and maximum stellar cluster mass are fixed to $M_{min}=2.5$ M$_\odot$ and $M_{max}=6.3\times 10^4$ M$_\odot$ respectively. To be consistent with \cite{Piskunov_GalSCGlobSurvIV_2018}, we also fix $M_{min}$ to the same value, which should correspond to binary systems or small brown dwarf aggregates. However, it is crucial to stress that when we will calculate the number of Galactic massive stellar clusters using Eq.~\ref{eq:NYMSC}, the minimum considered mass $M_{min}$ will be different from $2.5$ M$_\odot$. The choice of $M_{min}$ has a certain degree of arbitrariness. In \S~\ref{sec:DefAndTerm}, we defined as massive all those clusters with masses greater than 1000 M$_\odot$. In general, the choice of $M_{min}$ is made by considering a reasonably small mass to produce enough massive stars to make the existence of a collective cluster wind possible. We will check a posteriori in \S~\ref{sec:PopYMSCasCR} if this choice is reasonable. 

\begin{figure}[ht]
\begin{center}
\includegraphics[width=0.8\textwidth]{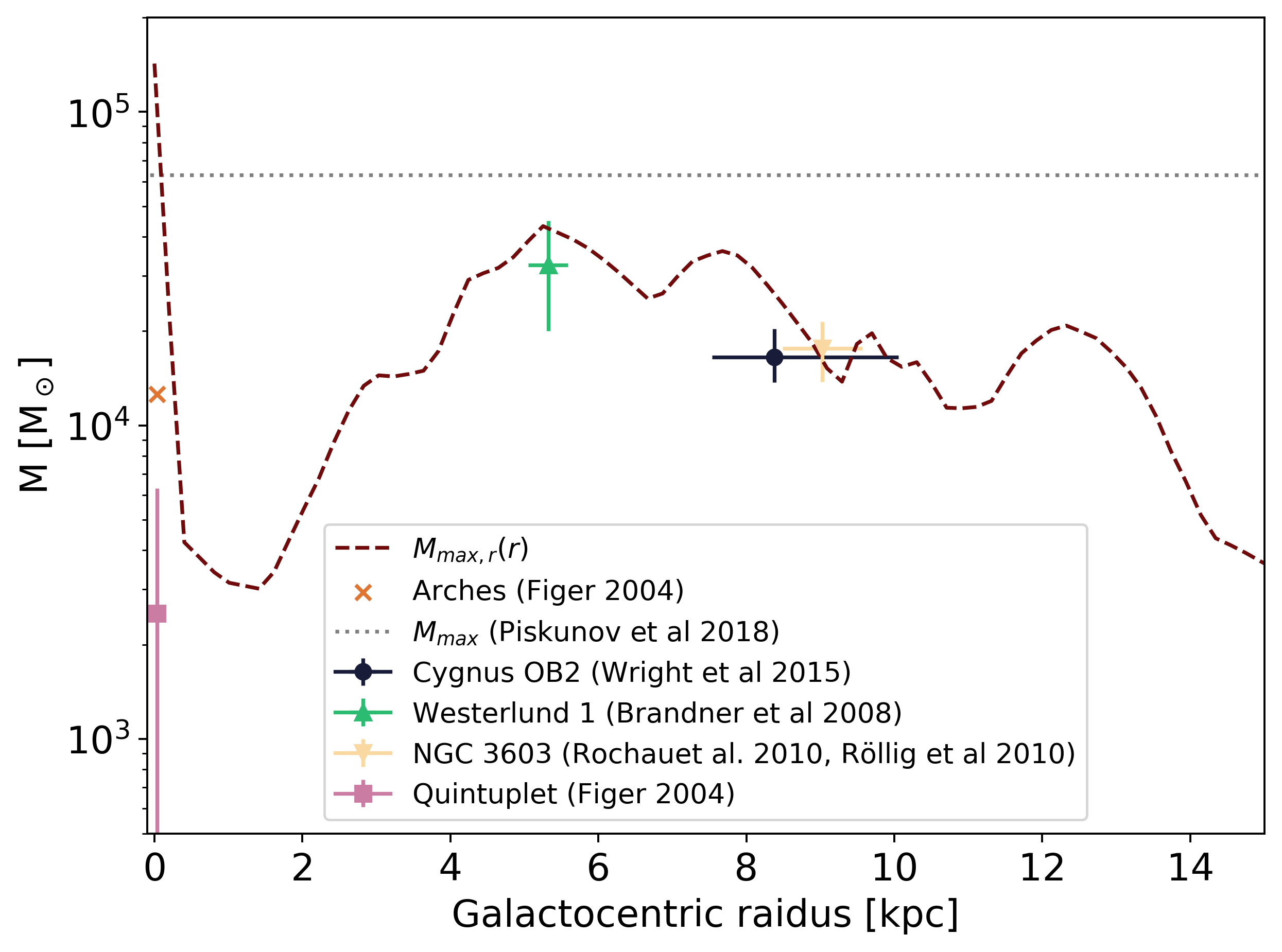}
\caption{Maximum cluster mass as a function of the galactocentric radius (dashed line). The normalization of $M_{max, r}$ at the Galactic Center is calculated using Eq.~\ref{eq:YMSCsMmax}. For comparison purposes, we show the mass of several of the most massive YMSCs, such as Cygnus OB2 \citep{Wright_MassiveStarPopCygOB2_2015}, Westerlund 1 \citep{Brandner_Wd1Mass_2008}, NGC 3603 \citep{Rochau_NGC3603Mass_2010, Rollig_NGC3603MassAndDist_2011}, Archer and Quintuplet \citep{Figer_ArchesQuintupletMass_2004}. The maximum mass used by \cite{Piskunov_GalSCGlobSurvIV_2018} is also reported (dotted line).}
\label{fig:YMSCMmaxVsR}
\end{center}
\end{figure}

The choice of $M_{max}$ in \cite{Piskunov_GalSCGlobSurvIV_2018} is dictated by the most massive star cluster observed in the Milky Way. However, observations of Milky Way star clusters are biased by extinction. Hence, we decided to make a consistency check based on observations of other closeby galaxies. Observational pieces of evidence show how the luminosity of YMSCs in spiral and dwarf galaxies correlates with the measured value of the star formation rate (SFR) in their host galaxy \citep{Weidner_YMSCsMaxM_2004}: the higher the SFR, the higher the luminosity of the brightest (hence the most massive) YMSCs. The inferred relation between the SFR and $M_{max}$ is \citep{Weidner_YMSCsMaxM_2004}:
\begin{equation}
\label{eq:YMSCsMmax}
M_{max}=k_{ML} \left( \frac{SFR}{1 \rm \ M_\odot yr^{-1}} \right)^{0.75 \pm 0.03} \times 10^{6.77 \pm 0.02} \ \rm M_\odot
\end{equation}
where $k_{ML}$ is the cluster mass-to-light ratio which depends on the cluster age. For clusters with ages less than 10 Myr one has $k_{ML}=0.0144$ \citep{Smith_YMSCskML_2001}. Considering that the Milky Way SFR is $\sim 2$ M$_\odot$ yr$^{-1}$ \citep{Elia_MWSFR_2022}, the expected maximum mass is $M_{max}\approx 1.4 \times 10^5$ M$_\odot$. Note that this value must be interpreted as an upper limit on a Galactic scale. However in different places of the Milky Way, it is reasonable to think that the maximum mass of a YMSC correlates with the total amount of gas. Eventually, the work performed by \cite{Pflamm_YMSCsMmaxVsR_2008} has proven this relation. In their paper, \cite{Pflamm_YMSCsMmaxVsR_2008} show how the H$\alpha$ emission in disk galaxies (tracing the population of short-lived massive stars) is characterized by a cut-off at some galactocentric distances, which can be smoothly explained by introducing a cluster mass function with $M_{max, r}$ dependent on the total gas surface density:
\begin{equation}
\label{eq:YMSCMmaxVsR}
M_{max, r}(r)=M_{max} \left[\frac{\Sigma_{gas}(r)}{\Sigma_{gas}(r=0)} \right]^\delta
\end{equation}
where $\delta=3/2$ that can be inferred by observations. Fig.~\ref{fig:YMSCMmaxVsR} shows the evaluation of $M_{max, r}(r)$. We estimate $M_{max, r}(r)$ considering the radial gas distribution implemented by \cite{Strong_GalpropDiffGamma_2000} in the GALPROP code \citep{Strong_Galprop_2009}. The radial gas profile accounts for both molecular and atomic gas phases. Interestingly, the maximum masses at different galactocentric radii are perfectly consistent with the measured mass of some of the most massive Galactic YMSCs. One can see that for radii larger than 2 kpc, the maximum mass is roughly constant, with variations by a factor of a few. 

It is worth underlining that if we consider a radius-dependent maximum mass, Eq.~\ref{eq:ClusterIMF} must necessarily be modified by replacing $M_{max}$ with $M_{max}(r)$. This will ultimately produce a radial dependence of the parameters $k_1$ and $k_2$ which can be then obtained by requiring the normalization of $f(M)$ at each galactocentric radius. Clearly, by doing so, $f(M)$ will become dependent on $r$, making invalid the hypothesis of factorization for $\xi_{SC}$. Given that the maximum mass does not vary much with galactocentric distance, we will use, for the rest of the work, the approximation of a constant mass equal to the value adopted by \cite{Piskunov_GalSCGlobSurvIV_2018} ($M_{max}=6.3\times 10^4$ M$_\odot$).

As stated at the beginning of this section, \cite{Piskunov_GalSCGlobSurvIV_2018} provide also an expression for the cluster formation rate $\psi (t)$ inferred from the present population of local clusters. Comparison with data shows that several functional forms are allowed. A useful form adopted by \cite{Piskunov_GalSCGlobSurvIV_2018} is:
\begin{equation}
\label{eq:ThCFR}
\psi(t)=\left [ A+B \exp \left(C \frac{T_p-T_f}{T_p} \right) \right ] \rm \ Myr^{-1} kpc^{-2}
\end{equation}
where $T_p=4.8$ Gyr is the present time, considering as initial reference point the age of the oldest observed cluster, and $T_f$ is the time at which the cluster was formed, such that the age of the cluster can be calculated as $t=T_p-T_f$. Eq.~\ref{eq:ThCFR} is such that the present SFR is given by $\psi_{now}=A+B$, while SFR at $t=T_p$ is $\psi_0=A+B e^C$. The coefficients $A$, $B$ and $C$ are obtained by fitting the observed cluster age distribution. Similarly to $x_1$ and $x_2$ in $f(M)$, the coefficients $A$, $B$, and $C$ also depend on the clusters capability to survive. Still considering the previously mentioned underfilled Roches lobe case, the values of $A$, $B$ ad $C$ are -0.55, 0.57, and 1 respectively. Using Eq.~\ref{eq:ThCFR} one can readly see that the cluster formation rate is practically constant in the last 10 Myr, as the value of $\psi(t)$ changes only by a few percent in the 1--10 Myr range:
\[
\frac{\psi (T_p-10\ \textrm{Myr})-\psi(T_p)}{\psi(T_p)} \simeq 0.06
\]
This is also well confirmed by observations, as shown in Fig. 1 of \cite{Piskunov_GalSCGlobSurvIV_2018}. Hence, we can consider the cluster formation rate as constant in the last 10 Myr.

The average value of the cluster formation rate ($\bar{\psi}$) is a crucial parameter as in the end it will be the one setting the normalization of $\xi_{SC}$. In order to determine the value of $\bar{\psi}$, we rely on the work by \cite{Lamers_SFRinSC_2006} who determined from observations the SFR in local clusters ($SFR_{SC}$) for masses between $M_-=100$ M$_\odot$ and $M_+=3 \times 10^4$ M$_\odot$. Such a rate is $\sim 350$ M$_\odot$ Myr$^{-1}$ kpc$^{-2}$. The result of \cite{Lamers_SFRinSC_2006} is also compatible with a more recent estimate done by \cite{Bonatto_SFRinSC_2011}, who found a SFR in local clusters of $790\pm 160$ M$_\odot$ Myr$^{-1}$ kpc$^{-2}$ after considering clusters within a mass range of $M_-=10$ M$_\odot$ and $M_+=7.5 \times 10^4$ M$_\odot$. Starting from $SFR_{SC}$, it is possible to infer $\bar{\psi}$ as:
\begin{equation}
\label{eq:PsiBarFromSFR}
\bar{\psi}= \frac{SFR_{SC}}{\int_{M_-}^{M_+} M f(M) dM} \ .
\end{equation}
This leads to average cluster formation rates of $\bar{\psi}_{LG}\approx 1.3$ Myr$^{-1}$ kpc$^{-2}$ and $\bar{\psi}_{BB}\approx 1.8$ Myr$^{-1}$ kpc$^{-2}$ for the works of \cite{Lamers_SFRinSC_2006} and \cite{Bonatto_SFRinSC_2011} respectively. When simulating the Galactic population of YSMCs, we will consider the value $\bar{\psi}_{BB}$ because it is the most recent one (however, given the uncertainties, the two values are consistent with one another).

\subsection{Cluster spatial distribution}
\label{subsec:YMSCGalDistr}
In the previous section we have determined the cluster formation rate in the solar neighborhood. However, the cluster formation rate is expected to vary across the Galactic disk according to the density of giant molecular clouds (GMC) from which stellar clusters originate. It is then reasonable to presume that the distribution of GMCs well traces the Galactic position of YSMCs, which in turn closely follow the spiral arm structure of the Milky Way. We spatially distribute the synthetic stellar clusters following a two-step procedure:
\begin{enumerate}
\item We start by generating stellar clusters with a galactocentric radial distribution following that of GMCs (under the assumption of an isotropic angular distribution) and assuming an exponential altitude distribution similar to the observed gas profile.
\item Afterward, based on its radial and angular position, we associate every synthetic YMSC to a specific Galactic structure, i. e. spiral arm, galactic bar, etc.   
\end{enumerate}
For the radial distribution of GMCs and the modeling of the Milky Way spiral structure, we rely on the materials and results of the work by \cite{Hou_MWStructure_2014}, who fitted spiral arms models simultaneously using the distribution of observed HII regions, masers, and GMCs.

\cite{Hou_MWStructure_2014} supply a complete catalog of GMCs, which reports, for each cloud, the galactic position (along with kinematic distance), kinematic velocity, size and mass. For a large portion of the GMCs, the kinematic ambiguity is resolved. We then recompute the kinematic distance of every cloud with disentangled kinematic ambiguity using the code developed by \cite{Wenger_KinDistCode_2018}, which calculates distances using two possible approaches: a Montecarlo method or the conventional estimate using the Galactic rotation curve (see \S~\ref{subsec:CygOB2ISM}). We adopt the classical rotation curve based method using the state of the art curve provided by \cite{Reid_GalRotCurve_2019}. A small fraction of GMCs in the catalog also have distances calculated using more reliable methods such as parallax. For those specific cases, the latter value is used.

\begin{figure}[ht]
\begin{center}
\includegraphics[width=0.8\textwidth]{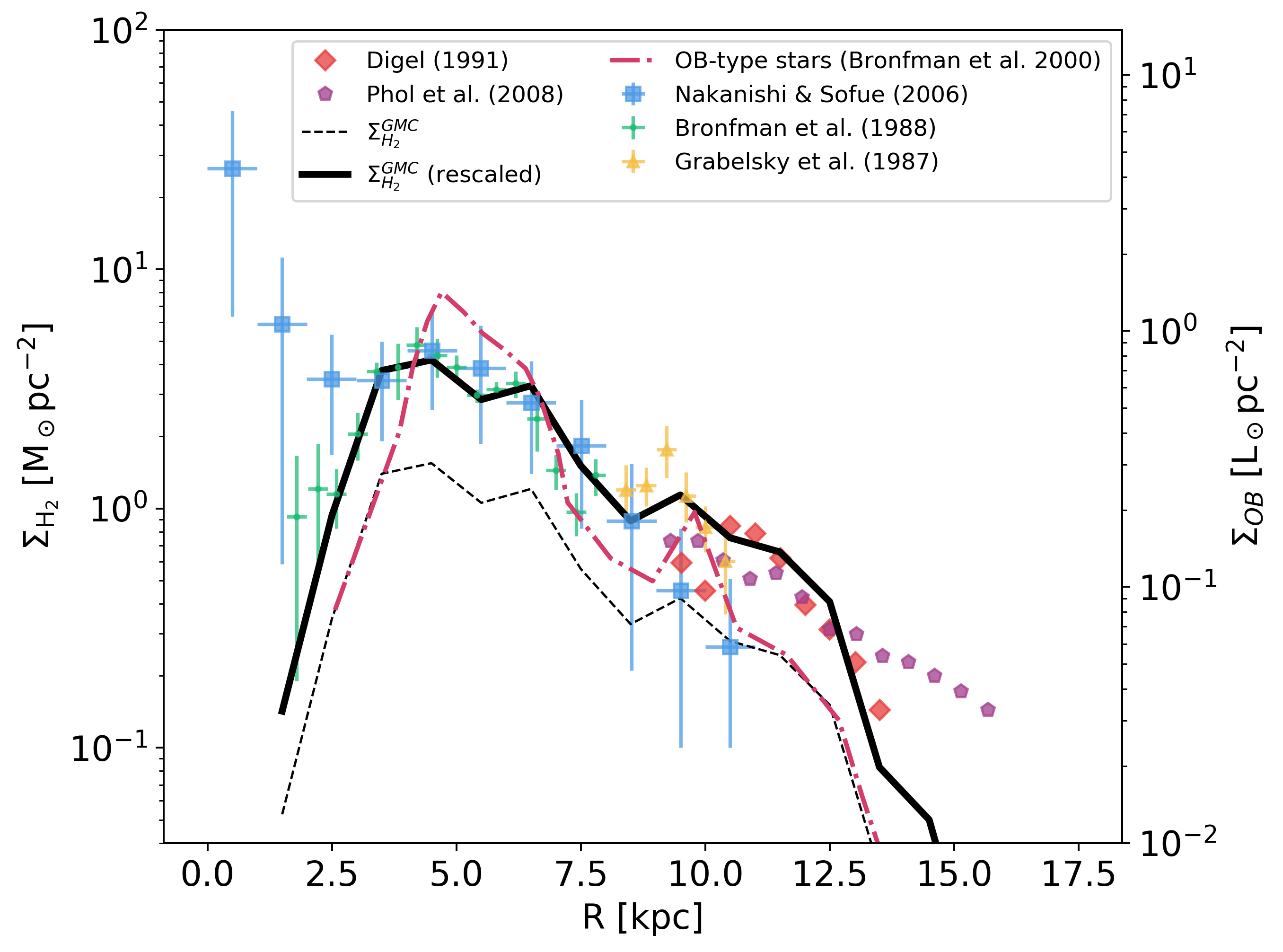}
\caption{Comparison of surface mass density of molecular gas enclosed in GMC (thin dashed line), calculated using \cite{Hou_MWStructure_2014} catalog, and mass density of diffuse molecular gas (see text for the references). The thick solid line is the surface mass density rescaled for comparison purposes. The dash-dotted line referring to the right axis is the FIR surface luminosity calculated from embedded massive stars \citep{Bronfman_OBStarDistrib_2000}.}
\label{fig:GMCgasDistrib}
\end{center}
\end{figure}

Once the location is known, we can estimate the radial distribution of GMCs, which we express in terms of the surface mass density of molecular gas ($\Sigma_{H_2}^{GMC}$). To do so, we average the total GMCs mass in 18 rings with constant width of 1 kpc and radii spanning in the interval 0--18 kpc, and centered on the Galactic center. The result is shown in Fig.~\ref{fig:GMCgasDistrib}, where the surface mass density of molecular material in GMCs is compared to several measurements of the diffuse molecular gas distribution. Note that the overall profile is remarkably compatible with the results from different works \citep{Grabelsky_MCinCarina_1987, Bronfman_MolGasSolCir_1988, Digel_MCinOutGal_1991, Nakanishi_H2MWDistrib_2006, Pohl_MW3DMolGasDistrib_2008}, except for two aspects. First, the normalization, which is found to be lower by a factor $\sim 3$. This is expected as we are considering only molecular gas within GMC, while the other results account for all the diffuse molecular gas. Second, the trend towards the Galactic Center, where data by \cite{Nakanishi_H2MWDistrib_2006} indicate an increase of the density, which in parallel seems not consistent with the findings of \cite{Bronfman_MolGasSolCir_1988}. Nevertheless, the profile is also in good agreement with the radial distribution of far-infrared emission detected from embedded OB stars \citep{Bronfman_OBStarDistrib_2000}, emphasizing the robustness of the hypothesis that YMSCs and molecular gas.

At this point we can formally define the radial distribution of YMSCs as:
\begin{equation}
\label{eq:YMSCRadDistr}
\rho(r)=\frac{\Sigma_{H_2}^{GMC}(r)}{\Sigma_{H_2}^{GMC}(r=8.5 \rm \ kpc)}
\end{equation}
Note that by doing so, the normalization of $\xi_{SC}$ at the sun position is given by the local observed cluster formation rate $\bar{\psi}_{LG}$, as $\rho(r)$ is normalized at the Sun position.  

\begin{figure}[ht]
\begin{center}
\includegraphics[width=0.8\textwidth]{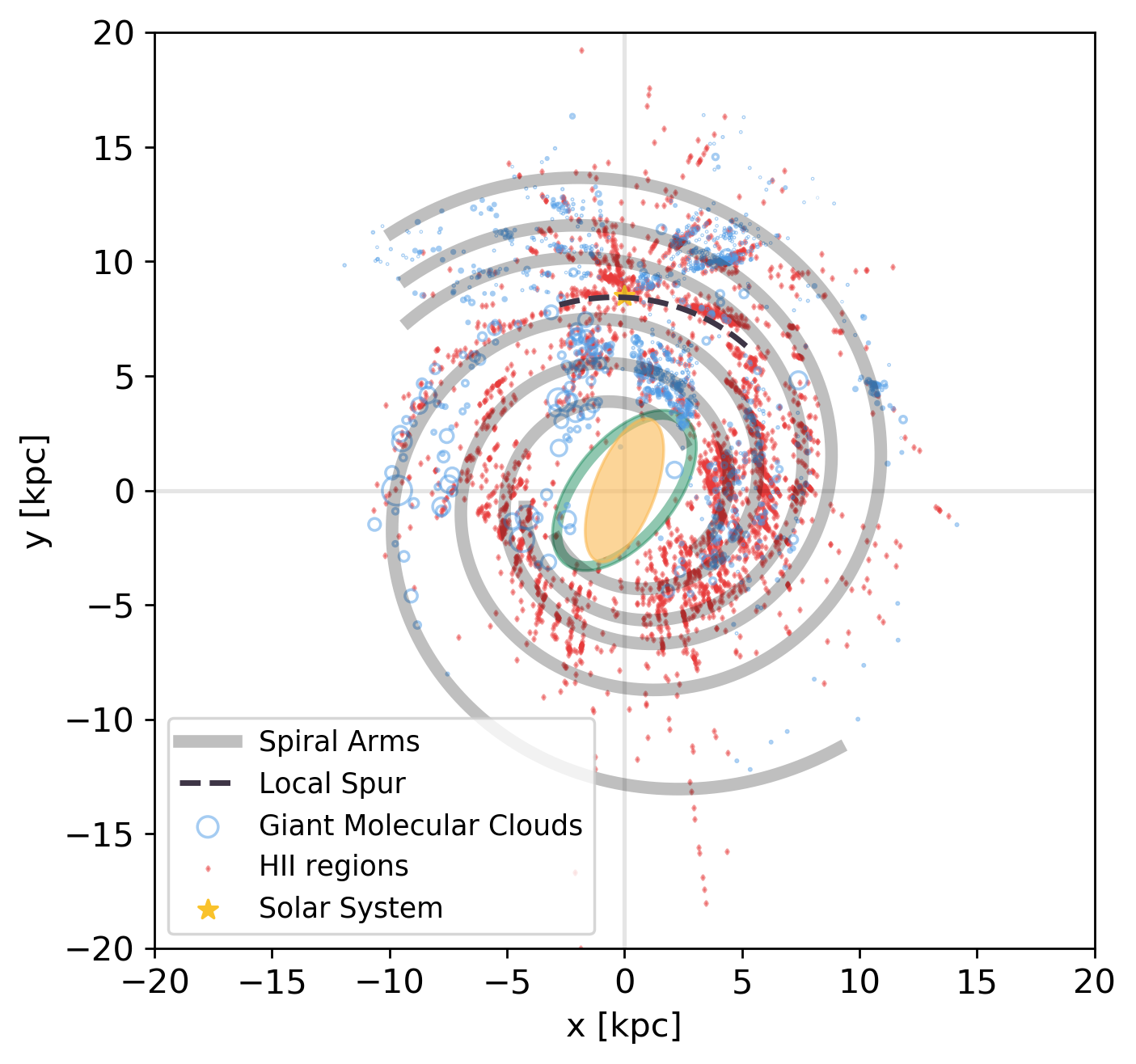}
\caption{Milky Way structure used in this work. The spiral arms (thick continuous gray lines) are modeled following the work of \cite{Hou_MWStructure_2014}. The dashed black line represents the Local Spur. Yellow and green ellipses represent the Galactic Bar \citep{Churchwell_GalBarGlimpse_2009} and the Near and Far 3kpc arms \citep{Green_NF3kpcArms_2011}, respectively. Also shown in the overlay are the positions of the GMCs (blue points, with sizes proportional to the GMC mass) and HII regions (red points) from the catalog provided by \cite{Hou_MWStructure_2014}. The yellow star marks the position of the Solar System.}
\label{fig:MWSpiralStruct}
\end{center}
\end{figure}

Once the radial distribution is known, we allocate stellar clusters following the Milky Way observed morphology\footnote{In the work presented, as we will describe in Section 3, we will only consider two regions of the Galactic plane. However, we still decided to simulate the distribution of clusters rigorously over the entire galaxy, especially in view of future more detailed work.}. Unfortunately, the Milky Way spiral structure still remains nowadays a matter of debate. Within their work, \cite{Hou_MWStructure_2014} model the Galaxy by considering two possible scenarios, one containing three and one containing four spiral arms. They furthermore consider two different functions for the spiral arms: logarithmic and polynomial-logarithmic functions. Here we use the 4-arm model with simple logarithmic function as it is on average the case returning the best fit to the spiral arm tracers. The logarithmic spiral arm, are defined as:  
\begin{equation}
\label{eq:LogSpiral}
\ln \left( \frac{r}{R_i} \right) = \left(\frac{\theta-\theta_i}{1 \rm \ rad} \right) \tan \Psi_i
\end{equation}
where $R_i$,$\theta_i$, and $\Psi_i$ are parameters inferred from the fit procedure to the position of HII regions, masers and GMCs. In addition to the 4 spiral arms, we also include the Local Spur. Table~\ref{tab:SpiralArmPar} shows the numerical values of the parameters in Eq.~\ref{eq:LogSpiral}.
\begin{table}
\begin{center}
\begin{tabular}[c]{l c c c c c}  
\toprule \toprule
Parameters  & Arm 1 & Arm 2 & Arm 3 & Arm 4 & Local Spur \\
\midrule
R$_i$ [kpc] & 3.27 & 4.29 & 3.58 & 3.98 & 8.16\\
$\Psi_i$ [$^\circ$] & 9.87 & 10.51 & 10.01 & 8.14 & 2.71\\
$\theta_i$ [$^\circ$] & 38.5 & 189 & 215.2 & 320.1 & 50.6\\
\bottomrule %\bottomrule
\end{tabular}
\caption{Parameters values used to calculate Milky Way spiral arms \citep{Hou_MWStructure_2014}.}
\label{tab:SpiralArmPar}
\end{center}
\end{table}  
On top of the spiral structure, we also take into account the structure of the innermost region. Here we consider the presence of the Galactic bar and the Near 3 kpc and Far 3 kpc arms. The first is modeled as an ellipse with an aspect ratio of 10:4 (length:width) having a half-length of 3.3 kpc and an inclination angle of 70$^\circ$, defined counterclockwise with respect to the positive direction of the x-axis (or equivalently, 20$^\circ$ clockwise from the Galactic Center - Solar System connecting line, see Fig.~\ref{fig:MWSpiralStruct}) \citep{Churchwell_GalBarGlimpse_2009}. The Near 3 kpc and Far 3 kpc arms are similarly modeled using an ellipse, with a semi-major axis of 4.1 kpc, and an aspect ratio of 0.54 (a semi-minor axis of 2.2 kpc). The ellipse major axis orientation is 52$^\circ$ counterclockwise with respect to the positive direction of the x-axis \citep{Green_NF3kpcArms_2011}. Fig.~\ref{fig:MWSpiralStruct} shows the resulting structure composed with the observed position of different tracers used by \cite{Hou_MWStructure_2014}. The positions of HII regions are recalculated using the same approach employed for the GMCs. 
  
Let us now describe in detail the implemented process to allocate the synthetic YMSCs in the Galaxy. For every j-th cluster, we start by randomly generating its radial ($R_j$) and angular coordinates ($\theta_j$). Radial distances are extracted considering the probability distribution $\rho(r)$ given by Eq.~\ref{eq:YMSCRadDistr}, while the angular coordinate is chosen assuming a uniform distribution. Afterward, based on the values of R$_j$ and $\theta_j$, we associate the YMSC to a specific structure, following the criteria listed in Tab.~\ref{tab:MWDistribCriteria}. Namely, depending on the criterion met, we randomly select one specific structure among the included ones. The minimum value of $r$ allowed for each arm corresponds to the position where the arm connects to the inner region of the Milky Way.

\begin{table}[]
\begin{center}
\begin{tabular}{l|c|c|}
\cline{2-3}
                                    & $R$                      & $\theta$                                                                                                                                        \\ \hline
\multicolumn{1}{|l|}{Spiral Arm 1}  & $R \geq 3.27$ kpc        & \begin{tabular}[c]{@{}c@{}}$\theta<360^\circ$ \\ (or $50^\circ>\theta>110^\circ$\\  if $7.59\ \textrm{kpc} <R<9.17\ \textrm{kpc}$)\end{tabular} \\ \cline{2-3} 
\multicolumn{1}{|l|}{Spiral Arm 2}  & $R \geq 4.29$ kpc        & \begin{tabular}[c]{@{}c@{}}$\theta<360^\circ$ \\ (or $50^\circ>\theta>110^\circ$\\  if $7.59\ \textrm{kpc} <R<9.17\ \textrm{kpc}$)\end{tabular} \\ \cline{2-3} 
\multicolumn{1}{|l|}{Spiral Arm 3}  & $R \geq 3.58$ kpc        & \begin{tabular}[c]{@{}c@{}}$\theta<360^\circ$ \\ (or $50^\circ>\theta>110^\circ$\\  if $7.59\ \textrm{kpc} <R<9.17\ \textrm{kpc}$)\end{tabular} \\ \cline{2-3} 
\multicolumn{1}{|l|}{Spiral Arm 4}  & $R \geq 3.98$ kpc        & \begin{tabular}[c]{@{}c@{}}$\theta<360^\circ$ \\ (or $50^\circ>\theta>110^\circ$\\  if $7.59\ \textrm{kpc} <R<9.17\ \textrm{kpc}$)\end{tabular} \\ \cline{2-3} 
\multicolumn{1}{|l|}{Local Spur}    & $7.59$ kpc $<R<9.17$ kpc & $50^\circ \leq \theta \leq 110^\circ$                                                                                                           \\ \cline{2-3} 
\multicolumn{1}{|l|}{NF 3kpc / Bar} & $R< 4.29$ kpc            & $\theta<360^\circ$                                                                                                                              \\ \hline
\end{tabular}
\caption{Criteria in radius and angle implemented to choose which structure to associate a given YMSC with.}
\label{tab:MWDistribCriteria}
\end{center}
\end{table}

Note that the angular coordinate $\theta_j$ is used only to check whether the YMSC should be associated with the Local Spur. Once the cluster is placed in a given arm, the coordinate $\theta_j$ is recalculated by inverting Eq~\ref{eq:LogSpiral}:
\begin{equation}
\theta_j= \left(\frac{\tan \Psi_i}{1 \ \rm rad} \right )^{-1} \ln \left( \frac{R_j}{R_i} \right) + \theta_i \ .
\end{equation}
In case there is an association with Near 3 kpc and Far 3 kpc arms or the Galactic Bar, the association method is slightly more involved. To be precise, we first check which structure to associate the cluster with, following a minimum distance criterion. After that, if the cluster is associated with the Galactic Bar we check whether its position is actually located within the Bar. If not, $\theta_j$ is varied until its position falls within the Galactic Bar. Note that, given the specific geometry of the Near and Far arms and the shape of the Galactic Bar, there are certain areas where YMSCs end up being closer to the Bar (and thus associated with it) but there are no values of $\theta_j$ such that the cluster can be moved within the Bar. In this specific case, we associate the cluster with the Near 3 kpc and Far 3 kpc arms.

When a cluster is associated with the Near 3 kpc and Far 3 kpc arms, we change its coordinates so as to have them within the ellipse that defines these structures. We do so by replacing the cluster coordinates with the ones of the nearest point of the ellipse. It should be emphasized that this procedure induces a distortion in the starting radial distribution. However, the deviation found is totally negligible for the final result.

Once all clusters are associated we proceed to perturb their positions according to a Gaussian distribution, following the fundamental concept that both Spiral Arms and Near and Far structures possess an intrinsic thickness. For spiral arms, we extract a non-constant radial fluctuation, which increases with the galactocentric distance such that the probability $\mathcal{P}$ of having a certain fluctuation $\Delta R_j$ is \citep{Faucher_IsolatedRadioPulsar_2006}:
\begin{equation}
\mathcal{P}(\Delta R_j) = \frac{1}{2 \pi (0.07 R_j)} e^{- \frac{\Delta R_j^2}{2 (0.07 R_j)^2}} \ .
\end{equation}
As for the clusters belonging to the Near and Far arms, we extract from a gaussian probability distribution the galactic x and y coordinates with a spread of $\sigma_x = \sigma_y = 0.1667$ kpc, such that the probability at $3 \sigma$ returns a scattering compatible with the observed radial thickness of 0.5 kpc.

Finally, after having set the position of all YMSCs in the Galactic plane, we generate the vertical coordinate ($z$) following the observed gas distribution profile, i.e. an exponential distribution ($\rho(z)$) with a characteristic spread of 100 pc \citep{Strong_GalpropDiffGamma_2000}:
\begin{equation}
\rho(z)=\exp \left(- \frac{z}{100 \ \rm pc} \right )
\end{equation} 

\subsection{Wind luminosity of star clusters}
To model the wind from each YMSC, first, its stellar population must be generated following the description given in \S~\ref{sec:ModStarPop}. Knowing the wind luminosity ($L_{\star, i}$) and mass loss rate ($\dot{M}_{\star, i}$) of each i-th star of the cluster, then, the luminosity ($L_w$) and mass loss rate ($\dot{M}$) of the collective cluster wind are readily obtained as:
\begin{subequations}
\begin{equation}
L_w=\sum_i L_{\star, w}
\end{equation}
\begin{equation}
\dot{M}=\sum_i \dot{M}_{\star, i}
\end{equation}
\end{subequations}
With these parameters, one can also estimate the wind speed as:
\begin{equation}
v_w=\sqrt{\frac{2 L_w}{\dot{M}}}
\end{equation}

\subsection{Statistical properties of the YMSC synthetic population}
\label{sec:PopYMSCasCR}
\begin{figure}[h]
\begin{center}
\includegraphics[width=0.8\textwidth]{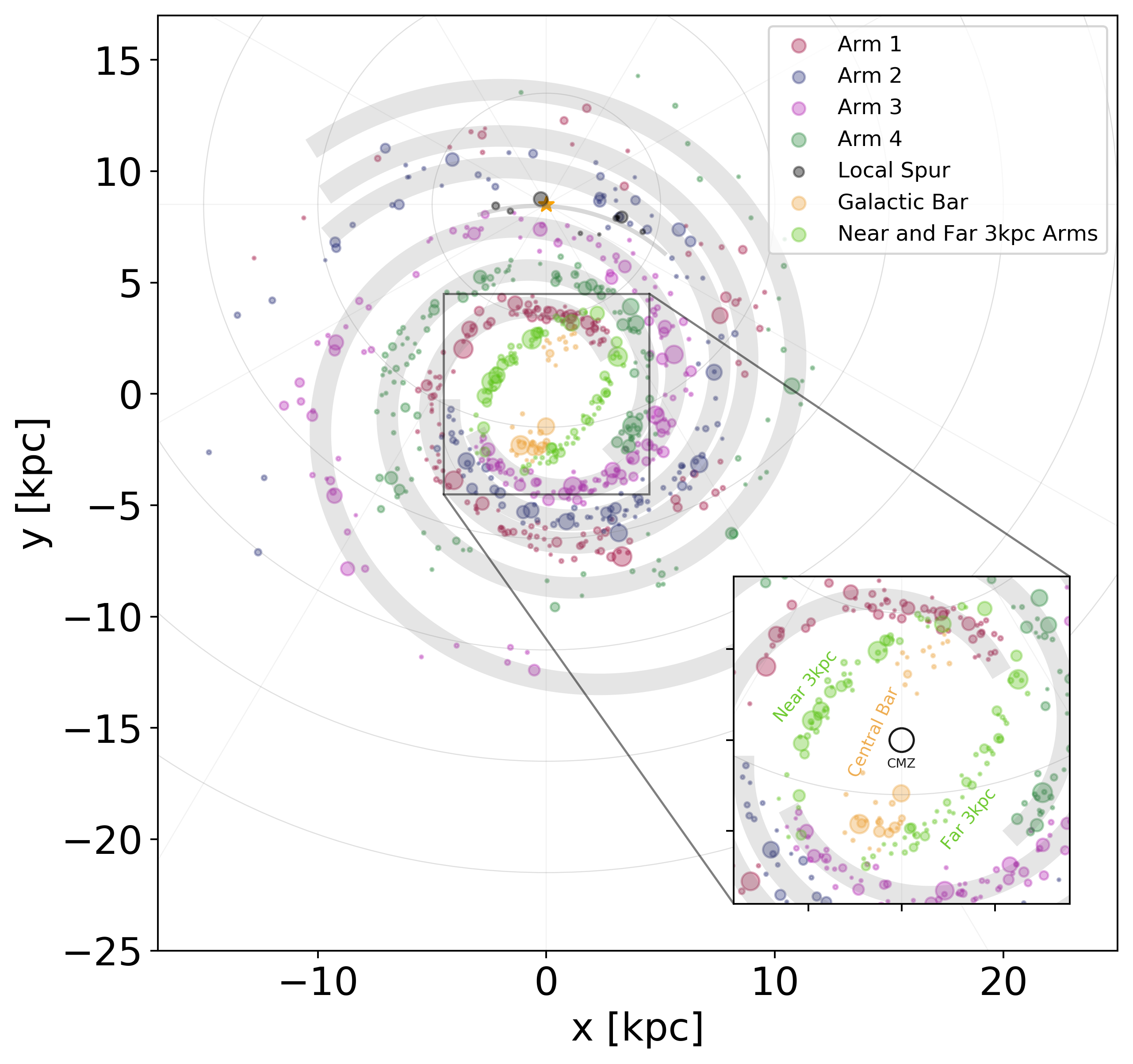}
\caption{Spatial distribution of a single realization of the YMSC Galactic population. Different colors correspond to different Galactic structures. The size of the dots is directly proportional to the mass of the clusters. The yellow star marks the Sun position.}
\label{fig:YMSCGalDistrib}
\end{center}
\end{figure}

With all the ingredients in our hands, we can finally generate a synthetic population of Galactic YMSCs and examine their general properties. After normalizing the cluster distribution function using Eq.~\ref{eq:PsiBarFromSFR} with the most updated value of the star formation rate \citep{Bonatto_SFRinSC_2011}, we obtain from Eq.~\ref{eq:NYMSC} a total number of 747 YMSCs. Fig.~\ref{fig:YMSCGalDistrib} shows the resulting spatial distribution for a specific realization of the Galactic cluster population. Unless differently stated, we will use this specific synthetic population for the rest of the chapter. 

\begin{figure}[H]
\begin{center}
\includegraphics[width=\textwidth]{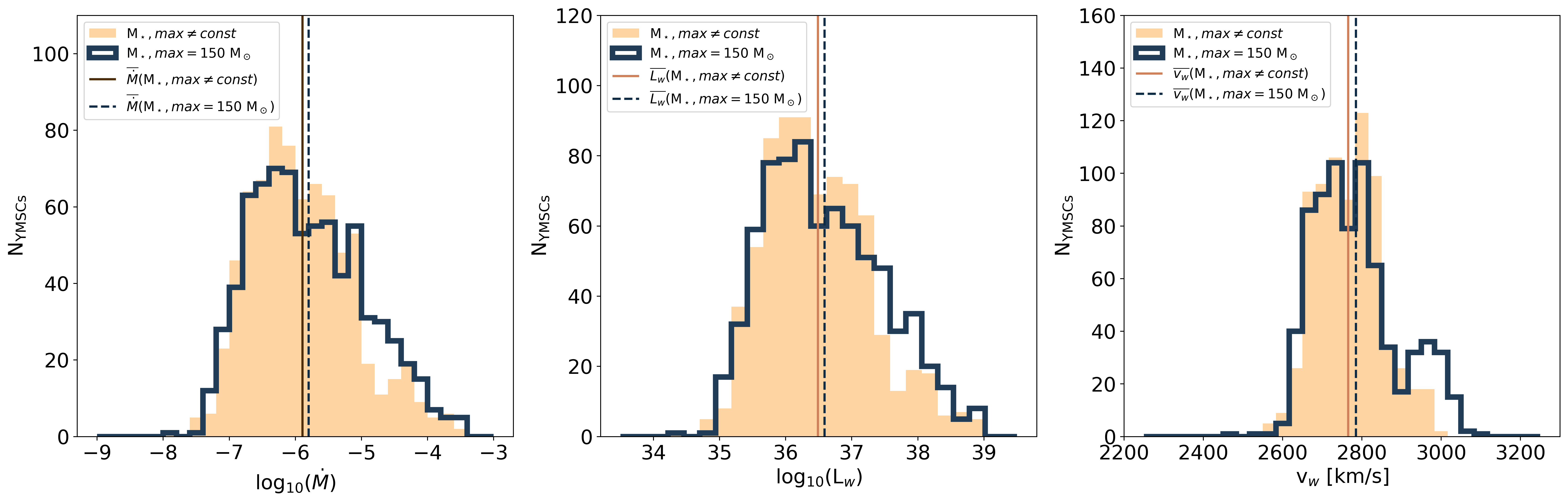}
\caption{Distribution of the mass loss rate (left panel), wind luminosity (central panel), and collective cluster wind speed (right panel) for the cases of constant (step histograms) and non-constant (filled histograms) maximum stellar masses. Continuous and dashed lines represent the average values for the two different cases (see plot legend).}
\label{fig:YMSCsLwMdotVw}
\end{center}
\end{figure}

The main parameters that determine the effectiveness of a cluster as a particle accelerator are the wind luminosity, along with the mass loss rate and the cluster wind speed. Fig.~\ref{fig:YMSCsLwMdotVw} shows the distribution of such values considering both the case when the maximum stellar mass depends on the cluster mass and the case when it is fixed to 150 M$_\odot$. Noticeably, although the latter case produces, as expected, distributions with more pronounced high-value tails, the parameters average values are almost unchanged and equals to $\overline{\dot{M}} \approx 10^{-6}$ M$_\odot$ yr$^{-1}$, $\overline{v_w}\approx 2800$ km s$^{-1}$ and $\overline{L_w} \approx 3 \times 10^{36}$ erg s$^{-1}$. 

\begin{figure}[H]
\begin{center}
\includegraphics[width=\textwidth]{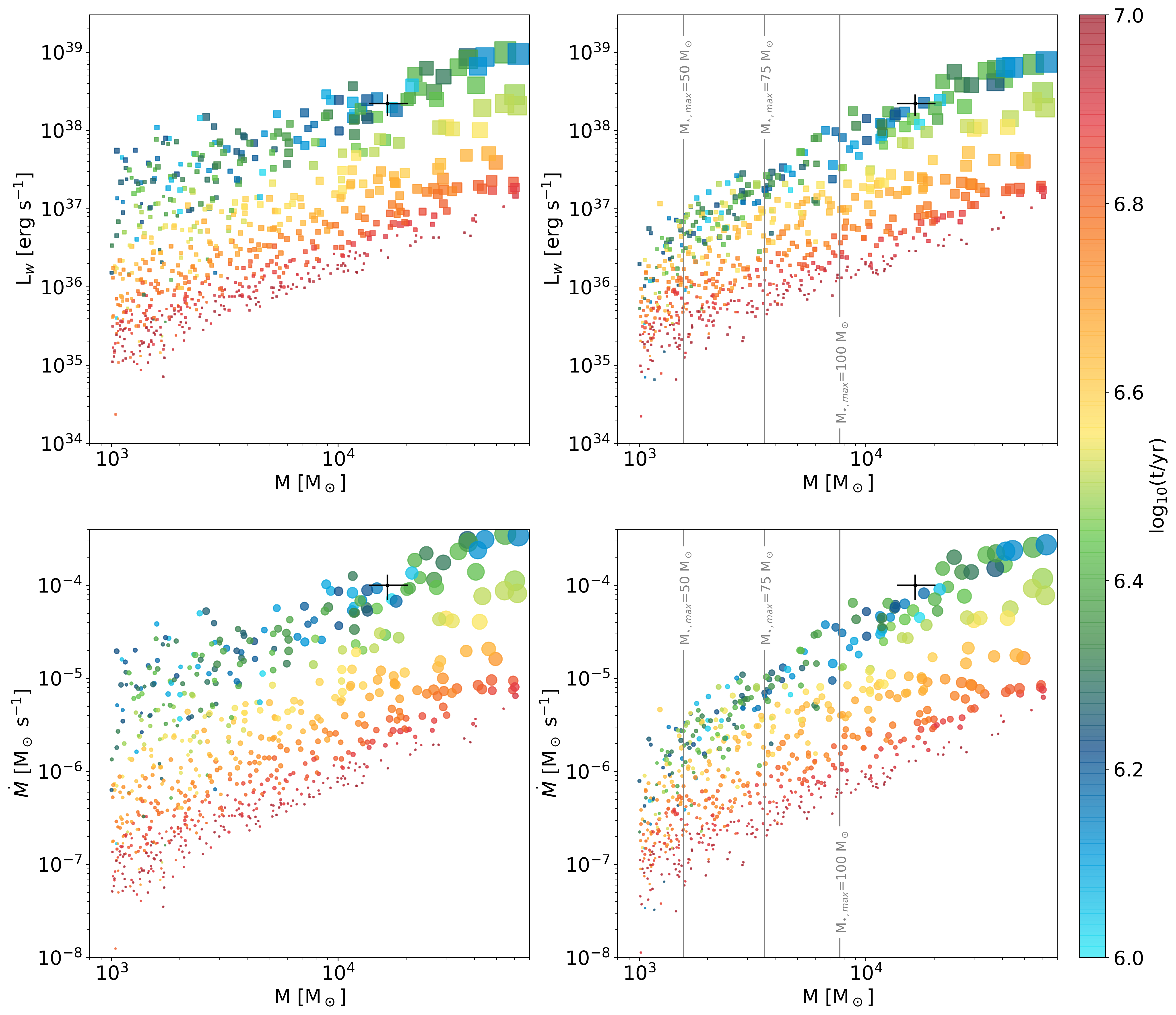}
\caption{Wind power (top row) and mass loss rate (bottom row) as a function of the cluster mass for the synthetic population shown in Fig.~\ref{fig:YMSCGalDistrib}. The left column shows the case where the maximum stellar mass ($M_{\star, max}$) is fixed to 150 M$_\odot$, while the right column is for a maximum stellar mass dependent on the cluster mass.
In all the plots, the marker size is proportional to the number of stars with $M_\star>20$ M$_\odot$. For comparison purposes, the properties of Cygnus OB2 along with their uncertainties obtained in \S~\ref{subsec:CygOB2LwMdot} are indicated with a black cross\citep{Wright_MassiveStarPopCygOB2_2015}.}
\label{fig:LwMdotVsM}
\end{center}
\end{figure}

\begin{figure}[H]
\begin{center}
\includegraphics[width=\textwidth]{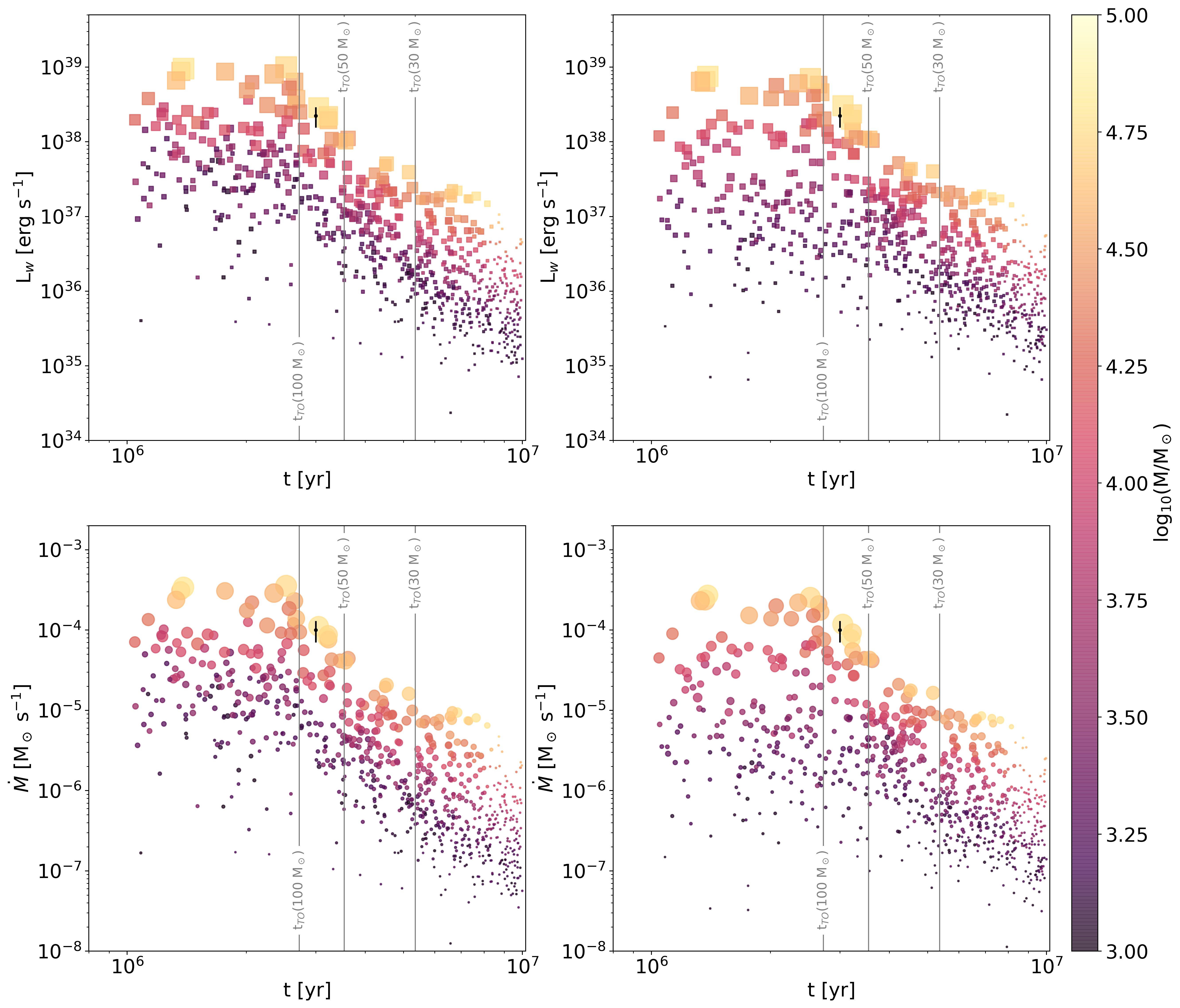}
\caption{Same as Fig.~\ref{fig:LwMdotVsM}, but the wind luminosity and mass loss rate are plotted as a function of the cluster age. Grey continuous line show for comparison the turn-off time of stars with masses of 100, 50, and 30 M$_\odot$.}
\label{fig:LwMdotVst}
\end{center}
\end{figure}

For the sake of completeness, we also show the trend of wind luminosity and mass loss rate as a function of mass (Fig .\ref{fig:LwMdotVsM}) and cluster age (Fig .\ref{fig:LwMdotVst}). As can be easily guessed, both wind power and mass loss rate are roughly proportional to the cluster mass, as the content of massive stars increases with the latter. Similarly, after $\sim 2$ Myr, the two values start to decrease with increasing cluster age, since the number of massive stars decreases as they start exploding as supernovae.

An interesting point worth outlining is the difference between the two cases of high stellar mass cutoff under analysis. A constant maximum stellar mass of 150 M$_\odot$ is able to produce clusters characterized by higher wind powers even for low cluster masses.

Along with the physical properties of the stellar clusters, it is worth investigating the dimensions of the associated wind-blown bubbles. To calculate the size of the TS ($R_{TS}$) and the forward shock ($R_b$), we need to estimate the interstellar medium density surrounding our YMSCs (see Eq.~\ref{eq:Rts} and Eq.~\ref{eq:Rbubble} respectively). As we stated in \S~\ref{subsec:YMSCBirth}, YMSCs are likely to be still partially (or completely) embedded in their parent GMC, so the average medium density ($\rho_0$) is expected to be higher than the average density of the interstellar medium.

\begin{figure}[H]
\begin{center}
\includegraphics[width=0.7\textwidth]{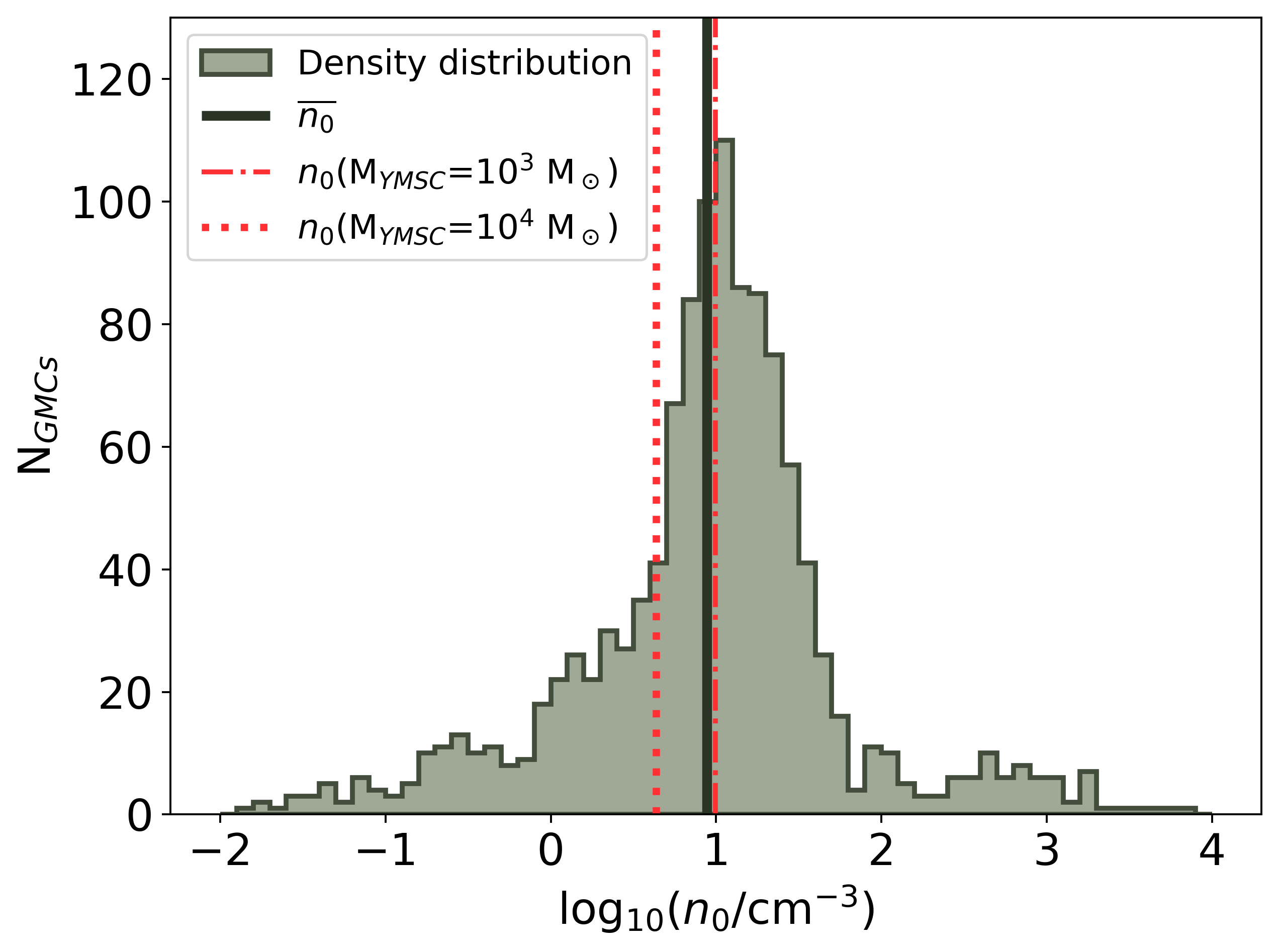}
\caption{Particle density distribution in GMCs derived from the catalog provided by \cite{Hou_MWStructure_2014}. The solid line shows the distribution average value, while the dotted and dash-dotted lines are the particle density obtained for clusters with masses of 10$^3$ and 10$^4$ M$_\odot$ respectively (see text).}
\label{fig:GMCn0}
\end{center}
\end{figure}

To compute a rough estimate of $\rho_0$, we can use again the GMC catalog provided by \cite{Hou_MWStructure_2014}, which additionally reports the mass and the angular size of the clouds. Considering the cloud distances calculated in \S~\ref{subsec:YMSCGalDistr}, from the angular size we can estimate an average cloud physical radius, from which, assuming spherical geometry, we infer the average gas density. Fig.~\ref{fig:GMCn0} shows the distribution of the particle number density for the clouds in the catalog of \cite{Hou_MWStructure_2014}, characterized by an average value of $\sim 8.76$ cm$^{-3}$, with a spread of $_{-9.96}^{+31.62}$ cm$^{-3}$.

\begin{figure}[ht]
\begin{center}
\includegraphics[width=0.8\textwidth]{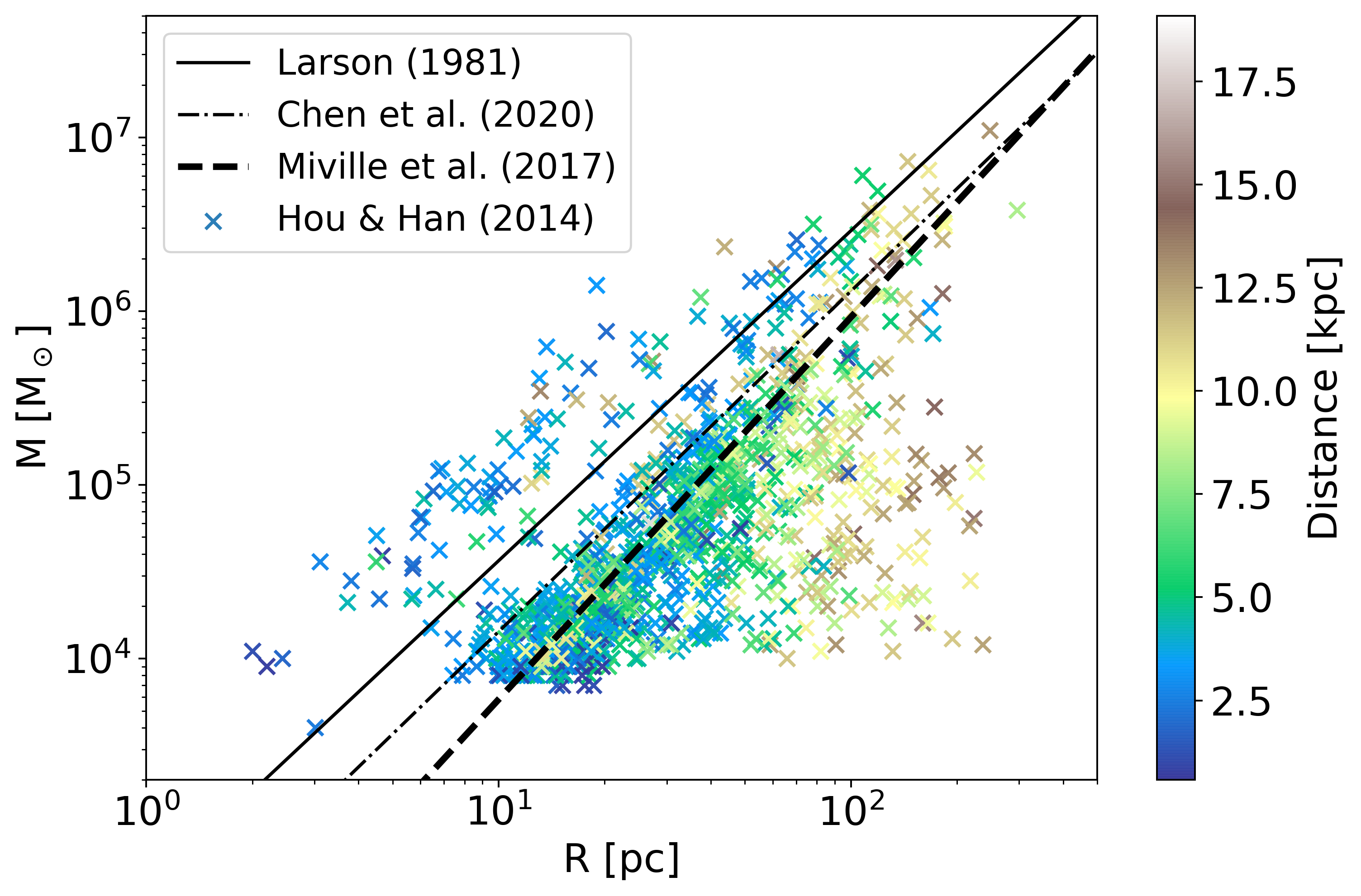}
\caption{GMC masses vs radii for the considered sample \citep{Hou_MWStructure_2014}. The solid line shows the first relation obtained by \cite{Larson_TurbulenceMCs_1981}. The dot-dashed and dashed lines are observed relations provided by \cite{Chen_MmcVsRmc_2020} and \cite{Miville_MmcVsRmc_2017} respectively. The color scale associated to the symbols represents the cloud distances from the Sun.}
\label{fig:GMCsMvsR}
\end{center}
\end{figure}

There is a second less direct way to infer the clouds density, that we can use as double check. In principle, the environmental density surrounding a given cluster, which is linked to the GMC mass, should be correlated to the cluster mass. In \S~\ref{subsec:YMSCBirth} we saw that for a given star formation efficiency $\epsilon_{SFE}$ (usually of the order of a few percent), the mass of a cluster is linked to the mass of the parent cloud as:
\begin{equation} 
M_{GMC}=M_{YMSC} \left( \frac{1}{\epsilon_{SFE}}-1 \right) \ .
\end{equation}
The density of the cloud as a function of the cluster mass is then:
\begin{equation}
\label{eq:AvgGMCDens} 
n_0=\frac{M_{YMSC}}{\frac{4}{3} \pi R_{GMC}^3} \left( \frac{1}{\epsilon_{SFE}}-1 \right).
\end{equation}
where $R_{GMC}$ is the radius of the cloud, which also depends on the GMC mass. Several mass-radius relations for the clouds have been proposed over the decades. In general, all relations are parametrized as power laws:
\begin{equation}
R_{GMC}=R_0 \left( \frac{M_{GMC}}{M_0} \right)^\alpha
\end{equation} 
with $R_0$ and $M_0$ as normalization constants. Fig.~\ref{fig:GMCsMvsR} shows some of the clouds mass-radius relations taken from the literature \citep{Larson_TurbulenceMCs_1981, Miville_MmcVsRmc_2017, Chen_MmcVsRmc_2020} and compared with mass and radius of the sample of GMCs from \cite{Hou_MWStructure_2014}. If we consider the relation provided by \cite{Miville_MmcVsRmc_2017} ($M_0=36.7$ M$_\odot$, $R_0=1$ pc, $\alpha=0.454$), which seems to be the one best reproducing the observed sizes and masses, using Eq.~\ref{eq:AvgGMCDens} (assuming $\epsilon_{SFE}=0.01$) leads to densities of $\sim 10$ cm$^{-3}$ and $\sim 4$ cm$^{-3}$ for $M_{YMSC}=10^3$ M$_\odot$ and $M_{YMSC}=10^4$ M$_\odot$ respectively. These densities are fully compatible with the estimated mean value of 10 cm$^{-3}$ obtained above. Taking this into account, and given that by changing the mass of the cluster by an order of magnitude the density only varies by a factor of $\sim$2, we can safely consider for our purposes a constant density equal to $\rho_0=10 m_p$ cm$^{-3}$ for all YMSCs.

\begin{figure}[ht]
\begin{center}
\includegraphics[width=0.9\textwidth]{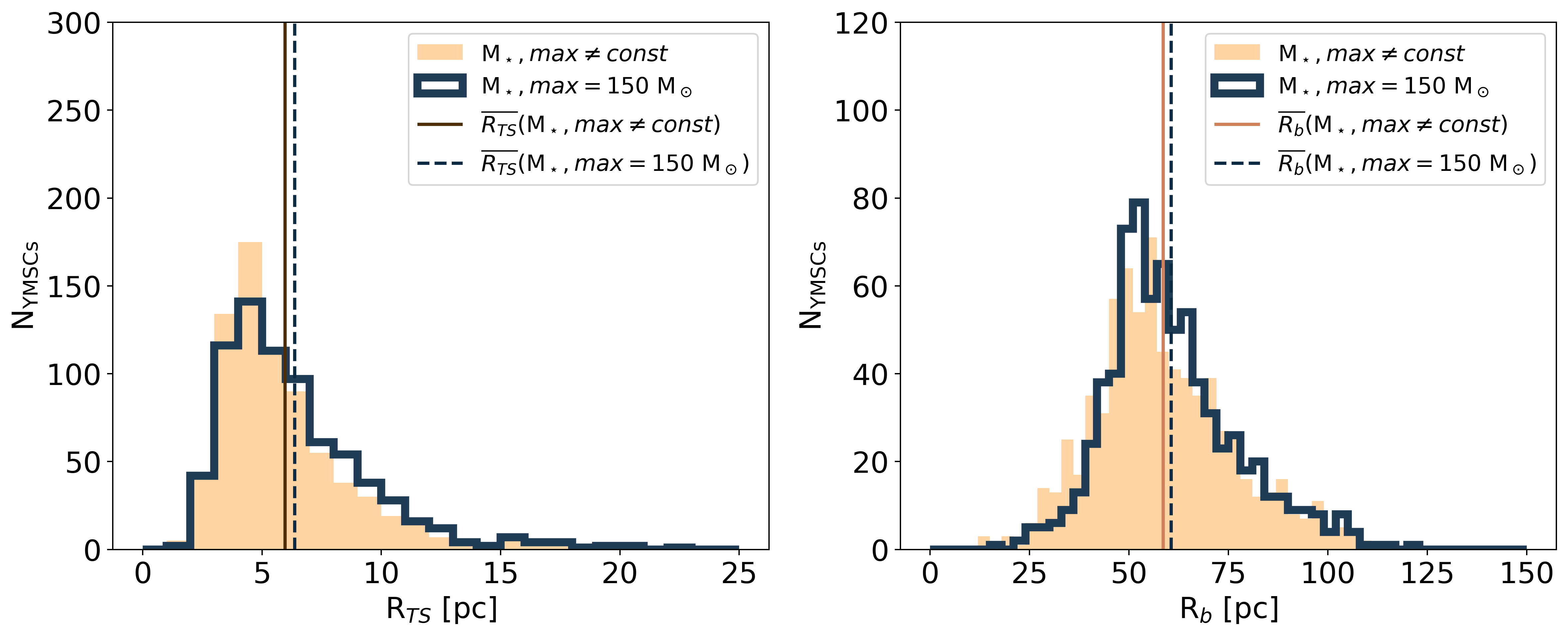}
\caption{Size distribution of the TS and forward shock for both cases under analysis, namely with maximum stellar mass fixes (step histogram) and dependent on the cluster mass (filled histogram). Solid and dashed lines represent the distributions average values for the filled and step histograms respectively.}
\label{fig:SyntRbRTS}
\end{center}
\end{figure}

Once the environmental density is set, we can readily calculate the sizes of the wind-blown bubbles associated with our synthetic population of YMSCs. Fig.~\ref{fig:SyntRbRTS} shows the resulting distribution. The distribution peaks at $R_{TS}=4$ pc and $R_b=50$ pc and is not Gaussian but has a long tail extending towards higher values. The average values are $\overline{R}_{TS}=6$ pc and $\overline{R}_b=60$ pc. Again, changing the maximum stellar mass does not significantly affect the distributions.

%---------------------------------------------------------

\section{Cosmic Ray distribution in YMSCs}
\label{sec:YMSCasCRfact}
Once the YMSC distribution is determined, we have all the ingredients to compute the CR distribution. Once again, we rely on the model of hadronic CR acceleration at the wind TS developed by \cite{Morlino_2021} and described in \S~\ref{subsec:WindTSAccModel}.

\begin{figure}[ht]
\begin{center}
\includegraphics[width=0.9\textwidth]{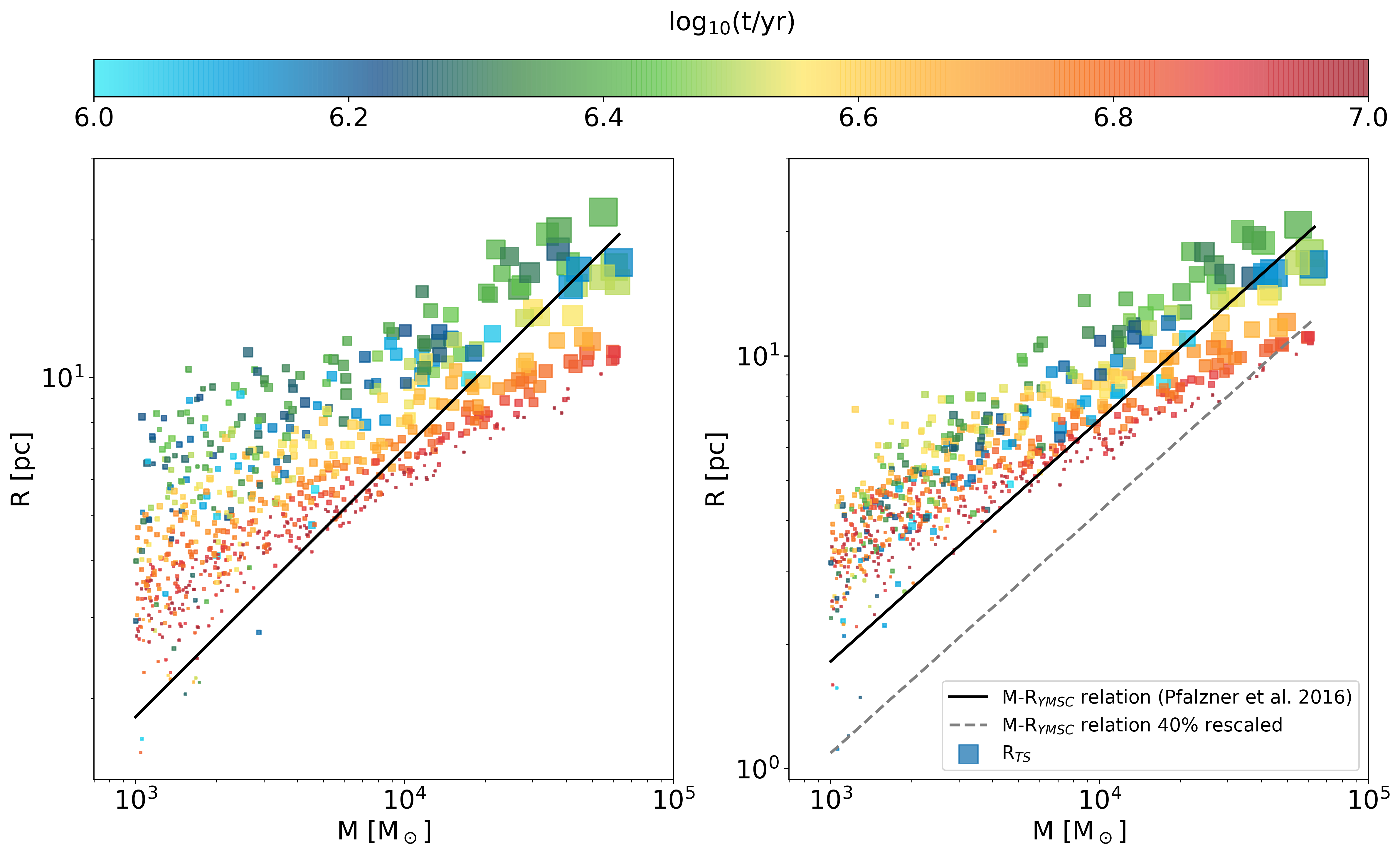}
\caption{Termination shock radius as a function of the cluster mass. Continuous line shows the cluster mass-half mass radius relation provided by \cite{Pfalzner_YMSCsMRR_2016}. The left panel is for the case of a synthetic Galactic cluster population realized with a fixed maximum stellar mass (150 M$_\odot$), while the right panel shows the case with maximum stellar mass dependent on the cluster mass.}
\label{fig:RtsVsMymsc}
\end{center}
\end{figure}

In \S~\ref{subsec:YMSCBirth}  and \S~\ref{subsec:WindTSAccModel} we emphasized that the validity of this model is bound to the existence of a collective cluster wind. We recall that the general rule for this to happen is that the average distance between stars has to be smaller than the size of the TS. When this condition is satisfied, we call the stellar cluster "compact". In the opposite cases we speak of "loose" clusters, defined as the ones where the collective TS will not be formed. An approximate way to distinguish between compact and loose stellar clusters is by using the half mass radius, which is, however, a non trivial quantity to estimate from observations. After a comprehensive analysis of the literature, \cite{Pfalzner_YMSCsMRR_2016} found a power-law relation of the form:
\begin{equation}
\label{eq:RymscVsM}
R_{YMSC}=\left(\frac{M}{359 \rm \ M_\odot} \right)^{0.585} \rm \ pc.
\end{equation} 
Fig.~\ref{fig:RtsVsMymsc} shows Eq.~\ref{eq:RymscVsM} compared to the sizes of $R_{TS}$ of our synthetic population plotted as a function of the cluster mass. Noticeably, for a significant fraction of YMSCs the TS radius is always larger than the cluster size. This condition seems to be violated by some fraction of clusters with masses larger than 10$^4$ M$_\odot$. This would lead one to think that for some very massive clusters the formation of a collective wind could be suppressed. However, one still needs to take into account the effect of mass segregation, which tends to aggregate the most massive stars (which are the largest contributors to the creation of the collective wind) toward the center of the cluster. In 90\% of the generated synthetic YMSCs, stars with masses greater than 10 M$_\odot$ contribute $\sim 90$\% of the cluster wind power. It follows therefore that, instead of the half-mass radius, one should rather consider an effective radius $R_{\rm eff}$ defined such that it contains a significant fraction of stars above 10 M$_\odot$. If we fix this fraction to 90\%, defining the distribution of stars within a cluster as a function of the mass and radius as $\Xi (M_\star, r) \equiv \frac{dN_\star}{dM_\star dr}$, the effective radius can be found by requiring:
\begin{equation}
\int_{10 \rm M_\odot}^{150 \rm M_\odot} \int_0^{R_{\rm eff}} \Xi (M_\star, r) dM_\star dr = 0.9 \ ;
\end{equation}
where we have assumed $\int \int \Xi(M_\star, r) dM dr = 1$. Unfortunately, the function $\Xi(M_\star, r)$ turns out to be difficult to estimate. However, we can calculate $R_{eff}$ as the containment radius of the fraction of cluster mass due to stars with $M_\star > 10$ M$_\odot$. Such a fraction is:
\begin{equation}
\frac{\int_{10 \ \rm M_\odot}^{150 \ \rm M_\odot} M_\star f(M_\star) dM_\star }{\int_{0.08 \ \rm M_\odot}^{150 \ \rm M_\odot}  M_\star f(M_\star) dM_\star } \simeq 0.2
\end{equation}
Assuming that the stellar mass distribution is $dM_\star/dr \propto r^{-\alpha}$ one has $R_{eff} < R_{half \ mass}$ (unless $\alpha \approx 3$). If $\alpha$ is such that $R_{eff} \lesssim 0.4 R_{half \ mass}$ (with $R_{eff} = 0.4 R_{half \ mass}$ if $\alpha=2$), then, almost all stellar clusters turn out to have a TS more extended than the region containing the most massive stars. $R_{eff}$ estimated in this way is reported in Fig.~\ref{fig:RtsVsMymsc} with a dashed line. Summarizing, the conclusion of our calculation is that all YMSCs in our synthetic population can generate a TS.

Now we can proceed to the calculation of the CR distribution both in terms of morphological and spectral shape. As illustrated in Ch.~\ref{ch:CygOB2}, for an individual cluster the CR distribution will depend on two factors, namely the injection efficiency and the type of plasma turbulence that will determine the diffusion properties in the system. For particle injection at the TS, we assume a fixed spectral slope $s=4$, typical of strong shocks, and acceleration efficiency ($\epsilon_{CR}$) equal to 10\% of the wind luminosity. Notice that the final result on the $\gamma$-ray emission will scale almost linearly with $\epsilon_{CR}$. Regarding the plasma turbulence, given the results obtained in Ch.~\ref{ch:CygOB2}, we again decide to investigate the two scenarios characterized by the Kraichnan and Bohm-like diffusion. 

\begin{figure}[ht]
\begin{center}
\includegraphics[width=\textwidth]{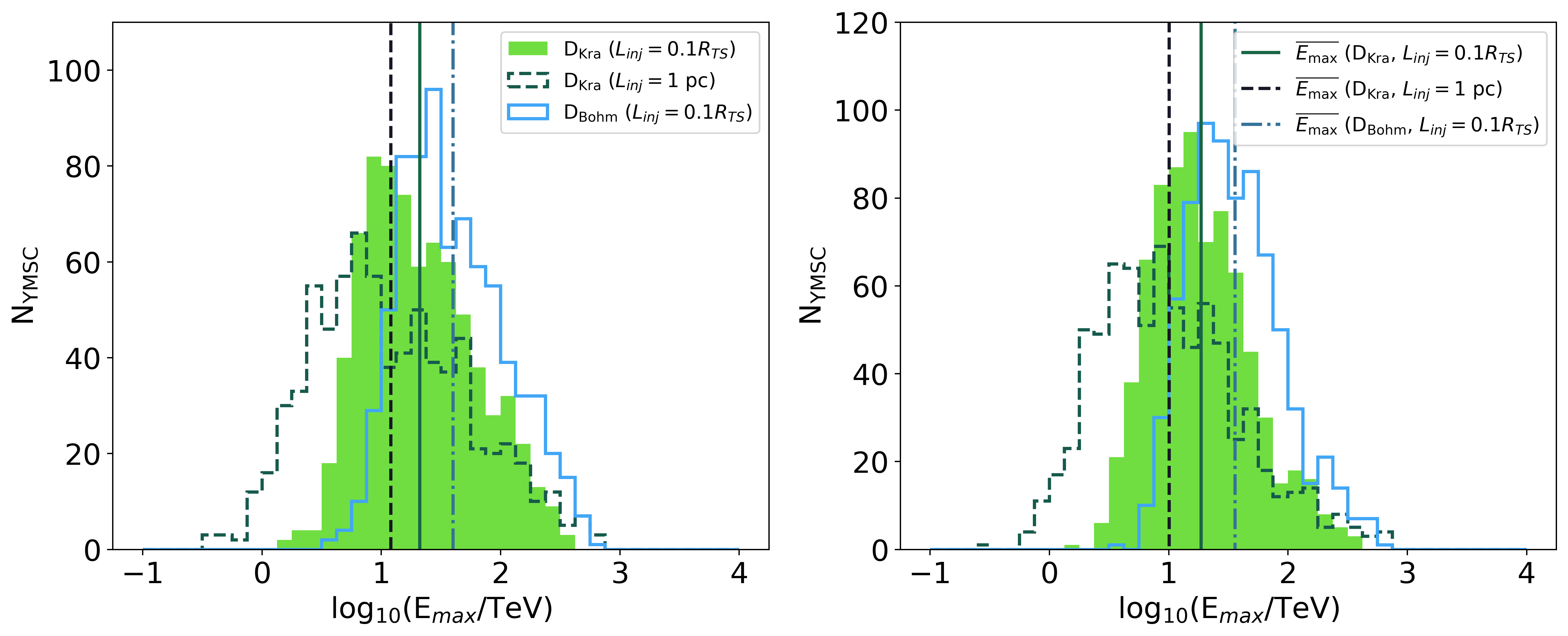}
\caption{Distribution of maximum energies obtained in the case where the maximum stellar mass is set to 150 $M_\odot$ (left panel) or dependent on the mass of the host cluster (right panel). Filled and step-dashed histograms represent maximum energy in the case of Kraichnan turbulence injected at a characteristic length scale of 0.1 $R_{TS}$ and 1 pc respectively. Step-continuous histogram reports instead the maximum energy distribution for Bohm-like diffusion.  Vertical lines show the distribution mean value (legend on the right panel).}
\label{fig:Emax_K_B_YMSC}
\end{center}
\end{figure}

A relevant aspect worth investigating is the distribution of the maximum particle energy reached in our synthetic population of CR factories. The maximum energies for the Kraichnan and Bohm cases are described by Eq.~\ref{eq:EmaxKraLw} and Eq.~\ref{eq:EmaxB}, respectively. We assume once more that a 10\% of the clusters wind luminosity is converted into turbulent magnetic field ($\eta_B=0.1$). Note that in the case of Kraichnan turbulence, the maximum energy also depends on the injection scale of the turbulence, which is totally unknown and hard to estimate. Consequently, we decide to further split the Kraichnan scenario into two subcases, defined by an injection length scale fixed to 1 pc or to 10\% of $R_{TS}$ respectively. The same also holds for the Bohm case. However here the dependence is not on a single length scale but rather on the interval of length scales for which the turbulence power is uniformly injected. We assume in this case that turbulence is constantly injected from 10\% of $R_{TS}$ to $10^{-5}$ pc. The results are shown in Fig.~\ref{fig:Emax_K_B_YMSC}. We can see that, when the maximum stellar mass as fixed to 150 M$_\odot$ the distribution of $E_{max}$ is slightly more skewed towards high values. However, the mean values for the maximum energy are basically the same, namely $E_{max}^{Kra}(L_{inj}$=1 pc) $\approx 24$ TeV, $E_{max}^{Kra}(L_{inj}=0.1 R_{TS}) \approx 37$ TeV and $E_{max}^{Bohm} \approx 36$ TeV. 

Noticeably, no PeVatrons are found. This is true both when considering Kraichnan and Bohm like turbulence, but also in all considered scenarios with different turbulence injection scales and maximum stellar masses. In fact, we found that the fraction of clusters with $E_{max}>100$ TeV is 13\% and 7\% for the Bohm and Kraichnan cases respectively. For $E_{max}>500$ TeV these fractions reduce to 1\% for the Bohm and 0.5\% for the Kraichnan case, corresponding to $\sim 7\--5$ clusters in total. These values do not vary a lot when considering the various scenarios under analysis, i.e. different turbulence injection scales for the Kraichnan case different maximum stellar mass. With such a small expected number of PeVatrons, the statistical fluctuations are non negligible, and multiple realizations of the Galactic population of YMSCs are needed to asses the contribution of YMSCs to CRs at energies above 100 TeV.
%An interesting consequence of the different shapes of the distribution tails at high energies is the resulting fraction of PeVatrons. Clearly, Bohm-like diffusion is the one providing the largest fractions of PeVatrons, with percentage values of $\sim 22$\% and $\sim 15$\% for the constant and not constant stellar maximum mass cases respectively. For Kraichnan turbulence these numbers are significantly reduced. In the most optimistic scenario where the maximum stellar mass is fixed at 150 M$_\odot$, the percentage of YMSC begin PeVatrons drops to $\sim 1$\%.

%Finally, in Fig.~\ref{fig:EmaxVsTageVsM} we show the maximum energy as a function of mass and age of our synthetic stellar cluster population. In fact, this plot does not add any new information, however, it returns a clear view of the acceleration capabilities of stellar clusters: the younger and more massive a given cluster is, the higher the particle energy it can produce.

%\begin{figure}[ht]
%\begin{center}
%\includegraphics[width=0.9\textwidth]{FIGURES/Chapter_3/EmaxVsTageVsM.png}
%\caption{Maximum particle energy (in color scale) as a function of the cluster age and mass for the Kraichnan (left panel) and Bohm (right panel) cases. We are considering the most favorable scenario where the maximum stellar mass is fixed as 150 M$_\odot$.}
%\label{fig:EmaxVsTageVsM}
%\end{center}
%\end{figure}

%---------------------------------------------------------
\section{$\gamma$-ray emission from single sources}
Now we are ready for the final step which is the estimate for the $\gamma$-ray emission from YMSCs. As already stated, we will consider only the hadronic mechanism of $\pi_0$ production and decay (see \S~\ref{subsec:RadMechanisms}).

Since we are not interested in the analysis of individual sources, but rather in the study of the total diffuse emission of the population, we perform a simplified calculation of $\gamma$-ray emission from a single cluster. More precisely, we model the emission from a single source without accounting for its morphology, namely by considering only the total flux obtained from all particles confined within the wind-blown bubble and assuming that the emission is uniformly distributed in the projected disk in the sky. Following this approach, the $\gamma$-ray flux coming from the i-th cluster will be:
\begin{equation}
\label{eq:PhiGammaYMSC}
\phi_{\gamma, i}(E_\gamma)=\frac{c n_{i}}{4 \pi d_i^2} \int F_{CR}(E_p) \frac{d \sigma(E_p, E_\gamma)}{dE_p} dE_p 
\end{equation}
where $n_{i}$ is the target medium numerical density, $d_i$ is the distance from the Sun of the i-th YMSC, $\sigma$ is the cross section for $\gamma$-ray production from $p\--p$ interaction \citep{Kafexhiu_SigmaPi0Gamma_2014}, and $F_{CR}(E_p)$ is defined as:
\begin{equation}
F_{CR}(E_p)=4 \pi \int_0^{R_{b, i}} r^2 f_{CR}(r, E_p) dr
\end{equation}
with $f_{CR}(r, E_p)$ calculated using Eq.~\ref{eq:fCRNoSea} after the appropriate transformation from particle momentum to energy. Note that by using Eq.~\ref{eq:fCRNoSea}, we are again accounting only for the contribution of freshly accelerated particles without considering the emission from Galactic CRs penetrating in the wind-blown bubble. 

In Eq.~\ref{eq:PhiGammaYMSC}, the density of the target medium $n_{i}$ remains a difficult parameter to estimate, as it can vary depending on the evolutionary conditions of the wind bubble. In general, one could consider the density obtained from the evaporation rate of the swept-up shell (see Eq.~\ref{eq:ShellMdot}) in addition to the material injected by the stellar winds. However, for the sake of simplicity and with the aim of proving a considerable but solid upper limit for the $\gamma$-ray flux\footnote{Under the assumption of a fixed acceleration efficiency of the particles.}, we will consider here the simplest case of a constant target density equal to that of the surrounding environment, i.e., $n_0=10$ cm$^{-3}$. %Finally, we would like to emphasize that, by not taking into account source morphology, we are implicitly assuming that all the observed flux from a given YMSC comes from a disk region with size equal to the projection of $R_b$. 

%---------------------------------------------------------
\section{Diffuse emission and comparison with data}
Once the flux of all YMSCs is obtained, we calculate the diffuse $\gamma$-ray emission of a given area of the sky by simply summing up all the contributions of the sources included in that region. This approach is computationally convenient. However, the final result may be slightly distorted as we are not excluding the emission from those regions of the wind bubbles that do not fall in the considered sky area. The opposite situation is also not considered, namely the possibility of including the contribution of portions of bubbles that fall within the region of interest whose centroid is instead outside the area of the sky under analysis. However, we expect the two contributions to cancel each other out, so that on average the final estimation does not differ much from the true result.

\begin{figure}[H]
\begin{center}
\includegraphics[width=0.75\textwidth]{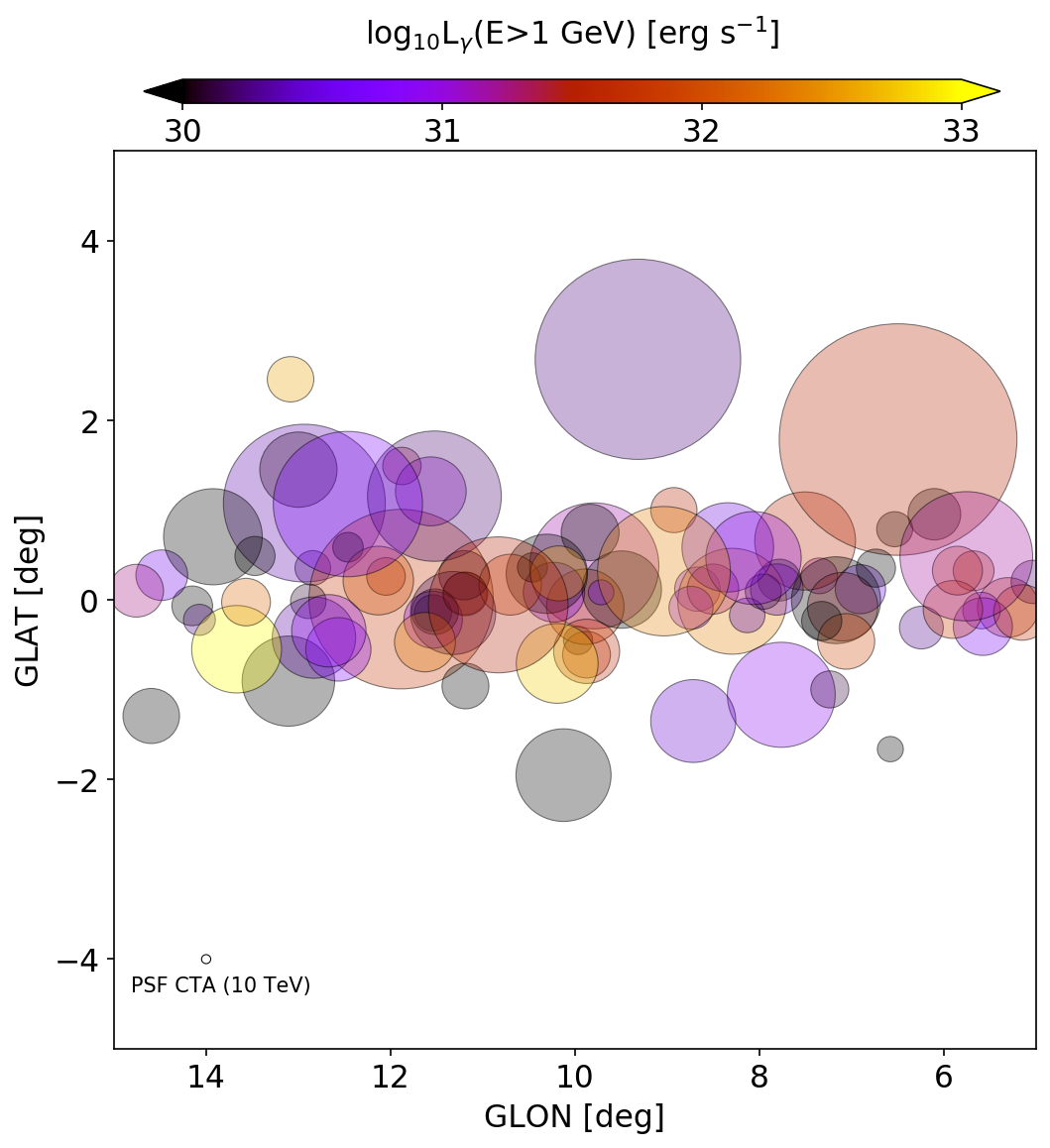}
\caption{Region of interest (\texttt{ROI1}) for the computation of the $\gamma$-ray diffuse emission using the work by \cite{Yang_FermiDiffuse_2016}. Circles represent the wind-blown bubbles from the synthetic YMSC. The color scale reports the integrated $\gamma$-ray luminosity above 1 GeV considering the Kraichnan scenario with maximum stellar mass as a function of the cluster mass. For comparison, we report the CTA point spread function at 10 TeV \citep{CTA_Performances}.}
\label{fig:ROI}
\end{center}
\end{figure}

\begin{figure}[H]
\begin{center}
\includegraphics[width=0.75\textwidth]{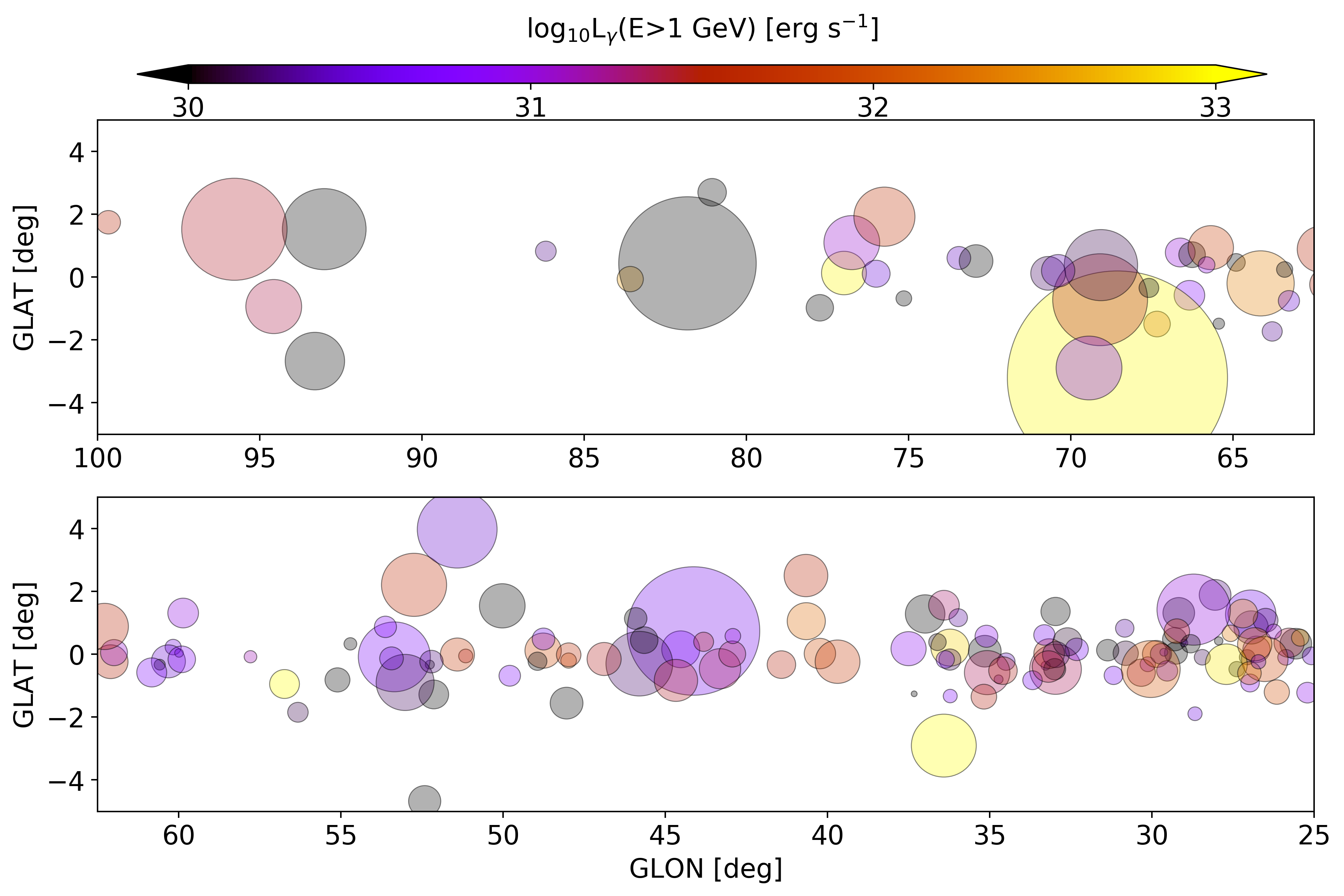}
\caption{Same as Fig.~\ref{fig:ROI}, but for the region of interest defined between $100^\circ<l<25^\circ$ and $-5^\circ<b<5^\circ$ (\texttt{ROI2}).}
\label{fig:ROI2}
\end{center}
\end{figure}

We select the regions from which to extract the $\gamma$-ray emission considering four different works. The first is the one by \cite{Yang_FermiDiffuse_2016}, who supply the diffuse $\gamma$-ray spectra in several areas of the Galactic Plane using Fermi-LAT observations. Among the provided area, we select the one between $15^\circ<l<5^\circ$ and $-5^\circ<b<5^\circ$. Fig.~\ref{fig:ROI} shows the selected region of interest and the synthetic YMSC within it. For simplicity, we will refer from now on to this region as \texttt{ROI1}. The remaining three studies that we considered are \cite{Hunter_EgretDiffuse_1997}, \cite{Bartoli_ArgoDiffuse_2015}, and \cite{Amenomori_TibetASgammaDiffuse_2021}. These works report the diffuse $\gamma$-ray emission within a region spanning $100^\circ<l<25^\circ$ and $-5^\circ<b<5^\circ$, employing data from EGRET, ARGO, and Tibet-AS$\gamma$. Fig.~\ref{fig:ROI2} shows the latter region and the synthetic YMSCs within it. We will refer to this region from now on as \texttt{ROI2}.

\begin{figure}[h]
\begin{center}
\includegraphics[width=0.9\textwidth]{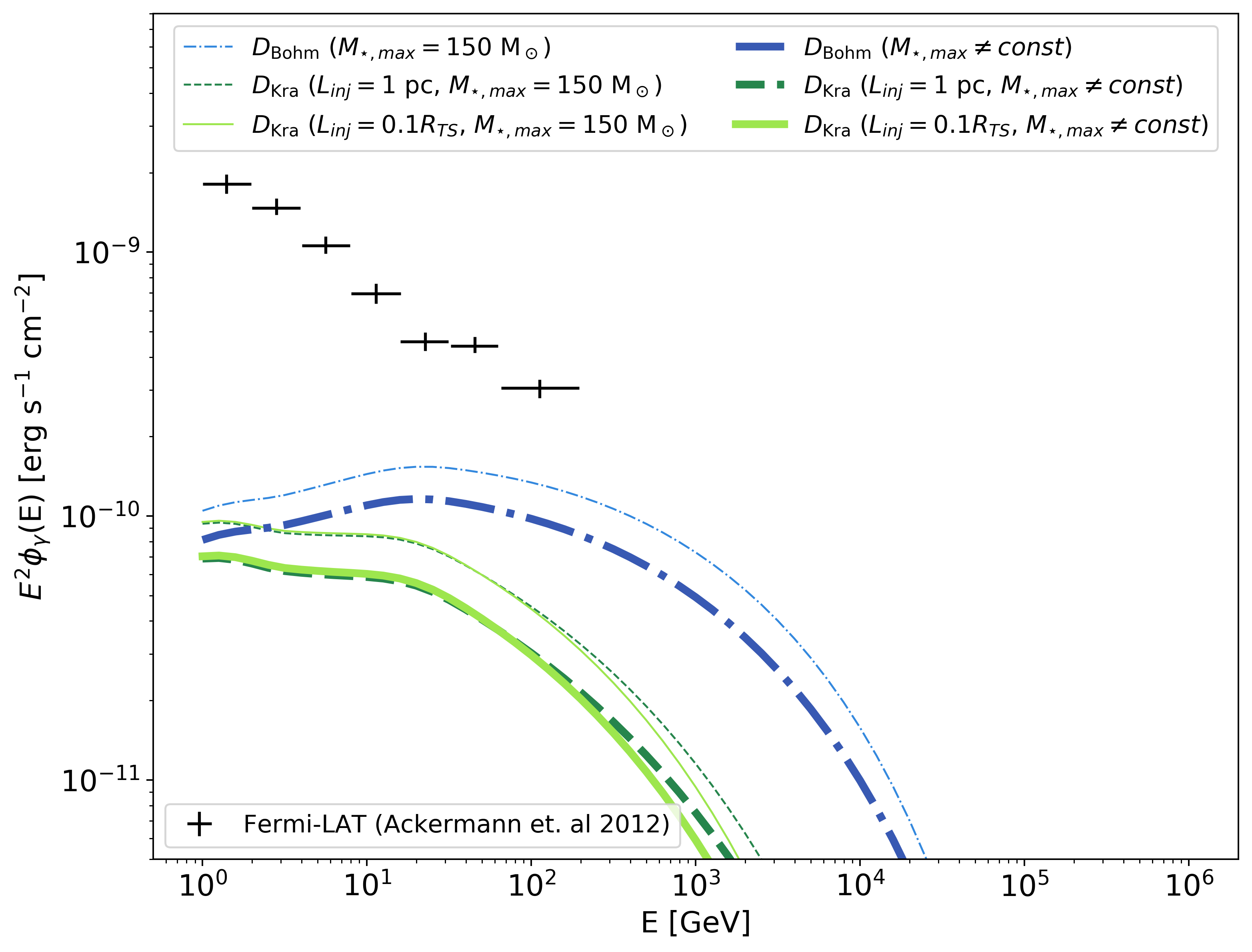}
\caption{Diffuse $\gamma$-ray emission from the synthetic population of YMSCs compared to Fermi-LAT observations in \texttt{ROI1} (sky region shown in Fig.~\ref{fig:ROI}). Thin and thick lines represent the spectra after considering constant and cluster mass-dependent maximum stellar masses. Dash-dotted lines are the spectra obtained in the case of Bohm-like turbulence. Solid and dashed lines are instead the spectra calculated for Kraichnan diffusion after considering $L_{inj}=1$ pc and $L_{inj}=01 R_{TS}$ respectively.}
\label{fig:SED_Diffuse_Gamma}
\end{center}
\end{figure}

As already mentioned, in this work we consider just one realization of the Galactic YMSC population and calculate $\gamma$-ray emission on top of it. Fig.~\ref{fig:SED_Diffuse_Gamma} shows the $\gamma$-ray spectral energy distributions compared with observations in \texttt{ROI1} for all the cases under analysis. There are several things worth discussing. The most important aspect is that the flux obtained from all the cases analyzed lies below the Fermi-LAT data by a factor of $\sim 2\--3$ for the Bohm scenarios and $\sim 10$ for the Kraichnan cases at $E_\gamma>10$ GeV. This is a remarkable result, as it implies that the way we modeled the YMSCs does not lead to a scenario inconsistent with observations and yet shows that YMSCs could be an important contributors of the diffuse emission in this band. A second result to underline is that, changing recipe for the maximum stellar mass does not lead to significant changes in the overall emission, only resulting in a normalization shift by a factor of $\sim 1.5$. This is due to a greater number of massive stars in the lower-mass star clusters, which produce a higher wind power and thus higher flux normalization. As we will see in short, the shift in amplitude is relatively small, as the contribution to the diffuse $\gamma$-ray is mainly due to the most massive stellar clusters.

Remarkable, and again not unexpected, turns out to be the difference between models using different diffusion coefficients. Bohm-like diffusion is capable of producing $\gamma$-ray emission up to very high energy, with a cut-off in the spectrum appearing at $\sim 1$ TeV, contrary to the Kraichnan case where the emission start to fade at $\sim 100$ GeV. Concerning the Kraichnan case, having a constant injection scale fixed to 1 pc produces a spectrum with a cut-off at higher energy compared to the case where the turbulence injection occurs at 10\% of $R_{TS}$, although, in the end, the overall spectra do not differ too much. 

Regarding the cut-off position in the $\gamma$-ray spectrum, we stress once more time the fact that the energy at which it appears is less than the expected 0.1 $E_{max}$ of accelerated particles.  As already discusses in \S~\ref{subsec:SpectAnalysis}, the maximum energy calculated in \S~\ref{subsec:WindTSAccModel} does not report the exact location of the cutoff in the injected particles spectrum, but is usually shifted to lower energy due to the spherical geometry of the system.

\begin{figure}[h]
\begin{center}
\includegraphics[width=0.9\textwidth]{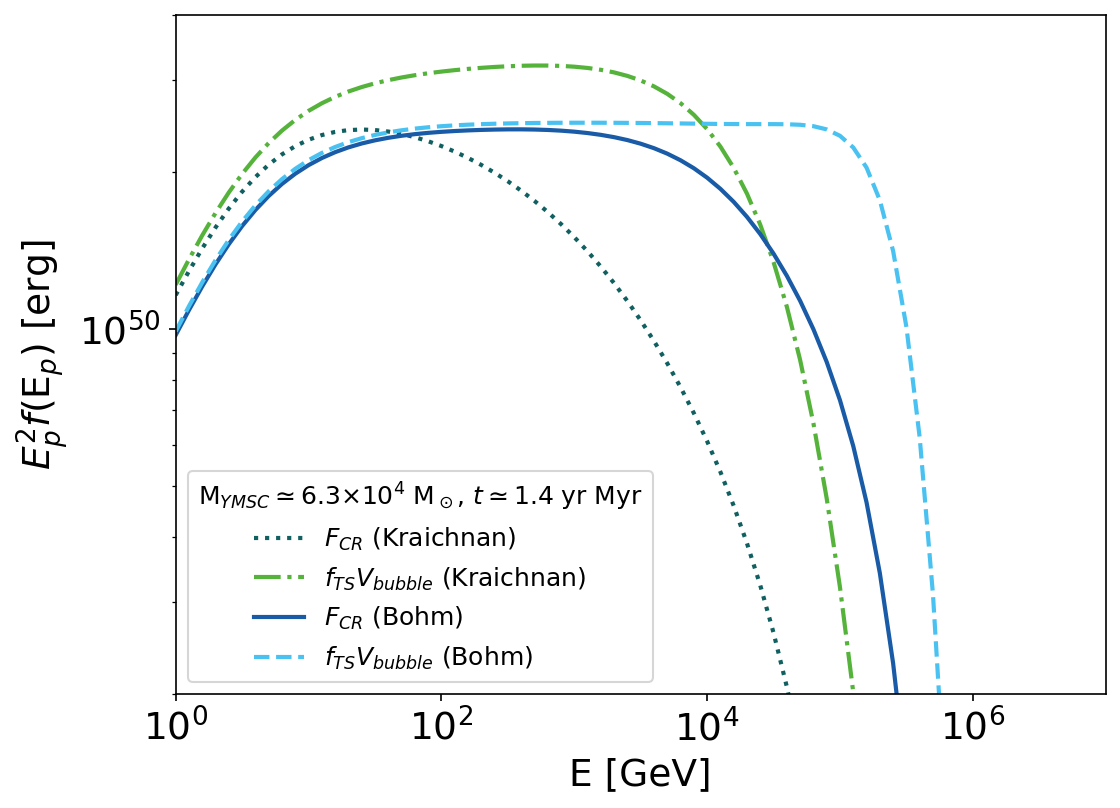}
\caption{CR spectra ($F_{CR}$) used to compute the $\gamma$-ray radiation (dotted and continuous lines for Kraichnan and Bohm cases respectively) compared to hypothetical spectra where the propagation mechanism is totally advection-dominated (dash-dotted and dashed lines for Kraichnan and Bohm cases respectively). All curves reported refer to the most massive stellar cluster in our synthetic population.}
\label{fig:FCRvsfTS}
\end{center}
\end{figure}

In general, it is important to note that the Krainchnan case returns an overall spectrum that is softer than the Bohm model, despite the same spectral slope at the acceleration site. This is a direct consequence of the interplay between advection and diffusion, whose relative importance differs in the two cases under analysis. To show the importance of diffusion, we can compare the full solution with the idealized case completely dominated by advection, which has $f_{CR}$ inside the bubble equal to $f_{TS}$. As an example, in Fig.~\ref{fig:FCRvsfTS} we consider the case of the most massive YMSCs of our synthetic population, and assuming the scenario where the maximum stellar mass is fixed to 150 M$_\odot$. One can readily note that the particle spectrum in the Kraichnan model is much softer than in the ideal situation of pure advection, which is characterized by the expected power law $f_{TS}\propto E_p^{-2}$. In the case of Bohm-like diffusion, advection is expected to dominate over diffusion up to very high energy. Therefore, the CR spectrum does not differ much from the ideal situation of pure advection (except close to the cut-off energies). As a consequence, the observed $\gamma$-ray spectrum is harder than the one obtained considering Kraichnan-like diffusion. 

\begin{figure}[h]
\begin{center}
\includegraphics[width=0.9\textwidth]{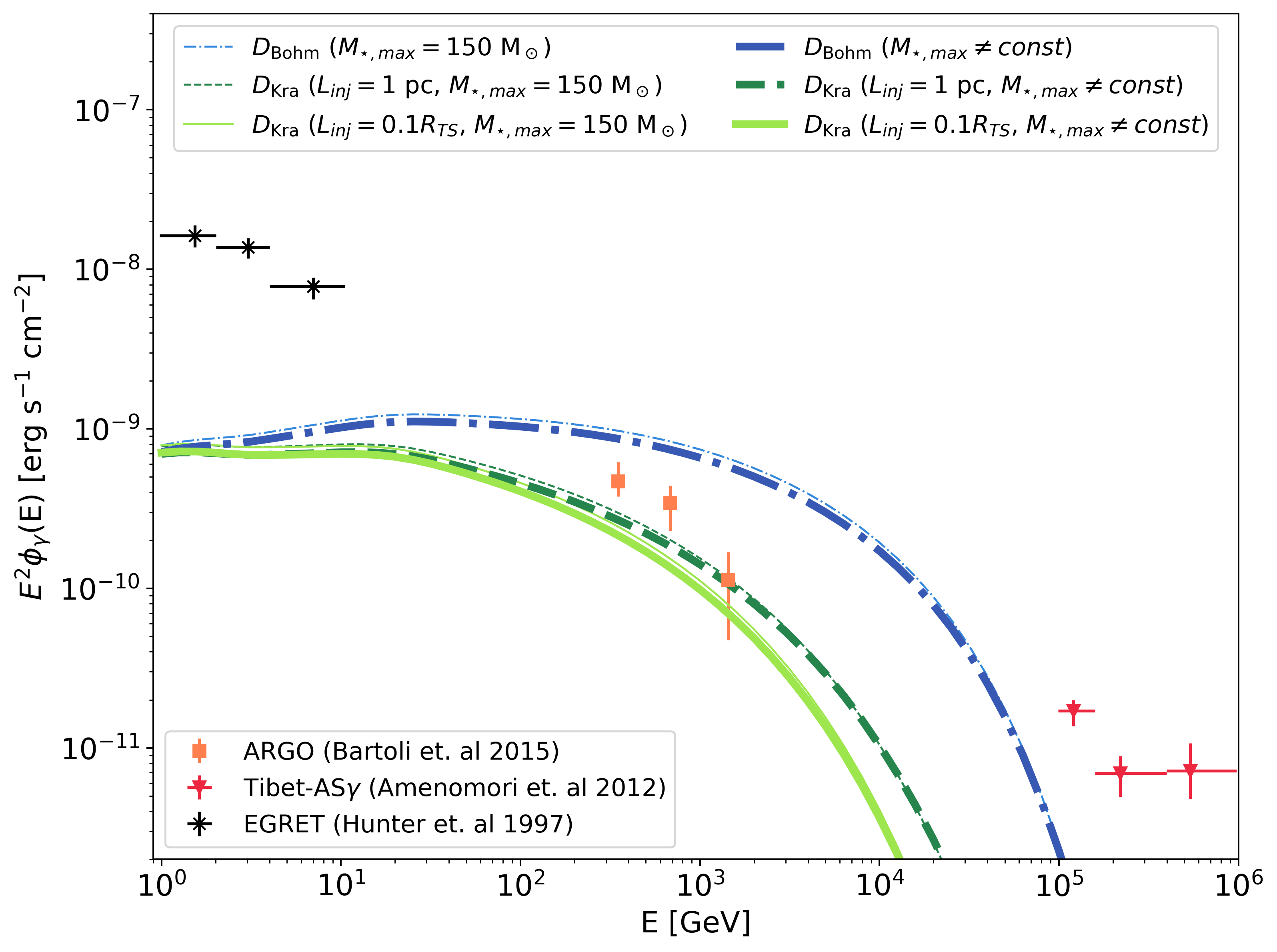}
\caption{Same as Fig.~\ref{fig:SED_Diffuse_Gamma}, but the spectral energy distributions are calculated considering \texttt{ROI2} (sky region shown in Fig.~\ref{fig:ROI2}). The crosses, square and triangles marks the emission observedy by EGRET \cite{Hunter_EgretDiffuse_1997}, ARGO \cite{Bartoli_ArgoDiffuse_2015} and Tibet-AS$\gamma$ \cite{Amenomori_TibetASgammaDiffuse_2021} respectively.}
\label{fig:SED_Diffuse_Gamma_ROI2_10}
\end{center}
\end{figure}

Fig.~\ref{fig:SED_Diffuse_Gamma_ROI2_10} shows the diffuse $\gamma$-ray spectrum in \texttt{ROI2} compared to EGRA, ARGO and Tibet-AS$\gamma$ observations. Similarly to \texttt{ROI1}, the emission at low-energy ($E_\gamma<10$ GeV), in all the considered cases under analysis, is an order of magnitude below the observed spectrum. On the contrary, above $E_\gamma \gtrsim 300$ GeV the flux for the Bohm case, regardless the maximum stellar mass considered, is found to be higher than the observed spectrum by Argo by a factor $\sim 3$ between $0.2$ TeV $\lesssim E_\gamma \lesssim 1$ TeV and $\sim 6$ at $E_\gamma \approx 1 \-- 2$ TeV. For the Kraichnan case, the flux is instead below the observed flux by a factor of $\sim 2$ at $0.2$ TeV $\lesssim E_\gamma \lesssim 1$ TeV, while at $E_\gamma \approx 1 \-- 2$ TeV the expected flux matches the observations. This is true independently on the chosen turbulence injection scale and the maximum stellar mass in the cluster. This is indeed an intriguing result, indicating that the non-resolved diffuse emission from YMSCs between a few hundreds of GeV and a few TeV is likely not negligible and possible even dominant.

\begin{figure}[h]
\begin{center}
\includegraphics[width=0.9\textwidth]{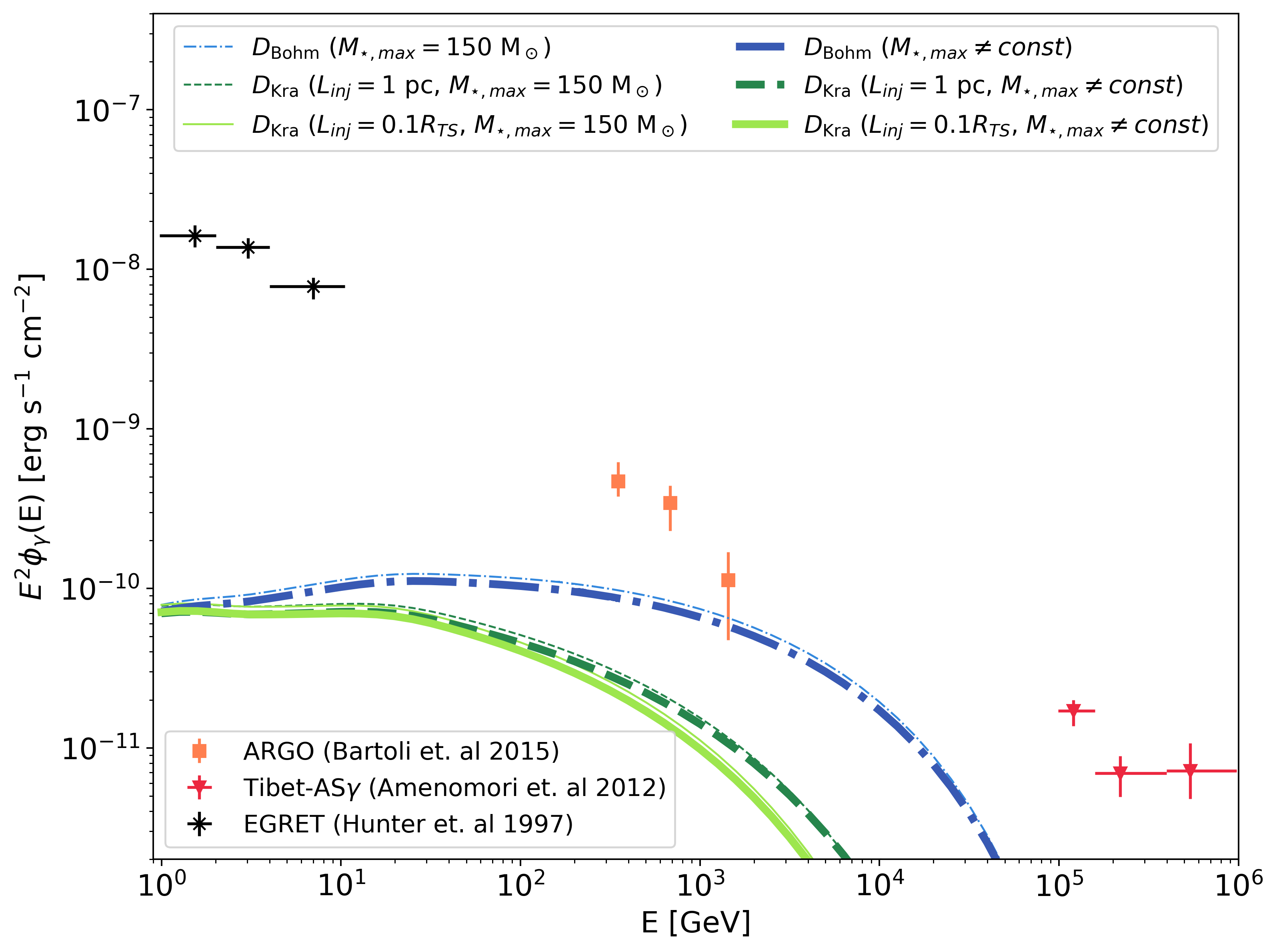}
\caption{Same as Fig.~\ref{fig:SED_Diffuse_Gamma_ROI2_10}, but the spectral energy distributions are calculated considering an efficiency of CR acceleration of $\epsilon_{CR}=0.1$.}
\label{fig:SED_Diffuse_Gamma_ROI2_1}
\end{center}
\end{figure}

We would like to emphasize that the outcomes we obtained were based on the assumption of high efficiency and high target density. It is possible to adjust these two parameters to reduce the flux and achieve consistency with Argo data, even in the Bohm scenario. Fig.~\ref{fig:SED_Diffuse_Gamma_ROI2_1} shows the same spectra after assuming an efficiency of CR acceleration of $\epsilon_{CR}=0.01$. Concerning the emission at very-high energy ($E_\gamma>100$ TeV), even considering the most optimistic scenario of a Bohm-like diffusion, the expected flux is significantly below the observations of Tibet-AS$\gamma$. As already discus, at these energies however, the number of contributing cluster is so low that no conclusion can be derived based on one single realization of the galactic population of YMSCs.

\begin{figure}[h]
\begin{center}
\includegraphics[width=\textwidth]{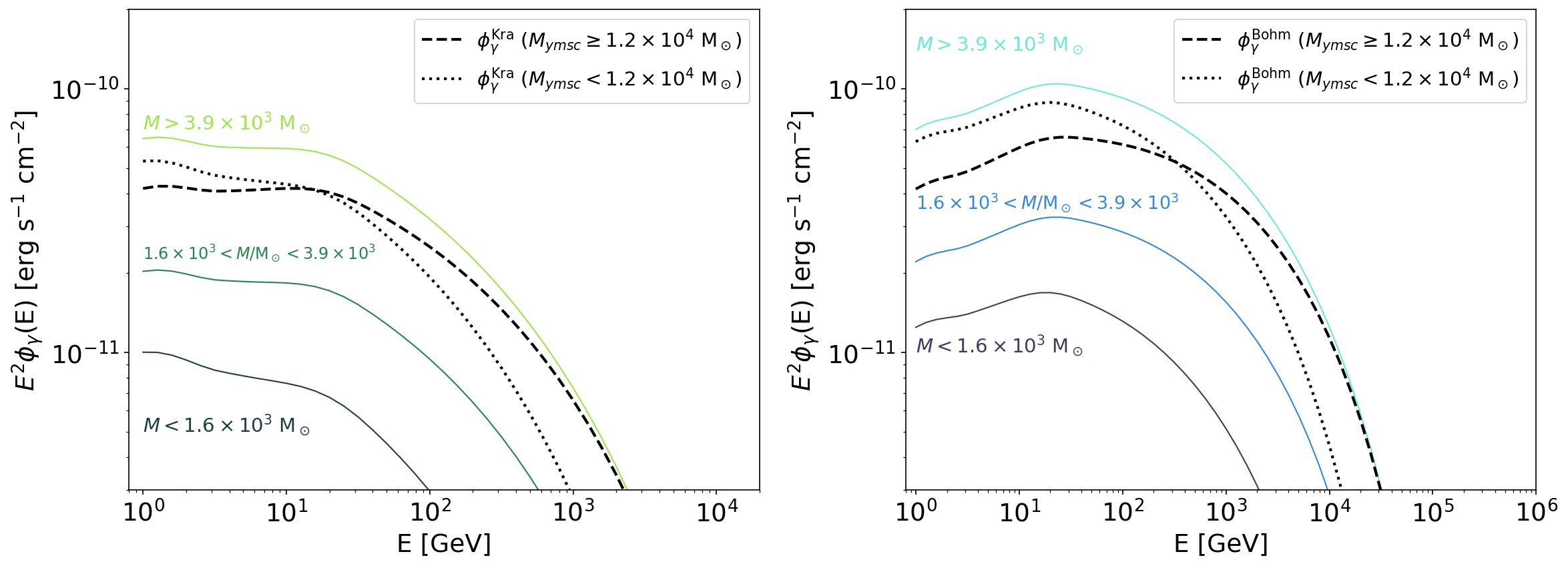}
\caption{Diffuse $\gamma$-ray emission from clusters in three diverse mass ranges (see legend in each panel). The dashed and dotted lines are the contribution of clusters above and below $10^4$ M$_\odot$ respectively.}
\label{fig:GammaSpectVsMass}
\end{center}
\end{figure}

Finally, for the sake of completeness, one last crucial aspect should be discussed. One may wonder whether the total $\gamma$-ray emission is mainly contributed by the most numerous and less massive stellar clusters or by the less numerous and most massive ones. The answer is the latter. In fact, the total $\gamma$-ray luminosity differentiated with respect to the cluster mass is:
\begin{equation}
\label{eq:dLgammadMsc}
\frac{dL_\gamma}{dM}= L_\gamma(M) \frac{dN_{YMSC}}{dM}
\end{equation}
where $\frac{dN_{YMSC}}{dM}\propto M^{-1.53}$ is the cluster mass function and $L_\gamma(M)$ is the $\gamma$-ray luminosity for a given cluster with mass $M$. As the $\gamma$-ray luminosity is dominated by the emission at low energy, one can write: 
\begin{equation}
\label{eq:LgammaProp}
L_\gamma (M) \approx \frac{4 \pi}{3} R_b^3 n_0 f_{TS} \sigma_{pp\rightarrow\gamma} \propto f_{TS} R_b^3 n_0
\end{equation}
where $\sigma_{pp\rightarrow\gamma}$is the integrated cross section for hadronic $\gamma$-ray emission. The distribution at the TS is given by Eq.~\ref{eq:fTS}, which scale as:
\begin{equation}
\label{eq:fTSPropto}
f_{TS} \propto \epsilon_{CR} n_1 v_w^2 \propto \dot{M} R_{TS}^{-2} v_w \epsilon_{CR}
\end{equation}
where in the last step we rewrote $n_1$ as $n_1=\dot{M}/4 \pi R_{TS}^2 v_w$. Knowing from Eq.~\ref{eq:Rbubble} and Eq.~\ref{eq:Rts} that:
\begin{subequations}
\begin{equation}
R_b \propto L_w^{1/5} n_0^{-1/5} t^{3/5}
\end{equation}
\begin{equation}
R_{TS} \propto \dot{M}^{1/2} v_w^{1/2} L_w^{-1/5} n_0^{-3/10} t^{2/5} \ ,
\end{equation}
\end{subequations}
and considering Eq.~\ref{eq:fTSPropto}, we can rewrite Eq.~\ref{eq:LgammaProp} as:
\begin{equation}
\label{eq:LgammaProp2}
L_\gamma (M) \propto L_w t n_0 \ .
\end{equation}
The wind luminosity has a quasi-linear dependence on the cluster mass $L_w \propto M$, with this acknowledged and using Eq.~\ref{eq:LgammaProp2}, Eq.~\ref{eq:dLgammadMsc} finally reads:
\begin{equation}
\frac{dL_\gamma}{dM} \propto M^{-0.5} t n_0 \ .
\end{equation}
If we then consider a wide interval of masses, ranging between some $M_{min}$ and $M_{max}$ with $M_{min} \ll M_{max}$, the total $\gamma$-ray luminosity will be:
\begin{equation}
L_\gamma \propto \int_{M_{min}}^{M_{max}} M^{-0.5} \propto M_{max}^{0.5}
\end{equation}
hence, dominated by the most massive stellar clusters. The above estimates can be verified plotting the $\gamma$-ray contribution of our synthetic population by mass intervals for \texttt{ROI1}. This is shown in Fig.~\ref{fig:GammaSpectVsMass}, where we have chosen three different mass intervals that contain the same number of stellar clusters, namely $M<1.6\times 10^3$ M$_\odot$, $1.6\times 10^3<M/ \textrm{M}_\odot <3.9\times 10^3$ and $M>3.9\times 10^3$ M$_\odot$. Even considering the most extreme case of constant maximum stellar mass, for both Kraichnan and Bohm cases the low mass clusters do not contribute much to the total diffuse emission: the contributions of the three different mass intervals in the total $\gamma$-ray flux at $\sim 10$ GeV is 9\%, 21\% and 70\% respectively. A different way to express the same concept is that 50 \% of the total $\gamma$-ray flux is contributed by stellar clusters with $M>1.2 \times 10^4$ M$_\odot$, which are only $\sim$20\% of the total ensemble (see dashed and dotted black lines in Fig.~\ref{fig:GammaSpectVsMass})

%---------------------------------------------------------

\section{Final remarks}
In this chapter we described the method for generating a synthetic galactic population of YMSCs. The procedure is complex and requires the knowledge or the assumption of several ingredients, starting from how we model stars within each cluster and ending with the rules for generation of the population itself. The final goal was to obtain an estimate of the diffuse $\gamma$-ray emission from the combination of multiple unresolved sources under the assumption of particle acceleration at the cluster wind TS, and to compare this emission with observations available in the literature. To this purpose, we considered two different regions of interest.

The first one, named \texttt{ROI1}, is defined as $15^\circ<l<5^\circ$ and $|b|<5^\circ$. Here, the measurement of the $\gamma$-ray diffuse emission is provided by the Fermi-LAT telescope. We obtain for a single realization of the galactic YMSC population a diffuse $\gamma$-ray emission consistent with observations. To be more specific, the contribution from stellar clusters in \texttt{ROI1} is a factor $2\--3$ below the data at $E_\gamma \gtrsim 10$ GeV considering the Bohm scenarios. For the Kraichnan case, the expected flux is a factor $\sim 10$ below the observed flux at $E_\gamma \gtrsim 10$ GeV. These result remains unchanged when considering different maximum stellar masses in the clusters, or different injection scales of the magnetic turbulence.

In the second region analyzed, named \texttt{ROI2}, and defined by $100^\circ<l<25^\circ$ and $|b|<5^\circ$, observations of the diffuse $\gamma$-ray emission are provided by EGRET, ARGO and Tibet-AS$\gamma$. We found here that the expected emission in the range $\sim 0.3\--1$ TeV is comparable with the observed flux from ARGO. More precisely, the expected emission overshoot by a factor $\sim 2\--3$ the observations by ARGO when considering the Bohm scenario, while is consistent at 1 TeV when considering the Kraichnan case. In general, full consistency can be recovered for the Bohm case by changing the product between $n_0$ (assumed to be 10 cm$^{-3}$) and $\epsilon_{CR}$ (assumed to be 0.1). We conclude that the emission at $\sim 1$ TeV from YMSCs is not negligible and that the observed diffuse emission could be even dominated by stellar clusters. 

%The final goal was to obtain an estimate of the diffuse $\gamma$-ray emission from the combination of multiple unresolved faint sources, and to compare this emission with observations available in the literature. Under the assumption of a particle acceleration model at the wind TS, we obtain for a single realization of the galactic YMSC population diffuse $\gamma$-ray emission consistent with observations from the Fermi-LAT telescope, meaning that the estimated $\gamma$-ray emission never over predicts observations. The emission remains true even considering various situations, such as different maximum stellar masses in the clusters, or different ways to model CRs distribution inside the wind-blown bubble. To be more specific, the contribution from stellar clusters can range from a maximum of $\sim 30\%$  at $E_\gamma \gtrsim 10$ GeV, up to a minimum of $\sim 90\%$ at $E_\gamma \approx 1 GeV$ for the most optimistic scenario where the particle diffusion inside the bubble is Bohm-like. Given our results in Ch.~\ref{ch:CygOB2} on Cygnus OB2, we have seen that the Bohm-like scenario is disfavored. Hence we do expect that the $\gamma$-ray contribution from stellar cluster will be something in between the Bohm and the Kraichnan cases. Whether this condition will also be valid in other regions of the sky will be investigated.

Beyond results concerning the $\gamma$-ray emission, our study contains several novelties from the point of view of stellar clusters. In fact, the generation of a population of YMSCs is already per se an interesting result, as it allows one to study the average properties of the Galactic clusters. For example, we found mean values of cluster wind luminosity and mass loss rate of $\overline{L_w} \approx 3 \times 10^{36}$ erg s$^{-1}$ and $\overline{\dot{M}} \approx 10^{-6}$ M$_\odot$ yr$^{-1}$ respectively. Furthermore, our analysis revealed that the synthetic population of YMSCs contains 157 clusters with masses exceeding $10^4$ M$_\odot$, which is approximately ten times larger than the number of Galactic clusters observed within this same mass range \citep{Zwart_YMSC_2010}.

Although highly simplified, the modeling of stellar parameters ultimately reproduces properties of winds that are fairly consistent with observations for high-mass stars.

From the point of view of YMSCs as particle accelerators, there are two main interesting aspects to emphasize. First, the validity of the acceleration model at the wind TS. Considering the effect of mass segregation, the establishment of a collective cluster wind is a likely scenario, making the model of particle acceleration we consider a physically motivated choice. A second important finding is, the distribution of the maximum energy of freshly accelerated particles obtained for the two considered types of plasma turbulence spectra. We found mean maximum energies of $E_{max}^{Kra}\approx$ 10--20 TeV, and $E_{max}^{Bohm} \approx 35$ TeV for Kraichnan and Bohm-like diffusion. No PeVatrons are found in this specific realization of the Galactic population, for both the considered models of particle diffusion.

The main limitation of our work is neglecting the contribution of SN explosions inside stellar clusters. For an age of $\gtrsim 3$ Myr we know that those events occur and they will probably dominate the energetics of the bubble \citep{Vieu_MSC+SNR_2022}. In this respect, our result for the $\gamma$-ray flux should be regarded as a lower limit in that SN explosions are likely to enhance the production of CRs. The reason why we have neglected SN explosions is because the modeling of particle acceleration ath the SNR shock propagating inside the wind bubble is not very well developed. One of the uncertainty is related to the SN shock Mach number, which is related to the particle acceleration efficiency by the shock. Let us consider, for example, the first SN exploding in the star cluster. The resulting SNR will reach the Sedov phase soon after the interaction with the wind TS. The Mach number of the shock in the hot bubble is $\mathcal{M}=v_{sh}/c_s$, where $v_{sh}$ is the SN shock speed and $c_s=\sqrt{\gamma k_B T_2/m_p}$ is the sound speed, with $\gamma$ as the adiabatic index. Neglecting possible effects of cooling, the temperature in the bubble is determined by the TS, hence, $k_B T_2=3/16 (m_p v_w^2)$, so that the Mach number is:
\begin{equation}
\mathcal{M}=\frac{v_{sh}}{c_s}=\frac{v_{sh}}{\sqrt{\gamma \frac{3}{16} v_w^2}}= \sqrt{\frac{16}{5}} \frac{v_{sh}}{v_w} \simeq 4 \left(\frac{v_{sh}}{5000 \ \rm km s^{-1}} \right ) \left(\frac{v_w}{2000 \ \rm km s^{-1}} \right )^{-1}
\end{equation}
Hence we see that standard values for $v_{sh}$ and $v_w$ give $\mathcal{M}\simeq a \ few$, which is not enough to generate efficient particle acceleration. The effect of cooling in the bubble could change the conclusion. However, subsequent SNRs will expand inside a material heated by the fist SN, hence again, the value of the Mach number should be estimated carefully.

\subsection{Future prospects}
The results presented so far are the outcome of work still in progress, for which some are to be improved, possibly leading to additional and more significant conclusions. 

\begin{figure}[h]
\begin{center}
\includegraphics[width=0.9\textwidth]{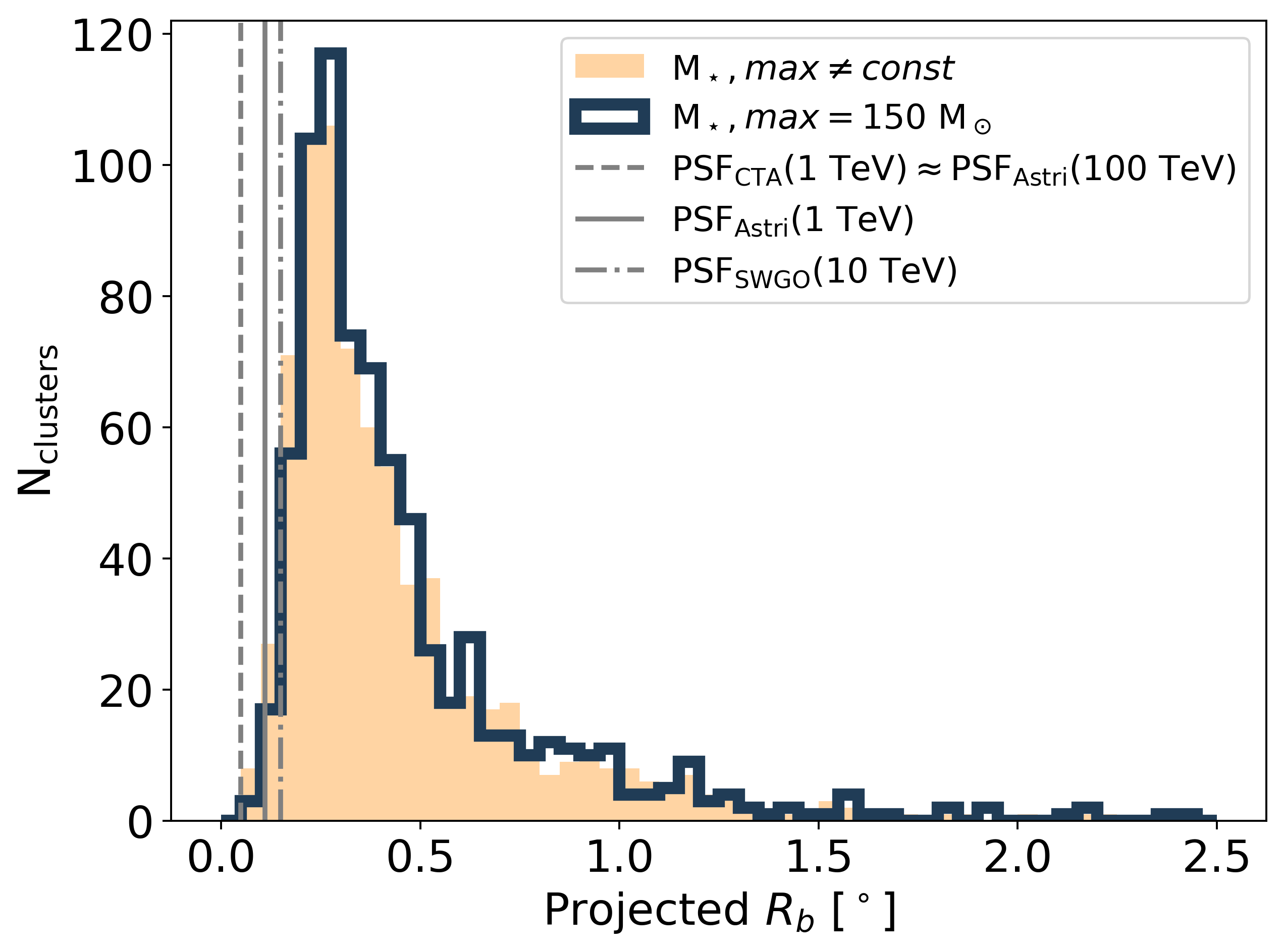}
\caption{Projected bubble radii compared to the point spread function of new generation $\gamma$-ray telescopes: Cherenkov Telescope Array (dashed line), Astri Mini Array (solid line), and the Southern Wide-field Gamma-ray Observatory (dot-dashed line). Filled and step histograms report the distribution of the projected radii for constant and nonconstant maximum stellar mass respectively. }
\label{fig:ProjRb}
\end{center}
\end{figure}

Some of these aspects are listed below:
\begin{enumerate}
\item From the point of view of stellar physics, rather than using empirical relations like we did here, we could include robust stellar models. In principle, we could generate tables of stars with different masses and ages, which can then be used to populate stellar clusters according to their age and masses. The upside of this approach is that stellar evolution would be taken into account, eventually accounting for evolutionary stages outside the main sequence, which are not currently considered. The downside is the computation time. 

\item Concerning the modeling of the star cluster population, to confirm the robustness of our stellar cluster population simulation method, lower-mass stellar clusters could be generated so that we could have a comparison with cluster surveys obtained from Gaia. In general, the number of locally generated clusters should be consistent with those observed. In addition, comparison with the cluster population of other Milky Way like galaxies can be performed.

\item In terms of modeling particle acceleration, several additional aspects could be considered. First, we could account for different injection spectral indexes after estimating for each cluster the termination shock Mach number. At lower energies, the contribution of second-order Fermi acceleration may prove to be non-negligible and should accordingly be considered. We do expect second order Fermi acceleration to affect particles up to several GeV if the magnetic turbulence is strongly enhanced.

\item For the calculation of the $\gamma$-ray emission, the estimation of the average target density remains a largely unbound parameter. For this purpose, a comparison with results obtained by considering a density dependent on the characteristics of the wind-blown bubble would be interesting. In addition, the inclusion of leptons and the subsequent calculation of their contribution to the diffuse $\gamma$-ray emission could prove to be and interesting exercise, possible leading to increasing constrains on the electron to proton fraction.  

\item An additional check that should be made is with the observed diffuse neutrino flux.

\item Finally, a relevant aspect which we want to investigate is the estimate of the number of detectable sources for the new generation of $\gamma$-ray telescopes, such as the Cherenkov Telescope Array, the ASTRI Mini Array, and the Southern Wide-field Gamma-ray Observatory. This specific task is non-trivial, and it will be covered in a future work. The main difficulty relies on the fact that stellar clusters are extended sources. This can be readily verified by looking at Fig.~\ref{fig:ProjRb}, where the projected sizes of the cluster bubbles are compared to the point spread functions of the future $\gamma$-ray telescopes. It should be noted that a considerable part of the YMSCs is characterized by an extension of 2--3 times the telescope point spread functions, and the analysis of extended sources, especially when located in crowded regions of the sky, is a particularly difficult and challenging task. In fact, we will need instrument response functions (IRFs) for extended sources, which at the moment are not available.
\end{enumerate}

%% file: CHAPTERS/chapter_4.tex
%---------------------------------------------------------
\chapter{Ionization of molecular clouds close to YMSCs}
%---------------------------------------------------------
\lettrine{A}{long} with high-energy particles, YMSCs also produce a sub-GeV population of CRs. The existence of these low energy particles is of fundamental interest for two reasons. From an observational point of view, the ability of low-energy particles to penetrate deep within the dense core of molecular clumps, inducing ionization of the dense material and triggering the generation of specific complex molecular compounds, makes them the perfect tool to probe the presence of freshly accelerated particles, in parallel with $\gamma$-ray observations. From the point of view of fundamental physics process, the induced ionization of low-energy CRs is a possible feedback mechanism that regulates the collapse of molecular clumps, making star formation less efficient. In fact, the molecular cloud interior can only be ionized by CRs, as UV and X-rays photons are efficiently absorbed at the cloud periphery \citep{Phan_W28_2020}. Moreover, it has been shown that ionization is mainly produced by CRs with energy $1\ \rm MeV<E\lesssim 100 \rm MeV$ \citep{Padovani_MCIon_2009}. The larger the ionization degree, the stronger the coupling of the plasma with the magnetic field, whose pressure works against the gravitational collapse.

Very young star clusters are often found to be close to or surrounded by the fragmented material of the parental GMC. As the cluster develops a wind-blown bubble, it is natural to expect some gas clumps to be encompassed within the expanding hot gas bubble. For these clumps, the ionization rate can be significantly different from the one induced by the low-energy population of galactic CRs. This is because at these energies, galactic CRs are swept away from the expanding motion of the hot bubble, and the continuos advection prevents any attempts of penetration. This effect is indeed similar to the shielding effect of solar wind on galactic CRs for energies $\lesssim 1$ GeV. Ionization is hence only provided by the population of freshly accelerated particles from the YMSC, whose spectrum can differ from the galactic standard one. 

In this chapter, we present the estimation of the ionization rate for molecular clumps found within a YMSC wind-blown bubble. Having in mind the result obtained in Ch.~\ref{Ch:chap3} we compute the ionization rate for an average Galactic YMSC. Afterward, we apply our calculation to the specific case of DR21, a molecular cloud observed in the close proximity of Cygnus OB2. For this case, the employed CR spectra used for the calculation of the ionization rate will be the ones found in Ch.~\ref{ch:CygOB2} that best reproduce the observed $\gamma$-ray emission. We will, hence, again consider the two extreme propagation cases, and we will discuss whether the measurement of the ionization rate can be used to constrain the propagation models, and under what circumstances.

The chapter is structured as follows: first, we describe the method employed to estimate the CR spectrum inside a molecular cloud accounting for the propagation of low-energy particles. Afterward, we apply this method to a synthetic YMSC considering the mean properties obtained in Ch.~\ref{Ch:chap3}. Finally, we calculate the ionization for parameters representative of DR21, and we compare the obtained result with archival data related to HCO$^+$ observations. 

%---------------------------------------------------------
\section{Penetration of low-energy CRs in a molecular cloud}
\label{sec:PartPenInMC}
\begin{figure}[h]
\begin{center}
\includegraphics[width=0.9\textwidth]{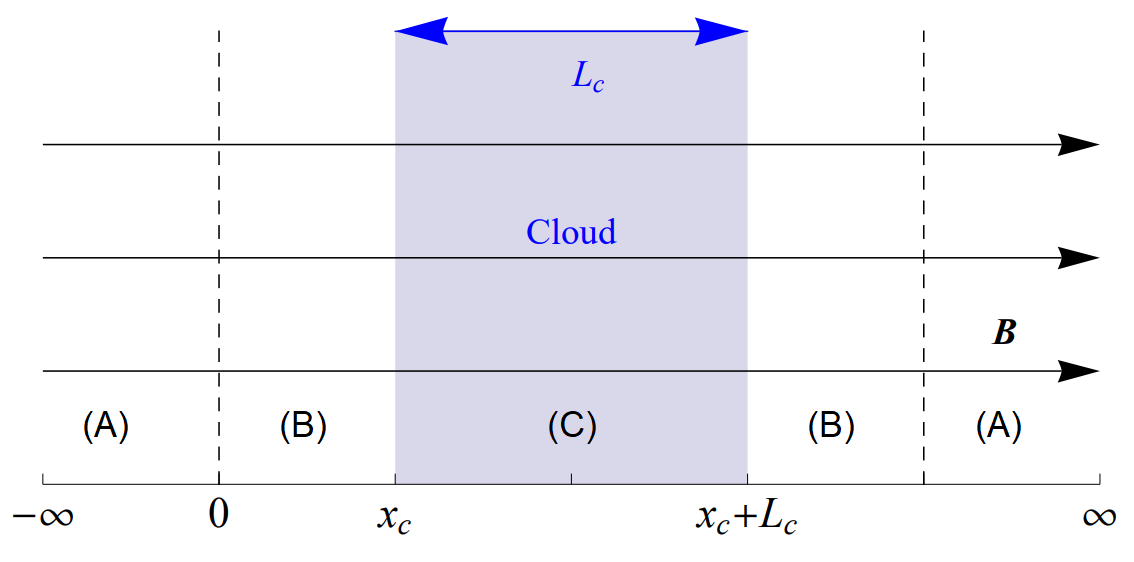}
\caption{Scheme of the simplified one-dimensional model employed to describe the geometry of the cloud.}
\label{fig:MCscheme}
\end{center}
\end{figure}

As CRs penetrate within a molecular cloud (MC), their spectrum is modified by the combination of propagation and energy loss effects. To obtain the final particle spectrum in a MC, we use the simple analytic prescription proposed by \cite{Morlino_CRsPenetrationMC_2015}. Let us, hence, consider a MC encompassed within a wind-blown bubble produced by a YMSC. The cloud is characterized by a size $L_c$ and a density $n_c$. Let us furthermore assume that the MC is completely permeated by the bubble magnetic field $B_2$, which we consider as spatially constant and aligned along the x-axis, such that the cloud can be schematized as one-dimensional (see Fig.~\ref{fig:MCscheme}). The MC density is higher with respect to the bubble density $n_2$. We then assume that the transition between the dense cloud environment and the hot, fully ionized, bubble material happens within a thin layer of thickness $x_c$. 

According to this picture, the system can be divided into three main regions:
\begin{itemize}
\item[(A)] A zone far away from the MC that we identify as the whole hot bubble downstream of the cluster wind TS. Here, the CR distribution ($f_2$) is provided by Eq.~\ref{eq:FDownstream} and is unaffected by the presence of the cloud\footnote{This statement is true unless a large fraction of the bubble volume is filled with MCs.}. In this one-dimensional problem, the downstream region is set at $x<0$ and $x>L_c+x_c$.
\item[(B)] A transition zone positioned at the cloud borders between $0<x<x_c$ and $L_c+x_c<x<L_c+2x_c$ where the distribution of CRs is affected by the presence of the MC.
\item[(C)] The cloud itself, defined between $x_c<x<L_c+x_c$.
\end{itemize}

To obtain the particle spectra within the MC, we assume that CRs propagate along the magnetic field lines crossing the cloud, hence, we neglect perpendicular diffusion, assuming that is far slower than parallel \citep{Morlino_CRsPenetrationMC_2015}. As explained in \S~\ref{subsec:MCIonization}, low-energy particles in a dense environment undergo severe energy losses due to ionization. Consequently, particles loose energy and migrate to lower energy, such that at a given momentum $p$, one has $f_2(p)>f_c(p)$, where we have defined $f_c(p)$ as the average CR spectrum within the cloud. Eventually, a negative spatial gradient will appear in the region (B), which in turn triggers the onset of streaming instabilities. The grown of streaming instabilities generates Alfvén waves with speed $v_A=B_2/\sqrt{4 \pi m_p n_2}$, that propagate towards the cloud direction. This will cause CRs escaping from the MC to be advected back into the cloud from region (B).

The particle distribution in region (B) can be obtained by solving the transport equation (see Eq.~\ref{eq:CRsTransportEq}). In the specific case of a steady state system, under the assumption of a spatially constant diffusion coefficient, the transport equation in (B) reads:
\begin{equation}
\label{eq:TranspEqIn(2)}
v_A \frac{\partial f}{\partial x} = D_{B} \frac{\partial^2 f}{\partial^2 x}
\end{equation}
where $D_{B}$ is the diffusion coefficient in region (B). Note that we are not considering energy losses. This is reasonable as the bubble density is expected to be relatively low.

If propagation in this region is mediated by the scattering with Alfvén waves, then simple dimensional analysis implies $D_B \sim v_A x_c$. As a consequence, for $x<x_c$ one has $x<D_B/v_A$, so that in Eq.~\ref{eq:TranspEqIn(2)} the diffusion term dominates. Under these circumstances, the solution is:  
\begin{equation}
f=-[f_2-f(x_c^-)]\frac{x}{x_c}+f_2
\end{equation}
where $f(x_c^-)$ is the value of the distribution calculated immediately outside of the MC. The former approximate solution can be then used to calculate the flux of CRs penetrating in the MC. This can be done by integrating Eq~\ref{eq:TranspEqIn(2)} between $-\infty$ and $r_c$:
\begin{equation}
-2 D_B \frac{\partial f}{\partial x} \bigg |_{x_c^-}+ 2 f(x_c^-) v_A = 2 f_2 v_A \ ,
\end{equation}
where the factor 2 accounts for both sides of the cloud. As emphasized by \cite{Morlino_CRsPenetrationMC_2015}, the validity of the previous equation does not require the presence of the streaming instability, but rather applies to the more general cases of Alfvén waves propagating towards the cloud. This is a common situation, as Alfvén waves cannot come from the cloud itself due to the damping induced by the dense and largely neutral environment. Note that the absence of Alfvén waves within the cloud implies that inside it particles move balistically along the magnetic field lines, so that the CR distribution can be considered approximately constant in space. 

Knowing the incoming particle flux, we can readily estimate the spectrum of CRs within the cloud by imposing an equilibrium between the incoming flux and the particle removal rate due to energy losses: 
\begin{equation}
\label{eq:FluxLossBalance}
2 f_2(p)v_A=\frac{L_c}{p^2} \frac{\partial}{ \partial p} [ \dot{p} p^2 f_c(p)]
\end{equation}
Noticeably, if energy losses are negligible, Eq.~\ref{eq:FluxLossBalance} leads to the trivial solution $f_c(p)=f_0(p)$. Using once again dimensional analysis, the condition for energy losses to be neglected is:
\begin{equation}
\label{eq:Iota}
\iota (p) \equiv \frac{v_A \tau_{loss}(p)}{L_c/2} \leq 1
\end{equation}
where $\tau_{loss}(p)=-p/\dot{p}$ is the energy loss time scale, which can be approximated as a power law in momentum \citep{Morlino_CRsPenetrationMC_2015}:
\begin{equation}
\label{eq:TauLossIon}
\tau_{loss}(p)= 1.46 \times 10^5 \left(\frac{p}{0.1 m_p c} \right)^{2.58} \left(\frac{n_c}{1 \ \rm cm^{-3}}\right)^{-1} \rm \ yr .
\end{equation}
The condition reported in Eq.~\ref{eq:Iota} can be phenomenologically interpreted as follow: as the particle propagation within the cloud is ballistic, the crossing time for a given CR can be simply defined as $\tau_{cross}=L_c/v_p$, where $v_p$ is the velocity of the particle with momentum $p$. The condition can be then rewritten as $\tau_{loss}/\tau_{cross}<v_p/v_A$, which means that energy losses are only important if the particle crosses several times the molecular cloud (of the order of $v_p/v_A$).

Using Eq.~\ref{eq:TauLossIon} in Eq.~\ref{eq:Iota}, we can estimate the energy above which ionization losses are negligible, which is:
\begin{equation}
E_{br}\simeq 70 \left(\frac{v_A}{100 \ \rm km s^{-1}} \right)^{-0.78} \left(\frac{N_c}{3 \times 10^{21} \ \rm cm^{-2}} \right)^{0.78} \ \rm MeV
\end{equation}
where $N_c$ is the column density of the MC. At energies above $E_{br}$, the CR spectrum in the cloud is the same as the one in region (A). On the other hand, below $E_{br}$, ionization losses start to be important, and the CR spectrum is modified. In this regime, the spectrum is obtained by integrating Eq.~\ref{eq:FluxLossBalance}. The final spectrum of particles within the MC can then be written as:
\begin{equation}
\label{eq:fc}
f_c(p) = 
\begin{cases}
f_2(p)  & \text{for } E \geq E_{br}  \\
f_2(p_{br}) \left(\frac{p}{p_{br}} \right)^{-0.42} \left\{ 1 - \frac{\iota(p)}{s-3} \left(\frac{p}{p_{br}} \right)^{-2.58} \left[1- \left(\frac{p}{p_{br}} \right)^{3-s} \right]  \right\} & \text{for } E < E_{br} \ .
\end{cases}
\end{equation}
where $f_2(p)$ is given by Eq.~\ref{eq:FDownstream}. We further assume that the particle spectrum in the bubble is well represented by a power-law, $f_2 \propto p^{-s}$. The latter assumption is fairly well motivated for the following reason: for the energy ranges we are interested in (E<10 GeV), particle propagation in the bubble is totally advective, hence, the spectrum is the same as the one at the injection site, which is a power-law for $p \ll p_{max}$. Note that Eq.~\ref{eq:fc} is a simplified analytical approximation of the true solution, which is also provided by \cite{Morlino_CRsPenetrationMC_2015}. However, it remains a good approximation for the purpose of this work.

For the sake of completeness, we underline that, in principle, the distribution of particles in the bubble has a radial dependence (see \S~\ref{sec:CygOB2fCR}). However, we stress again that, since we are interested in the low-energy regime, advection is the primary propagation process, and this produces a constant distribution in radius. Thus, ultimately, $f_2$ can be considered constant within the bubble to a good approximation.
%---------------------------------------------------------

%---------------------------------------------------------
\section{Ionization of molecular clouds embedded in the wind-blown bubble of a YMSC}
\label{sec:IonSynthYMSC}
As the wind-blown bubble develops around a newly born YMSC, it may happen that a molecular clump, formed due to the fragmentation of the parental GMC, is embedded within the hot shocked plasma. Let us then consider the case of a compact dense clump, with size $L_c=1$ pc and column density of $N_c=10^{23}$ cm$^{-2}$, close to an average galactic YMSC. We assume the cluster younger than $t_{age}\sim 3$ Myr, so that no supernova should have exploded within it yet. We furthermore fix the wind luminosity and mass loss rate to $L_w = 3 \times 10^{36}$ erg s$^{-1}$ and $\dot{M} = 10^{-6}$ M$_\odot$ yr$^{-1}$ respectively. These values correspond to the average wind luminosity and mass loss rate of an average Galactic YMSC, as esitmated in \S~\ref{subsec:YMSCGalDistr}.

We can compute the CR spectrum within the clump using Eq.~\ref{eq:fc}. To obtain the particle spectrum in the bubble, $f_2$, we assume the scenario of Kraichnan-like propagation, as it represents an intermediate case between the two most extreme regimes of particle propagation due to Kolmogorov or Bohm-like turbulence. Once again we assume that a fraction $\eta_B=0.1$ of the wind power is converted into a turbulent magnetic field, which is injected at a characteristic length scale of $L_{inj}=0.1 R_{TS}$. Under these conditions, the magnetic field in the bubble is $B_2=12$ $\mu$G.

\begin{figure}[h]
\begin{center}
\includegraphics[width=0.9\textwidth]{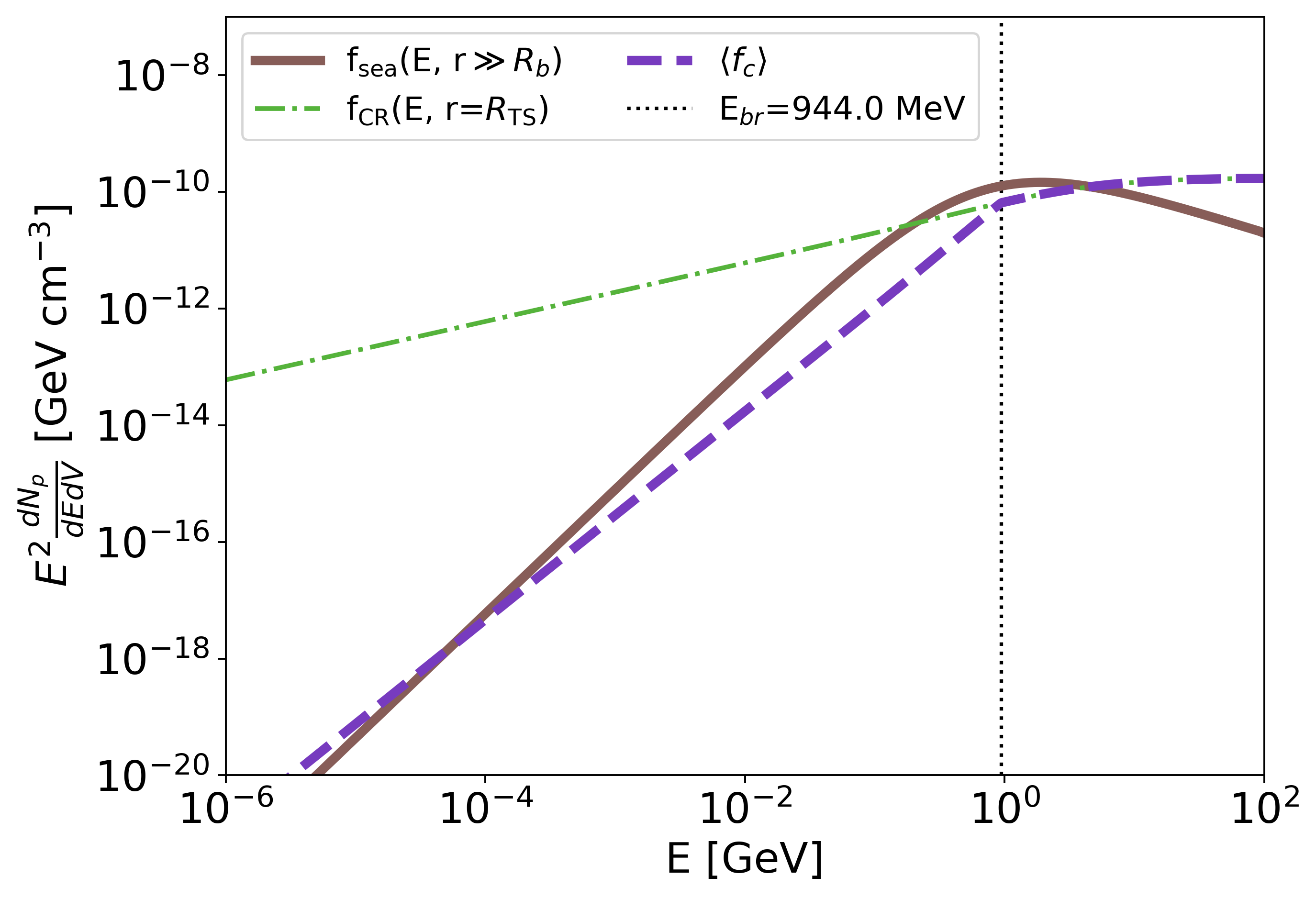}
\caption{Comparison between the CR spectrum in a molecular clump embedded within the wind-blown bubble (dashed line) and the particle spectrum in the bubble (dot-dashed line). The spectrum of the Galactic CR sea obtained from Voyager measurement is also reported (solid line). The vertical dotted line denotes the value of $E_{br}$ at which ionization-induced energy losses in the clump start to dominate.}
\label{fig:fcSynthYMSC}
\end{center}
\end{figure}

Furthermore, we calculate the density in the bubble as:
\begin{equation}
n_2=\frac{\dot{M}_{shell} t_{age}}{\frac{4 \pi}{3} R_b^3} \approx 0.05 \ \rm cm^{-3}
\end{equation}
where $\dot{M}$ is the mass of the shell evaporating in the bubble, which we estimate using Eq.~\ref{eq:ShellMdot}. Fig.~\ref{fig:fcSynthYMSC} shows the CR spectrum in the clump compared with the spectrum of the injected particles and the Galactic CR sea as measured by the Voyager spacecraft \citep{Cummings_Voyager1LECRs_2016}. It can easily be seen that below $E_{br}$, the particle spectrum is significantly different from that of the bubble, which, due to advection, is practically the same as that at the TS. The spectrum also differs from that of the CR sea. Therefore, as anticipated, the ionization rate will also be different. 

For simplicity, to estimate the ionization rate $\zeta_{H_2}$, we only consider the effect of ionization induced by primary protons. This can be calculated using Eq.~\ref{eq:IonRateIons}, that we report here for convenience:
\begin{equation}
\zeta_{H2}= \sum_k \int_{I(H_2)}^{E_{max}} c f_k(E_k) [1 + \phi_k(E_k)] \sigma_k^{\rm ion}(E_k) dE_k  + 
\int_0^{E_{max}} c f_p(E_p) \sigma_p^{\rm e.c.}(E_p) dE_p \ .
\end{equation}
We ignore the effect of primary electrons (Eq.~\ref{eq:IonRateEle}), as the content of accelerated electron inside stellar clusters has not been established yet. The obtained ionization rate is $\zeta_{H_2} \approx 2.2 \times 10^{-18}$ s$^{-1}$ , which is a factor $\sim 5$ lower than the Spitzer value $\zeta_{Spitzer}\approx 10^{-17}$ s$^{-1}$ \citep{Spitzer_IonizationMC_1968}. 

\begin{figure}[h]
\begin{center}
\includegraphics[width=0.9\textwidth]{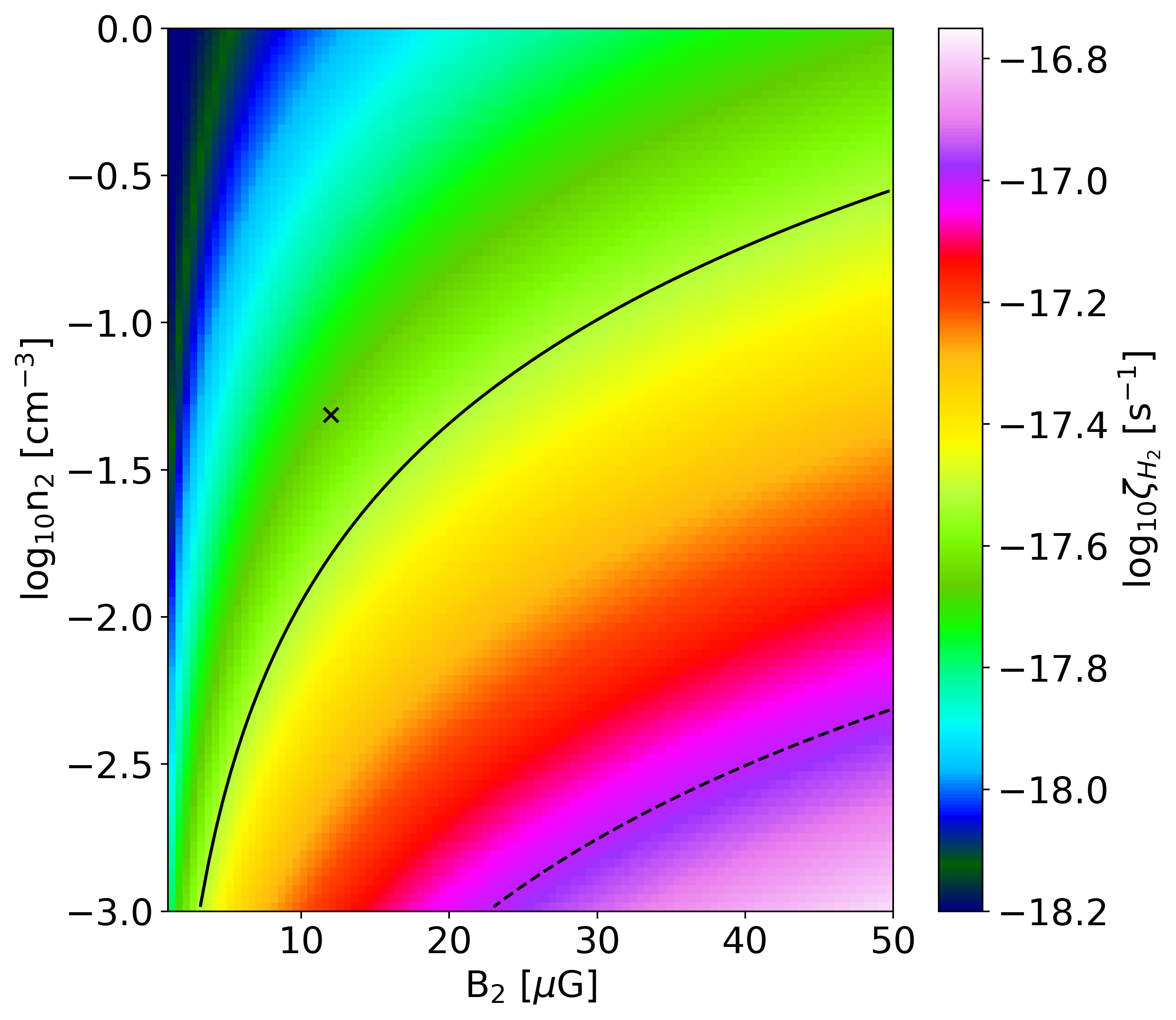}
\caption{Ionization rate induced by primary protons as a function of bubble density and magnetic field for a clump with $N_c=10^{23}$ cm$^{-2}$. The dashed line indicates the standard Spitzer value, while the cross marks our results obtained for values of $B_2$ and $n_2$ consistent with the YMSC properties (see text). The solid line indicates the ioniazation rate as estimated by \cite{Phan_IonRateDiffMC_2018}.}
\label{fig:ZetaSynthYMSC_MC22}
\end{center}
\end{figure}

A crucial aspect to emphasize is that the obtained ionization rate is strongly affected by the position of the break, namely by the energy at which ionization losses start to dominate. As described in \S~\ref{sec:PartPenInMC}, this depends on the number of times a particle crosses the molecular clump under consideration, which is related in turn to the value of the Alfvén velocity, which depends on the bubble magnetic field and density, both of which can easily be different from the expected values. When estimating the bubble density we assumed that the material within it is composed of the evaporated cold shell medium\footnote{The contribution of the wind material is negligible.}. However, shell fragmentation and turbulent mixing with the external interstellar medium can increase this value to a higher density. Concerning the magnetic field in the bubble, we are assuming conversion of the wind power into a turbulent magnetic field everywhere in the bubble. This is an approximate assumption, and the magnetic field fluctuation could have an additional spatial dependence. Following these concerns, we show in Fig.~\ref{fig:ZetaSynthYMSC_MC22} how the ionization rate changes as a function of the magnetic field and bubble density assuming that they are free variables. Noticeably, the value is always lower than $\zeta_{Sptizer}$ except for very low density and high magnetic fields. 

For the sake of completeness, we recall that the Spitzer value should be considered only as a reference value, as it is calculated without accounting for the physics of CR propagation within the cloud. \cite{Phan_IonRateDiffMC_2018} provide a better estimation using a refined version of the approach proposed by \cite{Morlino_CRsPenetrationMC_2015}. They found a ionization rate that is lower than $\zeta_{Spitzer}$ (see solid black line in Fig.~\ref{fig:ZetaSynthYMSC_MC22}), and closer to, but still higher than, our results, where the density of the bubble and the magnetic field are fixed to $n_2=0.05$ cm$^{-3}$ and $B_2=12$ $\mu$G, implying $\zeta_{H_2} \approx 9.2 \times 10^{-18}$. 

\begin{figure}[h]
\begin{center}
\includegraphics[width=0.9\textwidth]{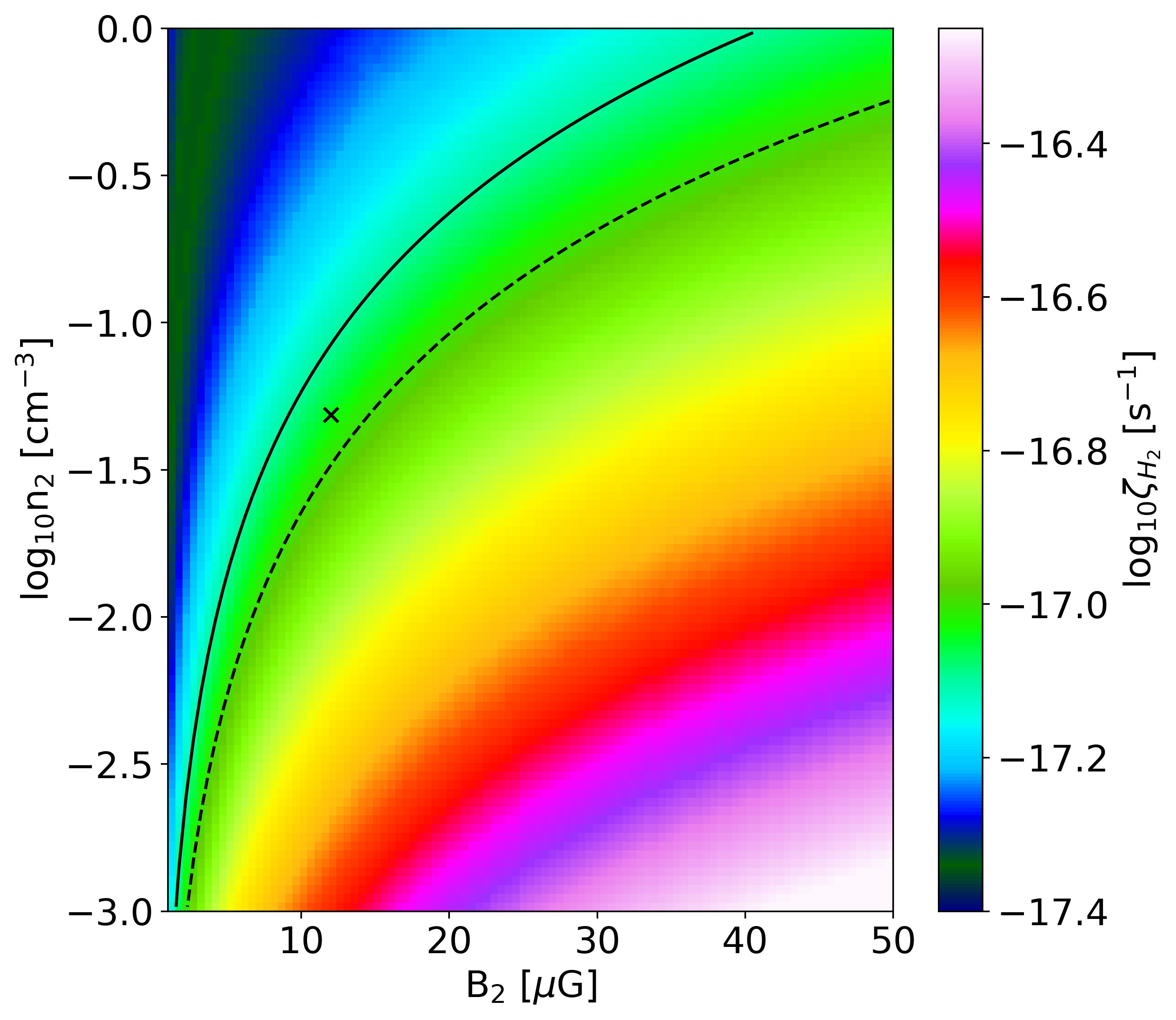}
\caption{Same as Fig.~\ref{fig:ZetaSynthYMSC_MC22}, but for a MC with column density $N_c=10^{22}$ cm$^{-3}$.}
\label{fig:ZetaSynthYMSC_MC23}
\end{center}
\end{figure}

Finally, we additionally calculate the ionization rates in the case of a molecular clump with $N_c=10^{22}$ cm$^{-2}$. In this case, the ionization rate estimated by \cite{Phan_IonRateDiffMC_2018} is much closer to the standard Spitzer value, and our result for fixed $n_2$ and $B_2$ lies between the two numbers (see Fig.~\ref{fig:ZetaSynthYMSC_MC23}). 

%---------------------------------------------------------

%---------------------------------------------------------
\section{Comparison with reality: The case of DR21}
Because Cygnus OB2 (Cyg OB2) lies within the Cygnus-X star formation complex, it represents the perfect test bed for studying the ionization rate in molecular clouds near a YMSC. Indeed, there are plenty of MCs that in projection are located close to the cluster, although the uncertainties in the position along the line of sight prevent a conclusive association. Nevertheless, a relevant fraction of these clouds is likely in close proximity of the cluster, as they show to be bright at 8 $\mu$m.
\begin{landscape}
\begin{figure}
\begin{center}
\includegraphics[width=1.5\textwidth]{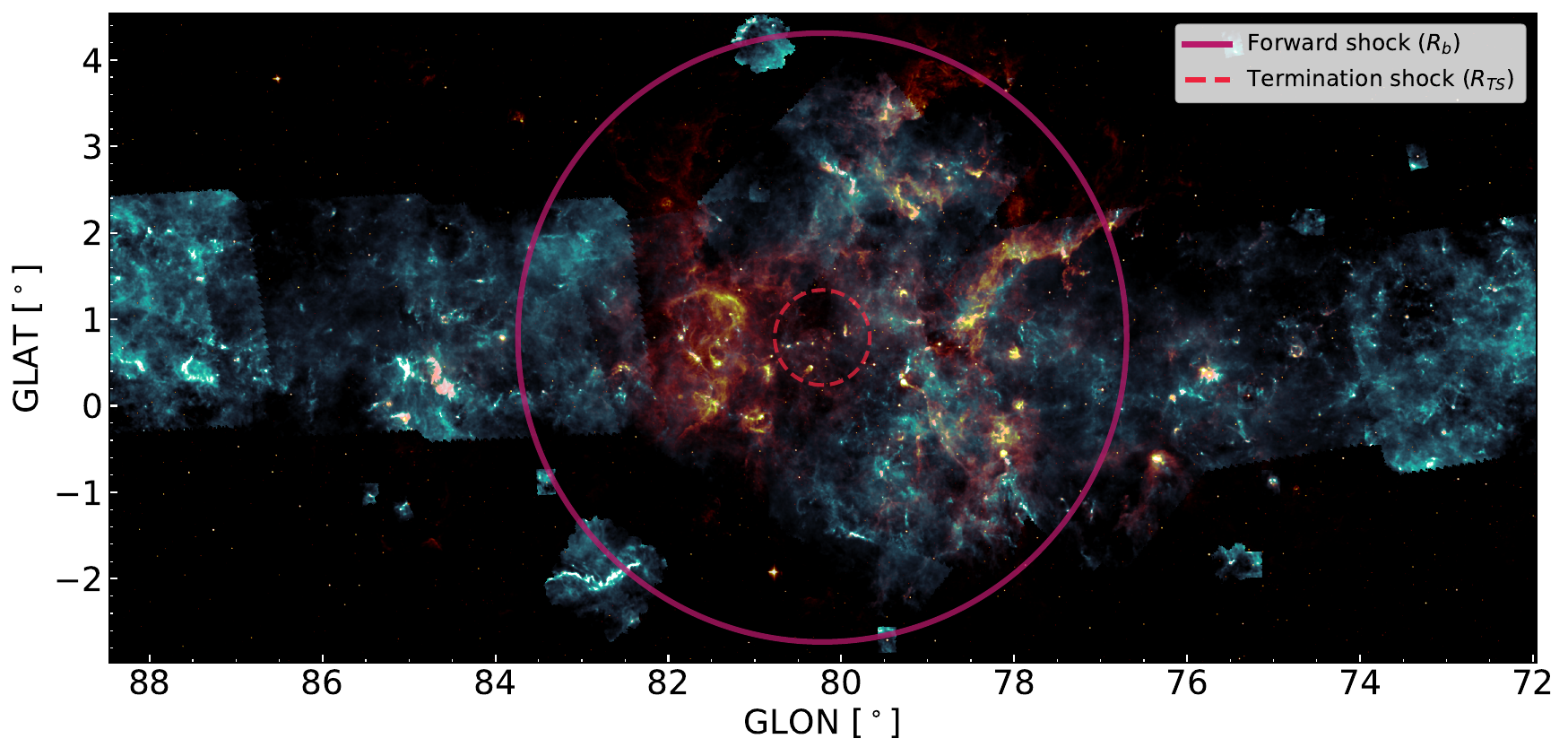}
\caption{Combination of observation at 500 $\mu$m and 8 $\mu$m from the Herschel SPIRE and MSX telescopes of the Cygnus X star-forming complex. Radiation at 500 $\mu$m (light blue scale) indicates emission from cold dust, while the 8 $\mu$m (orange scale) traces photon-dominated regions.}
\label{fig:CygX_8-500mu}
\end{center}
\end{figure}
\end{landscape}
Radiation at 8 $\mu$m should trace the emission from hot dust in photon dominated regions \citep{Rolling_PDRinNGC3603_2011}, heated by the ionizing light of massive stars, suggesting then the vicinity to an ionizing source (see Fig.~\ref{fig:CygX_8-500mu}). In addition, some clouds, such as DR18 and DR20NW, are characterized by an elongated shape pointing toward the direction of Cyg OB2 \citep{Schneider_CygnusXSFR_2006}, suggesting a direct interaction with the wind. More precisely, the vicinity of the DR20 complex is also confirmed by precise distance measurement with maser parallaxes \citep{Rygl_MaserParallCygX_2012}: the distance is estimated to be 1.46 kpc, very close to the estimated distance of Cyg OB2 (see \S~\ref{sec:CygOB2cluster}). In addition to DR20, again through maser distance measurements, two massive MCs are observed in the vicinity of Cyg OB2: DR21 and W75N. 

The MC DR21, positioned at 1.5 kpc, is of great interest for the scope of this work as the ionization rate has been measured by \cite{Hezareh_IonRateDR21_2008} based on HCO$^+$ observations (see \S~\ref{subsec:MCIonization}). Noticeably, the estimated ionization rate, $\zeta_{H_2}^{DR21}\approx 3.1 \times 10^{18}$ s$^{-1}$, is found to be lower than the one expected for clouds with the same column density \citep{Padovani_MCIon_2009}. This fact makes the case of DR21 even more interesting, given the result obtained in \S~\ref{sec:IonSynthYMSC} where a reduced ionization rate was shown for molecular clumps in interaction with a cluster wind-blown bubble.

Having in mind the results obtained in Ch.~\ref{ch:CygOB2}, we can estimate the ionization rate in DR21 by using the particle spectra that best reproduce the observed $\gamma$-ray emission, considering both the cases of Kraichnan and Bohm-like propagation. Following the same approach used in \S~\ref{sec:IonSynthYMSC} for the case of an average Galactic YMSC, we readily obtain the ionization rates of $\zeta_{H_2}\approx 1.88 \times 10^{-18}$ s$^{-1}$ and $\zeta_{H_2}\approx 1.08 \times 10^{-18}$ s$^{-1}$ for the Kraichnan and Bohm cases respectively. Noticeably, both values are in fairly good agreement with the measured ionization rate of DR21. 

Note that the particle spectra we are using are calibrated on $\gamma$-ray emission under the assumption of a constant particle density in the bubble equal to $n_2 \approx 8$ cm$^{-3}$. As we have already pointed out in \S~\ref{subsec:SpectAnalysis}, this density is perhaps too high (which we recall, causes a low acceleration efficiency). As mentioned in the previous section, the density in the bubble for estimating the ionization rate turns out to be a crucial parameter, so an erroneous estimate of the density is likely to produce a wrong result. In terms of the $\gamma$-ray spectrum, the efficiency and the density in bubble are totally degenerate parameters. This means that the same $\gamma$-ray spectrum can be re-obtained by keeping constant the product $n_2 \epsilon_{CR}$ while varying both. Following this reasoning, we additionally consider the case where the CR efficiency is fixed to $\epsilon_{CR}=0.1$, and the bubble density is changed accordingly. All other parameters are fixed to their previous values. The resulting ionization rates are $\zeta_{H_2}\approx 5.98 \times 10^{-17}$ s$^{-1}$ and $\zeta_{H_2}\approx 6.27 \times 10^{-18}$ s$^{-1}$ for the Kraichnan and Bohm cases respectively.

\begin{figure}
\begin{center}
\includegraphics[width=0.95\textwidth]{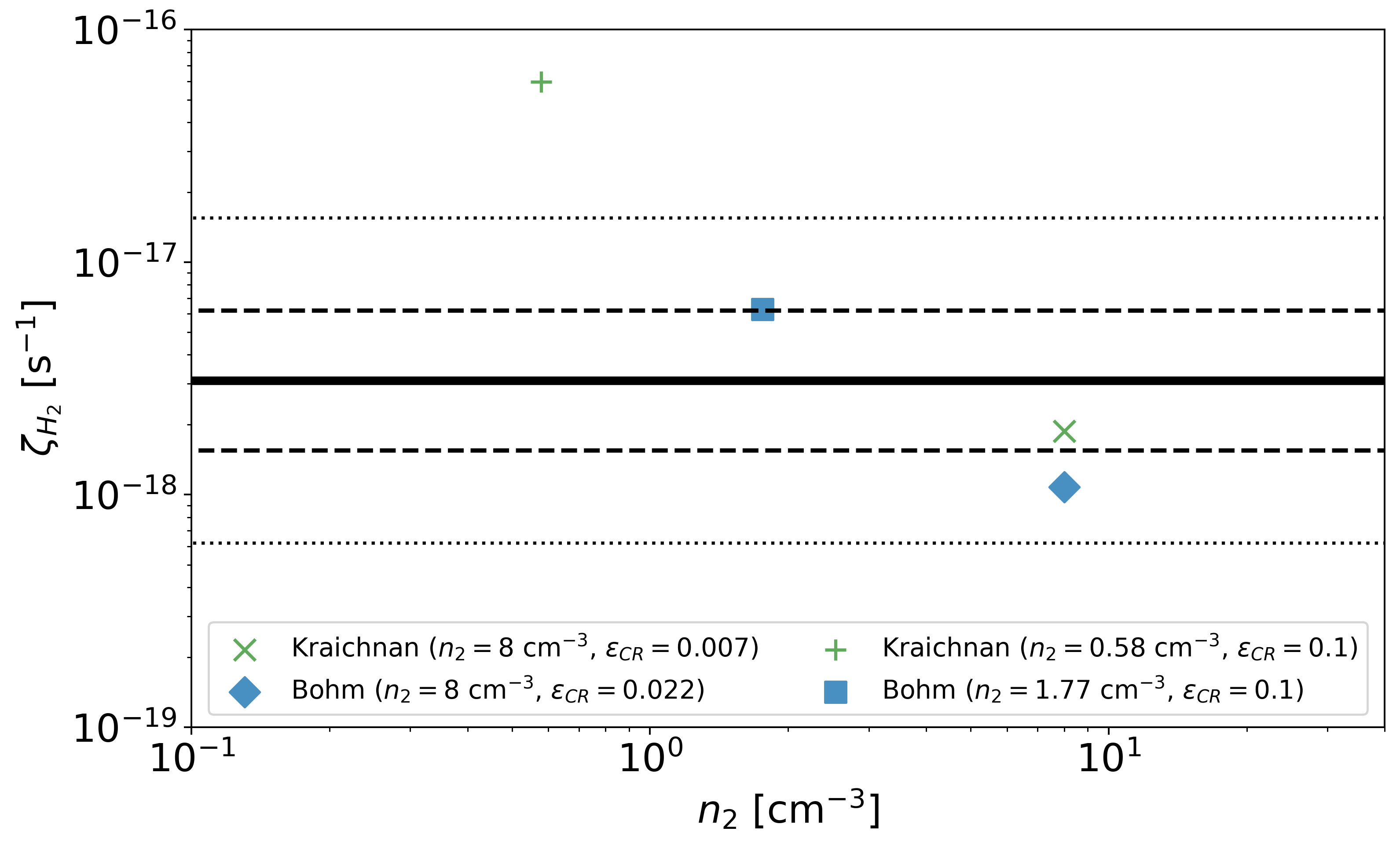}
\caption{Ionization rates estimated for DR21.  Cross and X points report the ionization rates for the Kraichnan case for fixed $n_2$ and $\epsilon_{CR}$ respectively. The same holds for the Bohm scenario, denoted by square and diamond markers respectively. The solid line indicates the ionization rate measured by \cite{Hezareh_IonRateDR21_2008}. Dashed and dotted lines report the uncertainty bars for a factor of 2 and 5 respectively. }
\label{fig:IonRateDR21}
\end{center}
\end{figure}

Fig.~\ref{fig:IonRateDR21} summarized the results obtained so far. \cite{Hezareh_IonRateDR21_2008} state that their measured ionization rate is correct within a factor "of a few", we hence report in Fig.~\ref{fig:IonRateDR21} also their value with uncertainty bars of factors 2 and 5. Interestingly, all values are roughly in agreement with the exception of the Kraichnan case with $\epsilon_{CR}=0.1$, which returns a ionization rate of more than one order of magnitude higher than the measured one. This is a remarkable result, showing that the measurement of the ionization rate of closeby MCs, combined with $\gamma$-ray observations, can be used to constrain (and in some cases even exclude) the models of CR diffusion in these sources.

%---------------------------------------------------------

%---------------------------------------------------------
\section{Preliminary conclusions and future prospectives}
The study of the ionization rate in clouds close to a YMSC is a relevant aspect to understand the physics of CRs in these sources. First, the measurement of the ionization rate, coupled with $\gamma$-ray observations, provides a unique method to constrain particle propagation in the vicinity of a stellar cluster, returning in additions valuable information on the CR spectrum at low energy. Secondly, understanding how YMSCs can affect the population of CRs in their neighborhood is a crucial aspect to better comprehend the process of star formation. Eventually, the enhancement (the suppression) of the content of CRs within the wind-blown bubble can induce negative (positive) feedback for star formation. 

The work presented in this chapter is split into two distinct parts. In the first half of the chapter, we estimated the ionization rate for a molecular cloud in close proximity of a typical Galactic YMSC. We fixed the wind luminosity and mass loss rate of the stellar cluster to the average values obtained in \S~\ref{subsec:YMSCGalDistr} during the study of the Galactic population of stellar clusters. We considered two possible column densities for the molecular cloud: $N_c=10^{23}$ cm$^{-2}$ and $N_c=10^{22}$ cm$^{-2}$, and we found a ionization rate always lower than the Spitzer value. We additionally compared our ionization rate with the more accurate estimates given by \cite{Phan_IonRateDiffMC_2018}, who calculated the ionization rate induced by the population of Galactic CRs using a robust modelization of particle propagation within the cloud. We found our ionization rates in good agreement with the value predicted by \cite{Phan_IonRateDiffMC_2018}, for both considered column densities. Finally, we provided predictions for the ionization rate considering  both the magnetic field and particle density in the bubble as free variables ranging between 1--50 $\mu$G and 10$^{-3}$--1 cm$^{-3}$ respectively.

In the second part of the chapter, we focused on the specific case of the molecular cloud DR21, which is located close to the YMSC Cygnus OB2.
To infer the ionization rate of DR21, we considered the particle distributions obtained in Ch.~\ref{ch:CygOB2} that best fit the observed $\gamma$-ray emission. More precisely, we used the particle distributions obtained assuming Kraichnan and Bohm-like diffusion. We then compared the resulting ionization rate with archival measurements obtained through observations of HCO$^+$ and found our result consistent with observations. However, we noticed that our estimates are probably biased. 

In fact, the employed particle distributions are obtained after fitting the $\gamma$-ray observations with the underlying assumption that the density in the bubble is 8 cm$^{-3}$, which is the average density in a $\pm 400$ pc region around Cygnus OB2. This number is too high and difficult to justify unless invoking extreme scenarios of material mixing and intense mass evaporation rate from the cold shell. We then recomputed the ionization rate using a lower particle density in the bubble, which we fix to $n_2=0.58$ cm$^{-3}$ and $n_2=1.77$ cm$^{-3}$ for the Kraichnan and Bohm cases respectively. The particle distributions are accordingly changed by setting the efficiency for CR acceleration to $\epsilon_{CR}=0.1$ such that the $\gamma$-ray spectrum remains unchanged and so does the consistency with observations. The resulting new ionization rates are in agreement with observations only when considering a Bohm-like diffusion. This would tents to favor a scenario where particle diffusion is provided by a diffusion coefficient with sharper energy dependence than for the Kraichnan case.

The entire work carried out in this chapter is still in its early stage, and there is room for significant improvements. At present, there are two main limitations to the presented work. The first is the absence of the contribution to the ionization from primary electrons. Currently, the electron distribution within the bubble remains unconstrained. Consequently, no reliable guesses can be made for their contribution. The second is the uncertainty in the diffusion coefficient in the bubble. This prevents us from making a robust modelization and systematic estimation of the ionization rate in clouds close to YMSCs. Therefore, it is currently not possible to make a conclusive statement regarding this particular feedback on star formation by YMSCs. 

To overcome these limitations, systematic radio and X-ray observations of YMSCs could be used to constrain the magnetic field intensity within the wind-blown bubbles. This would lead to better estimates of the ionization rates, in parallel to a better modelization of the particle diffusion in these systems. X-ray and radio observations could also provide precious constrains on the electrons distribution. We aim to investigate these aspects with the upcoming data from the MeerKAT radio telescope and the Galactic Plane survey carried out by eROSITA.

%% file: CHAPTERS/Conclusions.tex
%---------------------------------------------------------
\chapter{Conclusions}
%---------------------------------------------------------

\lettrine{Y}{oung} massive stellar cluster (YMSCs) represent at present one of the most intriguing sources in the panorama of high-energy astrophysics. The environment shaped by the winds of multiple young massive stars provides ideal conditions for particle acceleration, making these objects potential Galactic CR accelerators. In this manuscript, we analyzed the capability of YMSCs to produce CRs assuming the scenario where particle acceleration exclusively take place at the termination shock of the cluster wind. We performed the investigation by means of a comprehensive modelization of the hadronic $\gamma$-ray emission and by estimating the ionization rate in nearby molecular clouds induced by the injected low-energy particles. To model the CR distribution in the scenario we considered, we employed the model developed by \cite{Morlino_2021}, which describes both particle acceleration at the wind termination shock and particle propagation within the wind-blown bubble of the stellar cluster.

In the first part of the work, we focused on the specific case of the YMSC Cygnus OB2. We tried to interpret the diffuse $\gamma$-ray emission detected by the Fermi-LAT, HAWC, and ARGO experiments in terms of hadronic emission generated by a population of freshly accelerated particles, whose distribution is obtained following the model of \cite{Morlino_2021}. When modeling the particle distribution within the wind-blown bubble, we considered three possible scenarios corresponding to three different diffusion regimes: Kolmogorov, Kraichnan and Bohm like. We found that a Kolmogorov like diffusion in the system cannot reproduce the observed $\gamma$-ray spectrum, as the wind luminosity required to fit the $\gamma$-ray flux at very-high energy is more than one order of magnitude higher than the one inferred from the observed population of massive stars ($L_w \simeq 1.5 \-- 3 \times 10^{38}$ erg s$^{-1}$). Similarly, also Kraichnan like diffusion requires a wind luminosity higher than the one inferred from the stellar population. However, differently from the Kolmogorov case, the discrepancy is reduced and amount to a factor of $\sim 4$. When considering Bohm like diffusion, we found that the $\gamma$-ray spectrum can be explained with a wind luminosity consistent with our estimations, if $\sim 30\%$ of the wind power is converted into turbulent magnetic field. As the wind luminosity required to reproduce the $\gamma$-ray emission decreases when considering increasingly harder energy dependencies for the diffusion coefficient (at the cost of increasing by a factor of a few the fraction of wind power in magnetic turbulence), we concluded that, if acceleration proceeds according to our model, the most plausible scenario is the one where the turbulence spectrum has a spectral index that lies between the Kraichnan and Bohm cases.

In addition to the $\gamma$-ray spectrum, the observations of Fermi-LAT and HAWC provide also the radial profile of the emission in a circular region of 2.2$^\circ$ (corresponding to $\sim 54$ pc for an assumed distance of 1.4 kpc) centered on Cygnus OB2. We then compared the expected radial profile for our best-fit cases to the observed morphology. We limited this analysis only to the Kraichnan and Bohm scenarios. We found that both cases return a flat morphological profile in the same region. This profile is consistent with HAWC observations, but not with the Fermi-LAT ones, which are characterized by a centrally peaked morphology. From the point of view of our model, a flat morphology is the result of advection as the dominant propagation mechanism. Diffusive transport in a region of $2.2^\circ$ (corresponding to a physical radius of $\sim 54$ pc) is expected to start dominating only at energies above $\sim 100$ TeV. The reason for the peaked morphology observed by Fermi-LAT remains not fully understood. One possibility is that the increasing $\gamma$-ray luminosity towards the center of the system is due to inverse Compton emission by a population of accelerated leptons confined in a thin shell around the termination shock. This scenario will be investigated in a forthcoming paper, where a robust modelization of electrons acceleration and diffusion in stellar clusters, along with an estimate of their $\gamma$-ray emission, will be provided. 

Given the obtained results, we conclude that the model of particle acceleration at the wind termination shock of Cygnus OB2 is capable to reproduce the observed $\gamma$-ray emission, and that particle diffusion in the system is likely the result of a scenario in which the power spectrum of the magnetic turbulence has a spectral index in between Kraichnan and Bohm predictions. In the near future, it will be possible to confirm or reject these conclusions through a number of different tests, among which is the analysis of the $\gamma$-ray spectrum at different distances from Cygnus OB2, and the study of the $\gamma$-ray emission from nearby molecular clouds. In this regard, joint observations with the MAGIC and LST-1 telescopes are currently ongoing for the molecular clouds DR21 and W75N. \\

In the second part of the manuscript, we analyzed the $\gamma$-ray emission expected from a simulated population of Galactic YMSCs. Given that, within the model of particle acceleration and propagation in YMSCs we assume, the $\gamma$-ray radiation is expected to be mostly produced by particles confined within the wind-blown bubble, the emission of single clusters is expected to be on average extended ($\gtrsim 0.1^\circ$) and possibly difficult to distinguish from the diffuse. Therefore, we focused on the diffuse emission produced by the collective contribution of multiple sources, rather than examining the properties of each source individually.

To do so, we simulated a population of Galactic YMSCs following a cluster distribution function tuned on several observational parameters, such as mass and age distribution of local clusters, observed star forming rate in clusters, and the Galactic radial distribution of giant molecular clouds. To estimate the cluster wind properties, we simulated for each cluster a synthetic population of stars. When simulating the cluster stellar population, we considered two potential scenarios. In the first one, we assumed that the maximum mass of stars generated by each cluster is fixed as 150 M$_\odot$. In the second scenario, we assumed that the maximum stellar mass is a function of the cluster mass. After generating the star population, for each star we modeled the associated wind using again empirical relations based on the stellar parameters, such as bolometric luminosity, radius and temperature. We estimated these parameters starting from the stellar mass and using empirical relations. In addition to our main goal of estimating the diffuse $\gamma$-ray emission, the simulation of a synthetic Galactic population of YMSCs provides already per se an interesting results. We found that the average wind mass loss rate and wind power of stellar clusters are $\overline{\dot{M}} \approx 10^{-6}$ M$_\odot$ yr$^{-1}$ and $\overline{L_w} \approx 3 \times 10^{36}$ erg s$^{-1}$ respectively. The average size of the wind-blown bubble is found to be $\overline{R}_b\approx 60$ pc, so that the mean projected size of the $\gamma$-ray emission is $\approx 0.25^\circ$. These average values do not vary much when we considered different maximum stellar masses.

When calculating the CR distribution in each cluster, we considered two different diffusion regimes, Kraichnan and Bohm like. We further consider subcases for the Kraichnan regime, assuming two possible injection scales of the magnetic turbulence, 1 pc and 10\% of the termination shock radius. In all scenarios, and independently of the considered maximum stellar mass, we found no cluster able to produce maximum particle energies above 1 PeV. The fraction of YMSCs accelerating particles above 100 TeV is found to be $7\--13 \%$, while above 500 TeV it reduces to $\sim 0.5\--1 \%$, which corresponds to $3\--6$ clusters. With such a small number of clusters statistical fluctuations are not negligible, and multiple realizations of the galactic population of YMSCs are required to obtain a robust estimate of the expected number of PeVatron clusters.

After modeling the CR distribution in all YMSCs, we calculated the diffuse $\gamma$-ray emission in two specific regions of the Galactic plane, the first one defined by $15^\circ<l<5^\circ$ and $|b|<5^\circ$ (named \texttt{ROI1}) and the second one by $100^\circ<l<25^\circ$ and $|b|<5^\circ$ (named \texttt{ROI2}). The diffuse $\gamma$-ray spectrum in \texttt{ROI1} is available from Fermi-LAT observations, while in \texttt{ROI2}, data from EGRET, Argo and Tibet-AS$\gamma$ are considered. We found that in \texttt{ROI1} the computed $\gamma$-ray spectrum is below the data by a factor of $\sim 2\--3$ for the Bohm scenario and $\sim 10$ for the Kraichnan case at $E_\gamma>10$ GeV. In \texttt{ROI2}, we obtain a diffuse emission that overshoots the Argo observations by a factor $\sim 3$ between $0.2$ TeV $\lesssim E_\gamma \lesssim 1$ TeV and $\sim 6$ at $E_\gamma \approx 1 \-- 2$ TeV when considering the Bohm scenario. The spectrum in the Kraichnan case is instead below the data by a factor of $\sim 2$ at $0.2$ TeV $\lesssim E_\gamma \lesssim 1$ TeV, while accounting for most of the observed flux at $E_\gamma \approx 1 \-- 2$ TeV. With these numbers, we conclude that the observed diffuse emission at $\sim$ TeV energies could be largely provided by YMSCs. We also note that full consistency with observations in the Bohm scenario can be obtained by reducing the average target density (assumed to be 10 cm$^{-3}$) or the efficiency in CR acceleration (assumed to be 10\%).

In general, the obtained $\gamma$-ray spectra do not vary much when considering different maximum stellar masses in clusters, and different turbulence injection scales for the Kraichnan case. However, the spectra obtained when considering a Kraichnan and Bohm-like diffusion are significantly different. Using a Kraichnan diffusion coefficient returns a diffuse $\gamma$-ray emission characterized by a cut-off at energies of $\sim 100$ GeV, while the emission in the Bohm scenario starts to fade at energies above 1 TeV. At lower energies the spectrum for the Bohm case is found to be harder than for Kraichnan. This is a direct consequence of the interplay between advection and diffusion, whose relative importance differs in the two cases under analysis. In the Kraichnan case, diffusion starts to dominate at energies lower than in the Bohm case. This produces a decrease of the number of emitting particles in the wind blown bubble which causes a softening of the $\gamma$-ray spectrum. Lastly, we explored how the diffuse $\gamma$-ray emission varies with the cluster mass. Our analysis revealed that the less numerous and most massive clusters are the primary contributors to the diffuse emission.

Several additions and improvements to the current status of the work are possible and foreseen. At present, the mains missing ingredients in the second part of the presented work are the contribution to particle acceleration by supernova explosions and the contribution to $\gamma$-ray emission from accelerated leptons. Both aspect are still poorly understood and lack a firmly established theoretical framework. Regarding the simulation of the galactic population of YMSCs, our future plans include conducting a comprehensive consistency check of the generated population by comparing it to the observed population of clusters in Milky Way-like galaxies. Furthermore, multiple realizations of the Galactic population will have to be considered. \\

In the third and last part of the manuscript, we moved our analysis to lower energies, and we investigated the ionization rate induced by sub-GeV particles in molecular clouds close to YMSCs. Young clusters are expected to be surrounded by dense molecular clumps, which can be either remains of the cluster parent molecular cloud or generated by the fragmentation of the swept-up shell created by the expansion of the wind bubble. We considered a situation in which one of these clumps finds itself embedded within the wind blown bubble. We modeled the propagation of low energy CRs following the approach of \cite{Morlino_CRsPenetrationMC_2015}, where particles in the cloud move balistically along the magnetic field while experiencing energy losses due to direct ionization. Following this method, the particle spectrum in the cloud is equal to the one in the wind blown bubble for energies larger than a certain value $E_{br}$, which describes the energy below which losses start to be relevant. Energy losses are relevant when particles cross several times the molecular cloud, and this condition is related to the Alfvèn velocity in the bubble, hence, it depends on the value of plasma density and magnetic field strength in the bubble. For $E<E_{br}$, the spectrum is harder as it is modified by energy losses. 

As a first trial, we considered a molecular clump of size $L_c=1$ pc with two possible values of the column density $N_c=10^{22}$ cm$^{-2}$ and $N_c=10^{23}$ cm$^{-2}$. We assumed the case in which the clump is close to a YMSC of age less than 3 Myr whose wind luminosity and mass loss rate are equal to the average values estimated from our analysis of the Galactic population of clusters, i.e. $\overline{L_w} \approx 3 \times 10^{36}$ erg s$^{-1}$ and $\overline{\dot{M}} \approx 10^{-6}$ M$_\odot$ yr$^{-1}$. When calculating the CR distribution in the wind blown bubble, we considered the scenario where particle propagation is governed by Kraichnan like diffusion. Assuming for the magnetic field in the bubble a value such that the magnetic turbulence power is $10\%$ of the wind luminosity, and considering a density in the bubble consistent with the mass evaporated from the swept up shell, we found a ionization rate of $\zeta_{H_2} \approx 2.2 \times 10^{-18}$ s$^{-1}$ for $N_c=10^{23}$ cm$^{-2}$. This is lower than the Spitzer value by almost one order of magnitude. Even when compared with more recent estimates, presented in \cite{Phan_IonRateDiffMC_2018}, our ionization rate is still lower by a factor of a few.

When considering a lower column density of the cloud, we found a ionization rate of $\zeta_{H_2} \approx 9.2 \times 10^{-18}$, which is consistent within a factor of a few with both the Spitzer value and the estimates given by \cite{Phan_IonRateDiffMC_2018}. We noted that the ionization rate obtained can substantially differ by varying the energy at which the losses become important. As this value depends on the magnetic field and density inside the bubble, which are highly uncertain parameters, we also made predictions for the ionization rate by taking into account a broad range of magnetic field and density values. In general, low (high) magnetic field and high (low) density produced lower (higher) ionization rates than the Spitzer value.

After studying the case of a generic cluster, we focused on the specific case of Cygnus OB2 and estimated the expected ionization rate for the nearby molecular cloud DR21. To this end, we used the CR distributions in Cygnus OB2 calculated for a Kraichnan and Bohm like diffusion that best fit the observed $\gamma$-ray emission. We found the ionization rates to be $\zeta_{H_2,\ K}\approx 1.88 \times 10^{-18}$ s$^{-1}$ and $\zeta_{H_2,\ B}\approx 1.08 \times 10^{-18}$ s$^{-1}$ for the Kraichnan and Bohm cases respectively. These values are consistent within a factor of a few with the ionization rate estimated by observations of the HCO$^+$ and equal to $\zeta_{H_2}^{DR21}\approx 3.1 \times 10^{-18}$ s$^{-1}$ \citep{Hezareh_IonRateDR21_2008}. 

These ionization rates are obtained considering a density in the bubble of $8$ cm$^{-3}$, which is the average density observed near Cygnus OB2 and inferred from HI and CO observations. This value of density is considerably high.
Given the significant variation observed in the ionization rate by varying the density within the bubble, we also investigated a scenario where the density is kept constant at $0.58$ cm$^{-3}$ and $1.77$ cm$^{-3}$ for the Kraichnan and Bohm scenarios respectively. These values are chosen so that the $\gamma$-ray spectrum remains unchanged if the efficiency for CR acceleration is fixed to $10\%$ of the wind luminosity. The chosen densities are in better agreement with the expected value assuming that the material inside the bubble is composed of the mass evaporated from the swept-up shell. We found for this new scenario ionization rates of $\zeta_{H_2}\approx 5.98 \times 10^{-17}$ s$^{-1}$ and $\zeta_{H_2}\approx 6.27 \times 10^{-18}$ s$^{-1}$ for the Kraichnan and Bohm cases respectively. The ionization rate for the Kraichnan case is more than one order of magnitude higher than the value inferred from observations. This seems to favor the Bohm case, or in general, a diffusion coefficient in the system with an energy dependence stronger than the Kraichnan case. 

The work presented in this last part is still at an early stage of development. Nevertheless, the results obtained are promising and underline the importance of having a parallel modelization of the $\gamma$-ray emission together with the ionization rate induced by low-energy CRs. Indeed, the combination of these two pieces of information can provide a powerful consistency check for any model of CR acceleration and propagation. Furthermore, the capability of understanding the ionization rate of clouds close to YMSCs is crucial to assess the relative importance of the feedback channels that govern the star formation process.

At present, the main limitation of this work is given by the absence of the contribution to the ionization rate induced by primary leptons. As a future step, we plan to utilize X-ray and radio data from Galactic surveys by eROSITA and MeerKAT to conduct a systematic analysis of the environment surrounding YMSCs. This will enable us to obtain a more reliable estimation of the magnetic field within the wind-blown bubbles of stellar clusters, which will lead to even more solid estimates of the ionization rate.

Now more than ever the modelization of YMSCs as CR accelerators and $\gamma$-ray sources is of primary importance. In the coming years, thanks to the new generation of $\gamma$-ray observatories, such as CTA, ASTRI Mini Array, and SWGO, the importance of YMSCs as high-energy sources is bound to increase. The CTA and ASTRI observatories will perform deep surveys of the Galactic plane, which, combined with extended surveys in other bands, such as those of MeerKAT and eROSITA, will provide a valuable set of information for the systematic study of these sources. Eventually, multi-band analyses will be the key to study particle acceleration in YMSCs and evaluate their contribution to the generation of Galactic CRs. All these studies will hopefully soon bring us closer to the solution of the century-long enigma of the origin of CRs. In parallel, a campaign of observations and investigation of the ionization rate in molecular clouds in close proximity to YMSCs would provide a significant cross-check for the presence of accelerated particles. These observations could also provide insights into the star formation mechanism and improve our understanding of feedback mechanisms associated with the presence of freshly accelerated CRs in the vicinity of clusters.

%% file: CHAPTERS/appendix_A.tex
%\chapter*{Appendix}
%\addcontentsline{toc}{chapter}{Appendix}
%\markboth{APPENDIX}{}

\appendix
\chapter*{Appendix}

\markboth{Appendix}{}
\addcontentsline{toc}{chapter}{Appendix}
\renewcommand{\thesection}{A.\arabic{section}}

\section{$\gamma$-ray emission as a tracer of CRs}
\label{subsec:RadMechanisms}
The detection of high-energy radiation from a given object in the sky is direct proof of the presence of energetic particles. CRs may produce $\gamma$-rays through different possible radiation processes that depend on the type of emitting particles. Energetic hadrons interacting with the ISM matter may undergo the following nuclear interactions: 
\[
\begin{aligned}
p+p \rightarrow p+p+\pi^0 \\
p+p \rightarrow p+n+\pi^+ \\
p+p \rightarrow p+p+\pi^++\pi^-
\end{aligned}
\]
where $p$ are protons and $\pi$ are pions\footnote{It is worth noting that charged pions are also created as a result of nuclear interactions. These mesons can then decay, producing neutrinos that propagate straight to Earth. Neutrino astronomy can therefore be considered as an additional probe of the presence of CRs \citep{Spiering_NeutrinoAtro_2012}, with the fundamental and important difference from $\gamma$-ray astronomy that neutrinos can only be created via a hadronic interaction channel, so the detection of neutrinos is a smoking gun for the presence of hadronic CRs. Unfortunately, given the low interaction cross-section of neutrinos and the low fluxes expected at high-energy, this branch of astronomy remains affected by low statistics and hence particularly challenging, requiring massive detectors and very long exposure times.}. The production of neutral pions is the main channel to generate hadronic $\gamma$-ray emission, as these subsequent decays according to $\pi^0 \rightarrow \gamma + \gamma$. Clearly, the creation of neutral pions is only possible if the energy of a proton is sufficiently high. The proton energy threshold ($E_{th}$) can be obtained through a few straightforward calculations of particle kinematic \citep{Rybicki_RadiativeProcess_1986}, and is:
\begin{equation}
E_{th}=2m_\pi c^2 + \frac{m_\pi^2}{2m_p}c^2 \approx 280 \ \rm MeV
\end{equation}
where $m_\pi$ is the pion mass.

\begin{figure}
\begin{center}
\includegraphics[width=0.8\textwidth]{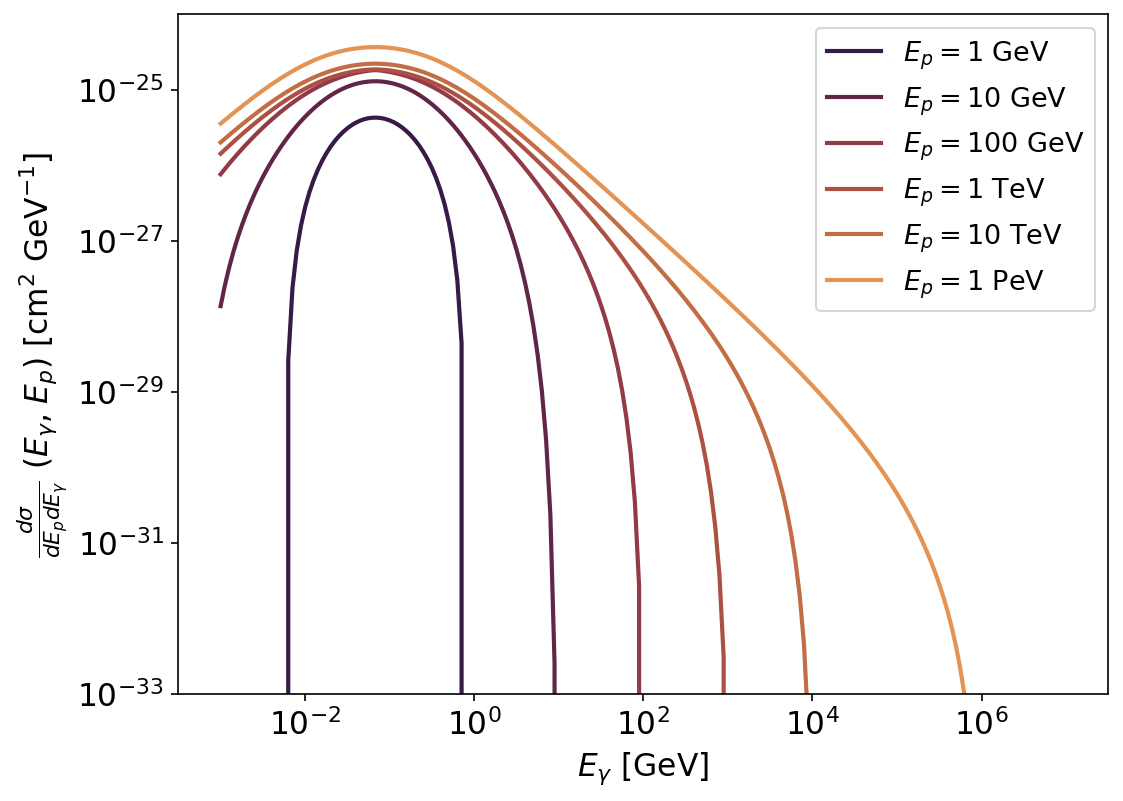}
\caption{Differential cross section for $\gamma$-ray production for different protons energy of 1 GeV, 10 GeV, 100 GeV, 1 TeV, 10 TeV, and 1 PeV.  The cross-section is calculated using the recipe of \cite{Kafexhiu_SigmaPi0Gamma_2014}, considering the specific case where the trend at energies higher than 100 GeV is obtained using the Montecarlo SIBYLL.}
\label{fig:CrossSection}
\end{center}
\end{figure}

Given a distribution of CRs ($f_{CR}$), it is possible to calculate the expected $\gamma$-ray flux ($\phi_\gamma$) from $\pi^0$ production within a certain volume $V$ as:
\begin{equation}
\label{eq:GammaFluxGeneral}
\phi_\gamma(E_\gamma)=\frac{1}{4 \pi d^2} \int \int_V c f_{CR}(E_p, r) n(r) \frac{d \sigma (E_p, E_\gamma) }{dE_p} dE_p dV, 
\end{equation}
where $d$ is the distance of the considered volume, $\sigma$ is the cross-section for $\gamma$-ray production from $\pi_0$ decay, $n$ is the number density of the target medium, and $E_p$ with $E_\gamma$ are the CR kinetic energy and the $\gamma$-ray energy respectively. \cite{Kafexhiu_SigmaPi0Gamma_2014} provides an analytical prescription for the cross-section, based on empirical fits of published results from p-p interactions at particle colliders and outcomes of Montecarlo predictions based on the Standard Model. The full expression results rather cumbersome, but it can be briefly described as the product of two contributing terms:
\begin{equation}
\frac{d\sigma}{dE_\gamma}(E_p, E_\gamma)= A(E_p) \times F(E_p, E_\gamma)
\end{equation}
where $A(E_p)=\rm max (d\sigma_\pi/dE_p)$ is the maximum value of the pion production cross section, while $F(E_p, E_\gamma)$ is a term describing the spectrum of produced $\gamma$-ray as a function of the proton energy. Fig.\ref{fig:CrossSection} shows the cross section as a function of $E_\gamma$ for different values of the parent proton energy. Notice that a proton is able to produce a $\gamma$-ray with a maximum energy of $\sim 0.1 E_p$. This is particularly important as the detection of $\gamma$-rays with a certain energy $E'_\gamma$ implies the presence of hadrons with energies of at least $\sim 10 E'_\gamma$.

Leptons produce $\gamma$-rays through the Inverse Compton (IC) process. This mechanism consists in the scattering of high energy electron with a low energy photon, resulting from one side in an average energy gain for the photon, and an average energy loss for the electron. If in the electron rest frame the photon energy is $h \nu \ll m_e c^2$, with $h$, $\nu$ and $m_e$ as the Planck constant, photon frequency and electron mass respectively, then, the process is mediated by Thomson cross section ($\sigma_T$). If, instead, $h \nu \simeq m_e c^2$ the cross-section decrease and is described by the Klein-Nishina formula \citep{Klein_KNCrossSection_1929}.

In case of an electron colliding with an isotropic monochromatic radiation field with frequency $\nu_0$, under the assumption of Thomson regime, the average frequency ($\nu_c$) of the upscattered photons and the radiative power emitted by the electron are \citep{Rybicki_RadiativeProcess_1986}:
\begin{equation}
\label{eq:ICFreq}
\nu_c=\frac{4}{3}\gamma^2 \nu_0
\end{equation}
and
\begin{equation}
\label{eq:ICPower}
\frac{d E_{IC}}{d t}=\frac{4}{3} \sigma_T \beta^2  \gamma^2 U_{ph}^2
\end{equation}
where $\gamma$ is the electron relativistic Lorentz gamma factor, and $U_{ph}$ is the radiation field energy density. In general, the outgoing photon will not have a single characteristic frequency, but rather a possible spectrum of frequencies, which makes the calculation of the IC spectrum non-trivial. However, one can with good accuracy assume that all photons are upscattered with the specific frequency given by Eq. \ref{eq:ICFreq}, in which case the IC spectrum from a given volume $V$ can be written as:
\begin{equation}
\label{eq:ICflux}
\phi_{IC}(E_\gamma)=\frac{1}{4\pi d^2} \int \int_V \frac{d E_{IC}}{d t} f_e(E_e, r) \delta(\nu-\nu_c) dE_e dV
\end{equation}
where $f_e$ is the distribution of emitting electrons and $E_e$ is the energy of the electrons. In the case of the Klein-Nishina regime, Eq. \ref{eq:ICPower} cannot be used, and the previous expression is no longer valid. The complete treatment becomes then quite involved, with the full solution in this regime given by \cite{Blumenthal_FFICSync_1970}. Yet, in the specific case where the target radiative field is represented by a black body, a simple analytical approximation is still possible in any scattering regime. This solution has been developed by \cite{Khangulyan_ICKNApprox_2014} and consists of approximating the energy losses as:
\begin{equation}
\frac{d E_{IC}}{d t}=q_1(T_{rad})G_{iso}(E_e, T_{rad})
\end{equation}
with
\begin{equation}
G_{iso}(E_e, T_{rad})=\left(\frac{m_e c^2}{k_B T_{rad}} \right)^2 \frac{c_{iso} Y(E_e, T_{rad})}{1+\frac{12}{\pi^2}c_{iso}Y(E_e, T_{rad})} \ln \left(1+ \frac{\pi^4 Y(E_e, T_{rad})}{135 c_{iso}} \right) \ ,
\end{equation}
\begin{equation}
q_1(T_{rad})=\frac{4}{3}\sigma_T c U_{ph} \frac{135}{16 \pi^4} \ ,
\end{equation}
and
\begin{equation}
Y(E_e, T_{rad})=\frac{4 E_e k_B T_{rad}}{(m_e c^2)^2} \,
\end{equation}
where $T_{rad}$ is the black body temperature, $k_B$ is the Boltzmann constant, and $c_{iso}=4.62$. The IC flux is then easily obtained by using Eq.\ref{eq:ICflux} with the new expression for the emitted power.

\section{Ionization of molecular clouds by low energy CRs}
\label{subsec:MCIonization}
Differently from ionizing radiation, low-energy CRs can penetrate deep within the core of gas clumps, providing ionization of clouds, possibly heating the dense cold gas \citep{Galli_CRHating_2015}, and indirectly inducing chemical reactions in the ISM, generating complex molecular compounds \citep{Dalgarno_CRsISMChem_2006}. As YMSCs are expected to be surrounded by the dense molecular envelope of the parent GMC, we will focus through this section on the ionization of molecular clouds (MC). The ionization processes in MCs have been exhaustively investigated by \cite{Padovani_MCIon_2009}. There is a plethora of interactions that can lead to the ionization of $H_2$\footnote{MCs also harbor a fraction of helium, which can be likewise ionized (see \cite{Padovani_MCIon_2009})}, namely, proton-induced ionization\footnote{Also nuclei may induce ionization but, for the sake of simplicity, as hadron induced process, we will only consider ionization generated by protons.}:
\begin{subequations}
\begin{gather}
p_{CR}+H_2 \rightarrow p_{CR} + H_2^+ + e \label{eq:pDirIonReac} \\
p_{CR}+H_2 \rightarrow  H + H_2^+ \label{eq:pElCaptIonReac} \\
p_{CR}+H_2 \rightarrow p_{CR} + H + H^+ + e \label{eq:pDissIonReac} \\
p_{CR}+H_2 \rightarrow p_{CR} + 2H^+ + 2 e \label{eq:pDoubleIonReac}
\end{gather}
\end{subequations}
or ionization induced by CR electrons:
\begin{subequations}
\begin{gather}
e_{CR}+H_2 \rightarrow e_{CR} + H_2^+ + e \label{eq:eDirIonReac} \\
e_{CR}+H_2 \rightarrow  e_{CR} + H + H^+ + e \label{eq:eDissIonReac} \\
e_{CR}+H_2 \rightarrow e_{CR} + 2H^+ + 2 e \ . \label{eq:eDoubleIonReac}
\end{gather}
\end{subequations}
The molecular hydrogen ion ($H_2^+$) production rate from the sole contribution of protons and electrons through \textit{direct ionization} (Eq. \ref{eq:pDirIonReac} and Eq. \ref{eq:eDirIonReac}) and \textit{electron capture} (Eq. \ref{eq:pElCaptIonReac}) processes can be calculated as:
\begin{equation}
\label{eq:IonRateIons}
\zeta_{H2}= \sum_k \int_{I(H_2)}^{E_{max}} c f_k(E_k) [1 + \phi_k(E_k)] \sigma_k^{\rm ion}(E_k) dE_k  + 
\int_0^{E_{max}} c f_p(E_p) \sigma_p^{\rm e.c.}(E_p) dE_p
\end{equation}
where the index $k$ account for the considered CR species (electrons or protons), $I(H_2)=15.603$ eV is the ionization potential of H$_2$, and $\sigma^{\rm ion}$ and $\sigma^{\rm e.c}$ are the direct ionization and electron capture cross sections respectively (see \cite{Padovani_MCIon_2009} and references therein). The quantity $\phi_k(E_k)$ is a correction factor accounting for the ionization induced by a population of secondary electrons created by direct ionization, and can be calculated as: 
\begin{equation}
\phi_k(E_k)=\frac{1}{\sigma_k^{\rm ion}(E_k)} \int_{I(H_2)}^{E_{max}} \mathcal{P}(E_k, E'_e) \sigma^{\rm ion}_e(E'_e)dE'_e
\end{equation}
with the term $\mathcal{P}(E_k, E'_e)$ describing the probability that a secondary electron with energy E'$_e$ is created during a primary ionization by a particle with energy E$_k$. To estimate, instead, the electron production rate, one also needs to account for the contribution of \textit{dissociative ionization} (Eq. \ref{eq:pDissIonReac} and Eq. \ref{eq:eDissIonReac}) and \textit{double ionization} (Eq. \ref{eq:pDoubleIonReac} and Eq. \ref{eq:eDoubleIonReac}) processes:
\begin{equation}
\begin{split}
\label{eq:IonRateEle}
\zeta_{e}= \sum_k \int_{I(H_2)}^{E_{max}} c f_k(E_k) [1 + \phi_k(E_k)] \sigma_k^{\rm ion}(E_k) dE_k & \\
+ \sum_k \int_{E_{\rm diss.ion.}}^{E_{max}} c f_k(E_k) [1 + \phi_k(E_k)] \sigma_k^{\rm diss.ion.}(E_k) dE_k \\
+ 2 \sum_k \int_{E_{\rm doub.ion.}}^{E_{max}} c f_k(E_k) [1 + \phi_k(E_k)] \sigma_k^{\rm doub.ion.}(E_k) dE_k
\end{split}
\end{equation}
where $\sigma^{\rm diss.ion.}$ and $\sigma^{\rm doub.ion.}$ are the cross sections for the dissociative and double ionization process. While we mention them for completeness, generally, these two ionization processes can be safely neglected in comparison to the direct ionization, as the cross section are a factor 10$\--$100 below $\sigma^{\rm ion}$.

Creation of $H_2^+$ ions in a dense environment can trigger an intricate chain of chemical gas phase based reactions with the formation of complex molecules, such as for example, DCO$^+$ and HCO$^+$ ($D$ stands for the deuterium isotope). The detection of these species through molecular radio emission lines can be used to assess the value of $\zeta_{H2}$ \citep{Caselli_IonFracMC_1998, Vaupre_W28IonRate_2014}. In a steady-state regime, the abundances of DCO$^+$ and HCO$^+$ are set by the following main reactions \citep{Guelin_DCO+_1977, Caselli_IonFracMC_1998}:
\begin{subequations}
\label{eq:ChemNet}
\begin{gather}
k_{CR}+H_2 \xrightarrow[]{\zeta_{H2}} k_{CR} + H_2^+ + e \\
H_2^++H_2 \xrightarrow[]{\kappa_{H_2^+}}  H_3^+ + H \\
H_3^++CO \xrightarrow[]{\kappa_H} HCO^+ + H_2 \\
HCO^+ + e \xrightarrow[]{\beta'} CO + H \\
H_3^+ + e \xrightarrow[]{\beta} 3H \text{ (or $H_2+H$)}\\
H+H \xrightarrow[]{\kappa'} H_2 \\
H_3^++HD \xrightleftharpoons[\kappa_f^{-1}]{\kappa_f} H_2D^+ + H_2  \\
H_2D^+ + CO \xrightarrow[]{\kappa_D} DCO^+H_2 \\
DCO^+ + e \xrightarrow[]{\beta'} CO+D  \\
H_2D^++e \xrightarrow[]{\kappa_e} 2H+D \text{ (or $H_2+D$ or $HD+H$)} \\
H+D \xrightarrow[]{\kappa''} HD \\
H_2D^+ + CO \xrightarrow[]{\kappa'_D} HCO^+ + H_2\\
H_3^+ + D \xrightleftharpoons[\kappa_f'^{-1}]{\kappa_f'} H_2D^+ + H  \\
CO^+ + HD \xrightarrow[]{\kappa_{CO^+}} DCO^+ + H
\end{gather}
\end{subequations}
where the parameters appearing over the arrows denotes the creation (or destruction) rates for each chemical compound (see \cite{Vaupre_W28IonRate_2014}). The ionization rate $\zeta_{H2}$ can then be analytically expressed in terms of the abundance ratios $R_D=[DCO^+]/[HCO^+]$ and $R_H=[HCO^+]/[CO]$ \citep{Wootten_ElecAbundance_1979, Guelin_IonStateMC_1982}:
\begin{equation}
R_D=\frac{[DCO^+]}{[HCO^+]}\simeq \frac{1}{3} \frac{x(H_2D^+)}{x(H_3^+)} \simeq
\frac{1}{3} \frac{\kappa_f x(HD)}{\kappa_e+x(e)+\delta+\kappa^{-1}_f/2}
\end{equation}
\begin{equation}
\label{eq:RatHCO+CO}
R_H=\frac{[HCO^+]}{[CO]} = \frac{\kappa_H x(H_3^+)}{\beta' x(e)} \simeq
 \frac{\kappa_H}{x(e)[2 \beta x(e)+ \delta]\beta'} \frac{\zeta_{H2}}{n_H}
\end{equation}
where $\kappa_f$, $\kappa_e$, $\kappa_H$, $\beta$, $\beta'$, are the reaction rates\footnote{Note that the reaction coefficients actually mediate collisional processes, and are consequently (some of them) dependent on the kinetic temperature of the gas \citep{Caselli_IonFracMC_1998}.} occurring in the chemical networks of Eq.\ref{eq:ChemNet}, $\delta \approx \delta_{H_3^+} \approx \delta_{H_2D^+}$ is the total destruction rate of H$_3^+$ or H$_2$D$^+$ due to reactions with neutral species such as CO and O, and $x(\mathcal{X})=n_\mathcal{X}/n_H$ denotes the fractional abundance of a given specie $\mathcal{X}$ with number density $n_\mathcal{X}$. Inversion of Eq. \ref{eq:RatHCO+CO} gives the expression for $\zeta_{H2}$ \citep{Caselli_IonFracMC_1998}:
\begin{equation}
\zeta_{H2}= [2 \beta x(e)+ \delta]\frac{R_H x(e) \beta' n_H}{\kappa_H} \simeq \left[ 7.5 \times 10^{-4} x(e) + \frac{4.6 \times 10^{-10}}{f_D} \right] x(e) n_H R_H
\end{equation}
where $f_D$ is the depletion factor of C and O, defined such that $1/f_D$ is the fraction of C and O in the gas phase, and
\begin{equation}
x(e)= \left[ \frac{\kappa_f x(HD)}{3 R_D} - \delta \right] \simeq \frac{2.7 \times 10^{-8}}{R_D} - \frac{1.2 \times 10^{-6}}{f_D}
\end{equation}
Assuming we know the depletion factor, $\zeta_{H2}$ is easily obtained as $R_D$ and $R_H$ can be readily estimated from observation through the ratios of molecular rotational lines. 

As a final remark, it is essential to emphasize that the value of $\zeta_{H2}$ does not return any information on the spectral shape of the CR distribution since the spectral information is lost in the integral in Eq. \ref{eq:IonRateIons}. However, one could use low-energy $\gamma$-ray emission to retrieve the CR spectrum, in this way, the value of the ionization rate can be employed as a cross-check to have a comprehensively self-consistent estimate of the CR distribution.

%% file: CHAPTERS/references.tex
%\begin{spacing}{1}

%\addcontentsline{toc}{chapter}{References}

\bibliographystyle{mnras}
\bibliography{Bibliography}

%\end{spacing}